\documentclass[%
reprint,
superscriptaddress,
frontmatterverbose, 
nofootinbib,
amsmath,amssymb,
aps,
prd,
prstab,
prstper,
floatfix,
]{revtex4-2}

\usepackage{graphicx}% Include figure files
\usepackage{dcolumn}% Align table columns on decimal point
\usepackage{bm}% bold math
\usepackage{hyperref}% add hypertext capabilities
\usepackage[mathlines]{lineno}% Enable numbering of text and display math
\usepackage[caption=false]{subfig}
\usepackage{ulem} 
\usepackage{comment}
\usepackage{orcidlink}
\usepackage{multirow}

\usepackage{tikz}
\usetikzlibrary{positioning, shapes.geometric, calc, fit}
\usetikzlibrary{arrows.meta}
\usepackage[%showframe,%Uncomment any one of the following lines to test 
scale=0.7, marginratio={1:1, 2:3}, ignoreall,% default settings
text={7in,10in},centering,
margin=0.24in, %1.5 %0.35
total={6.5in,8.75in}, top=1.2in, left=0.6in,right = 0.6in, includefoot, %left=0.9
]{geometry}
\usepackage{setspace}
\hypersetup{
	colorlinks=true, linkcolor=blue, %NavyBlue, %filecolor=magenta,      
    urlcolor=purple, %blue
	citecolor= cyan, %hyperindex=true, 
    linktocpage = true
}
\usepackage{hyperref}
\usepackage{url}
\usepackage{natbib}
\usepackage{standalone}

\usepackage{enumitem}
\usepackage{array} % for better column handling
\usepackage{multirow}
\usepackage{booktabs} % for cleaner lines
\usepackage{amssymb}

\usepackage{xcolor}
\usepackage{pifont}
\newcommand{\cmark}{\textcolor{green}{\checkmark}}

\begin{document}

\preprint{APS/123-QED}

%\title{A feature-driven machine learning approach for binary neutron star and neutron star-black hole early-inspiral warnings in the third generation gravitational wave observatories}

%\title{A feature-driven machine learning approach for BNS and NSBH early-inspiral warnings in the third generation GW observatories}

\title{\texttt{GW-FALCON}: A Novel Feature-Driven Deep Learning Approach for Early Warning \\ Alerts of BNS and NSBH Inspirals in Next-Generation GW Observatories}

%Low Mass

\author{\large Grigorios Papigkiotis \orcidlink{0009-0008-2205-7426}}
\email{gpapigki@auth.gr}
\affiliation{\large Department of Physics, Aristotle University of Thessaloniki,\\ Thessaloniki 54124, Greece}
\author{\large Georgios Vardakas \orcidlink{0000-0003-1352-2062}}
\email{gvardakas@auth.gr}
\affiliation{\large Department of Physics, Aristotle University of Thessaloniki,\\ Thessaloniki 54124, Greece}
\author{\large Nikolaos Stergioulas \orcidlink{0000-0002-5490-5302}}
\email{niksterg@auth.gr}
\affiliation{\large Department of Physics, Aristotle University of Thessaloniki,\\ Thessaloniki 54124, Greece}

\date{\today}

\begin{abstract}

Next-generation gravitational-wave (GW) observatories such as the Einstein Telescope (ET) and Cosmic Explorer (CE) will detect binary neutron star (BNS) and neutron star–black hole (NSBH) inspirals with high signal-to-noise ratios (SNRs) and long in-band durations, making systematic early-warning alerts both feasible and scientifically valuable. Such pre-merger triggers are essential for coordinating the rapid detection of electromagnetic follow-up, enabling searches for short gamma-ray bursts, kilonovae, and afterglows from the immediate pre-merger phase through the minutes to hours following coalescence. Detections during the inspiral phase would enable earlier alerts, allowing for faster electromagnetic follow-up and more comprehensive multimessenger coverage from merger onward. In this work, we introduce \texttt{GW-FALCON}, a novel feature-driven deep learning framework for early-time detection between GW signal+noise and noise-only data in next-generation detectors. Instead of feeding raw time series to deep convolutional or more complex neural network architectures, we first extract a large set of statistical, temporal, and spectral quantities from short observational time windows using the TSFEL library. The resulting fixed-length feature vectors are then used as input to compact feed-forward artificial neural networks (ANNs) suitable for low-latency operation. We demonstrate the method using simulated BNS and NSBH inspiral waveforms injected into colored Gaussian noise generated from the ET and CE design power spectral densities. More specifically, we train separate ANNs on feature sets extracted from partial-inspiral windows characterized by different maximum instantaneous frequencies. Within this setup, early-warning triggers could be issued from tens to hundreds of seconds before merger. Across all detector configurations and training and test datasets employed, the resulting classifiers achieve high accuracy and detection efficiency, with ET-like networks typically reaching test accuracies of order $90\%$ and CE-like ones exceeding $97\%$ at low false-alarm probability (at most a few percent for ET and well below the percent level for CE). The models exhibit the expected dependence of detection efficiency on partial-inspiral SNR, with CE configurations achieving near-perfect detection efficiency at lower SNR than ET due to their higher sensitivity and more informative feature representations. To the best of our knowledge, this work presents the first comprehensive feature-based, deep-learning GW detection framework for Next-generation GW observatories, connecting feature extraction from strain time series data to robust signal–noise classification within a setup that can be readily extended to real data and to more advanced neural network architectures.

\end{abstract}

\maketitle

\section{Introduction}
Gravitational waves (GWs), as subtle distortions in the fabric of spacetime's curvature,
%---gravitational waves (GWs)---
serve as a powerful tool to investigate the dynamics of the universe’s most extreme astrophysical phenomena, such as compact binary coalescences (CBCs). The direct observation of these CBC signals has established a cornerstone of modern astrophysics.
%provide a unique window for investigating the dynamics of the universe’s most extreme astrophysical phenomena. 
Over the past few years, the global network of ground-based gravitational wave detectors has recorded an increasing number of confident GW signals. So far, four observing runs have been concluded. More specifically, during the first three observing runs (O1-O3) conducted by the LIGO-Virgo Collaboration \cite{aasi2015advanced, acernese2014advanced}, and later joined by KAGRA \cite{akutsu2021overview}, roughly 90 events were confidently identified and published in the associated short-duration (transient) GWTC catalogs \cite{abbott2019gwtc,abbott2021gwtc,abbott2021gwtc_2,abbott2023gwtc}. The majority of these detections correspond to binary black hole (BBH) mergers, with a smaller fraction arising from binary neutron star (BNS) and neutron star-black hole (NSBH), respectively. Additional events were reported in the OGC \cite{nitz20191,nitz20202,nitz20213,nitz20234}, IAS \cite{olsen2022new,mehta2025new,wadekar2023new}, PyCBC\_KDE \cite{kumar2024optimized}, and AresGW catalogs \cite{koloniari2025new}. In addition, the IAS event catalog has recently been extended to include more BBH mergers identified in the LIGO-Virgo O3 data through GW searches that incorporate higher-order harmonics \cite{wadekar2025new}. A recent re-analysis in Ref. \cite{williams2025beyond} performed coherent Bayesian parameter estimation for a large subset of these additional catalogs candidates, finding that a majority is consistent with BBH mergers.

Following the previous observing periods, the LIGO-Virgo-KAGRA (LVK) global network scheduled and completed the O4 observing run, organized into three consecutive phases. Recently, regarding the first part of the fourth observing run (O4a), the collaboration released the GWTC-4.0 catalog \cite{abac2025gwtc, abac2025gwtc_2, abac2025gwtc_3}, announcing 128 new credible GW events with astrophysical probability $p_{\mathrm{astro}} \geq 0.5$, consistent with signals from BBH and NSBH binaries, thereby increasing the total number of reported confident detections to 218. All the necessary data associated with each GW candidate are publicly available through the Gravitational Wave Open Science Center (GWOSC) \cite{gwosc}. It should be highlighted that notable O4a detections include an NSBH merger with a BH within the lower mass gap and a particularly massive BBH candidate. The former \cite{ligo2024observation} suggests a potentially higher rate of that kind of mergers with electromagnetic counterparts, while the latter \cite{abac2025d} likely originates from the most massive binary source observed to date, featuring large component spins and systematic differences depending on the employed waveform models.

%A landmark in astronomy was achieved in 2017 with GW170817, the first BNS merger detected by the LIGO-Virgo detectors and observed through both gravitational and electromagnetic signals, marking the beginning of the era of multi-messenger astronomy (MMA) with GWs \cite{abbott2017gw170817, coulter2017swope, Abbot2017multi,abbott2017gravitational, cowperthwaite2017electromagnetic, soares2017electromagnetic,nicholl2017electromagnetic,margutti2017electromagnetic,margutti2017electromagnetic,chornock2017electromagnetic, abbott2019properties}.
In 2017, a significant milestone in astronomy was achieved with GW170817, the first BNS merger detected by the LIGO-Virgo detectors and simultaneously observed through both gravitational and electromagnetic signals, marking the beginning of the era of multimessenger astronomy (MMA) with GWs \cite{abbott2017gw170817, coulter2017swope, Abbot2017multi,abbott2017gravitational, cowperthwaite2017electromagnetic, soares2017electromagnetic,nicholl2017electromagnetic,margutti2017electromagnetic,margutti2017electromagnetic,chornock2017electromagnetic, abbott2019properties}.  In particular, the detection of the short $\gamma$-ray burst GRB170817A \cite{goldstein2017ordinary}, observed $1.7 \ \mathrm{s}$ after the coalescence by the Fermi Gamma-Ray Burst Monitor (Fermi-GBM) \cite{meegan2009fermi} and the INTEGRAL satellite \cite{savchenko2017integral}, and reported about six minutes later by the low-latency single-detector identification of a coincident GW signal in the Advanced LIGO Hanford data, provided the first direct evidence linking BNS mergers to short GRBs \cite{goodman1986gamma}.
%More specifically, the $1.7 \ \mathrm{s}$-detection after the coalescence of the short $\gamma$-ray burst signal GRB170817A \cite{goldstein2017ordinary} by the Fermi Gamma-Ray Burst Monitor (Fermi-GBM) \cite{meegan2009fermi} and the INTEGRAL satellite \cite{savchenko2017integral}, followed approximately six minutes later by the low-latency identification of the trigger in LIGO-Virgo data, provided the first direct evidence linking BNS mergers to short GRBs \cite{goodman1986gamma}. 
The candidate event was rapidly disseminated through a $\gamma$-ray Coordinates Network (GCN) notice, and a subsequent rapid re-analysis of Hanford, Livingston, and Virgo data confirmed a highly significant, coincident detection \cite{abbott2017gw170817}. Following the Fermi-GBM and LIGO-Virgo alerts, a comprehensive multi-wavelength observational effort was conducted across the whole electromagnetic (EM)  spectrum, resulting in the discovery of the kilonova associated with the merger, designated as AT 2017gfo, and powered by the radioactive emission of r-process heavy nuclei synthesized in the ejecta \cite{Abbot2017multi,perego20172017gfo,mccully2017rapid}. Following the first multimessenger detection, subsequent BNS and NSBH mergers observed during the O3 and O4a runs did not yield associated electromagnetic counterparts due to the limited sky localization accuracy \cite{abbott2023gwtc,abac2025gwtc,abac2025gwtc_2}.

Within the framework of MMA, the joint detection of the GW signal and the associated electromagnetic counterparts from the GW170817 BNS merger has fundamentally advanced our understanding of the Universe. This landmark observation provides critical insights into the origin of heavy elements through r- and s-process nucleosynthesis, allows independent determinations of the Hubble constant to probe existing cosmological tensions, and offers stringent tests of general relativity (GR) such as the propagation speed of GWs \cite{cowperthwaite2017electromagnetic, soares2019first, fishbach2019standard, berti2018extreme, abbott2019tests,liu2020measuring,palmese2024standard,ligo2017gravitational_hubble, di2021realm, abdalla2022cosmology, gianfagna2024potential}. Looking forward, upgraded or next-generation ground-based detectors, including the LIGO A+ \cite{LIGO_A_plus}, LIGO-India \cite{saleem2021science}, Voyager \cite{adhikari2020cryogenic}, Virgo nEXT \cite{garufi2024advanced,abac2025gwtc}, NEMO \cite{ackley2020neutron}, Einstein Telescope (ET) \cite{hild2011sensitivity,maggiore2020science,abac2025science,branchesi2023science}, and Cosmic Explorer (CE) \cite{abbott2017exploring,reitze2019cosmic,evans2021horizon}, will significantly expand observational capabilities by offering unprecedented sensitivity over a broader frequency range. These facilities are expected to increase the detection rate of CBCs by orders of magnitude, enabling far more systematic and detailed investigations of their properties than currently possible with existing detectors \cite{abbott2023population, abac2025gwtc_pop, abac2025all}.

In general, GWs from CBCs typically comprise three main stages: inspiral, merger, and ringdown. A central aspect of MMA is the time delay between the GW detection and source localization. When the signal remains in the detector’s sensitive band for a sufficiently long duration, it becomes possible to identify its presence and estimate its sky position before merger, thereby enabling early warnings that improve the chances of successful EM counterparts observations \cite{cannon2012toward}. The duration of a signal within a detector’s sensitive band is primarily determined by the detector’s low-frequency cut-off and the masses of the binary components. Notably, premerger alerts from BNS and NSBH systems have recently drawn considerable interest, as they enable rapid EM and astroparticle follow-ups \cite{sachdev2020early,tsutsui2022early,nitz2020gravitational,nitz2018rapid,chan2018binary,li2022exploring,kovalam2022early,magee2022observing,chaudhary2024low,kang2022electromagnetic,banerjee2023pre}. Since the emitted GW radiation during the binary's inspiral phase enters the sensitive band of ground-based interferometers well before coalescence, timely identification of that signal is crucial to provide EM facilities with sufficient lead time to respond \cite{magee2021first,chaudhary2024low}. The response time, however, varies significantly across telescopes, which further underscores the need for early detection during the inspiral stage. 

As the binary's GW waveform passes through the detector, the associated signal-to-noise ratio (SNR) accumulates progressively: while the low-frequency, low-amplitude early inspiral part of the signal contributes only modestly, it grows substantially as the system evolves toward merger, entering the detectors’ most sensitive frequency band and becoming more readily observable. Once identified, such signals can trigger early-warning alerts, facilitating coordinated EM observations over a broad range of frequencies. Considering that BNS and NSBH mergers are the dominant progenitors of short GRBs with multi-wavelength EM afterglows \cite{granot2018off,sarin2022linking,hendriks2023gravitational, gaspari2025binary}, predictions vary widely across the x-ray, optical, and radio bands \cite{granot2018off, saleem2018rates,hayes2023unpacking}. The dominant source of variation in these estimates arises from the poorly constrained BNS and NSBH distributions and merger rates, compounded by the limited number of so far observed events \cite{abbott2021population,abbott2023gwtc,abbott2023population,abac2025gwtc,abac2025gwtc_2,abac2025gwtc_3,abac2025gwtc_pop}.

%Considering that BNS and NSBH mergers are the dominant progenitors of short GRBs with associated x-ray, optical and radio emissions \cite{granot2018off,sarin2022linking,hendriks2023gravitational, gaspari2025binary}, and adopting a top-hat jet model, joint detection rates with the LVK global network are predicted to range from 0.02-27 per year in x-rays, 0.01-19 per year in optical, and 0.02-25 per year in radio at design sensitivity consistent with a three-detector analysis \cite{granot2018off, saleem2018rates,hayes2023unpacking}. Above all, the dominant source of variation in these estimates arises from the so far poorly constrained BNS merger rate \cite{abbott2023gwtc,abbott2023population,abac2025gwtc,abac2025gwtc_2,abac2025gwtc_3,abac2025gwtc_pop}.

The deployment of third-generation detectors, such as ET and CE \cite{hild2011sensitivity,punturo2010einstein,abbott2017exploring,maggiore2020science,abac2025science,reitze2019cosmic,branchesi2023science,evans2021horizon}, is expected to significantly enhance GW detection capabilities. By extending the sensitive frequency band, reducing instrumental noise, and improving strain sensitivity across both low- and high-frequency ranges, these observatories will substantially increase the candidate signal's accumulated SNR in the detector's band. As a result, these improvements will enable the detection of longer inspiral signals, raise the annual number of observed events to the order of tens of thousands (depending on astrophysical rate uncertainties and the detection threshold; e.g., $N_{\mathrm{BBH}}(\mathrm{SNR}>10) = 9.5^{+5.5}_{-2.8}\times10^{4} \ \mathrm{yr}^{-1}$ and $N_{\mathrm{BNS}}(\mathrm{SNR}>10) = 4.7^{+7.2}_{-3.5}\times10^{5} \ \mathrm{yr}^{-1}$ for a network of CE+CE+ET detectors \cite{iacovelli2022forecasting, kalogera2021next, gupta2023characterizing}),
%$\sim8\times 10^{4}$ for BBHs \cite{iacovelli2022forecasting} and $\sim 7\times 10^{4}$ for BNSs \cite{kalogera2021next})
and significantly improve early-warning and sky-localization performance, providing crucial lead time for coordinated multimessenger follow-ups across the entire EM spectrum \cite{chan2018binary,abac2025science,li2022exploring,tsutsui2022early,nitz2020gravitational,chaudhary2024low,reitze2019cosmic}.

%Advances in the sensitivity of second-generation LVK interferometers, and in particular the deployment of third-generation facilities such as the ET and CE \cite{maggiore2020science,abac2025science,reitze2019cosmic}, are expected to substantially increase the accumulated SNR by extending the detectors’ sensitive frequency band, reducing instrumental noise, and enhancing strain sensitivity across both low- and high-frequency ranges. These improvements will enable the detection of longer inspiral signals, raise the annual number of observed events to the order of thousands, and significantly enhance early-warning and sky-localization capabilities, providing essential lead time for coordinated multimessenger follow-ups across the EM spectrum \cite{chan2018binary,abac2025science,li2022exploring,tsutsui2022early,nitz2020gravitational,chaudhary2024low,iacovelli2022forecasting,kalogera2021next,reitze2019cosmic}.

The standard methodology for GW detection relies on {\it matched filtering} \cite{maggiore2008gravitational}, in which a large bank of template waveforms is constructed and correlated with detector data across the sensitive frequency band to extract possible signals hidden in noise. While highly effective and established, matched filtering is computationally demanding as it typically employs the entire bank of theoretical waveforms \cite{cannon2012toward,usman2016pycbc,dal2014implementing,messick2017analysis}
. To achieve low-latency detection performance, pipelines such as GstLAL \cite{messick2017analysis, sachdev2019gstlal, hanna2020fast, cannon2021gstlal}, PyCBC and PyCBCLive \cite{biwer2019pycbc,nitz2018rapid,allen2005chi, dal2014implementing, usman2016pycbc, nitz2017detecting,davies2020extending, nitz2022dfinstad, kumar2024optimized}, MBTAOnline \cite{adams2016low,aubin2021mbta,allene2025mbta}, SPIIR \cite{chu2022spiir}, as well as wavelet-based searches with cWB \cite{klimenko2016method, drago2021coherent, klimenko2021cwb, klimenko2011localization} have been developed and deployed for online analyses, providing rapid identification of candidate triggers. However, such pipelines face computational bottlenecks in true real-time operation, particularly for sources with complex dynamics such as precessing or non-aligned spins. Unmodeled search methods offer a complementary alternative, but their sensitivity varies across source classes, underscoring the need for further development. A detailed overview of the low-latency efforts undertaken by the LVK collaboration during the previous observing runs (O1-O4a) can be found in Refs. \cite{abbott2019gwtc,abbott2021gwtc,abbott2021gwtc_2,abbott2023gwtc,davis2021ligo,abac2025gwtc,abac2025gwtc_2,abac2025gwtc_3,ligo2025all}.

In contrast, early-warning alerts require the use of only premerger waveform information. Recent developments have demonstrated that matched filtering restricted to the inspiral phase can deliver such alerts. More specifically, Ref. \cite{sachdev2020early} presented a GstLAL-based pipeline that computes the matched-filter statistics, false-alarm rate, and sky localization using only the low-frequency portion of the signal, corresponding to the early inspiral. This method enables detections up to a minute before the final merger. In addition to GstLAL, PyCBCLive supports an early-warning search by employing a bank of truncated inspiral (e.g., TaylorF2 \cite{messina2017parametrized,messina2019quasi}) templates that discretely sample time before coalescence. Performance studies indicate this configuration can identify sufficiently loud, nearby BNS signals tens of seconds—up to about a minute—before merger (network- and sensitivity-dependent) \cite{nitz2020gravitational}. Likewise, the MBTA pipeline is operated online for CBC searches and alert production \cite{aubin2021mbta,allene2025mbta}. It also supports early-warning operation using inspiral-only matched filtering with templates truncated at low $f_{\max}$ \cite{allene2025mbta}. Under favorable conditions, this can produce premerger candidates tens of seconds before merger, particularly for BNS systems. Furthermore, early-warning times could reach hours to days for rare, strongly lensed BNS/NSBH events, where lensing time delays could provide advance notice of the arrival of subsequent lensed images in favorable cases \cite{magare2023gear,magare2025early}.

%, \sout{although the number of premerger triggers is smaller than detections from full-band analyses} \cite{nitz2020gravitational}.

Despite these advances, substantial challenges remain, especially with the advent of third-generation observatories, which are expected to detect hundreds of CBC signals per day for a network of CE+CE+ET detectors \cite{gupta2023characterizing}, thereby imposing unprecedented demands on low-latency data analysis and alert generation with the traditional matched-filtering pipelines. In recent years, machine learning (ML) and deep learning (DL) techniques have emerged as powerful tools in GW astronomy (see, e.g., Refs. \cite{cuoco2020enhancing,benedetto2023ai,zhao2023dawning,cuoco2025applications,stergioulas2024machine} 
for detailed reviews), due to their effectiveness across a diverse set of tasks and their computational efficiency, since most processing is performed during the training phase \cite{bishop2006pattern,goodfellow2016deep,prince2023understanding,lecun2015deep}. These methods provide a robust framework for data processing, pattern recognition, and GW analysis, encompassing artificial neural networks (ANNs), Bayesian neural networks (BNNs), convolutional neural networks (CNNs), deep residual networks (ResNets), Autoencoders, and other advanced architectures, with successful applications to CBC detection, burst searches, glitch classification, parameter estimation, sky localization, and synthetic data generation\cite{gabbard2018matching,george2018deep,gebhard2019convolutional,corizzo2020scalable,schafer2020detection,wang2020gravitational,krastev2020real,skliris2020real,lin2021detection,dodia2021detecting,marianer2021semisupervised,wei2021deep,alvares2021exploring,jadhav2021improving,chaturvedi2022inference,choudhary2022sigma,schafer2022one,barone2023novel,schafer2022training,andrews2022deepsnr,verma2024detection,aveiro2022identification,andres2023searches,alhassan2023detection, langendorff2023normalizing,dax2023neural,bini2023autoencoder,tian2024physics,murali2023detecting,bacon2023denoising,mcleod2022rapid,qiu2023deep,andres2024fast,wang2024rapid,fernandes2023convolutional,freitas2024comparison,yamamoto2023deep, beveridge2025novel,jadhav2023towards,tang2024deep,marx2025machine,sasaoka2024comparative,zelenka2024convolutional,gabbard2022bayesian,sasaoka2022localization,chatterjee2023premerger,chatterjee2023rapid,kolmus2022fast,chatterjee2019using,raza2024explaining,chan2024gwskynet, iess2020core, lopez2021deep, boudart2022machine, zevin2017gravity, soni2021discovering,mcginn2021generalised,lopez2022simulating, nousi2023deep, koloniari2025new, trovato2024neural,nagarajan2025identifying, wu2025advancing}.

A significant step in assessing the applicability of ML methods for realistic GW detection was the first Machine Learning Gravitational-Wave Mock Data Challenge (MLGWSC-1) \cite{schafer2023first}. This challenge provided a standardized benchmark to assess the sensitivity and efficiency of ML pipelines relative to traditional algorithms, using simulated BBH precessing injections into both Gaussian and real O3a detector noise. Among the evaluated methods, the Virgo-AUTH code (predecessor of AresGW), based on a 1-D ResNet architecture, emerged as the leading ML algorithm in the most demanding dataset with real noise. With additional improvements \cite{nousi2023deep} (named AresGW model 1), the code achieved, for the first time, sensitivity surpassing that of a standard (non-optimized) matched-filtering search in the $7-50 \ M_{\odot}$ components mass range, when tested on part of the O3a dataset from both LIGO detectors. In addition, some recent AresGW improvements \cite{koloniari2025new} include training and testing on O3 LIGO data from both detectors and producing a single ranking statistic capturing inter-detector correlations, which is further used to estimate the astrophysical probability $p_{\mathrm{astro}}$ of candidate events \cite{aubin2021mbta,allene2025mbta,farr2015counting,abbott2023gwtc,abac2025gwtc,abac2025gwtc_2}. The updated version of the code, designated as AresGW model 2, successfully recovers most previously identified events from the O3 observing run and enables, for the first time, the detection of eight new coincident, low-SNR, confident GW candidates with $p_{\rm astro} \geq 0.5$ using a network of LIGO detectors through a machine-learning-based approach. These eight new events were independently validated through a parameter estimation analysis provided in Ref. \cite{williams2025beyond}, with the resulting parameter posteriors closely matching those reported in Ref. \cite{koloniari2025new}. Furthermore, apart from AresGW and its subsequent improvements, several additional DL approaches have been introduced for BBH detection in real interferometer data, exploring diverse architectures, training strategies, and varying degrees of integration with traditional search pipelines (see, e.g., Refs. \cite{nagarajan2025identifying,marx2025machine, yamamoto2025hybrid, vandyke2025cobits, silver2025search}). %\gv{Optional: Additionall efforts have been appeared in harvard and japan.}
%\gp{China? (always china!)... etc}.

Furthermore, DL techniques have demonstrated significant potential for early-warning applications \cite{baltus2021convolutional,yu2021early,baltus2022convolutional,martins2025improving,alfaidi2024long}, and could enable the rapid identification of GW candidates and prompt estimation of their sky localization for real-time alerts \cite{dax2025real}. In this work, we present a DL approach for early alerts of BNS and NSBH mergers, 
exploiting the enhanced sensitivity and detection capabilities of third-generation gravitational-wave observatories, such as ET and CE \cite{hild2011sensitivity,punturo2010einstein,abbott2017exploring,maggiore2020science,abac2025science,reitze2019cosmic,branchesi2023science,evans2021horizon}. More specifically, rather than feeding long-duration time series directly to CNNs, ResNets, or more complex DL architectures, we suggested a novel feature-based framework for GW signal discrimination based on feed-forward ANNs, which we refer to as \texttt{GW-FALCON} (\texttt{G}ravitational-\texttt{W}ave Feature-based deep-learning \texttt{A}pproach for \texttt{L}ow-latency \texttt{C}lassificati\texttt{ON}). Features are extracted from the strain data using the Time Series Feature Extraction Library (TSFEL) \cite{barandas2020tsfel}. Therefore, observation windows are transformed from the time series domain into fixed-length feature vectors that summarize key signal properties through statistical, temporal, and spectral characteristics. Related work in the LVK context has also combined matched filtering with supervised ML by training an MLP on feature vectors derived from matched-filter trigger parameters, including the SNR, a signal-consistency statistic $\xi$, and template intrinsic parameters (masses and spins), to separate astrophysical signals from glitches \cite{lopez2025ameliorating}. In this proof-of-concept study, we compute such features for (i) foreground strain data, comprising simulated gravitational-wave signals embedded in Gaussian noise generated from theoretical power spectral densities (PSDs) for the CE and ET detectors, and (ii) background data consisting of noise-only realizations produced using the same PSD models. This way, these quantities are used as inputs to feed-forward ANNs trained for binary classification (signal+noise vs. noise-only). To the best of our knowledge, we introduce the first comprehensive feature-based gravitational-wave discrimination framework in third-generation GW observatories, providing a complete workflow from strain preprocessing and feature extraction to robust signal–noise classification, with the classifiers' output treated as a ranking statistic that is calibrated on background data to estimate a false-alarm rate per unit time.

Alongside this study aimed at third-generation observatories, we are pursuing parallel developments of the proposed feature-based DL framework for the upgraded instruments of the LVK network, including the Advanced LIGO A+ \cite{LIGO_A_plus} upgrade and the planned Virgo NExt configuration \cite{garufi2024advanced,abac2025gwtc}. These efforts aim to evaluate the proposed approach under the anticipated noise spectra of the upgraded detectors, and to assess whether a unified analysis workflow can support both low-latency early-warning operation and offline searches (including searches for BBH mergers) in networks spanning upgraded second-generation and third-generation facilities. In addition, we are benchmarking the method on more realistic, non-Gaussian data and plan to apply the \texttt{GW-FALCON} framework to archival LVK observations from previous observing runs in the context of offline searches. A comprehensive treatment of these directions, including multi-detector training strategies and the impact of detector glitches and data-quality artifacts, will be presented in forthcoming work.

The plan of the current paper is as follows. In Sec. \ref{sec:methodology}, we outline the physical and computational setup, including the loudness and time–frequency evolution of the inspiral signal, the in-band response and design noise power spectral densities of the ET and CE configurations, the corresponding antenna pattern functions, and the BNS/NSBH signal population waveforms adopted in our simulations. Sec. \ref{sec:datasets_feature_extraction} describes the construction of the supervised-learning datasets and the feature-extraction strategy, from the generation of detector data segments and observational time windows to the mapping of the associated time series windows into fixed-length feature vectors. Sec. \ref{sec:GWFDA_framework} introduces the \texttt{GW-FALCON} DL framework, covering the architecture of the early-warning ANNs, the training procedure, and the evaluation methodology. Next, in Sec. \ref{sec:results_discussion}, we present and discuss the performance of the trained ANN models for BNS and NSBH early-warning alerts. %, including applications to independent, more realistic data streams containing injections with parameters similar to existing GW events.
Sec. \ref{sec:summary_conclusions} provides a concise summary of the main results and draws the overall conclusions of this work. Finally, Appendices \ref{app:PI_SNR_distributions},
%\ref{app:tsfel_features},
\ref{app:Ann_training_prop}, \ref{app:classification_eval_measures}, and \ref{app: FAP_t_for_other_det_cases} provide supplementary material on the simulated event populations and detector models,
%the TSFEL feature sets,
the ANN models' training hyperparameters, and additional performance diagnostics that support the discussion in the main text.

%%%%%%%%%%%%%%%%%%%%%%%%%%%%%%%%%%%%%%%%%%%%%%%%%%%%%%%%%%%%%%%%%%%%%%%%%%%%%%%%%%%%%%%%%%%%%%%%%%%%%
%ML methods have also emerged as a promising tool for early-warning detection, enabling the identification of candidate GW events before merger [14,16]. Several studies have demonstrated that deep learning (DL) algorithms can achieve sensitivities comparable to traditional matched-filtering methods [29–34] and generalize across waveform families, e.g., by training on nonspinning templates and achieving high performance on precessing systems [29,30]. Some approaches utilize only a fraction of the inspiral waveform, for instance by classifying time-frequency maps derived from detector data with pretrained networks such as ResNet-50 [32].

%%%%%%%%%%%%%%%%%%%%%%%%%%%%%%%%%%%%%%%%%%%%
%Gravitational waves—tiny distortions of spacetime from compact binaries and other energetic sources—have opened a new window on the universe. Extracting these weak signals from noisy detector data is challenging and traditionally relies on matched filtering. Recently, machine learning (ML) has emerged as a powerful complement, promising faster and more efficient analyses. These advances are especially relevant for third-generation detectors such as the Einstein Telescope, where rapid and early alerts of binary neutron star mergers will be crucial for multi-messenger astronomy.

\section{\label{sec:methodology}Methodology}

\subsection{Loudness and Evolution of the GW Signal}

In CBC GW searches, the matched-filter SNR serves as a cumulative quantitative measure of the similarity between detector data and a modeled waveform template \cite{maggiore2008gravitational, nitz2018rapid}. Following the {FINDCHIRP} algorithm \cite{allen2012findchirp}, as implemented in PyCBC \cite{biwer2019pycbc,nitz2018rapid,allen2005chi, dal2014implementing, usman2016pycbc, nitz2017detecting,davies2020extending, nitz2022dfinstad, kumar2024optimized}, both the detector strain data $d(t)$ and the template approximate waveform $h(t;\vec{\theta} )$ with $\vec{\theta}$ parameters are first transformed into the frequency domain to enable efficient cross-correlation. In addition, the noise-weighted time-dependent inner product between two real functions $h(t;\vec{\theta} )$ and $d(t)$ can be expressed in the frequency domain as \cite{cutler1994gravitational}: 
\begin{align}
    \langle h,d \rangle(t) = 4\operatorname{Re}\left(\int_{f_{\mathrm{min}}}^{f_{\mathrm{max}}}  \frac{\tilde{h}^{*}(f;\vec{\theta})\tilde{d}(f)}{S_n(f)} df \right),
\end{align}
where $\tilde{h}^{*}(f;\vec{\theta})$ and $\tilde{d}(f)$ indicate, respectively, the Fourier-transform representations of the complex conjugate of the template and of the strain data, while $S_{n}(f)$ denotes the detector’s one-sided noise power spectral density (PSD) around the time of a candidate event. Also, in the above definition, $f_{\mathrm{min}}$ corresponds to the lower cutoff of the detector’s sensitive frequency band, whereas $f_{\mathrm{max}}$ is associated with the upper limit, typically defined by the Nyquist frequency, which is equal to half of the sampling rate. %%%%%%%%%%%%%%%%%%%%%%%%%%%%%%%%%%%%%%%
Then, the SNR time series is formulated as:
\begin{align}
    \rho(t) = \left(\frac{\langle h,d \rangle (t)} {\langle h,h \rangle}\right)^{1/2},
\end{align}
and the reported matched-filter output is its maximum over the analysis window $\mathcal{T}$,
\begin{align}
    \rho \equiv \max_{t\in\mathcal{T}} \rho(t).
\end{align}

%\begin{align}
%\rho = \left(4\operatorname{Re}\left(\int_{f_{\mathrm{min}}}%^{f_{\mathrm{max}}}  \frac{\tilde{h}^{*}(f)\tilde{d}(f)}{S_n(f)} df \%right)\right)^{1/2},
%\end{align}
%Based on the noise-weighted inner product:
%\begin{align}
%\rho = \left(4\operatorname{Re}\left(\int_{f_{\mathrm{min}}}^{f_{\mathrm{max}}}  \frac{\tilde{h}{*}%(f)\tilde{d}(f)}{S_n(f)} df \right)\right)^{1/2},
%\end{align}
%%%%%%%%%%%%%%%%%%%%%%%%%%%%%%%%%%%%%%55

%In practice, when a full inspiral–merger–ringdown template matches the data, $\rho(t)$ typically peaks near the coalescence time; for truncated (partial-inspiral) templates, the peak occurs earlier, near the end of the included inspiral segment. In noise-only data, $\rho_{\max}$ reflects the largest noise fluctuation within $\mathcal{T}$.

At this point, it is important to note that the matched-filter output $\rho$ depends on the signal’s phase at a chosen reference time, for example, when the signal first enters the detector’s sensitive frequency band \cite{maggiore2008gravitational,allen2012findchirp}. For a network of GW detectors, identified by an index $i = 1, \dots ,N$, the standard (incoherent) network SNR is defined as:
\begin{align}
\label{eq:net_snr}
\rho_{\mathrm{net}} = \left( \sum_{i=1}^{N} \rho_i^2 \right)^{1/2},
\end{align}
where $\rho_i$ denotes the SNR computed independently for each ground-based interferometer in the network. Most importantly, $\rho_{\mathrm{net}}$ provides a more reliable measure of signal significance than any single-detector SNR.

Above all, the SNR $\rho$ quantifies the degree of overlap between a template waveform $h(t;\vec{\theta})$ and the detector strain data $d(t)$, which contains both noise and a potential GW signal. In practice, a bank of precomputed templates is correlated with the data to identify the waveform that maximizes the SNR, yielding the best detection statistic under the assumption of stationary Gaussian noise and a well-modeled signal  \cite{maggiore2008gravitational, nitz2018rapid,biwer2019pycbc,nitz2022dfinstad, kumar2024optimized, cannon2012toward,usman2016pycbc,dal2014implementing,messick2017analysis,adams2016low, aubin2021mbta,allene2025mbta}. When only this type of noise is present, the SNR values fluctuate around a background mean value; however, the presence of a candidate GW signal produces a significant increase in the SNR. A candidate trigger is registered once this value exceeds a predefined detection threshold.

Although the matched-filtering method provides an efficient detection strategy under the assumption of stationary Gaussian noise, real interferometer data deviate significantly from this idealization. Non-Gaussian and non-stationary noise artifacts, such as transient glitches, can produce spurious peaks in the SNR time series that mimic true GW signals. To mitigate these effects, candidate events should be observed in coincidence across multiple detectors, thereby reducing the likelihood of false triggers. Additional signal-consistency tests, such as the $\chi^2$ time–frequency discriminator, are applied, and a reweighted SNR can then be defined to downweight triggers that are affected by noise transients and yield high SNR values inconsistent with the expected waveform morphology \cite{allen2005chi,babak2013searching,nitz2018rapid}.

For each candidate trigger, the detection confidence in low-latency searches is typically expressed in terms of the false-alarm rate (FAR), which quantifies how often random noise fluctuations and glitches yield a ranking statistic as high as that of a candidate event. The ranking statistic itself can be constructed as a multivariate statistic that includes different metrics, such as SNR and the $\chi^2$ signal-consistency test \cite{messick2017analysis,nitz2018rapid,cannon2021gstlal,usman2016pycbc,davies2020extending, nitz2022dfinstad, kumar2024optimized}. In the case of the DL approach, the ranking statistic is constructed directly from the network’s binary output \cite{koloniari2025new,cuoco2025applications,cuoco2020enhancing,trovato2024neural}. In parallel, each candidate should be accompanied by an astrophysical probability $p_{\mathrm{astro}}$, quantifying the confidence that the event is of astrophysical origin rather than a noise transient. In standard analyses, events with $p_{\mathrm{astro}} \geq 0.5$ are considered as confident GW candidates \cite{aubin2021mbta,allene2025mbta,farr2015counting,abbott2023gwtc,abac2025gwtc,abac2025gwtc_2}.
%\sout{In standard analyses, events with $p_{\mathrm{astro}}\geq 0.5$ are considered as confident GW candidates \cite{aubin2021mbta,allene2025mbta,farr2015counting,abbott2023gwtc,abac2025gwtc,abac2025gwtc_2}.}

The optimal SNR is obtained when the template waveform is correlated with itself \cite{maggiore2008gravitational}:
\begin{align}
\label{eq:opt_snr}
\rho_{\mathrm{opt}} = 2\left(\int_{f_{\mathrm{min}}}^{f_{\mathrm{max}}}  \frac{|h(f;\vec{\theta})|^2}{S_n(f)} df \right)^{1/2}.
%= (h|h)^{1/2} 
\end{align}
The value of $\rho_{\mathrm{opt}}$ quantifies the intrinsic signal strength or “loudness” as observed by the detector, serving as a benchmark for assessing detectability under ideal matched-filtering conditions with stationary Gaussian noise.
%which quantifies the intrinsic loudness of a signal relative to the detector’s noise power spectral density

%In early inspiral analyses, only a limited portion of the inspiral phase is considered. Consequently, the signal’s loudness is no longer characterized by the optimal SNR but instead by the partial-inspiral SNR (PI-SNR). This quantity follows the same definition as the optimal SNR, with the key distinction that the partial template $h_{\mathrm{PI}}$ represents only the fraction of the early inspiral waveform. In the frequency domain, this corresponds to replacing the upper integration limit $f_{\mathrm{max}}$ in Eq. (\ref{eq:opt_snr}) with the maximum frequency reached within the selected inspiral segment, typically within the tens of hertz range rather than extending to several kilohertz. The resulting PI-SNR value provides a reduced yet physically meaningful measure of the signal strength accumulated during the early inspiral phase,  which is critical for premerger detection and early warning studies.

In early inspiral analyses, only a limited portion of the inspiral waveform is taken into account. As a result, the signal strength is characterized not by the optimal SNR but by the partial-inspiral optimal SNR (PI SNR) \cite{baltus2021convolutional,baltus2022convolutional}. This quantity is defined analogously to the optimal SNR, with the key distinction that instead of the template $h$ it employs a partial template $h_{\mathrm{PI}}$, which represents only the portion of the early inspiral waveform containing the premerger information. This formulation provides a more relevant measure of signal strength during the inspiral phase, where timely detection is critical. In addition, in this definition, the upper frequency  $f_{\mathrm{max}}$ in Eq. (\ref{eq:opt_snr}) is replaced by the highest frequency reached within the partial inspiral, typically in the range of a few tens of $\mathrm{Hz}$. Above all, it is important to emphasize that as the CBC signal enters the detector’s sensitive band, the SNR accumulates gradually during the early inspiral phase but increases rapidly throughout the late inspiral and merger stages. As a result, for the PI SNR, the associated signal power is reduced compared to the merger, reflecting the weaker amplitude of the earlier phase.

Therefore, the resulting PI SNR provides a reduced yet physically meaningful measure of the signal strength accumulated during the early inspiral phase, which is essential for issuing early warning alerts and facilitating coordinated EM follow-up observations. 
%For a network of detectors, in direct analogy with the full-band case, the network PI-SNR can be defined accordingly by employing the individual detectors’ PI-SNRs $\rho_i^{\mathrm{PI}}$, each weighted by its own PSD.
%For a network of detectors, in direct analogy with the full-band case, the network PI-SNR is defined as the quadrature sum of the associated single-detector value $\rho_i^{\mathrm{PI}}$, each weighted with its detector’s own noise PSD. 
For a network of detectors, in direct analogy with the full-band case, the network PI SNR is defined as in Eq. (\ref{eq:net_snr}), replacing $\rho_i$ with the partial-inspiral values $\rho_{i,\mathrm{PI}}$, each computed using the corresponding detector’s noise PSD. This way, $\rho_{\mathrm{PI}, \mathrm{net}}$ provides a robust premerger detection metric for low-latency early-warning triggers and subsequent sky-localization.

%—typically below 50 Hz, rather than extending to several kilohertz—thereby yielding a reduced SNR value reflective of the partial waveform’s contribution.

The behavior of the PI SNR follows from the leading-order time-frequency evolution of a compact binary inspiral. In the post-Newtonian expansion, at the lowest order in velocity, the GW frequency $f(t)$ evolves as \cite{maggiore2008gravitational}:
\begin{align}
f(t) = \frac{1}{\pi}\left(\frac{G (1+z)\mathcal{M}_c}{c^3} \right)^{-5/8}\left(\frac{5}{251}\frac{1}{(t_c-t)} \right)^{3/8},
\end{align}
where G is the gravitational constant, $c$ is the speed of light, $z$ is the gravitational redshift, $\mathcal{M}_c = (m_1m_2)^{3/5}/(m_1+m_2)^{1/5}$ is the chirp mass of the source defined in terms of the component masses $m_1$ and $m_2$, while $t_c$ is the time of coalescence. Therefore, for a fixed chirp mass $\mathcal{M}_c$, detecting the signal $(t_c-t)$ seconds before merger is equivalent to observing it at the corresponding instantaneous maximum frequency $f(t)$. 
%In Fig. \ref{}, we present the time, frequency, and PI-SNR evolutions for GW waveforms associated with indicative BNS and NSBH mergers 
%In Fig. \ref{fig:pi_snr_evolution}, as an indicative example, we present the evolution of the GW strain, instantaneous frequency, and PI-SNR for indicative BNS and NSBH mergers \footnote{In each case, we show only the plus polarization of the associated GW waveform.}.

\begin{figure*}[!thb]
    %\resizebox{0.5\textwidth}{!}{
    \includegraphics[width=0.50\textwidth]{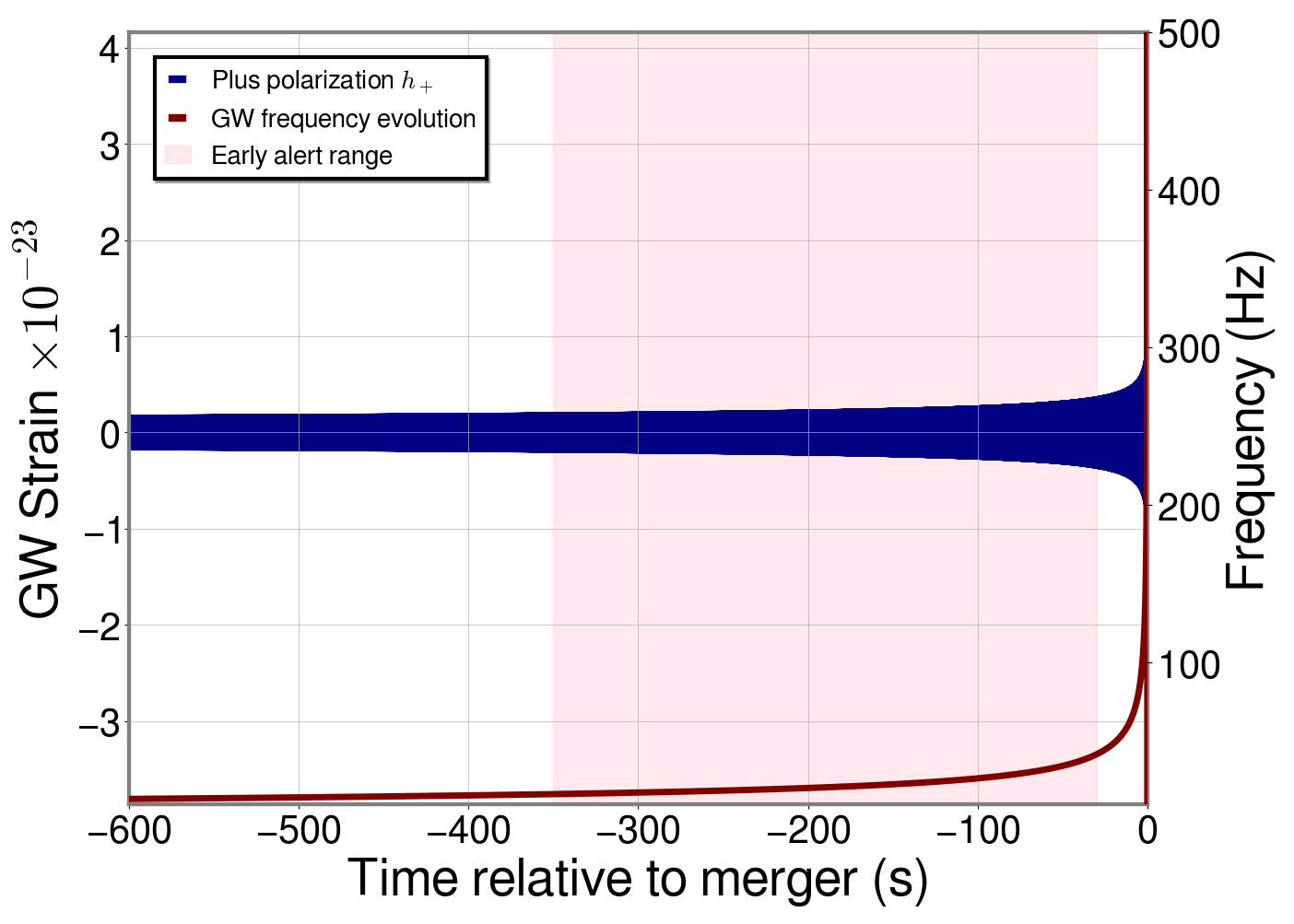}\hfill
    \includegraphics[width=0.50\textwidth]{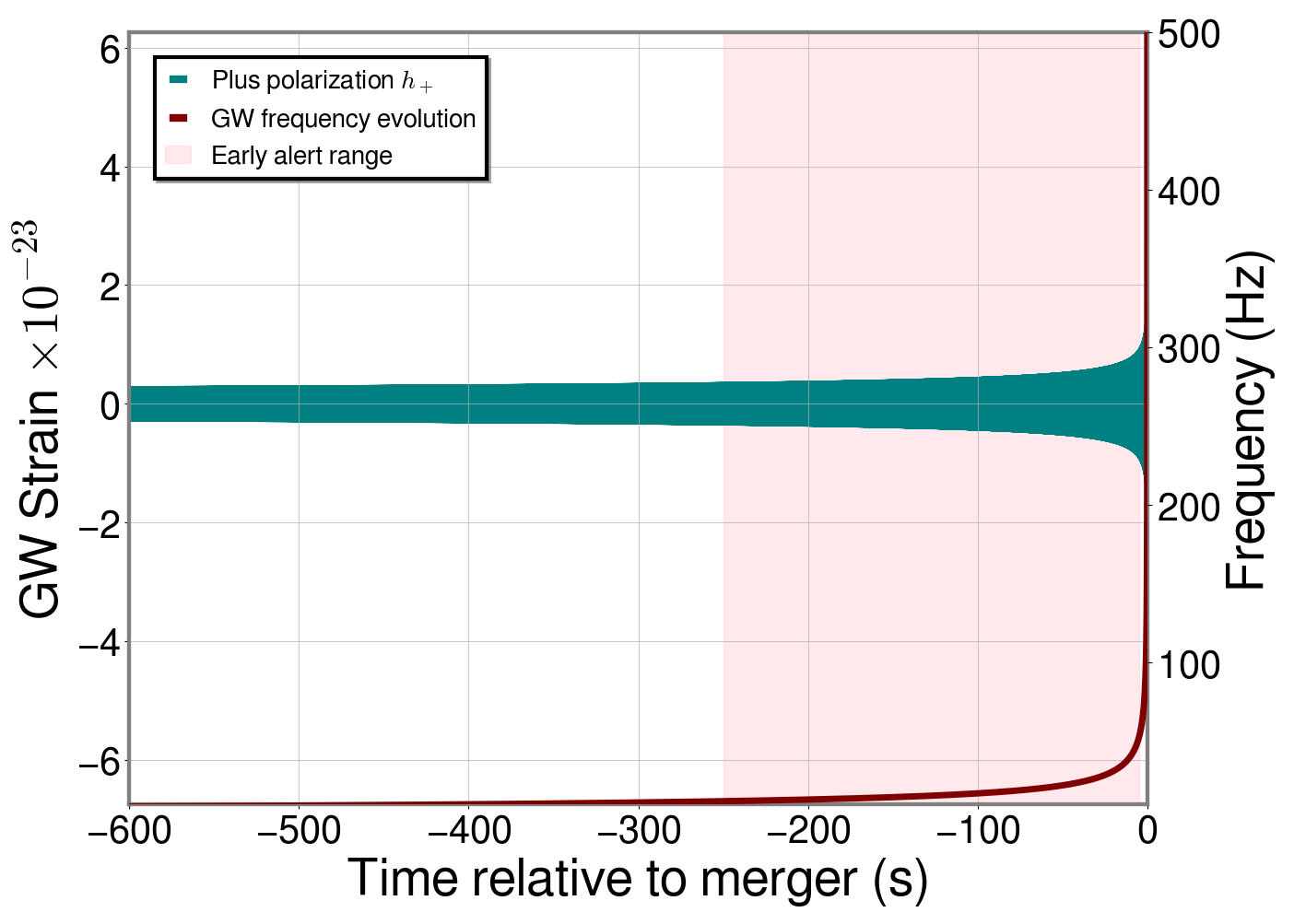}\hfill
    \caption{\label{fig:gw_strain_and_f_evolution} Left panel: Indicative plus polarization $h_{+}$ for a BNS GW template with component masses $m_{1} = 1.4\,M_{\odot}$ and $m_{2} = 1.0\,M_{\odot}$. 
    Right panel: Same as left panel for an NSBH GW template with component masses $m_{1} = 4.0\,M_{\odot}$ and $m_{2} = 1.5\,M_{\odot}$.
    In both panels, the shaded region marks the time interval during which early-warning alerts are expected with the methodology deployed in this work. The merger time is set to $t_{c} = 0 \ \mathrm{s}$. The corresponding instantaneous GW frequency, shown as a function of time relative to the merger, is also displayed in each panel.}
    %The presented data points correspond to the associated data within the test set for each EoS included in our catalog.
\end{figure*}

In Fig. \ref{fig:gw_strain_and_f_evolution}, we present the simulated GW strain and instantaneous frequency evolution for representative BNS (left panel) and NSBH mergers (right panel)\footnote{At this point, for simplicity, we show only the plus polarization of the associated GW waveforms.}.
%As we can observe, the time evolution of the amplitude and the frequency at merger will be different depending on the masses.
%PI-SNR depends on the fraction of the signal that is integrated and on the highest frequency reached within the observation window. Extending the observation window increases PI-SNR and improves detectability; however, premerger alerts require triggering as early as possible. This introduces a practical trade-off: longer integration yields higher PI-SNR and more robust detections, while triggering earlier enables prompt alerts. Therefore, a balance is required between achieving a sufficiently high PI-SNR and maintaining prompt detections.
The PI SNR depends on the fraction of the signal integrated and on the highest frequency reached within the observation window. Extending this window increases PI SNR and improves detectability, whereas premerger alerting favors triggers as early as possible. However, this fact introduces a practical trade-off: longer integration yields higher PI SNR and more robust detections, while earlier triggering has its own advantages. Therefore, a balance is required between achieving a sufficiently high PI SNR and achieving early detections.

\subsection{In-Band CBC Signal Presence and Detector PSD}

In the detector band, the observable duration of a CBC signal is mainly determined by the masses of binary components $m_1$ and $m_2$. This dependence is encoded by the associated chirp mass $\mathcal{M}_c$. During the early inspiral, where orbital velocities and strong-field effects are quite subdominant, the first-order in the post-Newtonian approximation (1PN) reliably captures the system's behavior.
Therefore, for binaries at cosmological distances and to the first order in velocity $\upsilon/c$, the duration of the GW signal is expressed as \cite{maggiore2008gravitational,samajdar2021biases}:
\begin{align}
\tau(s)\simeq 3 \left(\frac{M_{\odot}}{(1+z)\mathcal{M}_c}\right)^{\frac{5}{3}}\left[ \left(\frac{100 \ \mathrm{Hz}}{f_{\mathrm{low}}}\right)^{\frac{8}{3}} - \left(\frac{100 \ \mathrm{Hz}}{f_{\mathrm{high}}} \right)^{\frac{8}{3}} \right],
\end{align}
where $f_{\mathrm{low}}$ represents the lowest frequency cutoff of the detector sensitivity band, while $f_{\mathrm{high}}$ is the maximum frequency reached during the inspiral, typically taken as the frequency at the innermost stable circular orbit (ISCO) given by \cite{maggiore2008gravitational}:
\begin{align}
    f_{\mathrm{ISCO}} = \frac{1}{6\pi \sqrt{6}}\left(\frac{1}{1+z}\right)\frac{c^3}{GM_{\mathrm{tot}}},
\end{align}
where $M_{\mathrm{tot}} = m_1+m_2$ is the system's total mass. When the frequency $f_{\mathrm{low}}$ is maintained constant, an increase in the chirp mass $\mathcal{M}_c$ shortens the duration of the in-band signal, while systems with smaller $\mathcal{M}_c$ evolve more slowly and remain observable longer. This reflects the fact that heavier binaries pass through the detector band more rapidly, while lighter ones evolve more slowly and remain visible for an extended duration.

Furthermore, in the first order of the post-Newtonian approximation, the optimal SNR admits a compact expression \cite{maggiore2008gravitational,kastha2020imprints}:
\begin{align}
\label{eq:opt_snr_2}
\nonumber    \rho_{\mathrm{opt}} = &\frac{1}{2}\sqrt{\frac{5}{6}}\frac{1}{\pi^{2/3}}\frac{c}{d_{L}(1+z)^{1/6}}\left(\frac{G\mathcal{M}_c}{c^3}\right)^{5/6} \\ & \times  Q(\theta,\phi,\psi,\iota) \times I(M)^{1/2},
\end{align}
where $d_L$ is the luminosity distance, and $Q(\theta,\phi,\psi,\iota)$ is the geometric factor that encodes the source's sky position ($\theta,\phi$), polarization angle $\psi$, and inclination angle $\iota$ through the detector's antenna pattern functions. These angles describe the location and orientation of the CBC source with respect to the detector \cite{maggiore2008gravitational,schutz2011networks}. In addition, the frequency integral:
\begin{align}
I(M) = \int_{f_{\mathrm{min}}}^{f_{\mathrm{max}}}\frac{f^{-7/3}}{S_n(f)}df%\simeq \int_{f_{\mathrm{low}}}^{f_{\mathrm{ISCO}}}\frac{f^{-7/3}}{S_n(f)}df
\end{align}
is also a factor that quantifies how SNR accumulates across the detector's sensitive band. For the inspiral, the PI SNR retains the same form, with the upper integration limit equal to the highest frequency reached at the chosen time. Close to the merger, this limit is well approximated by the frequency at ISCO \cite{samajdar2021biases,kastha2020imprints}. Furthermore, from Eq. (\ref{eq:opt_snr_2}), it is evident that $\rho_{\mathrm{opt}} \propto \mathcal{M}_c^{5/6}/d_L$; thus, at fixed luminosity distance, lighter systems that correspond to lower chirp mass binaries attain smaller optimal SNRs, even though they are visible for longer. Keeping $d_L$ constant, besides the frequency integral $I(M)$, it should also be highlighted that the geometric factor $Q(\theta,\phi,\psi,\iota)$ is a significant quantity that directly modulates the optimal SNR.

It is important to emphasize that the main factor governing the detectability of a signal is the sensitivity of the associated interferometer, typically quantified by its one-sided noise PSD. The PSD varies with frequency and reflects the combined influence of the detector’s different noise sources. A lower PSD corresponds to higher matched-filter SNRs and, consequently, an increased detection range \cite{hild2011sensitivity, abbott2017exploring, buikema2020sensitivity, abbott2020guide, acernese2022virgo, capote2025advanced}. In this work, we generated colored Gaussian noise realizations based on the theoretical design-sensitivity PSD curves of the third-generation GW observatories, namely ET and CE \cite{punturo2010einstein, hild2011sensitivity,abbott2017exploring,maggiore2020science,abac2025science,reitze2019cosmic,branchesi2023science}. 
%With third-generation detectors in operation, detection rates would increase by orders of magnitude, shifting GW observations to survey scale with routine low-latency premerger alerts and sensitivity to sources at cosmological redshifts 
With third-generation detectors in operation, detection rates would increase by orders of magnitude. Gravitational-wave observing would become survey-scale, making low-latency premerger alerts routine and extending sensitivity to sources at higher cosmological redshifts \cite{maggiore2020science,reitze2019cosmic,iacovelli2022forecasting,kalogera2021next}. For our investigation, we used the ET-D and the wideband CE design sensitivity curves--\texttt{EinsteinTelescopeP1600143} and \texttt{CosmicExplorerWidebandP1600143}, respectively--to generate colored Gaussian noise and synthesize the simulated detector strain data \cite{punturo2010einstein,hild2011sensitivity,abbott2017exploring,hild2011sensitivity,maggiore2020science,abac2025science,reitze2019cosmic,evans2021horizon}. 
%Furthermore, we have to note that the specific PSDs used correspond to the official design configurations catalog distributed with the PyCBC package \footnote{For more information, analytical PSD generators appropriate to each use case are documented in the PyCBC module: \url{https://pycbc.org/pycbc/latest/html/pycbc.psd.html}.} \cite{biwer2019pycbc, nitz2022dfinstad}. 
%Furthermore, the specific PSDs used here correspond to an indicative choice of design-configuration sets distributed with PyCBC \footnote{For more information, analytical PSD generators appropriate to each use case are documented in the PyCBC module: \url{https://pycbc.org/pycbc/latest/html/pycbc.psd.html}.} \cite{biwer2019pycbc,nitz2022dfinstad} and are employed as a proof-of-concept that is utilized to demonstrate our methodology under representative third-generation sensitivities.
Furthermore, the PSDs used here are an indicative subset of the design-configuration sets available in PyCBC\footnote{For more information, analytical PSD generators appropriate to each use case are documented in the PyCBC module: \url{https://pycbc.org/pycbc/latest/html/pycbc.psd.html}.}. We adopt them as a proof of concept for generating strain data under representative third-generation sensitivities in order to demonstrate our methodology.

\begin{figure}[!thb]
    %\resizebox{0.5\textwidth}{!}{
    \includegraphics[width=0.47\textwidth]{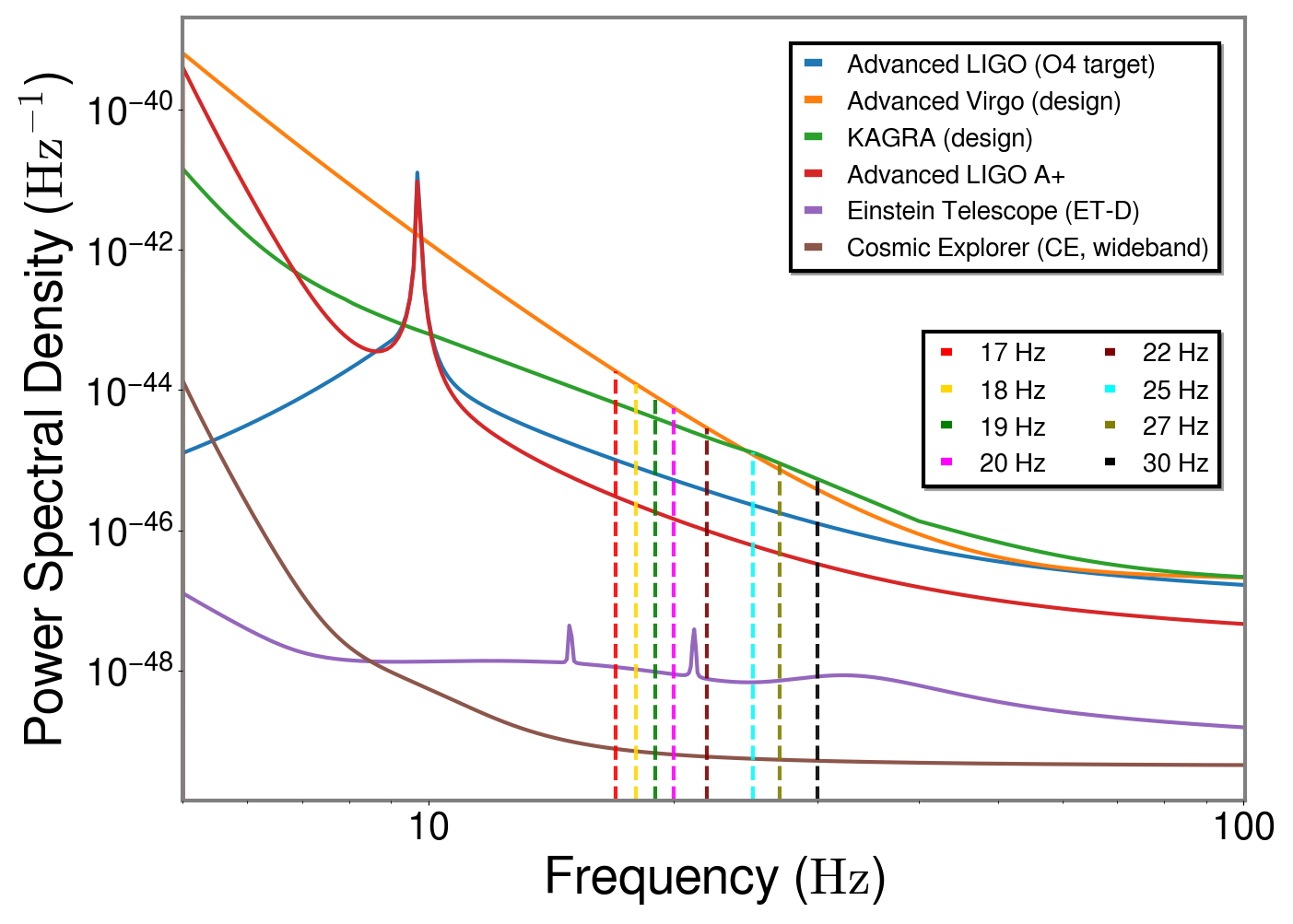}\hfill
    \caption{\label{fig:design_psds} Design-sensitivity (analytical) power spectral densities (PSDs) for next-generation LVK detectors and third-generation observatories. In this panel, we present the Advanced LIGO analytical PSD (O4 target), the Advanced LIGO A+, the Advanced Virgo (O4 target), the KAGRA (design), the Einstein Telescope (ET-D), and, finally, the Cosmic Explorer (CE, wideband) PSD as a function of frequency within the range $f\in[5,100] \ \mathrm{Hz}$. In addition, we show coloured vertical lines indicating benchmark frequencies that are useful for our investigation, as discussed in the following sections. In any case, for the chosen frequency band, the indicative noise curves illustrate the anticipated frequency-dependent improvements in strain sensitivity.}
    %The presented data points correspond to the associated data within the test set for each EoS included in our catalog.
\end{figure}

In Fig. \ref{fig:design_psds}, we present the analytical PSDs of the third-generation detectors used in this study, along with representative analytical noise curves associated with planned LVK upgrades\footnote{Here, we use benchmark design (theoretical) PSDs provided via PyCBC/LALSuite: \texttt{aLIGO175MpcT1800545} (Advanced LIGO), \texttt{AdVO4T1800545} (Advanced Virgo), \texttt{KAGRA128MpcT1800545} (KAGRA design), and \texttt{aLIGOAPlusDesignSensitivityT1800042} (Advanced LIGO A+ design). These design-sensitivity curves model the anticipated performance of next-generation LVK ground-based interferometers.} \cite{kagra2019kagra,abbott2020prospects}. The corresponding theoretical curves are shown in the frequency band $5-100 \ \mathrm{Hz}$. In our simulations, we approximate realistic detector conditions by generating colored Gaussian noise consistent with the associated PSDs. These stochastic noise realizations exhibit sample-to-sample fluctuations around the smooth spectra. However, to facilitate comparison across theoretical sensitivity designs, the figure displays only the smooth analytical PSDs that define the associated noise levels\footnote{Here, run-specific PSDs associated with O1-O4a LVK observing runs are not shown. These PSDs depend on various noise sources and display non-stationary and non-Gaussian characteristics. A detailed treatment of such measured PSD spectra would require additional context and lies beyond the present focus regarding the design sensitivities of next-generation GW observatories.}.
%In our simulations, we approximate realistic detector conditions by generating colored Gaussian noise from the associated PSDs, where the stochastic realizations produce sample-to-sample fluctuations around the smooth spectra.and are therefore not shown.
%As such, the corresponding colored Gaussian noise realizations
%extracted from design PSDs to generate data to approximate real conditions that 
%used in our simulations, which introduce stochastic, sample-to-sample fluctuations about these spectra, are not displayed. %For the same reason, we choose not to include measured, run-specific PSDs from the O1 through O4a observing runs, whose epoch-dependent variations would complicate a clean, design-focused comparison.

%In addition, to maintain a strictly design-focused comparison, we exclude from the figure the estimated, run-specific LVK PSDs derived from O1-O4a interferometer data, whose epoch- and configuration-dependent variability would obscure interpretation.

Both the ET and CE are designed to substantially exceed the CBC detection performance of the current LVK network and its expected near-term upgrades, with complementary sensitivity profiles. 
%The ET-D design sensitivity shown here achieves the highest noise reductions at low frequencies ($\lesssim 10 \ \mathrm{Hz}$), allowing the signal to enter the detector's band earlier, accumulate observable matched-filter SNR over a longer time, and thereby improve sensitivity to long-duration inspirals and early-warning prospects \cite{hild2011sensitivity, maggiore2020science,abac2025science}. 
The ET-D design sensitivity achieves the highest low-frequency noise reduction at frequencies $\lesssim 10 \ \mathrm{Hz}$, making CBC signals observable in the detector band earlier and capturing a longer part of the observed inspiral. Consequently, extending sensitivity to the $1-10 \ \mathrm{Hz}$ frequency band will substantially lengthen the in-band duration, with a signal at $1 \ \mathrm{Hz}$ remaining in-band for roughly five days as discussed in Ref. \cite{li2022exploring}. As a result, both the partial-inspiral matched-filter and optimal SNRs (accumulated up to a chosen cutoff time) rise more rapidly and can exceed detection thresholds sooner, whereas integration over the full frequency band yields larger optimal and matched-filter SNR values for a source at a fixed distance. This enhancement in sensitivity enables credible premerger alerts, since substantial SNR accrues at earlier times while maintaining controlled false-alarm rates. A trade-off remains between timeliness and robustness: waiting improves SNR and sky localization, whereas the enhanced low-frequency sensitivity of ET in the $1-10 \ \mathrm{Hz}$ band shifts this balance toward earlier alerts with comparable reliability and better coordination of multimessenger EM follow-up, compared to what is achievable with the wideband CE sensitivity in the same frequency range \cite{hild2011sensitivity,hu2023rapid,maggiore2020science,abbott2020guide,abac2025science}.

In addition, CE (wideband) design PSD provides broadband improvements from tens of hertz up to the kilohertz range, increasing reach to more distant sources and enhancing sensitivity to intermediate- and high-mass mergers as well as higher-frequency waveform content. For premerger alerts, this broadband response enables prompt early triggering across the detector's frequency band, particularly for higher-mass systems with short in-band durations, thereby complementing the ET’s early-warning capability \cite{reitze2019cosmic,evans2021horizon}. 
%Collectively, these expected reductions in the one-sided PSD associated with the operation of such third-generation ground-based observatories translate into higher SNR at fixed distance and, consequently, order-of-magnitude enhancements in detection range, premerger alerts, accessible redshift, and sensitive volume relative to current LVK instruments 
Overall, reductions in PSD anticipated for third-generation ground-based observatories would increase the candidate event's matched-filter SNR at fixed distance, improve the sky localization of GW events, and, in turn, yield order-of-magnitude enhancements in detection range, low-latency premerger alert capability, accessible redshift, and sensitive volume relative to the current LVK network and its associated future upgrades.
\cite{punturo2010einstein,evans2021horizon,hu2023rapid,chan2018binary,hild2011sensitivity,abbott2017exploring,maggiore2020science,abac2025science,reitze2019cosmic}.

\subsection{Detector's Response for an Incoming GW Signal}

Beyond the sensitivity gains of the proposed third-generation ground-based observatories, the detector morphology also warrants discussion. Initially, the ET has been proposed as an underground facility comprising three interferometers with $10 \ \mathrm{km}$ long arms arranged in an equilateral triangle. Each interferometer has an opening angle of $ 60^{\circ} $, and each will be rotated relative to the others by $ 120^{\circ}$. Candidate locations include Sardinia (Italy), the border region spanning the Netherlands, Belgium, and Germany, and locations within Germany. Besides the single-site triangular ET geometry, viable alternatives remain under study in the form of two-interferometer “$2$L” realizations built from co-sited $15 \ \mathrm{km}$ L-shaped interferometers. In this direction, two relative arm layouts are considered: a $2$L-parallel option with co-aligned arms, and a $2$L-$45^{\circ}$ alternative with the second facility rotated by $45^{\circ}$ with respect to the first. Additional variants are under consideration, including a triangular layout with $15 \ \mathrm{km}$ arms and two-interferometer configurations with $20 \ \mathrm{km}$ arms in either parallel or $45^{\circ}$ relative orientations (2L-parallel and $2$L-$45^{\circ}$) \cite{maggiore2020science,abac2025science,branchesi2023science}. It is worth noting that the final ET site configuration—whether a single triangular facility or a pair of co-sited interferometers within Europe—has not yet been decided, and the choice among candidate locations is expected to be made in the near future.

Among ET geometry variants, the single-site triangular geometry ($3 \times 10 \ \mathrm{km}$-arms) remains the most favorable single-observatory option, providing sufficient independent baselines to reconstruct the two gravitational-wave polarizations at a single site and offering a more nearly isotropic directional sensitivity across the sky. On the other hand, two-interferometer designs with co-aligned $15 \ \mathrm{km}$ L-shaped arms ($2$L-parallel geometry) seem to be disfavored due to limited orientation diversity, which reduces all-sky coverage. The misaligned configuration ($2$L-$45^{\circ}$) is more competitive, particularly in a network of detectors that includes a CE as a facility, where orientation associated with the multiple-detector diversity enhances sky localization (see e.g., Refs. \cite{maggiore2020science,abac2025science,branchesi2023science} for a detailed review).

Unlike the ET, the CE will have the conventional L-shaped geometry employed by second-generation LVK observatories. Current plans suggest two North America CE facilities with arm lengths of about $20 \ \mathrm{km}$ and $40 \ \mathrm{km}$, respectively. Further technical details and design considerations are provided in Refs. \cite{abbott2017exploring,reitze2019cosmic,evans2021horizon}. In a network of detectors, CE’s high strain sensitivity improves time-of-arrival precision and strengthens coherent localization, which combines data streams using the relative amplitudes and phases across sites (set by each detector’s antenna response) together with arrival times, thereby yielding more accurate sky position and enabling earlier, better-targeted multimessenger follow-up. In addition, although specific sites and configurations for the CE facilities in North America are still under study, they are expected to be finalised in the coming years.

Above all, the detector response to an incoming signal is not uniform and depends on the associated geometry of the instrument and the source position in the sky. Considering a spherical coordinate system centered on the detector, the antenna pattern functions of each interferometer of an ET-like triangular detector can be expressed as \cite{regimbau2012mock}:

\begin{align}
\label{eq:et_triangular_antennas}
F^{(1)}_{+}(\theta,\phi,\psi)
  \nonumber &= -\frac{\sqrt{3}}{4}\Big[(1+\cos^{2}\theta)\sin 2\phi \cos 2\psi \notag\\
  \nonumber &\qquad\quad + 2\cos\theta \cos 2\phi \sin 2\psi\Big],\\
F^{(1)}_{\times}(\theta,\phi,\psi)  
   \nonumber &= +\frac{\sqrt{3}}{4}\Big[(1+\cos^{2}\theta)\sin 2\phi \sin 2\psi \notag\\
   \nonumber &\qquad\quad -2\cos\theta \cos 2\phi \cos 2\psi\Big],\\
F^{(2)}_{+,\times}(\theta,\phi,\psi) \nonumber &= F^{(1)}_{+,\times}\left(\theta,\phi + \frac{2\pi}{3},\psi \right), \\
F^{(3)}_{+,\times}(\theta,\phi,\psi) &= F^{(2)}_{+,\times}\left(\theta,\phi - \frac{2\pi}{3},\psi \right).
\end{align}
At this point, $F^{(n)}_{+}$ and $F^{(n)}_{\times}$ denote the plus and cross polarization responses of the $n$-th interferometer. They depend on the GW source sky position $(\theta,\phi)$ in the detector frame and the polarization angle $\psi$. Additionally, the associated patterns for an L-shape detector, rotated by an angle $\zeta$ about the local vertical axis, take the form \cite{maggiore2008gravitational,schutz2011networks}:
\begin{align}
    \label{eq:et_L_antennas}
    \nonumber F_{+}(\theta,\phi,\psi;\zeta) =& +\frac{1}{2}(1+\cos^2\theta)\cos[2(\phi-\zeta)]\cos2\psi \\
    \nonumber&-\cos\theta\sin[2(\phi-\zeta)]\sin2\psi,\\
    \nonumber F_{\times}(\theta,\phi,\psi;\zeta) =& +\frac{1}{2}(1+\cos^2\theta)\cos[2(\phi-\zeta)]\sin2\psi \\
    &-\cos\theta\sin[2(\phi-\zeta)]\cos2\psi.
\end{align}
Building on this framework, for a CE L-shaped detector, we set the orientation angle
\(\zeta=0\), so the antenna pattern functions take the standard form as reported in Ref.  \cite{schutz2011networks}. On the other hand, for a co-sited $2$L-\(45^{\circ}\) ET
configuration with two L-shaped detectors \(A\) and \(B\) whose bisectors differ by \(45^{\circ}\),
we choose:
\begin{align}
    \zeta_A=\zeta_0,\qquad \zeta_B=\zeta_0+\frac{\pi}{4},  
\end{align}
and then define the associated antenna patterns as:
\begin{align}
    \label{eq:et_2L_45_antennas_2}
    \nonumber F^{(A)}_{+,\times}(\theta,\phi,\psi)=F_{+,\times}(\theta,\phi,\psi;\zeta_A)\\
    F^{(B)}_{+,\times}(\theta,\phi,\psi)=F_{+,\times}(\theta,\phi,\psi;\zeta_B),    
\end{align}
where, for simplicity, we adopt \(\zeta_0=0\).

As previously discussed, at lower frequencies, GW signals can remain in-band for extended durations. As the Earth rotates, the apparent sky position ($\theta,\phi$) evolves, rendering the associated antenna patterns $F_{+}$ and $F_\times$ explicitly time-dependent. Furthermore, as the signal remains in the detector's band, it will experience the Doppler effect due to the detector's relative motion with respect to the signal. However, as reported in Ref. \cite{chan2018binary}, this effect is not significant to the source's sky uncertainty estimation.

%through Eq. \ref{eq:et_L_antennas}
The response of a single detector to a passing GW with polarizations $h_{+}(t;\vec{\theta})$ and $h_{\times}(t;\vec{\theta})$ in the wave frame can be written in terms of the antenna pattern functions as \cite{schutz2011networks,abac2025science}:
%Given the detector's response, an incoming GW signal with polarizations $h_{+}$ and $h_{\times}$ in the source frame can be defined as:
\begin{align}
    \label{eq:h_det}
    h_{\mathrm{det}}(t;\vec{\theta}) = h_{+}(t;\vec{\theta})F_{+}(\theta,\phi,\psi) + h_{\times}(t;\vec{\theta})F_{\times}(\theta,\phi,\psi).
\end{align}
%In any case, this projection onto the arms of a detector maps the source-frame signal to its detector-frame representation that is ultimately observed. %can be observed in the detector frame.
This relation illustrates how the signal from the wave frame is projected onto the arms of the L-shape detector, converting it into a waveform representation that is hidden in the detector's strain noise. Furthermore, for a triangular ET-like detector, the response of the $k$-th interferometer ($k=1,2,3$) with antenna patterns as shown in Eq. (\ref{eq:et_triangular_antennas}) is given by \cite{abac2025science}:
\begin{align}
    \label{eq:h_det}
    h_{k}(t;\vec{\theta}) = h_{+}(t;\vec{\theta})F^{(k)}_{+}(\theta,\phi,\psi) + h_{\times}(t;\vec{\theta})F^{(k)}_{\times}(\theta,\phi,\psi).
\end{align}
Finally, for the co-sited $2$L-$45^{\circ}$ ET configuration, the GW strain has the same form, with $k$ indexing the two L-shaped observatories ($\ell\in \lbrace A, B\rbrace$) and the corresponding pattern functions given in Eq. (\ref{eq:et_2L_45_antennas_2}) with $\zeta_0 = 0$. Accordingly, whether ET is modeled as a triangular or as a $2$L-$45^{\circ}$ configuration, we treat it as a single observatory and compute the optimal SNR by combining the per-interferometer contributions via Eq. (\ref{eq:net_snr})-with 
$N=3$ for the former case or $N=2$ for the later. In either configuration, the same aggregation also applies to the quantification of the PI SNR used for premerger alerting.

Since at the time our study was completed, the final locations of the upcoming third-generation detectors ET and CE were unknown, {\it we adopted an agnostic approach and used the current locations of existing detectors as temporary placeholders}\footnote{Once the locations of ET and CE are decided, our analysis can be straightforwardly updated to reflect the final detector configurations.}. Therefore, for our study, we consider the following configurations. For the ET observatory: (i) a single-site triangular array with $10 \ \mathrm{km}$ arms at the Virgo site, Italy (longitude, latitude) = ($10.4 \ ^{\circ}\mathrm{E}$, $43.7\ ^{\circ}\mathrm{N}$); and (ii) a co-sited $2$L-$45^{\circ}$ layout in which observatory A is a $15 \ \mathrm{km}$ arms L-shaped interferometer at Virgo and detector B is a $15 \ \mathrm{km}$ L-shaped interferometer at the GEO600 site, Germany ($9.81 \ ^{\circ}\mathrm{E}$, $52.24 \ ^{\circ}\mathrm{N}$) \cite{willke2002geo}, with the B arms rotated by $45^{\circ}$ relative to A. For the CE configuration: (iii) an L-shaped detector at the Hanford site, USA ($-119.41 \ ^{\circ}\mathrm{E}$, $46.45 \ ^{\circ}\mathrm{N}$) with $20 \ \mathrm{km}$  arms; and (iv) an L-shaped detector at the Livingston site, USA ($ -90.77\ ^{\circ}\mathrm{E}$, $30.56 \ ^{\circ}\mathrm{N}$) with $40\ \mathrm{km}$ arms. In this framework, we expect only weak dependence on the exact detector locations within North America and Europe over a large ensemble of injected simulated signals $h_{\mathrm{det}}(t;\vec{\theta})$ \cite{li2022exploring}. Each simulated single-detector configuration is summarized in Table \ref{tab:det_network}. In all cases, we adopt a consistent notation to identify the participating third-generation observatories and the corresponding sites considered.

\begin{table}[!tbh]
\caption{\label{tab:det_network} Configurations of assumed third-generation GW detectors employed in this work. Columns list the detector configuration and notation, followed by the geographical locations of the assumed sites: Virgo, V1 ($10.4 \ ^{\circ}\mathrm{E}$, $43.7\ ^{\circ}\mathrm{N}$), GEO600, G1 ($9.81 \ ^{\circ}\mathrm{E}$, $52.24 \ ^{\circ}\mathrm{N}$), Hanford, H1 ($-119.41 \ ^{\circ}\mathrm{E}$, $46.45 \ ^{\circ}\mathrm{N}$), and Livingston, L1 ($ -90.77\ ^{\circ}\mathrm{E}$, $30.56 \ ^{\circ}\mathrm{N}$). The mark symbol indicates the site where the detector was placed. ET is a triangular detector array, while $2$L–$45^{\circ}$ ET configuration denotes two co-sited L-shaped detectors with the second rotated by $45^{\circ}$ relative to the first. In addition, CE-like configurations correspond to single L-shaped interferometers.}
\begin{ruledtabular}
\begin{tabular}{c|c|cccc}
\multirow{2}{*}{\textbf{Single-Detector}} & \multirow{3}{*}{\textbf{Notation}}& \multirow{3}{*}{\textbf{V1}}& \multirow{3}{*}{\textbf{G1}} & \multirow{3}{*}{\textbf{H1}}&\multirow{3}{*}{\textbf{L1}} \\
 \multirow{2}{*}{\textbf{Configuration}} &  &  &  &  &\\
 &  &  &  &  & \\

\hline
% &  &  & \\
ET & ET1 & \cmark & - & - &-\\
$2$L-$45^{\circ}$ ET & ET2 & \cmark & \cmark & - & -\\
CE-20km & CE1 & - & - & \cmark  & -\\
CE-40km & CE2 & - & - & - & \cmark\\
\end{tabular}
\end{ruledtabular}
%\vspace{4pt}
%\raggedright\footnotesize
\end{table}

\subsection{\label{sec:bns_nsbh_parms_distributions}BNS and NSBH Parameter Distributions}

Prior knowledge of the associated gravitational-wave signal characteristics is essential for guiding the overall design of our investigation. Focusing on the early-inspiral part of the signal, relevant for enabling premerger alerts, we generate BNS and NSBH waveforms using the SpinTaylorT4 approximant \cite{buonanno2003detecting}. We choose the source-frame component-mass priors to be uniformly distributed between $1$ and $3$ $M_{\odot}$ for BNS systems, covering all the possible mass range \cite{kiziltan2013neutron}. For NSBH systems, we adopt uniform distribution priors of $1-3 \ M_{\odot}$ for the NS and $3-20 \ M_{\odot}$ for the BH,  targeting the relatively low-mass BHs and incorporating the $3-5 \ M_{\odot}$ lower mass gap region \cite{ligo2024observation}. 

Building on this framework, it is also important to highlight that the generated waveform templates are restricted to nonspinning BNS and NSBH systems. In the early inspiral, spin effects are absent at the first post-Newtonian order. The first spin-orbit and spin-spin waveform phase corrections arise at 1.5PN and 2PN orders, respectively \cite{blanchet2024post}. Likewise, for tidal deformabilities, tidal effects first contribute to the inspiral phasing at 5PN and become informative mainly at higher frequencies near merger ($\gtrsim 100-400 \ \mathrm{Hz}$); hence, their impact in the lower-frequency premerger part of the signal is not taken into account \cite{chatziioannou2020neutron,flanagan2008constraining,dietrich2021interpreting}.

Consistent with the MLGWSC-1 challenge setup \cite{schafer2023first}, we do not sample sources uniformly in comoving volume. Instead, the generated GW waveforms are sampled according to the chirp distance rather than the luminosity distance. This approach increases the number of low-mass systems that can be observed, raising their detection rate. Notably, the chirp distance $d_c$ is defined in terms of the chirp mass $\mathcal{M}_{c}$ and the luminosity distance $d_L$ as
\begin{align}
    \label{eq:dc_relation}
    d_c = d_L\left(\frac{\mathcal{M}_{c,0}}{\mathcal{M}_c}\right)^{5/6},
\end{align}
where $\mathcal{M}_{c,0} = 1.4/2^{1/5} \ M_{\odot}$ is a fiducial value corresponding to an equal-mass $1.4-1.4 \ M_{\odot}$ neutron star binary. Therefore, once the chirp distance has been provided, the luminosity distance is estimated accordingly by inverting the chirp-distance relation in Eq. (\ref{eq:dc_relation}). Then, for each luminosity distance, we also derive the associated redshift $z$ from a fiducial flat $\Lambda$CDM cosmology\footnote{Redshift $z$ is obtained by inverting the luminosity-distance relation for the Planck 2018 flat \(\Lambda\)CDM cosmology using the \textsc{Astropy} Python library \cite{price2022astropy}.}. In this way, we further map the associated GW template representations from the source frame to the detector frame (e.g., $m_{1,2 \ \mathrm{det}} = (1+z)m_{\mathrm{1,2 \ source}}$), ensuring consistency with the cosmologically redshifted signals observed in the detector's strain data. We also adopt uniform sky position ($\theta,\phi$), drawing right ascension $\alpha$ uniformly within the range $[0,2\pi]$ and sampling declination $\delta$ via a uniform distribution in $\sin\delta \in [-1,1]$. In addition, we consider isotropic source orientation by sampling the inclination $\cos \iota $ uniformly within the range $[-1,1]$. Accordingly, the signal templates were distributed uniformly in the coalescence phase $\Phi_0$ and polarization angle $\psi$. A summary of the distributions employed for all BNS and NSBH GW parameters is provided in Table \ref{tab:inj_parmas_dist}.
\begin{table*}[!tbh]
\caption{\label{tab:inj_parmas_dist}Summary of GW parameter distributions employed for the early-inspiral part of the BNS and NSBH waveform injections.}
\begin{ruledtabular}
\begin{tabular}{ccccc}
\textbf{CBC Category} & \textbf{Parameter} & \textbf{Notation} & \textbf{Parameter Distribution} & \textbf{Constraint} \\
\hline
% &  &  & \\
BNS & \multirow{2}{*}{Source Masses} & \multirow{2}{*}{$m_1,m_2$} &
$ m_1,m_2 \in \mathcal{U}(1,3)\,M_\odot $ & $m_1>m_2$\\
NSBH &  &  & $m_1\in \mathcal{U}(3,20) \ M_{\odot}, \ m_2\in \mathcal{U}(1,3) \ M_{\odot}$ & --\\
\hline
% &  &  & \\
BNS & \multirow{2}{*}{Chirp Diastance} & \multirow{2}{*}{$d_c$} &
$d_c^{2} \in \mathcal{U}({3}^{2},\,{500}^{2})\ \mathrm{Mpc}^{2}$ & --\\
NSBH &  & &$d_c^{2} \in \mathcal{U}({10}^{2},\,{600}^{2})\ \mathrm{Mpc}^{2}$ & --\\
\hline
% &  &  & \\
BNS \& NSBH & Spins & $|\vec{\chi}_1|, |\vec{\chi}_2|$ &
$0,0$ & -- \\
BNS \& NSBH & Tidal Deformabilities & $\Lambda_1,\Lambda_2$ &
$0,0$ & -- \\
BNS \& NSBH & Inclination & $\iota$ &
$\cos\iota \in \mathcal{U}(-1,1)$ &--\\

BNS \& NSBH & Coalescence Phase & $\Phi_0$ &
$\Phi_0 \in \mathcal{U}(0,2\pi)$ &--\\

BNS \& NSBH & Declination & $\delta$ &
$\sin\delta \in \mathcal{U}(-1,1)$ &-- \\

BNS \& NSBH &Right Ascension & $\alpha$ & $\alpha \in \mathcal{U}(0,2\pi)$ & -- \\

BNS \& NSBH &Polarization & $\psi$ & $\psi \in \mathcal{U}(0,2\pi)$ & -- \\
\end{tabular}
\end{ruledtabular}
%\vspace{4pt}
%\raggedright\footnotesize
\end{table*}

\begin{figure}[!thb]
    %\resizebox{0.5\textwidth}{!}{
    \includegraphics[width=0.47\textwidth]{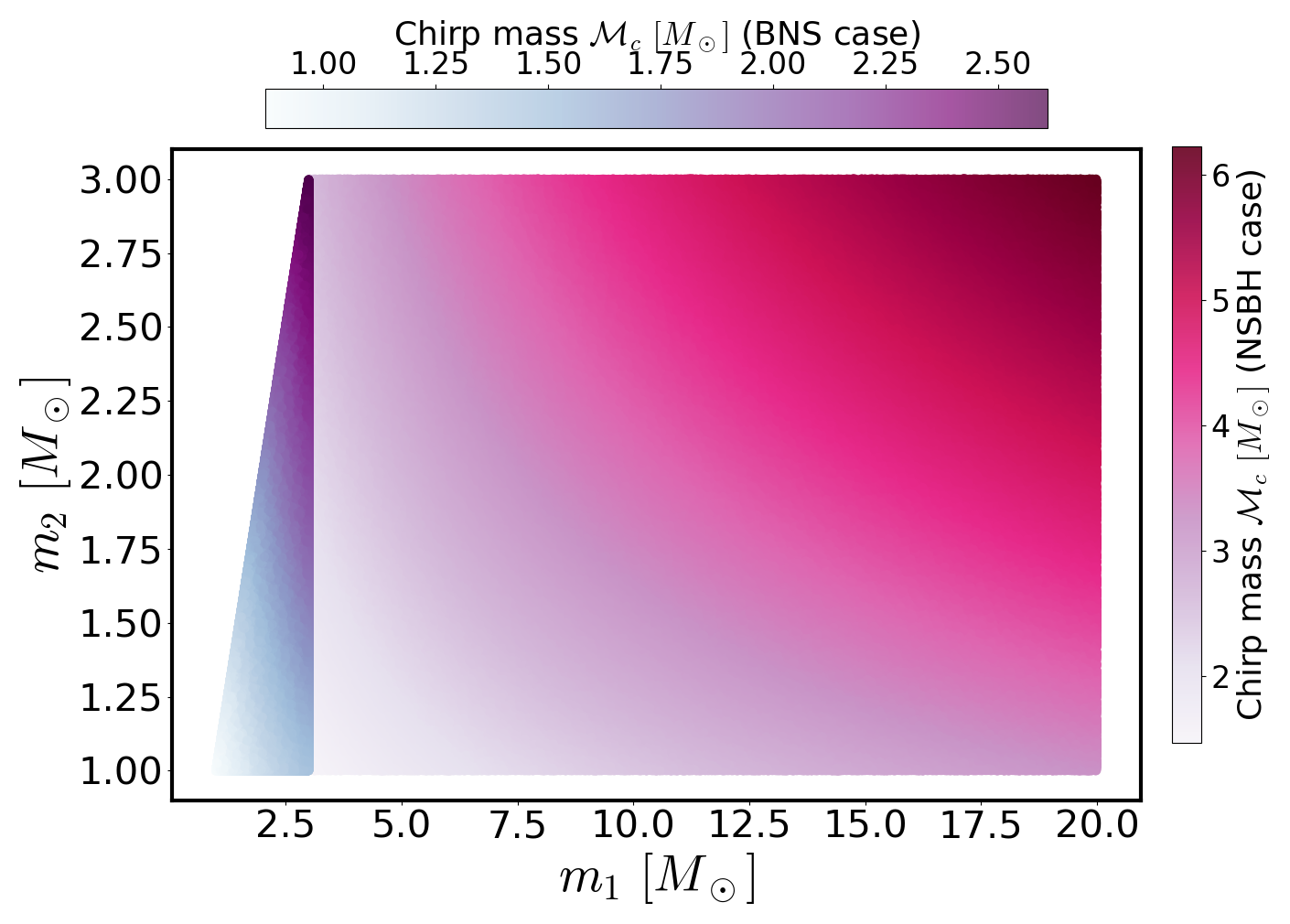}\hfill
    \includegraphics[width=0.47\textwidth]{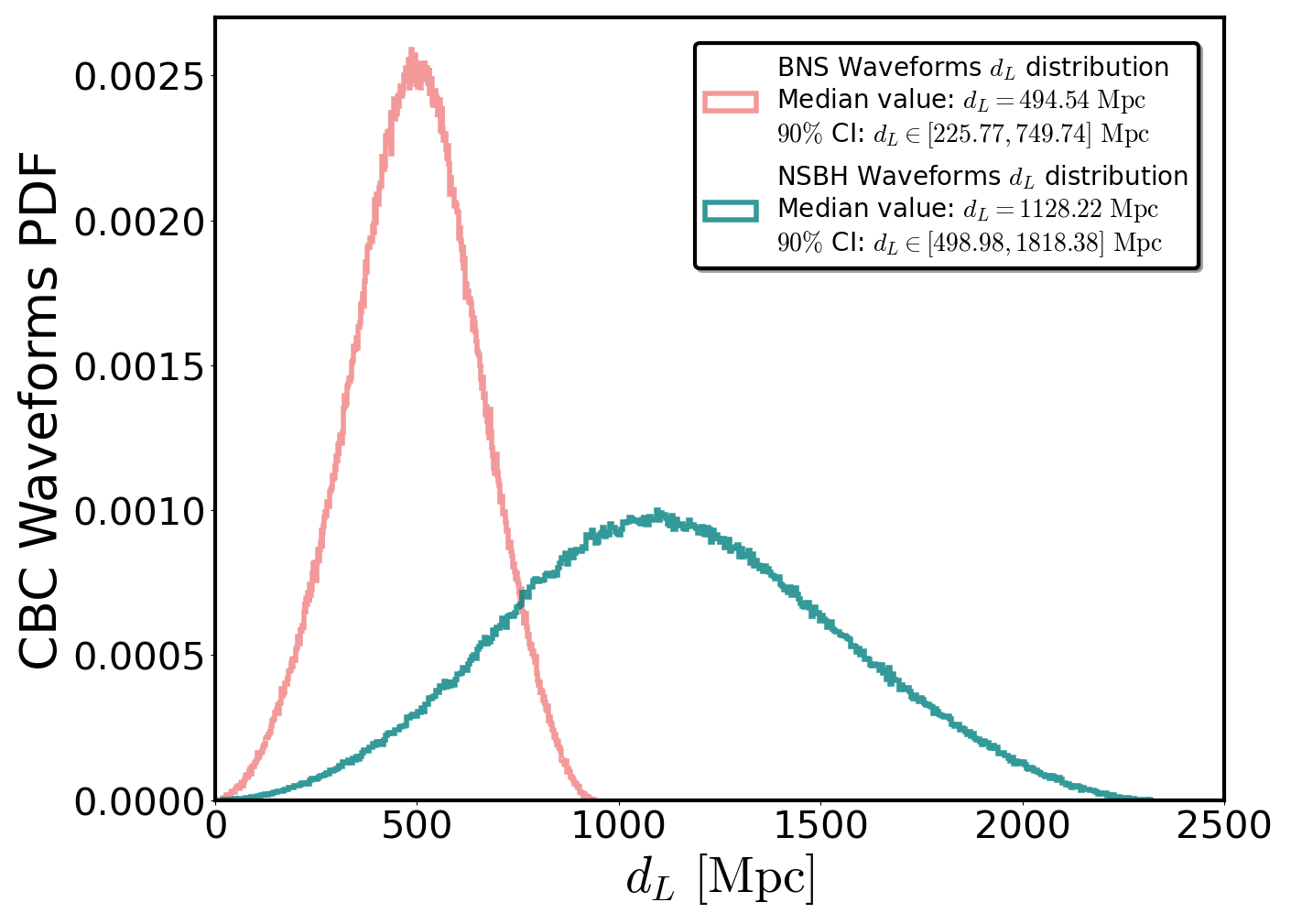}\hfill
    \caption{\label{fig:masses_and_d_L_params} Top panel: source frame $m_1-m_2$ masses for BNS and NSBH waveforms employed in this work, color-coded by the source frame chirp mass $\mathcal{M}_c$. Bottom panel: Luminosity-distance histogram distributions for the associated CBC populations considered. For each case, we also provide the median luminosity distance and the associated $90\%$ credible interval for the distributions illustrated.}
    %The presented data points correspond to the associated data within the test set for each EoS included in our catalog.
\end{figure}

Both the generated detector strain data and the constructed waveforms were sampled with a sampling frequency of $512 \ \mathrm{Hz}$. More specifically, regarding GW signals, we generated an ensemble of $10^6$ waveforms, encompassing both BNS and NSBH sources. In Fig. \ref{fig:masses_and_d_L_params}, the top panel shows the source-frame $m_1-m_2$ plane color-coded by chirp mass $\mathcal{M}_c$, while the bottom panel presents the corresponding luminosity-distance distributions for each CBC category considered.

The motivation for adopting this $d_L$ distribution for BNS systems was to restrict the sample to distances at which a kilonova could still be plausibly detectable in EM follow-up observations, given that kilonovae are relatively faint and rapidly evolving transients (see, e.g., Refs. \cite{metzger2020kilonovae, sagues2021detectability}). By contrast, if one sampled sources out to the full GW reach of third-generation detectors, the distribution of detectable events would be dominated by cosmological distances corresponding to luminosity distances of order $\sim 10-20 \ \mathrm{Gpc}$, i.e., tens of times farther than our EM-motivated cutoff \cite{abac2025science,maggiore2020science,abbott2017exploring,reitze2019cosmic,evans2021horizon}. For NSBH systems, EM counterparts are less certain and often expected to be weaker or absent altogether because producing bright emission typically requires tidal disruption of the neutron star outside the black hole, which occurs only in a restricted region of parameter space (see, e.g., Refs. \cite{kyutoku2021coalescence, barbieri2020electromagnetic}). Consequently, extending the NSBH distance distribution to $2\,\mathrm{Gpc}$ is optimistic for joint GW-EM observations; nevertheless, it remains useful for exploring a best-case multimessenger scenario, while noting that GW-only NSBH detections are intrinsically valuable for constraining formation channels and compact-object properties. This is also consistent with follow-up of GW200115, which reported no confirmed EM counterpart and placed constraining upper limits (see, e.g., Ref. \cite{dichiara2021constraints} for a review).

Lastly, it should be highlighted that waveforms are generated with a low-frequency cutoff of $12 \ \mathrm{Hz}$ for BNS and $8 \ \mathrm{Hz}$ for NSBH systems, respectively\footnote{BNS systems, being less massive, spend considerably more time in the detector band. A modestly higher low-frequency cutoff reduces the runtime with minimal information loss. On the other hand, NSBH signals are shorter, spending less time in the detector band.}. Under the employed mass parameters and the associated $f_{\mathrm{low}}$ settings, simulated BNS signals last from hundreds up to thousands of seconds, while NSBH signals are of order a few hundred seconds, with longer durations at lower chirp mass binaries.

\section{\label{sec:datasets_feature_extraction}Datasets and Feature Extraction}

\subsection{Data Segments and Employed Observational Time Windows}

Building on the preceding description, an extended ensemble of redshifted signal waveforms $h_{\mathrm{det}}(t;\vec{\theta})$ associated with BNS and NSBH mergers is employed as injections into simulated Gaussian noise segments generated from the ET-D and CE (wideband) design-sensitivity PSDs. Depending on the component masses, the employed detector's noise strain duration employed for the associated injections ranges from a few hundred to several thousand seconds. 
%Above all, since our goal is to develop an early-warning trigger identification methodology, we focus our analysis on the signal’s early inspiral phase.

Motivated by the development of an early-warning trigger identification methodology, we focus our analysis on the signal’s inspiral phase. Our central objective is to discriminate detector strain data containing specific pre-merger signal portions from noise-only data. For each detector configuration presented in Table \ref{tab:det_network}, we generated $8$ BNS and $6$ NSBH foreground datasets, respectively. At first glance, useful information about these datasets can be found in Table \ref{tab:inj_datasets}. Furthermore, we also generated an equal number of pure-noise datasets to serve as background. Therefore, each case consists of a balanced pair: one foreground dataset and its corresponding background one. Each set of data corresponds to an Observation Time Window (OTW) class with specific characteristics, which are defined and explained in detail below in the remainder of this section.

\begin{table*}
\caption{\label{tab:inj_datasets} Overview of the foreground injection datasets used in this work. For each foreground dataset $D_i$ (BNS and NSBH), we present the maximum and minimum reference frequencies, $f_{\mathrm{max}}$ and $f_{\mathrm{min}}$, together with the minimum, maximum, mean, and median time before merger $\Delta t = t_c - t_{\max}$. Each foreground dataset contains the associated portion of the CBC injection employed. The corresponding $60\,\mathrm{s}$ strain segments $D_i^{\star}$, from which the foreground samples are drawn, are also indicated, along with the high-pass and low-pass filter frequencies applied to the data.}
\begin{ruledtabular}
\begin{tabular}{c|c|cccccc|c|ccc}
%CBC Category & \multirow{3}{*}{Datasets} & \multirow{3}{*}{$f_{\mathrm{max}}$}& \multirow{3}{*}{V1}& \multirow{3}{*}{G1} & \multirow{3}{*}{H1}&\multirow{3}{*}{L1} \\

\multirow{2}{*}{\textbf{Injections}} & \textbf{Foreground} & $f_{\mathrm{max}}$ & $f_{\mathrm{min}}$&  Minimum & Maximum  & Mean & Median & $60 \ \mathrm{s}$ \textbf{Data} & \textbf{Highpass} & \textbf{Lowpass}    \\

  & \textbf{Samples} &  $(\mathrm{Hz}$) &  $(\mathrm{Hz}$) & $\Delta t$ $(\mathrm{s})$ & $\Delta t$ $(\mathrm{s})$ & $\Delta t$ $(\mathrm{s})$  & $\Delta t$ $(\mathrm{s})$ & \textbf{Segments} & \textbf{filter} $(\mathrm{Hz})$& \textbf{filter} $(\mathrm{Hz}$) \\
\hline
% &  &  & \\
 &   $D_{1}$ & $17.00$ & 13.75 & 51.90 & 412.68 & 136.39 & 124.98 & $D^{\star}_{1}$ &  $12.00$ &  $17.00$&\\
 &   $D_2$ & $18.00$ & 14.19 & 44.42 & 354.40 & 117.07& 107.26 & $D^{\star}_{2}$ &  $12.00$ &  $18.00$&\\
 &   $D_3$ & $19.00$ & 14.59 & 38.33 & 306.85 & 101.32 & 92.81 & $D^{\star}_{3}$ &  $12.00$ &  $19.00$ &\\
 \multirow{2}{*}{BNS}&  $D_4$ & $20.00$ & 14.95 & 33.30 & 267.62 & 88.32& 80.90&  $D^{\star}_{4}$&  $12.00$&  $20.00$&\\
 &   $D_5$ & $22.00$ & 15.57 & 25.62 & 207.55 & 68.41 & 62.65 &  $D^{\star}_{5}$&  $12.00$&  $22.00$&\\
 &   $D_6$ & $25.00$ & 16.31 & 17.92 & 147.55 & 48.53 & 44.43 &  $D^{\star}_{6}$&  $12.00$&  $25.00$&\\
 &  $D_7$ & $27.00$ & 16.69 & 14.40 & 120.12 & 39.44 & 36.09 &  $D^{\star}_{7}$&  $12.00$&  $27.00$&\\
 &   $D_8$ & $30.00$ & 17.14 & 10.62 & 90.60 & 29.66 & 27.13 & $D^{\star}_{8}$ &  $12.00$&  $30.00$&\\

\hline
&  $D_1$& $17.00$ & 8.33 & 19.79 & 178.37 & 45.19 & 39.53& $D^{\star}_{1}$& $12.00$& $17.00$&\\
\multirow{3}{*}{NSBH}&  $D_2$ & $18.00$& 8.42 & 18.54 & 154.17 & 40.40 & 35.53 & $D^{\star}_{2}$& $12.00$& $18.00$&\\

& $D_3$ & $19.00$& 8.55 & 17.49 & 134.94 & 36.49 &32.28 & $D^{\star}_{3}$& $12.00$& $19.00$&\\

&$D_4$ & $20.00$&8.73 & 16.64 & 119.13 & 33.27 & 29.59 & $D^{\star}_{4}$& $12.00$& $20.00$&\\
&$D_5$ & $22.00$& 8.96 & 15.34 & 94.87 & 28.35 & 25.49 & $D^{\star}_{5}$& $12.00$& $22.00$&\\
&$D_6$ & $25.00$& 9.15 & 14.03 & 70.90 & 23.44 & 21.41& $D^{\star}_{6}$& $12.00$& $25.00$&\\

\end{tabular}
\end{ruledtabular}
%\vspace{4pt}
%\raggedright\footnotesize
\end{table*}

%Within each CBC category investigated, the associated injection parameter distributions for the foreground data samples remain the same. 

For each injected waveform that belongs to the particular BNS or NSBH ensemble, the related set of data corresponds to a suitably partitioned segment that includes the early inspiral signal portion, characterized by a chosen instantaneous maximum frequency $f_{\mathrm{max}}$ as upper frequency limit. Therefore, relative to the underlying parameter distributions, the only variations across the datasets considered are the selected maximum instantaneous frequencies and the resulting optimal PI SNRs that follow from them. Additionally, the time instant $t_{\mathrm{max}}$ at which the signal's instantaneous frequency reaches the selected $f_{\mathrm{max}}$ defines the time before merger, measured relative to the time of coalescence $t_c$.

\begin{figure*}[!thb]
    %\resizebox{0.5\textwidth}{!}{
    \includegraphics[width=1.0\textwidth]{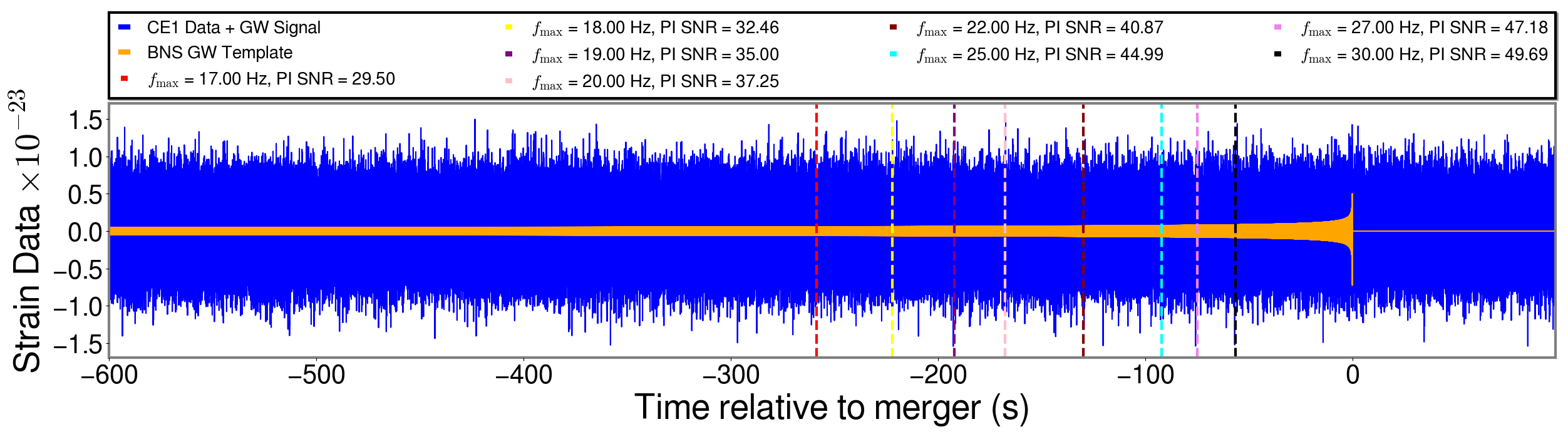}\hfill
    \caption{\label{fig:gw170817_f_crossings_example} CE1 strain data together with an injection with component source masses similar to the GW170817 BNS event ($m_1 = 1.46 \ M_{\odot}$, $m_2 = 1.27 \ M_{\odot}$), but located at a different luminosity distance, $d_L = 440\ \mathrm{Mpc}$. Colored vertical lines mark the time instants relative to the merger when the waveform’s instantaneous frequency reaches the corresponding $f_{\mathrm{max}}$ values employed. For each reference value of instantaneous maximum frequency, we also show the corresponding PI SNR, illustrating how the accumulated SNR increases in the detector's band as the system evolves toward merger. The time of coalescence corresponds to $t_c = 0 \ \mathrm{s}$.}
    %The presented data points correspond to the associated data within the test set for each EoS included in our catalog.
\end{figure*}

\begin{figure*}[!thb]
    \centering
    \subfloat[\label{fig:CEH_PI_snr_BNS} PI SNR histograms for BNS signals. ]{%
        \includegraphics[width=0.50\textwidth]{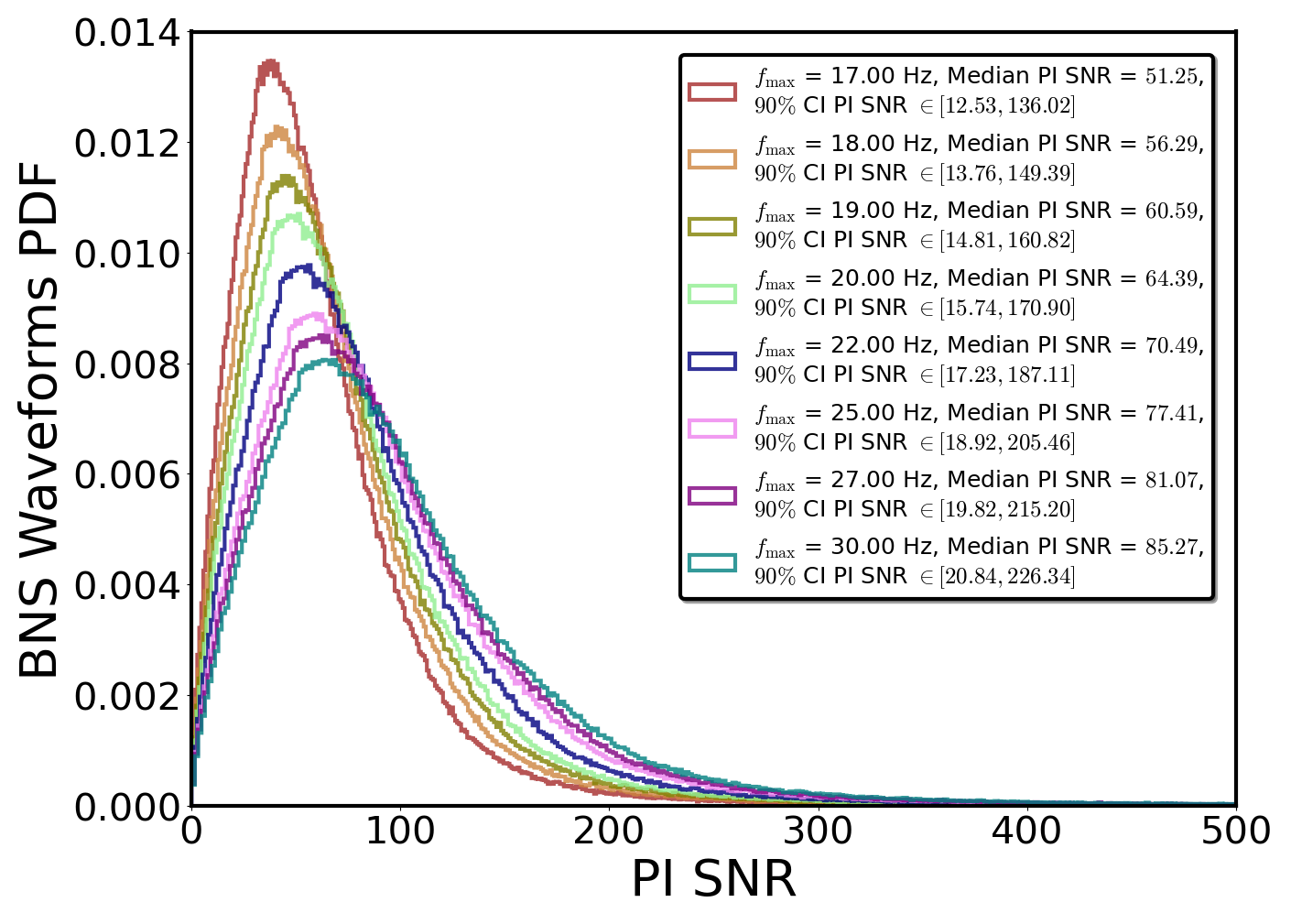}}
    \subfloat[PI SNR histograms for NSBH signals.]{%
        \includegraphics[width=0.50\textwidth]{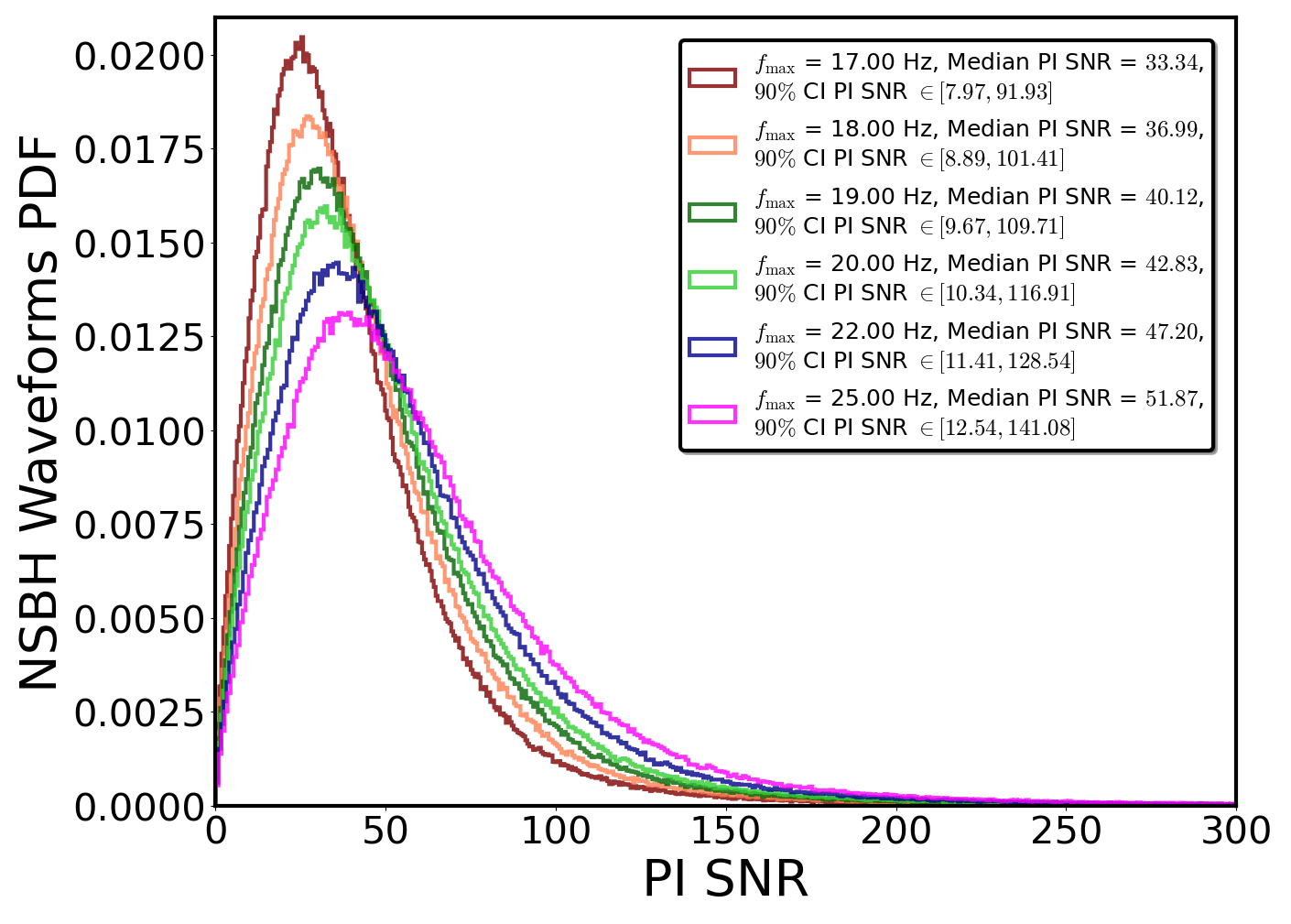}
        \label{fig:CEH_PI_snr_NSBH}}
    \caption{\label{fig:CEH_PI_SNR_distributions} CE1 Detector: PI SNR distributions for BNS (left panel) and NSBH (right panel) injections embedded in the CE1 detector's strain data. Each histogram is color-coded by the annotated maximum instantaneous frequency 
   $f_{\mathrm{max}}$. For each histogram, we report the median PI SNR, along with the corresponding $90\%$ credible interval, which indicates the central value and spread of the distribution, enabling a direct comparison of the typical signal strength and its variability across the different cases.
   %The optimal SNR distribution related to the time of merger $t_c$ is also shown in black. 
   }
\end{figure*}

In Fig. \ref{fig:gw170817_f_crossings_example}, we present a GW170817-like \cite{abbott2017gw170817} BNS waveform embedded in the CE1 detector's Gaussian noise. Each vertical colored dashed line indicates the time $t_{\mathrm{max}}$ at which the instantaneous signal frequency reaches the associated instantaneous frequency $f_{\mathrm{max}}$.
Furthermore, for each foreground dataset annotated by its maximum instantaneous frequency, the PI SNR histograms for the CE1 detector in the case of BNS and NSBH signals are shown in Fig. \ref{fig:CEH_PI_SNR_distributions}.
As PI SNR depends on the integrated signal content and the window’s upper instantaneous frequency, larger frequency values lead to higher PI SNR and better detectability. Consistent with this pattern, the demonstrated histograms shift toward lower values for the lower-frequency $f_{\mathrm{max}}$ signal portions, whereas higher-frequency ones yield larger PI SNR. For completeness, each panel shows the median PI SNR and the corresponding $90\%$ credible interval for the PI SNR distribution. Accordingly, for all other detector cases considered, the associated BNS and NSBH PI SNR distributions are provided in Appendix \ref{app:PI_SNR_distributions}.
%An accompanying table summarizes each detector’s distribution, reporting the maximum, minimum, mean, and median PI–SNR.

Furthermore, we consider distinct inspiral phases across the employed injected $h_{\mathrm{det}}(t;\vec{\theta})$ waveforms. For each injection, we selected a $60 \  \mathrm{s}$ foreground time window within the range $[t_{\mathrm{max}} - 50 \ \mathrm{s}, t_{\mathrm{max}} + 10 \ \mathrm{s}]$. %In each case, the instantaneous frequency of interest $f_{\mathrm{max}}$ lies within the chosen time segment.
This way, the chosen time segment encompasses the signal portion with the instantaneous frequency $f_{\mathrm{max}}$ under consideration.
%the instantaneous frequency $f_{\mathrm{max}}$ associated with the signal portion under consideration.
Then, for each data strain time series under consideration, we further whitened the associated $60 \  \mathrm{s}$ foreground segments corresponding to distinct portions of the inspiral using a PSD estimated from an independent $400 \  \mathrm{s}$ background-noise frame. For the PSD computation, we followed the standard practice in gravitational-wave data analysis. More specifically, the employed noise data were divided into $4 \  \mathrm{s}$ segments with $50 \%$ overlap, Hann-windowed, and transformed to the frequency domain. Welch’s method \cite{welch2003use,villwock2008application} was then applied, using the median across segments at each frequency bin to suppress the influence of the transient noise. The resulting PSD was interpolated onto the target analysis frequency grid and regularized via inverse spectrum truncation for smoothing \cite{allen2012findchirp}. Furthermore, the selected $400 \  \mathrm{s}$ background-noise segments used for whitening were independently produced across injections, using a different random seed in each instance. In parallel, we applied the same whitening procedure to noise-only detector-strain segments, yielding independently generated $60 \  \mathrm{s}$ whitened background-noise segments. 

\begin{figure*}[!thb]
    \centering
    \subfloat[\label{fig:tc_tmax_BNS} Relative times before merger for BNS signals. ]{%
        \includegraphics[width=0.50\textwidth]{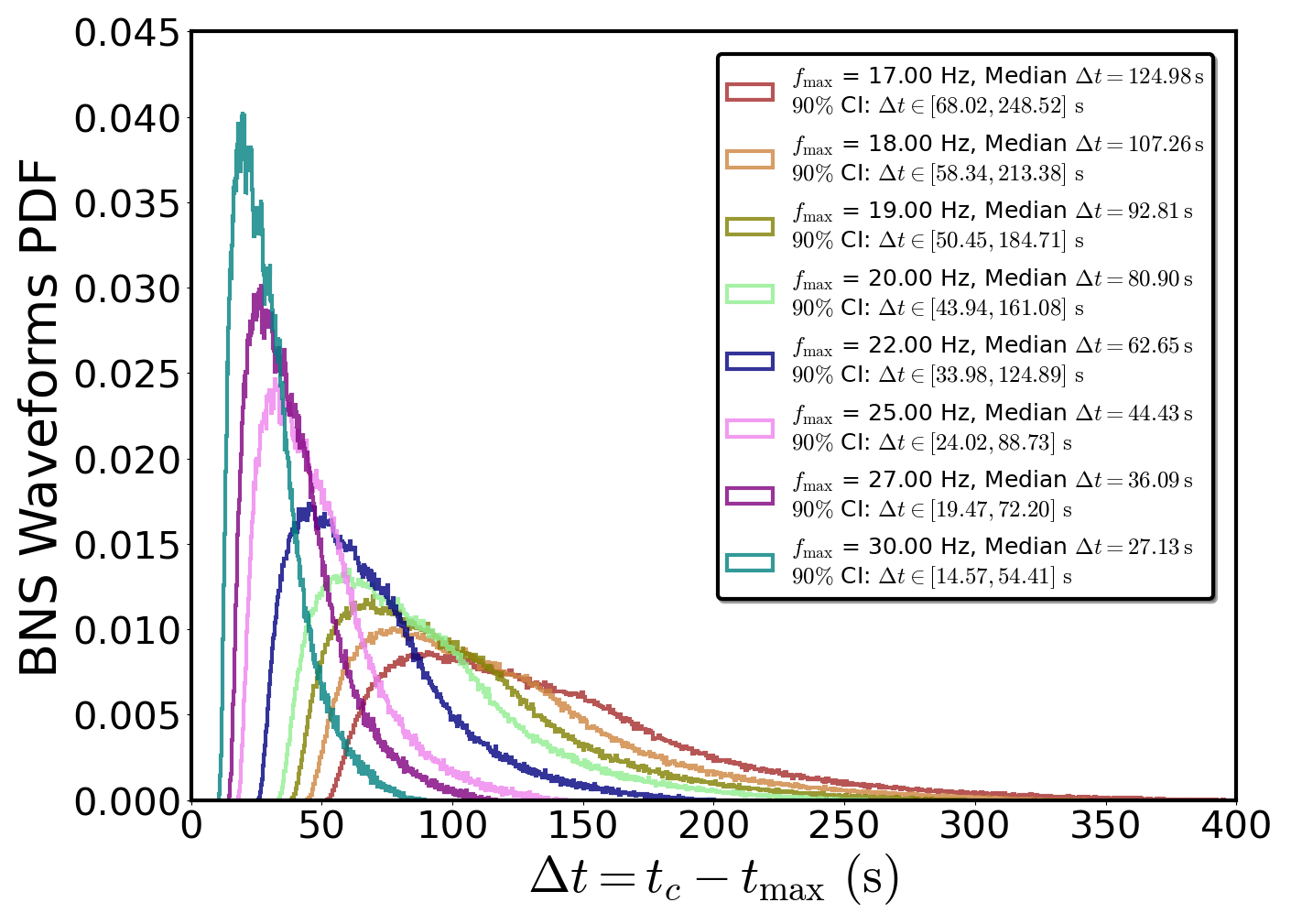}}
    \subfloat[Relative times before merger for NSBH signals.]{%
        \includegraphics[width=0.50\textwidth]{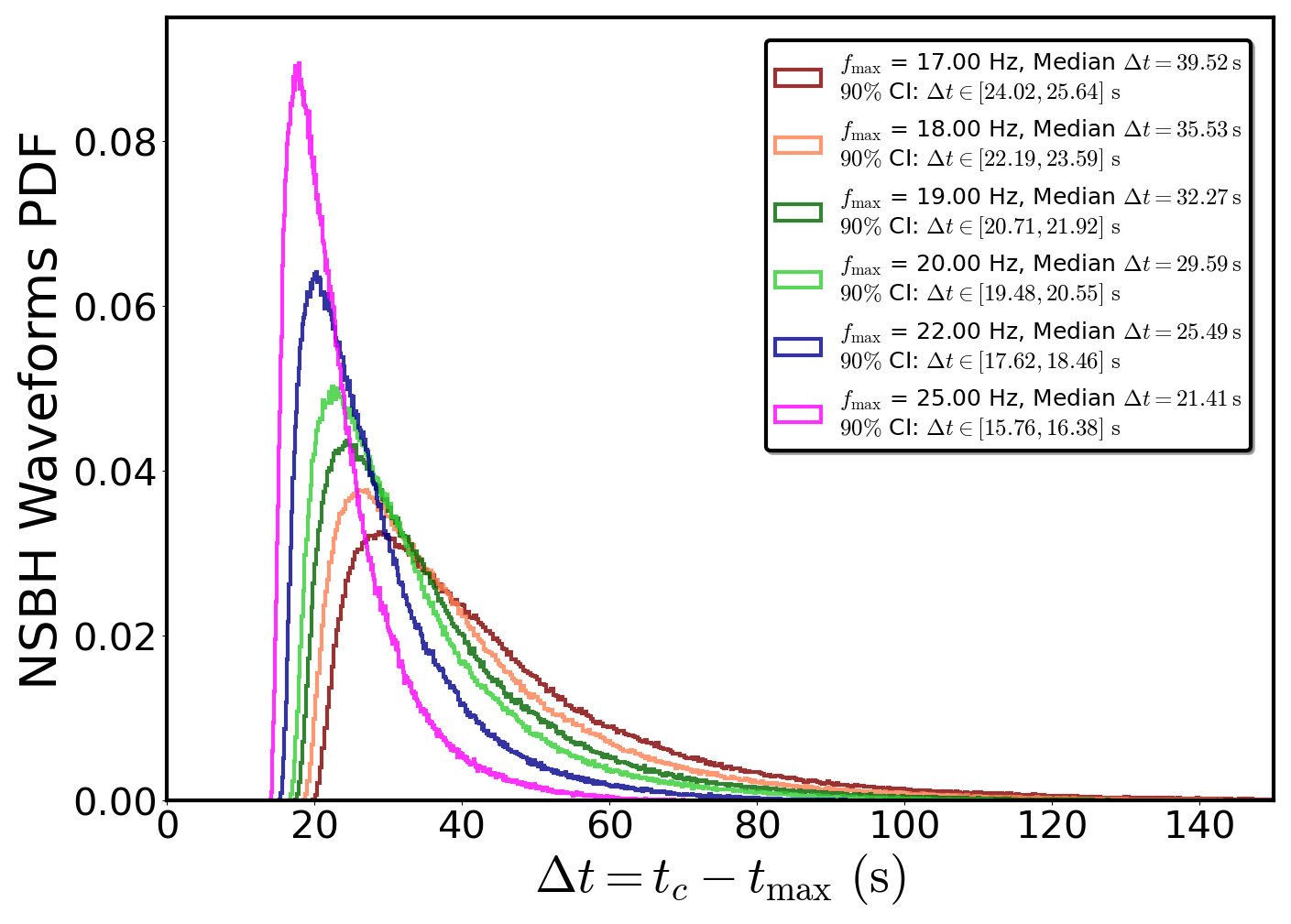}
        \label{fig:tc_tmax_NSBH}}
    \caption{\label{fig:delta_t_distributions} $ \Delta t =  t_c - t_{\mathrm{max}}$ time offset distributions for the BNS (left panel) and NSBH (right panel) waveform ensembles that were employed in this study. Each histogram is color-coded by the annotated time instant $t_{\mathrm{max}}$ corresponding to the inspiral's chosen maximum instantaneous frequency 
   $f_{\mathrm{max}}$. For each histogram, we report the median time offset, along with the corresponding $90\%$ credible interval, which indicates the central value and spread of the distribution, enabling a direct comparison of the typical times associated with instantaneous maximum frequency (prior to merger) and their variability across the chosen cases.}
\end{figure*}

After the whitening phase, all $60 \  \mathrm{s}$ foreground and background segments were bandpass-filtered using highpass and lowpass filters \cite{usman2016pycbc,nitz2018rapid}. The frequency cutoff settings applied to the foreground segments, defined as $D^{\star}_{i}$ and each selected to contain the signal portion with the chosen frequency $f_{\mathrm{max}}$ within the data, are summarized by each $i$-case  in the final columns of Table \ref{tab:inj_datasets}. Accordingly, each whitened background-noise realization was bandpass-filtered using the same cutoff frequencies as its paired foreground segment. Then, from each filtered foreground segment, we extract a $40 \  \mathrm{s}$ OTW whose upper bound matches the analysis upper frequency limit and whose window-specific signal's $f_{\mathrm{min}}$ depends on the injected waveform parameters. We adopt this OTW duration for both BNS and NSBH signal portions that we have used for early warning alert investigations. Above all, it is an empirical trade-off that balances signal coverage and latency across the considered waveform parameters. Based on that, it should be emphasized at this point that these OTWs constitute the foreground datasets $D_i$ listed in the corresponding column of Table \ref{tab:inj_datasets}. Alongside $f_{\mathrm{max}}$ and $f_{\mathrm{min}}$, for each foreground dataset $D_i$, we also provide the minimum, maximum, median, and mean times before merger ($\Delta t$), quantities directly informative for early-warning capability across the waveform models considered. In addition, in Fig. \ref{fig:delta_t_distributions} (left and right panels), we illustrate the associated histograms of the time offset $t_c - t_{\mathrm{max}}$ for the BNS and NSBH waveform ensembles used in this study. Notably, these distributions are essentially unchanged across detectors, since coincident arrival times differ by only a few milliseconds. As the selected instantaneous frequency increases,  $t_{\mathrm{max}}$ occurs closer to coalescence, and therefore, the relevant time before merger decreases. This pattern is consistent with the chirp evolution in the detector's band, where the instantaneous frequency rises during inspiral as the compact objects approach merger.
Additionally, corresponding bandpass-filtered time windows are obtained from background segments, yielding a matched background OTW for every foreground one defined by the injected signal’s $f_{\mathrm{max}}$ as upper frequency limit.

\subsection{From Observational Time Windows to Feature Space}

In total, we generated $10^6$ foreground OTWs and $10^6$ matched-pair background ones, with each segment window containing $40\ \mathrm{s}$ of time series data. Therefore, each foreground dataset $D_i$, as outlined in Table \ref{tab:inj_datasets}, is the collection of bandpass-filtered OTWs containing injections with the instantaneous frequency of interest as an upper limit, and a matched-pair background dataset $D^{\prime}_i$ with a collection of noise-only data segments filtered identically as the $D_i$ counterpart. Therefore, in each case, the paired time series data collections $D_i$ and $D^{\prime}_i$ are balanced, each containing $10^6$ OTWs in total. Based on that, we define the composite time series datasets as $X^{(s)}_{i} = [D^{(s)}_i, D^{(s)\prime}_{i}]$ with CBC source \(s \in \{\mathrm{BNS}, \mathrm{NSBH}\}\), for an $f_{\mathrm{max}}$ setting indexed by $i$. As presented in Table \ref{tab:inj_datasets}, we consider $i=1,\ldots,8$ for the BNS early-alert investigation and $i=1,\ldots,6$ for the associated NSBH case. This design provides a balanced binary classification setting (GW signal-portion + noise vs. noise-only) for each case under consideration \cite{cuoco2020enhancing,benedetto2023ai,zhao2023dawning,cuoco2025applications}.

In this framework, the dataset definition $X^{(s)}_{i}$ given above applies directly to the L-shaped CE1 and CE2 detector configurations. A schematic overview of the data-construction procedure, including foreground/background generation, whitening and bandpass filtering, OTW extraction, and the final concatenation into the time series dataset $X^{(s)}_{i}$, is shown in Fig. \ref{fig:dataset_flow}.

\begin{figure*}[!htb]
\centering
\begin{tikzpicture}[
  font=\small,
  node distance=8mm and 12mm
]
\usetikzlibrary{arrows.meta,positioning,shapes.geometric,fit}

\tikzset{
  proc/.style={
    draw, rounded corners, align=center,
    minimum width=40mm, minimum height=9mm,
    fill=blue!6
  },
  dataset/.style={
    draw, very thick, rounded corners, align=center,
    minimum width=46mm, minimum height=11mm,
    fill=yellow!10
  },
  merge/.style={
    draw, diamond, aspect=2, align=center,
    inner sep=1pt, fill=orange!10
  },
  note/.style={draw=none, align=left, font=\scriptsize},
  arrow/.style={-Latex, thick}
}

% Top: source choice
\node[proc] (source) {Choose CBC source\\ $s\in\{\mathrm{BNS},\,\mathrm{NSBH}\}$};

\node[proc, below=of source] (fmax) {Select $f_{\max}$ setting\\ (index $i$)};

% Move i-ranges next to the fmax cell
\node[note, right=3mm of fmax, anchor=west] (irange) {$\begin{aligned}
\text{BNS: }& i=1,\ldots,8\\
\text{NSBH: }& i=1,\ldots,6
\end{aligned}$};

% Left branch: foreground
\node[proc, below left=14mm and 20mm of fmax] (genF)
{Generate $10^6$ foreground samples\\ (strain data with injections)};
\node[proc, below=of genF] (filtF)
{Whiten + bandpass filtering\\ $\longrightarrow$ extract $40\,\mathrm{s}$ OTWs\\ with upper cutoff $f_{\max}$};
\node[dataset, below=of filtF] (Di)
{Final foreground OTWs\\ dataset $D^{(s)}_i$};

% Right branch: matched-pair background
\node[proc, below right=14mm and 20mm of fmax] (genB)
{Generate $10^6$ background samples\\ (noise-only)};
\node[proc, below=of genB] (filtB)
{Apply whitening and identical\\ bandpass filtering as for the foreground\\ datasets $\longrightarrow$ extract $40\,\mathrm{s}$ OTWs\\ with upper cutoff $f_{\max}$};
\node[dataset, below=of filtB] (Dip)
{Final background OTWs\\ dataset $D^{(s)\prime}_i$};

% Merge + final dataset + task
\node[merge, below=18mm of fmax] (m) {Concatenate};
\node[dataset, below=of m] (Xi)
{Composite balanced timeseries\\ dataset: $X^{(s)}_{i}=[D^{(s)}_i,\,D^{(s)\prime}_{i}]$};
\node[proc, below=of Xi] (task)
{Appropriate for binary classification\\ or further processing};

% Arrows
\draw[arrow] (source) -- (fmax);
\draw[arrow] (fmax) -- (genF);
\draw[arrow] (fmax) -- (genB);

\draw[arrow] (genF) -- (filtF);
\draw[arrow] (filtF) -- (Di);

\draw[arrow] (genB) -- (filtB);
\draw[arrow] (filtB) -- (Dip);

% Direct dataset-to-concatenate connections
\draw[arrow, shorten >=2pt, shorten <=2pt] (Di.east)  -- (m.west);
\draw[arrow, shorten >=2pt, shorten <=2pt] (Dip.west) -- (m.east);

\draw[arrow] (m) -- (Xi);
\draw[arrow] (Xi) -- (task);

% Dashed "paired" grouping
\node[
  draw, dashed, rounded corners, inner sep=17pt,
  fit=(Di)(Dip)(m)
] (pairbox) {};

\end{tikzpicture}
%\caption{Construction of balanced composite time-series datasets $X^{(s)}_{i}$ for each source type $s$ and $f_{\max}$ index $i$.}
\caption{\label{fig:dataset_flow} CE1 and CE2 detectors: compact workflow for constructing the balanced composite time-series datasets $X^{(s)}_{i}$ for each CBC source type $s$ and $f_{\max}$ index $i$. For each case, we generate $10^6$ foreground data segments $D^{\star}_{i}$ and $10^6$ noise-only ones $D^{\star\prime}_{i}$. Both streams undergo identical whitening and bandpass filtering, after which $40\,\mathrm{s}$ observational time windows (OTWs) are extracted with an upper analysis frequency limit set by $f_{\max}$. The resulting paired OTW collections $D^{(s)}_{i}$ and $D^{(s)\prime}_{i}$ are concatenated to form $X^{(s)}_{i}=[D^{(s)}_i,\,D^{(s)\prime}_{i}]$, yielding a balanced binary setting for further processing.}
\end{figure*}

On the contrary, for the ET1 configuration, which comprises of three interferometers denoted by $k = 1,2,3$ arranged on a triangular geometry, we should take into account the associated per-interferometer composite datasets $\tilde{X}^{(s)}_{i,k} = [D^{(s)}_{i,k}, D^{(s)\prime}_{i,k}]$. This way, the total ET1 time series dataset considered for case $i$ and source class $s$ is then the collection of the per-interferometer composites, \(X^{(s)}_{i}  = \{\tilde{X}^{(s)}_{i,1}, \tilde{X}^{(s)}_{i,2}, \tilde{X}^{(s)}_{i,3}\}\). In addition, for the ET2 configuration considered, which consists of two L-shaped detectors \(\ell \in \{A, B\}\) corresponding to V1 and G1, the definition of the composite datasets $\tilde{X}^{(s)}_{i,\ell} = [D^{(s)}_{i,\ell}, D^{(s)\prime}_{i,\ell}]$ is analogous. In this setting, the employed dataset, which includes time series data from the two co-sited detectors in a single-detector baseline, has the form \(X^{(s)}_{i}  = \{\tilde{X}^{(s)}_{i,A}, \tilde{X}^{(s)}_{i,B}\}\). Figure \ref{fig:dataset_flow_et} schematically summarizes the ET dataset assembly, with the top and bottom panels corresponding to the ET1 and ET2 cases, respectively.

\begin{figure*}[!htb]
\centering
\begin{tikzpicture}[font=\small]
\tikzset{
  proc/.style={
    draw, rounded corners, align=center,
    minimum width=120mm, minimum height=9mm,
    fill=blue!6
  },
  dataset/.style={
    draw, very thick, rounded corners, align=center,
    minimum width=120mm, minimum height=11mm,
    fill=green!10
  },
  smallset/.style={
    draw, thick, rounded corners, align=center,
    minimum width=36mm, minimum height=10mm,
    fill=yellow!8
  },
  arrow/.style={-Latex, thick}
}

% -------------------- TOP BLOCK: ET1 --------------------
\node[proc] (et1top) {ET1: per-interferometer composite datasets\\
Apply the procedure of Fig.~\ref{fig:dataset_flow} independently \\ for each $k\in\{1,2,3\}$};

\node[smallset, below=10mm of et1top, xshift=-44mm] (xk1)
{$\tilde{X}^{(s)}_{i,1}=[D^{(s)}_{i,1},\,D^{(s)\prime}_{i,1}]$};
\node[smallset, right=10mm of xk1] (xk2)
{$\tilde{X}^{(s)}_{i,2}=[D^{(s)}_{i,2},\,D^{(s)\prime}_{i,2}]$};
\node[smallset, right=10mm of xk2] (xk3)
{$\tilde{X}^{(s)}_{i,3}=[D^{(s)}_{i,3},\,D^{(s)\prime}_{i,3}]$};

\node[dataset, below=12mm of xk2] (et1X)
{Balanced ET1 time series dataset\\
$X^{(s)}_{i}=\{\tilde{X}^{(s)}_{i,1},\,\tilde{X}^{(s)}_{i,2},\,\tilde{X}^{(s)}_{i,3}\}$};

\draw[arrow] (et1top.south) -- ++(0,-3mm) -| (xk1.north);
\draw[arrow] (et1top.south) -- ++(0,-3mm) -| (xk2.north);
\draw[arrow] (et1top.south) -- ++(0,-3mm) -| (xk3.north);

\draw[arrow] (xk1.south) -- ++(0,-3mm) -| (et1X.north);
\draw[arrow] (xk2.south) -- (et1X.north);
\draw[arrow] (xk3.south) -- ++(0,-3mm) -| (et1X.north);

\node[
  draw, dashed, rounded corners, inner sep=8pt,
  fit=(et1top)(xk1)(xk2)(xk3)(et1X),
  %label={[font=\scriptsize]above left:{\textbf{ET1 handling}}}
] (boxET1) {};

% -------------------- BOTTOM BLOCK: ET2 --------------------
\node[proc, below=18mm of et1X] (et2top)
{ET2: per-detector composite datasets\\
Apply the procedure of Fig.~\ref{fig:dataset_flow} independently \\ for each $\ell\in\{A,B\}$};

\node[smallset, below=10mm of et2top, xshift=-23mm] (xA)
{$\tilde{X}^{(s)}_{i,A}=[D^{(s)}_{i,A},\,D^{(s)\prime}_{i,A}]$};
\node[smallset, right=10mm of xA] (xB)
{$\tilde{X}^{(s)}_{i,B}=[D^{(s)}_{i,B},\,D^{(s)\prime}_{i,B}]$};

\node[dataset, below=12mm of xA, xshift=23mm] (et2X)
{Balanced ET2 time series dataset\\
$X^{(s)}_{i}=\{\tilde{X}^{(s)}_{i,A},\,\tilde{X}^{(s)}_{i,B}\}$};

\draw[arrow] (et2top.south) -- ++(0,-3mm) -| (xA.north);
\draw[arrow] (et2top.south) -- ++(0,-3mm) -| (xB.north);

\draw[arrow] (xA.south) -- ++(0,-3mm) -| (et2X.north);
\draw[arrow] (xB.south) -- ++(0,-3mm) -| (et2X.north);

\node[
  draw, dashed, rounded corners, inner sep=8pt,
  fit=(et2top)(xA)(xB)(et2X),
  %label={[font=\scriptsize]above left:{\textbf{ET2 handling}}}
] (boxET2) {};

\end{tikzpicture}
\caption{\label{fig:dataset_flow_et} Extension of the composite-dataset definition to ET detector layouts considered. Top panel: For ET1, which comprises three interferometers $k\in\{1,2,3\}$ in a triangular geometry, the OTW construction workflow of Fig.~\ref{fig:dataset_flow} is applied independently per interferometer to form per-interferometer composite datasets $\tilde{X}^{(s)}_{i,k}=[D^{(s)}_{i,k},\,D^{(s)\prime}_{i,k}]$, and the total dataset for case $(s,i)$ is the collection $X^{(s)}_{i}=\{\tilde{X}^{(s)}_{i,1},\,\tilde{X}^{(s)}_{i,2},\,\tilde{X}^{(s)}_{i,3}\}$. Bottom panel: For ET2, consisting of two co-sited L-shaped detectors $\ell\in\{A,B\}$, the definition is analogous, yielding $X^{(s)}_{i}=\{\tilde{X}^{(s)}_{i,A},\,\tilde{X}^{(s)}_{i,B}\}$.}
\end{figure*}

Above all, time series data possess characteristic statistical distributions, temporal dependencies, and spectral content that can be used in a binary classification framework. Therefore, an investigation based on these characteristics, rather than on raw time series strain data, warrants further study. On this basis, for each OTW, whether foreground or background, we estimated statistical, temporal, and spectral features using the Time Series Feature Extraction Library (TSFEL) \cite{barandas2020tsfel}. TSFEL is an open-source Python package that aggregates a broad set of feature extractor functions across statistical, temporal, and spectral domains for one-dimensional time series data\footnote{For all experiments, we used the TSFEL library (version 0.2.0) in Python.}. 
%It implements functions across statistical, temporal, and spectral domains and provides a unified Application Programming Interface (API) for configuration and batch processing. 
Notably, it provides a deterministic mapping from a time series observational window to a fixed-length feature vector. For each OTW, the output dimension is fixed at $156$, yielding a feature vector \(\mathbf{x} = [x_1,\ldots,x_{156}]^{\top} \in\mathbb{R}^{156}\). In our analysis, we used the library’s full set of available features rather than a preselected subset. A systematic investigation of feature selection will be addressed in future work. %A comprehensive list and definitions of the constituent features are provided in Appendix \ref{app:tsfel_features}. 
This representation effectively moves the analysis from filtered strain time series data to a consistent feature space that is comparable across detectors, CBC source classes, and $f_\mathrm{max}$ settings.

Let \(\phi\) be the general transformation that maps each OTW to its corresponding feature vector $\mathbf{x}$. Regarding the CE1 and CE2 detector configurations, each composite dataset 
$X^{(s)}_{i,\mathrm{det}} = [D^{(s)}_{i,\mathrm{det}}, D^{(s)\prime}_{i,\mathrm{det}}]$ used per CE detector is mapped to the associated parameter space as:
\begin{align}
   \nonumber \mathbb{X}^{(s)}_{i,\mathrm{det}} = \phi\!\big(D_{i,\mathrm{det}}^{(s)}\big)\in\mathbb{R}^{N_{\mathrm{fg}}\times 156}\\
   \mathbb{X}^{(s) \prime}_{i,\mathrm{det}} = \phi\!\big(D_{i,\mathrm{det}}^{(s)\prime}\big)\in\mathbb{R}^{N_{\mathrm{bg}}\times 156},
\end{align}
where \(\mathrm{det} = \{\mathrm{CE1},\mathrm{CE2}\}\) points the detector configuration employed in each CE case, and $N_{\mathrm{bg}} = N_{\mathrm{fg}} = \mathcal{N} = 10^6$ is the number of OTWs employed for the corresponding detectors' backgrounds and foregrounds, respectively. Both matrices share {\it identical} parameters \((x_{1},\ldots,x_{156})\) with the same names and ordering, ensuring feature compatibility across foreground and background.
% for classification.
Then, to obtain a unified representation for the subsequent datasets, we combine the foreground and background balanced feature matrices into a single design matrix by vertical concatenation,
\begin{align}
   \mathbf{X}^{(s)}_{i,\mathrm{det}} = \begin{bmatrix}
    \mathbb{X}^{(s)}_{i,\mathrm{det}}\\[7pt]
    \mathbb{X}^{(s) \prime}_{i,\mathrm{det}}   
    \end{bmatrix}
    \in\mathbb{R}^{(N_{\mathrm{fg}}+N_{\mathrm{bg}})\times 156}.
\end{align}

This construction places all data in the same $156$-dimensional feature space representation, with a corresponding label vector defined as:
\begin{align}
{y}^{(s)}_{i,\mathrm{det}}=
\begin{bmatrix}
\mathbf{1}_{i,N_{\mathrm{fg}}\times1}\\[7pt]
\mathbf{0}_{i,N_{\mathrm{bg}}\times 1}
\end{bmatrix},
\end{align}
where entries labeled as $\mathbf{1}$ and $\mathbf{0}$ denote the associated foreground and background matrix categories, respectively. Combining the above, the final dataset for each CE detector has the form:
\begin{align}
  \mathcal{D}^{(s)}_{i,\mathrm{det}}=\big\{\,(\mathbf{X}^{(s)}_{i,\mathrm{det}},\,{y}^{(s)}_{i,\mathrm{det}})\,\big\},  
\end{align}
where $s$ denotes the source class, and $i$ indexes the associated instantaneous maximum $f_{\mathrm{max}}$ setting.
%%%%%%%%%

Furthermore, for the ET1 configuration (triangular layout with interferometers $k=1,2,3$) and the ET2 configuration (two co-sited L-shaped interferometers \(\ell\in\{A,B\}\)), we construct the corresponding per-interferometer feature matrices using the same $156$-feature formulation as described above. Then, the foreground and background features are concatenated columnwise across interferometers as,
\begin{align}
   \mathrm{ET1}: \ \mathbf{X}^{(s)}_{i,k} = \begin{bmatrix}
    \mathbb{X}^{(s)}_{i,k}\\[7pt]
    \mathbb{X}^{(s) \prime}_{i,k}   
    \end{bmatrix}
    \in\mathbb{R}^{(N_{k,\mathrm{fg}}+N_{k,\mathrm{bg}})\times 156}, \\
    \mathrm{ET2}: \ \mathbf{X}^{(s)}_{i,\ell} = \begin{bmatrix}
    \mathbb{X}^{(s)}_{i,\ell}\\[7pt]
    \mathbb{X}^{(s) \prime}_{i,\ell}   
    \end{bmatrix}
    \in\mathbb{R}^{(N_{\ell,\mathrm{fg}}+N_{\mathrm{\ell,bg}})\times 156}.
\end{align}
Accordingly, for each detector configuration, we assemble a unified design matrix by columnwise concatenation across interferometers and vertical stacking of foreground above background:
%Accordingly, for each detector, we form the unified design matrices that gather the concatenated per-interferometer features as:
\begin{align}
     \mathbf{X}^{(s)}_{i,\mathrm{ET1}}  = \begin{bmatrix}
    \mathbb{X}^{(s)}_{i,1} \ \ \mathbb{X}^{(s)}_{i,2} \ \ \mathbb{X}^{(s)}_{i,3}\\[7pt]
    \mathbb{X}^{(s)\prime}_{i,1} \ \ \mathbb{X}^{(s)\prime}_{i,2} \ \ \mathbb{X}^{(s)\prime}_{i,3}
    \end{bmatrix},  \\ 
     \mathbf{X}^{(s)}_{i,\mathrm{ET2}}  = \begin{bmatrix}
    \mathbb{X}^{(s)}_{i,A} \ \ \mathbb{X}^{(s)}_{i,B}\\[7pt]
    \mathbb{X}^{(s)\prime}_{i,A} \ \ \mathbb{X}^{(s)\prime}_{i,B}   
    \end{bmatrix},   
\end{align}
while the associated dataset-label vector, either for ET1 or ET2 detector configurations, is now taking the form: 
\begin{align*}
    {y}_i^{(s)}=\begin{bmatrix}
    \mathbf{1}_{i,\mathcal{N}\times1}\\[7pt]
    \mathbf{0}_{i,\mathcal{N}\times1}
    \end{bmatrix}.
\end{align*}
In this representation, the feature columns retain identical names and ordering across interferometers, preserving compatibility between foreground and background and ensuring consistency. Consequently, the final balanced datasets across detector configuration take the form:
\begin{align}
  \mathcal{D}^{(s)}_{i,\mathrm{ET1}}=\big\{\,(\mathbf{X}^{(s)}_{i,\mathrm{ET1}},\,{y}^{(s)}_{i,\mathrm{ET1}})\,\big\},  
\end{align}
and
\begin{align}
  \mathcal{D}^{(s)}_{i,\mathrm{ET2}}=\big\{\,(\mathbf{X}^{(s)}_{i,\mathrm{ET2}},\,{y}^{(s)}_{i,\mathrm{ET2}})\,\big\}. 
\end{align}
Each dataset in \(\mathbf{X}^{(s)}_{i,\mathrm{ET1}}\) has $468$ features, 
whereas those in \(\mathbf{X}^{(s)}_{i,\mathrm{ET2}}\) comprise  $312$ features, with the corresponding label given by the
entries of ${y}^{(s)}_{i}$. For completeness, Table \ref{tab:detector_datasets_definition} summarizes the datasets' notation adopted in the remainder of this work, while Fig. \ref{fig:generic_feature_construction} provides a schematic overview of the corresponding time-series-to-feature mapping and dataset assembly.
\begin{table}
\caption{\label{tab:detector_datasets_definition} Final datasets definition by detector configuration, for each source class $s\in\{\mathrm{BNS},\mathrm{NSBH}\}$ and chosen $f_{\max}$ setting indexed by $i$. For the BNS early-warning alert investigation, we have $8$ distinct datasets ($i=1,\ldots,8$), while for the associated NSBH we have $6$ ($i=1,\ldots,6$). Each $(i,s)$ dataset is balanced, consisting of an equal number of $\mathcal{N} = 10^6$ foreground and background data points.}
\begin{ruledtabular}
\begin{tabular}{ccc}
\multirow{2}{*}{\textbf{Detector}} & \textbf{Final Dataset} & \textbf{Number of Features}\\
& \textbf{for case} $(i,s)$ &  \textbf{/Dimensions}\\
\hline
%ET1 & $\mathcal{D}^{(s)}_{i,\mathrm{ET1}}$ & \([x_1,\ldots,x^{\prime}_1,\ldots, x^{\prime\prime}_1,\ldots x^{\prime\prime}_{156}]^{\top}\) \\
ET1 & $\mathcal{D}^{(s)}_{i,\mathrm{ET1}}$ & $468$ \\
%ET2 & $\mathcal{D}^{(s)}_{i,\mathrm{ET2}}$ & \([x_1,\ldots,x_{156},x^{\prime}_1\ldots x^{\prime}_{156} ]^{\top}\)  \\
ET2 & $\mathcal{D}^{(s)}_{i,\mathrm{ET2}}$ & $312$  \\
%CE1 & $\mathcal{D}^{(s)}_{i,\mathrm{CE1}}$ &  \([x_1,\ldots,x_{156}]^{\top}\) \\
CE1 & $\mathcal{D}^{(s)}_{i,\mathrm{CE1}}$ & $156$ \\
%CEL & $\mathcal{D}^{(s)}_{i,\mathrm{CE2}}$ & \([x_1,\ldots,x_{156}]^{\top}\)  \\
CE2 & $\mathcal{D}^{(s)}_{i,\mathrm{CE2}}$ & $156$  \\
\end{tabular}
\end{ruledtabular}
\end{table}
\begin{figure*}[!htb]
\centering
\begin{tikzpicture}[font=\small, node distance=8mm and 14mm]
\usetikzlibrary{arrows.meta,positioning,fit,calc}
\tikzset{
  head/.style={draw, rounded corners, fill=blue!6, align=center,
               minimum width=150mm, minimum height=8mm},
  proc/.style={draw, rounded corners, fill=blue!4, align=center,
               minimum width=52mm, minimum height=10mm},
  mat/.style={draw, very thick, rounded corners, fill=yellow!10, align=center,
              minimum width=88mm, minimum height=12mm},
  arrow/.style={-Latex, thick},
  note/.style={draw=none, font=\scriptsize, align=left}
}

\node[head] (H)
{Detector-agnostic feature construction from balanced time series datasets};

% Start: time-series dataset (generic)
\node[proc, below=9mm of H] (Xts)
{Time-series dataset $X^{(s)}_{i}$\\
(for a chosen detector configuration)};

\node[note, right=3mm of Xts, anchor=west] (chnote)
{Contains $C\in\{1,2,3\}$ balanced\\
data cases: depending on the\\
detector (each dataset contain \\ foreground + background\\ time series data)};

% Mapping to features
\node[proc, below=11mm of Xts] (phi)
{Apply feature map per detector\\
$\phi:\text{OTW}\mapsto \mathbf{x} \in \mathbb{R}^{d}$};

\node[note, right=3mm of phi, anchor=west] (fdim)
{feature dimension depends on the\\ configuration $d\in\{156,312,468\}$};

\draw[arrow] (Xts) -- (phi);

% Design matrix + labels combined
\node[mat, below=11mm of phi] (Xy)
{Feature-space dataset for binary classification\\[-1mm]
\scriptsize stack foreground over background; concatenate across detectors (when $C>1$):
\\
$\mathbf{X}^{(s)}_{i,C}\in\mathbb{R}^{(N_{\mathrm{fg}}+N_{\mathrm{bg}})\times d}$,\;
$y^{(s)}_{i,C}=
\begin{bmatrix}
\mathbf{1}_{i,N_{\mathrm{fg}\times1}}\\
\mathbf{0}_{i,N_{\mathrm{bg}\times1}}
\end{bmatrix}$\\[-1mm] };

\draw[arrow] (phi) -- (Xy);

% Final dataset
\node[mat, below=10mm of Xy] (Dfinal)
{Final dataset\\
$\mathcal{D}^{(s)}_{i,C}=\{(\mathbf{X}^{(s)}_{i,C},\,y^{(s)}_{i,C})\}$};

\draw[arrow] (Xy) -- (Dfinal);

% Optional examples note
\node[note, below=6mm of Dfinal] (ex)
{\scriptsize Notation: $C=1$ (CE-like detector), $C=2$ (ET2), $C=3$ (ET1).};

% Panel box
\node[draw, dashed, rounded corners, inner sep=8pt,
      fit=(H)(Xts)(chnote)(phi)(fdim)(Xy)(Dfinal)(ex)] {};

\end{tikzpicture}
\caption{Detector-agnostic schematic for mapping balanced foreground/background time series datasets to the feature space. Starting from $X^{(s)}_{i}$, which may contains $C\in\{1,2,3\}$ datasets depending on the detector configuration, each dataset is mapped via $\phi$ from OTWs to feature vectors. Foreground and background features are vertically stacked and (when $C>1$) concatenated across detectors to form the design matrix $\mathbf{X}^{(s)}_{i,C}$ in a detector-dependent $d$-dimensional feature space, together with the associated binary label vector $y^{(s)}_{i,C}$, yielding $\mathcal{D}^{(s)}_{i,C}$.}
\label{fig:generic_feature_construction}
\end{figure*}
In each case, the datasets are organized as feature matrices with corresponding labels, making them directly suitable for binary classification using standard machine-learning models, such as feed-forward ANNs.

\section{\label{sec:GWFDA_framework}\texttt{GW-FALCON} Deep-Learning Framework}

The transformation from one-dimensional time series data to a corresponding feature space motivates the use of a comparatively simple neural network model within a supervised learning framework, rather than the more complex DL architectures proposed in the literature to date \cite{cuoco2020enhancing,benedetto2023ai,zhao2023dawning,cuoco2025applications}. The goal is to perform a binary classification task that discriminates foreground from background feature data using the balanced datasets reported in Table \ref{tab:detector_datasets_definition}. Despite their architectural simplicity, feed-forward ANNs are well established as powerful learners that have significantly advanced the state of the art across a wide range of data-science applications, making them a natural choice for this investigation \cite{bishop2006pattern,goodfellow2016deep,prince2023understanding,lecun2015deep}. In this way, our approach primarily relies on the development of fully connected feed-forward ANN models for classification that are trained to learn discriminative representations of the input feature data. 

%\gv{On the contrary, classical ML methods, such as support-vector machines, random forests, or gradient-boosted trees, can in principle be applied to the present classification task. However, such methods often rely on the careful tuning of multiple hyperparameters, a process that can be nontrivial and sensitive to the specific dataset \cite{bishop2006pattern}. A common approach is to perform a grid search over the hyperparameter space, which rapidly becomes computationally expensive as the number of parameters and their ranges increase. On the other hand, feed-forward ANNs naturally capture nonlinear interactions among correlated features, which are essential for classification. Notably, ANNs scale efficiently with both dataset size and feature dimensionality, offering fixed memory and inference costs, whereas many classical approaches become memory- or computational time-limited in large, high-dimensional datasets \cite{prince2023understanding}.}

A natural question is whether more traditional machine-learning methods, such as support-vector machines, random forests, or gradient-boosted trees, could also be effective for the present classification task. While these approaches are in principle viable, they often require careful hyperparameter tuning, which can be nontrivial and sensitive to the dataset \cite{bishop2006pattern}. A common strategy is to perform a grid or random search over the hyperparameter space, which rapidly becomes computationally expensive as the number of parameters and the range of values increase. By contrast, feed-forward ANNs naturally capture nonlinear interactions among features, which are essential for the present classification problem. Moreover, ANNs tend to scale more favorably with both dataset size and feature dimensionality, maintaining relatively modest memory and inference costs, whereas many classical approaches can become memory- or time-limited when applied to large, high-dimensional datasets \cite{prince2023understanding}.

%for the supervised classification task. 

In general, the optimal functional form of the neural-network model cannot be specified a priori and must instead be inferred empirically from the data. Considering a dataset  $\mathcal{D} = \lbrace (\mathbf{X}_m, {y}_m)\rbrace^{n}_{m=1}$, consisting of $n$ samples, where \(\mathbf{X}_m \in \mathbb{R}^d\) denotes the input feature vector of dimension $d$ and \(y_m \in \{0,1\}\) is the corresponding binary class label. In this setting, our objective is to identify the most concise ANN model $\hat{\mathcal{Y}}(\theta) = \hat{\mathcal{F}}_{\theta}(\mathbf{X})$, with parameter vector $\theta$, that adequately discriminates the foreground from background feature data. For the binary classification task, the model output for sample $\mathbf{X}_m$ is interpreted as the predicted probability $p_\theta(\mathbf{X}_m) = \hat{\mathcal{F}}_{\theta}(\mathbf{X_m}) \in [0,1]$ that sample $m$ belongs to the foreground $({y}_m = 1)$ or background class $({y}_m = 0)$. The associated model's class assignment is obtained by applying a decision threshold of $0.5$ to this probability, namely
\begin{align}
 \hat{\mathcal{Y}}_m= 
\begin{cases}
1, & \text{if } p_\theta(\mathbf{X}_m) \ge 0.5,\\[4pt]
0, & \text{if } p_\theta(\mathbf{X}_m) < 0.5,
\end{cases}   
\end{align}
corresponding to the foreground and background classes, respectively. For the chosen model, the parameters $\theta$ are estimated by minimizing the Cross-Entropy\footnote{Further details regarding the Cross-Entropy loss function can be found at: \href{https://docs.pytorch.org/docs/stable/generated/torch.nn.CrossEntropyLoss.html}{https://docs.pytorch.org/docs/stable/generated/ torch.nn.CrossEntropyLoss.html}.} as a loss function\cite{bishop2006pattern,goodfellow2016deep}. 

To obtain the optimal parameter vector $\theta^\star$, we employ a gradient-based optimization procedure to minimize the chosen loss function. Minimizing this loss while ensuring that the model captures the relevant structure of the training data is essential for effective learning. The ultimate goal, however, is not merely to describe the training data, but to achieve good generalization performance \cite{bishop2006pattern,prince2023understanding}. A well-generalizing classifier provides accurate predictions for previously unseen samples drawn from the same underlying distribution, thereby reliably distinguishing foreground from background instances. This generalization capability distinguishes a robust model from an overfitted one that fails to deliver reliable predictions on new data. At this point,  it is important to note that generalization does not imply arbitrary prediction, but rather a reliable pattern recognition within the validated domain of the data on which the model was trained.

%For all the BNS and NSBH waveform injections considered in the discrimination between foreground signal features and background noise ones, we employ a common structure for the ANN architecture as a baseline. Regarding the detector configuration investigated and the associated dataset, summarized in a compact form in Table~\ref{tab:detector_datasets_definition}, the model's hyperparameters used for training are not necessarily the same. 

To implement the DL architecture that satisfies our requirements, we used the \href{https://pytorch.org/}{PyTorch} Library \cite{paszke2019pytorch}. For all BNS and NSBH waveform injections considered in the discrimination between foreground signal features and background noise ones, we employ a feed-forward neural network as our baseline classifier. 
Across the detector configurations considered, a separate ANN model is trained for each final dataset $\mathcal{D}^{(s)}_i$ presented in Table~\ref{tab:detector_datasets_definition}. The network architecture is identical in all cases, while the learned parameters and training hyperparameters (e.g., learning rate, batch size, number of epochs) are specific to each dataset.

Next, we describe the datasets employed for training and validation of the proposed classifiers. For each dataset $\mathcal{D}^{(s)}_i$, we first set aside a randomly selected subset of $1.6\times 10^{6}$  samples, comprising $8\times 10^{5}$ background and $8\times 10^{5}$ foreground examples. These subsets correspond to $80\%$ of the corresponding available data and are used for training. The remaining $2\times 10^{5}$ foreground and $2\times 10^{5}$ background examples ($20\%$ of the data) set aside for validation. A fixed random seed is employed in all cases to ensure that, across the different dataset cases $(s,i)$, feature representations derived from the same injected waveforms for the foreground data are assigned consistently to the corresponding data splits.

In our systematic investigation, we employed a feed-forward network for early warning alerts described by an input layer matching the dimensionality of the employed feature vector, followed by six hidden layers, denoted as $H_1,\ldots,H_6$, and a final output layer with two units. The number of neurons in each hidden layer is reported in Table~\ref{tab:network_hidden_layers_struct}. 
\begin{table}[!th]
  \caption{\label{tab:network_hidden_layers_struct}
  ANN hidden-layer structure. Each hidden neuron uses the non-linear $\mathrm{GeLU}$ activation function while dropout is applied after the first three hidden layers to mitigate overfitting.}
  \begin{ruledtabular}
      \begin{tabular}{cccc} 
        \textbf{Hidden} & \textbf{Number of} & \textbf{Activation} & \textbf{Dropout} \\
        \textbf{Layer} & \textbf{Neurons}& \textbf{Function} & \textbf{Rate} [$\%$] \\
        \hline
        $H_1$ & 100 &  $\mathrm{g(x)} =\mathrm{GeLU(x)}$ & $30$ \\
        $H_2$ & 75  & $ \mathrm{g(x)} =\mathrm{GeLU(x)}$ & $20$ \\
        $H_3$ & 50  &  $\mathrm{g(x)} =\mathrm{GeLU(x)}$ & $10$ \\
        $H_4$ & 25  & $ \mathrm{g(x)} =\mathrm{GeLU(x)}$ & $0$  \\
        $H_5$ & 15  & $ \mathrm{g(x)} =\mathrm{GeLU(x)}$ & $0$  \\
        $H_6$ & 10  & $ \mathrm{g(x)} =\mathrm{GeLU(x)}$ & $0$  \\
      \end{tabular}
  \end{ruledtabular}
\end{table}
Furthermore, we have used the GeLU activation function \cite{hendrycks2023gaussian},
\begin{align}
    \mathrm{g(x)} = \mathrm{x}\Phi(\mathrm{x}),
\end{align}
to introduce the non-linearity in each hidden layer. At this point,  $\Phi(\mathrm{x})$ corresponds to the cumulative distribution function for the Gaussian distribution. In addition, dropout regularization \cite{srivastava2014dropout} is applied between the first three hidden layers, with rates $30\%$, $20\%$, and $10\%$, respectively, to prevent co-adaptation of the employed neurons, and therefore mitigate overfitting. Finally, at the output layer, we use the softmax activation function \cite{goodfellow2016deep} to produce class probability estimates. Overall, the employed ANN architecture provides an efficient framework tailored to our classification task. Based on extensive experimentation, this design yielded more stable and consistent training behavior, well aligned with the statistical properties of the datasets considered.

%Here, 

Before training, we apply a standardization step to the feature data. Let $\mu_j$ and $\sigma_j$ denote the mean and standard deviation of the $j$-th feature, computed over the dataset of interest. The standardized feature vector $\tilde{\mathbf{X}}_m$ is then defined element-wise as,
\begin{align}
    \tilde{X}_{mj} = \frac{X_{mj} - \mu_j}{\sigma_j} \, ,
\end{align}
where $j = 1, \dots, d$ indexes the feature components and $d$ denotes the dimensionality of the feature vector. This transformation produces features with approximately zero mean and unit variance along each dimension, so that most components of the standardized feature vectors lie within a few standard deviations of zero (typically within a range of about $\pm 5\,\sigma_j$ for well-behaved features). Above all, feature standardization is essential to ensure that all features contribute on a comparable scale, preventing those with larger numerical ranges from dominating the learning dynamics. This, in turn, improves the conditioning of the optimization problem, typically leading to faster convergence and more stable overall performance of the DL classifier. %Nevertheless, the training time is linearly proportional to the number of epochs.
Furthermore, due to the non-linear nature of the optimization process, we adopted the Adamax optimizer \cite{kingma2015adam} during backpropagation to obtain the optimal set of $\theta^{\star}$ parameters. In each case, the model parameters $\theta$ were initialized using the Xavier uniform initialization algorithm to ensure stable gradient flow during early training phases \cite{pmlr-v9-glorot10a}. In total, we trained 56 models, covering all BNS and NSBH early-warning triggering cases considered in this work. Appendix \ref{app:Ann_training_prop} provides further details on the training hyperparameters, including the batch size, learning rate, and number of epochs for each ANN model.

In Fig. \ref{fig:tsfel_pipeline}, we illustrate the overall \texttt{GW-FALCON} framework underlying the employed ML methodology. The pipeline starts from time series OTWs of the detector strain data \(d(t)\); these windows are processed to extract feature vectors using the TSFEL library \cite{barandas2020tsfel}, which are then standardized and passed to an ANN classifier that produces a binary decision between GW signal-portion + noise and pure noise.
\begin{figure}[!t]
\centering
\begin{tikzpicture}[
    node distance=1cm,
    >=Latex,
    data/.style={
        draw,
        rounded corners,
        minimum width=5.0cm,
        minimum height=1.2cm,
        align=center,
        fill=blue!8
    },
    block/.style={
        draw,
        rounded corners,
        minimum width=5.0cm,
        minimum height=1.2cm,
        align=center,
        fill=green!8
    },
    clf/.style={
        draw,
        rounded corners,
        minimum width=5.0cm,
        minimum height=1.2cm,
        align=center,
        fill=orange!12
    },
    arrow/.style={
        ->,
        line width=0.8pt
    }
]

% Nodes (top to bottom)
\node[data] (ts) {Time Series OTW\\(e.g., detector's filtered strain data \\ $d(t)$ of specific duration)};
\node[block, below=of ts] (tsfel) {Feature Extraction:\\TSFEL API};
\node[block, below=of tsfel] (std) {Feature Standardization:\\Preprocessing Step};
\node[clf,   below=of std] (clf) {Deep Learning Model:\\ANN Classifier};
\node[data,  below=of clf] (out) {Model's Binary Output\\ GW Signal + Noise / Noise};

% Arrows (top to bottom)
\draw[arrow] (ts) -- (tsfel);
\draw[arrow] (tsfel) -- (std);
\draw[arrow] (std) -- (clf);
\draw[arrow] (clf) -- (out);

\end{tikzpicture}
\caption{Schematic flowchart representation of the \texttt{GW-FALCON} deep learning framework that is suggested in this work.}
\label{fig:tsfel_pipeline}
\end{figure}

Lastly, across all ANN models considered for signal detection, we quantified their performance and robustness on the associated training and test datasets using standard classification evaluation measures, which are widely used in the field of machine learning. More specifically, for the training and test subsets, we report standard classification metrics, including accuracy, precision, recall (detection efficiency), specificity, F1-score, Matthews correlation coefficient (MCC), false-alarm probability (FAP), and false-positive rate (FPR) \cite{bishop2006pattern, goodfellow2016deep, lecun2015deep, prince2023understanding}. In all cases, these quantities are estimated from the elements of the corresponding confusion matrix, that is, the numbers of true positives (TP), true negatives (TN), false positives (FP), and false negatives (FN). Together, they characterize the models’ ability to discriminate between signal+noise and noise-only examples. The employed quantities, along with their mathematical definitions, are presented in Appendix~\ref{app:classification_eval_measures}. In general, for a well-trained classifier, a slight degradation in performance on the test set relative to the training set is expected. For each ANN model considered in the remainder of this work, we therefore present the corresponding evaluation measures on both the training and test datasets to enable a direct comparison. In addition, as a future extension of this work, it would be natural to complement these global performance evaluation measures with local, feature-level interpretability of the models’ predictions, to quantify how individual features contribute to the ANN decisions, in analogy with the analysis presented in Ref. \cite{papigkiotis2025assessing}.

\begin{comment}
   %BCE
\[
\mathcal{L}_{\mathrm{BCE}}(\theta)
= -\frac{1}{n} \sum_{i=1}^{n} 
\Big[
y_i \log p_\theta(\mathbf{x}_i)
+ (1 - y_i)\, \log \big(1 - p_\theta(\mathbf{x}_i)\big)
\Big],
\]
where \(p_\theta(\mathbf{x}_i)\) denotes the model-predicted probability that sample \(i\) belongs to the foreground class.

For the chosen model, the parameters \(\theta\) are estimated by minimizing the cross-entropy loss for a two-class classification problem

, defined as
\[
\mathcal{L}_{\mathrm{CE}}(\theta)
= -\frac{1}{n} \sum_{i=1}^{n} 
\log p_{i,y_i},
\]
where \(y_i \in \{0,1\}\) denotes the true class label of sample \(i\), and \(p_{i,k}\) is the model-predicted probability that sample \(i\) belongs to class \(k\), with
\[
p_{i,k} = \frac{\exp\big(f_{\theta,k}(\mathbf{x}_i)\big)}{\sum_{j=0}^{1} \exp\big(f_{\theta,j}(\mathbf{x}_i)\big)}, 
\quad k \in \{0,1\},
\]
and \(f_{\theta,k}(\mathbf{x}_i)\) denoting the logit (pre-softmax output) associated with class \(k\).
 
\end{comment}

%which penalizes discrepancies between the true labels \(y_i\) and the predicted probabilities \(p_\theta(\mathbf{x}_i)\).

%%%%%%%%%%%%%%%%%%%%%%%%%%%%%%%%%%%%%%%%%%%%%%%%%%%%%%%%%%%%%%%%%%%%%%%%%%%%%%%%%%%%%%%%%%%%%%%%%%%%%%%%%%%%%%

\section{\label{sec:results_discussion}Results and Discussion}

In this section, we evaluate the performance of the proposed \texttt{GW-FALCON}-based early-warning pipeline. In particular, we assess the ANN models developed for early-warning alerts (EW-ANN models) for BNS and NSBH systems. Having constructed the $(i,s)$ datasets summarized in Table~\ref{tab:detector_datasets_definition}, we trained a separate ANN model for each dataset of interest. For the pattern-recognition task, we employed the feed-forward network architecture presented in Table \ref{tab:network_hidden_layers_struct}. As described in the previous section, for each dataset $\mathcal{D}^{(s)}_i$, we set aside $2\times10^{5}$ foreground and $2\times10^{5}$ background examples as a test set. The remaining $8\times10^{5}$ foreground and $8\times10^{5}$ background examples, corresponding to $80\%$ of the total dataset, were used for training. In each case, the training and test subsets are drawn from the same underlying data distribution, ensuring a consistent assessment of model performance and generalization.

We first discuss the detection characteristics of the resulting ANN configurations used for BNS and NSBH early-warning alerts on the datasets introduced in the previous sections. 
%We then turn to more realistic scenarios by applying the \texttt{GW-FALCON} approach, together with these trained classifiers, to independent time series data streams that contain signals similar to GW170817, GW190425, and GW230529. This second step serves as an illustrative validation step, allowing us to assess how the models perform in practice and to quantify the associated computational cost, in particular, the inference time relevant for low-latency alerts. {\bf see this again} \gp{TODO}
Then, we proceed to more realistic scenarios by applying the suggested approach to independent strain data containing simulated BNS and NSBH mergers. This second part serves as a practical validation step, allowing us to assess how the models perform in realistic conditions and to determine when the classifiers would issue triggers suitable for low-latency alerts.

\subsection{Performance of EW-ANNs for BNS mergers}

%Having introduced the datasets and the GWFDA framework, we now examine the performance of the ANN classifiers developed for third-generation GW observatories in the context of BNS early-warning alerts. Table \ref{tab:detector_dataset_metrics} summarizes the corresponding model performance on the training and test subsets for each dataset and each detector considered. More specifically, for each detector–dataset pair, a separate model is trained, and the table reports its behavior on the training and test sets in terms of accuracy, precision, recall, F1-score, specificity, Matthews correlation coefficient (MCC), false-alarm probability (FAP), and false-alarm rate (FAR).

\subsubsection{Global classification performance}

Having introduced the datasets and the \texttt{GW-FALCON} framework, we now examine the performance of the ANN classifiers developed for third-generation GW observatories in the context of BNS early-warning alerts. Table \ref{tab:detector_dataset_metrics} summarizes the corresponding model performance on the training and test subsets for each dataset and each detector considered. More specifically, for each detector-dataset pair, a separate model is trained, and the table reports its behavior in terms of accuracy, precision, recall, F1-score, specificity, Matthews correlation coefficient (MCC), false-alarm probability (FAP), and false-positive rate (FPR), evaluated on both the training and test sets.

\begin{table*}[!t]
\caption{\label{tab:detector_dataset_metrics}
Overall classification performance of the early-warning (EW) ANN models for each detector and dataset used in the BNS early-alerts investigation. For every detector–dataset pair, a separate EW-ANN is trained, and its performance on the training and test sets is reported for all evaluation measures defined in Appendix \ref{app:classification_eval_measures}.}
%Classification performance of the early-warning (EW) ANN models for each detector–dataset pair used in the BNS early-alerts investigation. For each model, validation (Val) and test (Test) metrics are reported for all evaluation measures defined in Appendix~\ref{app:classification_eval_measures}.

  \centering
  \resizebox{\textwidth}{!}{%
  %\begin{tabular}{lll*{16}{c}}
  \begin{tabular}{c|c|c|cc|cc|cc|cc|cc|cc|cc|cc}
    \toprule
    \multirow{2}{*}{Detector}& \multirow{2}{*}{Dataset Train and}& \multirow{2}{*}{EW-ANN}&
    \multicolumn{2}{c}{Accuracy [\%]} &
    \multicolumn{2}{c}{Precision [\%]} &
    \multicolumn{2}{c}{Recall [\%]} &
    \multicolumn{2}{c}{F1-score [\%]} &
    \multicolumn{2}{c}{Specificity [\%]} &
    \multicolumn{2}{c}{MCC [\%]} &
    \multicolumn{2}{c}{FAP [\%]} &
    \multicolumn{2}{c}{FPR [\%]} \\
    \cmidrule(lr){4-5}
    \cmidrule(lr){6-7}
    \cmidrule(lr){8-9}
    \cmidrule(lr){10-11}
    \cmidrule(lr){12-13}
    \cmidrule(lr){14-15}
    \cmidrule(lr){16-17}
    \cmidrule(lr){18-19}
    &Test Subsets&  
    & Train & Test    % Accuracy
    & Train & Test    % Precision
    & Train & Test    % Recall
    & Train & Test    % F1
    & Train & Test    % Specificity
    & Train & Test    % MCC
    & Train & Test    % FAP
    & Train & Test    % FAR
    \\
    \midrule
    \multirow{8}{*}{ET1}
      & $\mathcal{D}_{1, \mathrm{ET1}}$ & model 1 & 89.03 & 88.96 & 93.52 & 93.35 & 83.88 & 83.90 & 88.44 & 88.37 & 94.19 & 94.02 & 78.48 & 78.32 & 5.81 & 5.98 & 6.48 & 6.65 \\
      & $\mathcal{D}_{2, \mathrm{ET1}}$ & model 2 & 89.56 & 89.45 & 93.83 & 93.70 & 84.70 & 84.59 & 89.03 & 88.91 & 94.43 & 94.31 & 79.51 & 79.28 & 5.57 & 5.69 & 6.17 & 6.30 \\
      & $\mathcal{D}_{3, \mathrm{ET1}}$ & model 3 & 89.99 & 89.81 & 94.25 & 94.03 & 85.17 & 85.03 & 89.48 & 89.30 & 94.81 & 94.60 & 80.35 & 79.99 & 5.19 & 5.40 & 5.75 & 5.97 \\
      & $\mathcal{D}_{4, \mathrm{ET1}}$ & model 4 & 90.33 & 90.20 & 94.37 & 94.26 & 85.77 & 85.62 & 89.86 & 89.73 & 94.88 & 94.79 & 80.99 & 80.74 & 5.12 & 5.21 & 5.63 & 5.74 \\
      & $\mathcal{D}_{5, \mathrm{ET1}}$ & model 5 & 89.79 & 89.67 & 93.84 & 93.61 & 85.17 & 85.16 & 89.30 & 89.18 & 94.41 & 94.18 & 79.93 & 79.67 & 5.59 & 5.82 & 6.16 & 6.39 \\
      & $\mathcal{D}_{6, \mathrm{ET1}}$ & model 6 & 91.64 & 91.55 & 95.42 & 95.34 & 87.47 & 87.38 & 91.27 & 91.19 & 95.80 & 95.73 & 83.57 & 83.40 & 4.20 & 4.27 & 4.58 & 4.66 \\
      & $\mathcal{D}_{7, \mathrm{ET1}}$ & model 7 & 91.90 & 91.81 & 95.56 & 95.41 & 87.88 & 87.84 & 91.56 & 91.47 & 95.92 & 95.77 & 84.07 & 83.88 & 4.08 & 4.23 & 4.44 & 4.59 \\
      & $\mathcal{D}_{8, \mathrm{ET1}}$ & model 8 & 91.73 & 91.58 & 95.40 & 95.22 & 87.69 & 87.56 & 91.38 & 91.23 & 95.78 & 95.61 & 83.74 & 83.43 & 4.22 & 4.39 & 4.60 & 4.78 \\
    \midrule
    \multirow{8}{*}{ET2}
      & $\mathcal{D}_{1,\mathrm{ET2}}$ & model 1 & 89.54 & 89.47 & 94.03 & 93.31 & 84.45 & 84.41 & 88.98 & 88.90 & 94.63 & 94.52 & 79.50 & 79.34 & 5.37 & 5.48 & 5.97 & 6.09 \\
      & $\mathcal{D}_{2,\mathrm{ET2}}$ & model 2 & 90.18 & 90.10 & 94.68 & 94.55 & 85.14 & 85.11 & 89.66 & 89.58 & 95.22 & 95.10 & 80.77 & 80.61 & 4.78 & 4.90 & 5.32 & 5.45 \\
      & $\mathcal{D}_{3,\mathrm{ET2}}$ & model 3 & 90.46 & 90.33 & 94.75 & 94.55 & 85.67 & 85.60 & 89.98 & 89.85 & 95.25 & 95.12 & 81.30 & 81.03 & 4.75 & 4.93 & 5.25 & 5.45 \\
      & $\mathcal{D}_{4,\mathrm{ET2}}$ & model 4 & 91.06 & 90.98 & 95.13 & 95.10 & 86.55 & 86.41 & 90.64 & 90.54 & 95.57 & 95.54 & 82.45 & 82.29 & 4.43 & 4.46 & 4.87 & 4.90 \\
      & $\mathcal{D}_{5,\mathrm{ET2}}$ & model 5 & 90.27 & 90.14 & 94.79 & 94.71 & 85.22 & 85.04 & 89.75 & 89.61 & 95.32 & 95.21 & 80.95 & 80.71 & 4.68 & 4.75 & 5.21 & 5.29 \\
      & $\mathcal{D}_{6,\mathrm{ET2}}$ & model 6 & 92.13 & 92.07 & 95.86 & 95.73 & 88.08 & 88.05 &  91.79 &  91.74 & 96.20 & 96.07 & 84.53 & 84.42 & 3.80 & 3.93 & 4.14 & 4.27 \\
      & $\mathcal{D}_{7,\mathrm{ET2}}$ & model 7 & 92.47 & 92.41 & 96.23 & 96.19 & 88.41 & 88.31 & 92.15 & 92.08 & 96.53 & 96.50 & 85.22 & 85.10 & 3.47 & 3.50 & 3.77 & 3.81 \\
      & $\mathcal{D}_{8,\mathrm{ET2}}$ & model 8 & 92.30 & 92.26 & 96.03 & 96.00 & 88.25 & 88.19 & 91.97 & 91.93 & 96.35 & 96.33 & 84.88 & 84.80 & 3.65 & 3.67 & 3.97 & 4.00 \\
    \midrule
    \multirow{8}{*}{CE1}
      & $\mathcal{D}_{1,\mathrm{CE1}}$ & model 1 & 97.35 & 97.29 & 99.30 & 99.31 & 95.38 & 95.25 & 97.30 & 97.23 & 99.33 & 99.33 & 94.78 & 94.66 & 0.67 & 0.67 & 0.70 & 0.69 \\
      & $\mathcal{D}_{2,\mathrm{CE1}}$ & model 2&  97.49 &  97.45 & 99.34 & 99.35 & 95.62 & 95.53 & 97.45 & 97.40 & 99.36 & 99.38 & 95.05 & 94.98 & 0.64 & 0.62 & 0.66 & 0.65 \\
      & $\mathcal{D}_{3,\mathrm{CE1}}$ & model 3 & 97.61 & 97.57 & 99.35 & 99.36 & 95.84 & 95.75 & 97.56 & 97.52 & 99.37 & 99.39 & 95.27 & 95.20 & 0.63 & 0.61 & 0.65 & 0.64 \\
      & $\mathcal{D}_{4,\mathrm{CE1}}$ & model 4 & 97.65 & 97.62 & 99.41 & 99.43 & 95.87 & 95.79 & 97.61 & 97.58 & 99.43 & 99.45 & 95.37 & 95.30 & 0.57 & 0.55 & 0.59 & 0.57 \\
      & $\mathcal{D}_{5,\mathrm{CE1}}$ & model 5 & 97.76 & 97.68 & 99.47 & 99.46 & 96.03 & 95.89 & 97.72 & 97.64 & 99.49 & 99.48 & 95.58 & 95.43 & 0.51 & 0.52 & 0.53 & 0.54 \\
      & $\mathcal{D}_{6,\mathrm{CE1}}$ & model 6 & 97.82 & 97.76 & 99.65 & 99.62 & 95.98 & 95.89 & 97.78 & 97.72 & 99.66 & 99.64 & 95.71 & 95.60 & 0.34 & 0.36 & 0.35 & 0.38 \\
      & $\mathcal{D}_{7,\mathrm{CE1}}$ & model 7 & 97.82 & 97.79 & 99.57 & 99.56 & 96.07 & 95.99 & 97.79 & 97.75 & 99.58 & 99.58 & 95.71 & 95.63 & 0.42 & 0.42 & 0.43 & 0.44 \\
      & $\mathcal{D}_{8,\mathrm{CE1}}$ & model 8 & 97.83 & 97.77 & 99.61 & 99.60 & 96.04 & 95.92 & 97.80 & 97.72 & 99.63 & 99.61 & 95.73 & 95.60 & 0.37 & 0.39 & 0.39 & 0.40 \\
    \midrule
    \multirow{8}{*}{CE2}
      & $\mathcal{D}_{1,\mathrm{CE2}}$ & model 1 & 97.35 & 97.31 & 99.34 & 99.35 & 95.33 & 95.25 & 97.29 & 97.26 & 99.37 & 99.38 & 94.77 & 94.71 & 0.64 & 0.62 & 0.66 & 0.65 \\
      & $\mathcal{D}_{2,\mathrm{CE2}}$ & model 2 & 97.51 & 97.45 & 99.47 & 99.45 & 95.52 & 95.42 & 97.46 & 97.39 & 99.49 & 99.47 & 95.09 & 94.97 & 0.51 & 0.53 & 0.53 & 0.55 \\
      & $\mathcal{D}_{3,\mathrm{CE2}}$ & model 3 & 97.56 & 97.53 & 99.48 & 99.47 & 95.62 & 95.56 & 97.51 & 97.48 & 99.50 & 99.50 & 95.19 & 95.13 & 0.50 & 0.51 & 0.52 & 0.53 \\
      & $\mathcal{D}_{4,\mathrm{CE2}}$ & model 4 & 97.67 & 97.63 & 99.48 & 99.47 & 95.84 & 95.78 & 97.63 & 97.59 & 99.50 & 99.49 & 95.41 & 95.33 & 0.50 & 0.51 & 0.52 & 0.53 \\
      & $\mathcal{D}_{5,\mathrm{CE2}}$ & model 5 & 97.75 & 97.73 & 99.47 & 99.47 & 96.02 & 95.96 & 97.71 & 97.69 & 99.49 & 99.49 & 95.56 & 95.51 & 0.51 & 0.51 & 0.53 & 0.53 \\
      & $\mathcal{D}_{6,\mathrm{CE2}}$ & model 6 & 97.83 & 97.77 & 99.48 & 99.45 & 96.17 & 96.08 & 97.80 & 97.74 & 99.50 & 99.47 & 95.72 & 95.60 & 0.50 & 0.53 & 0.52 & 0.55 \\
      & $\mathcal{D}_{7,\mathrm{CE2}}$ & model 7 & 97.83 & 97.76 & 99.61 & 99.58 & 96.04 & 95.93 & 97.79 & 97.72 & 99.62 & 99.59 & 95.73 & 95.59 & 0.38 & 0.41 & 0.39 & 0.42 \\
      & $\mathcal{D}_{8,\mathrm{CE2}}$ & model 8 & 97.86 & 97.80 & 99.48 & 99.48 & 96.22 & 96.10 & 97.82 & 97.76 & 99.49 & 99.50 & 95.76 & 95.66 & 0.51 & 0.50 & 0.52 & 0.52 \\
    \bottomrule
  \end{tabular}
  }% end resizebox
\end{table*}

Across all ET- and CE-based configurations, the ANN models achieve high classification performance. For the ET1 and ET2 detectors, train and test-set accuracies typically lie between $89\%$ and $92\%$. The corresponding precision values are in the range $93\%$-$96\%$, recall values lie between about $84\%$ and $88\%$, and specificities are in the range $94\%$-$96.5\%$, with MCC values around $78\%$-$85\%$. Together, these evaluation measures indicate strong overall discrimination between GW signal+noise and pure noise examples. For the CE1 and CE2 detectors, the models' performance further improves, with train and test-set accuracies above $97\%$, precision and specificity values close to $99.5\%$, recall values around $95\%$-$96\%$, and MCC values near $95\%$. This behavior is consistent with the higher sensitivity of the CE1 and CE2 designs: their lower noise PSDs yield larger PI SNRs for the same injected signals, which enhances the separability of signal-related features from the background. As a result, the feature representations used as input to the EW-ANNs for the CE detectors are of higher quality, enabling the networks to learn more distinct patterns in the data and to discriminate more reliably between foreground and background examples than in the corresponding ET-based configurations.

Within each detector configuration, there is also a clear pattern across the datasets indexed by $i=1,\dots,8$. As we move from $\mathcal{D}_1$ to $\mathcal{D}_8$, the train and test-set accuracies and the other evaluation measures systematically improve. This behavior is consistent with the increase in PI SNR with $f_{\max}$ for the corresponding injections, since a larger fraction of the inspiral signal is accumulated within the analyzed window (see, e.g., Figs. \ref{fig:CEH_PI_SNR_distributions} and \ref{fig:other_PI_SNR_distributions} for a detailed review). The resulting louder signals lead to more informative feature representations for the EW-ANNs and, thus, facilitate better discrimination between background and foreground cases. 

In the ET1 and ET2 configurations, a slight degradation in the performance of EW-ANN model 5 is observed for dataset $\mathcal{D}_5$. This dataset corresponds to features extracted from OTWs with instantaneous maximum frequency $f_{\max} = 22\,\mathrm{Hz}$. The slightly reduced ANN performance on this dataset can be explained by the steep rise of the ET noise PSD within the OTW corresponding to this value of $f_{\max}$, which increases the noise contribution in the filtered time segment and results in quite less informative features for training (see, e.g., Fig. \ref{fig:design_psds} for a review). As a result, the improvement that would be expected from extending the analysis to higher $f_{\max}$ as the system approaches merger is compensated by the additional noise in this band.

Above all, the recall values reported in Table \ref{tab:detector_dataset_metrics} quantify the detection efficiency for the GW signal+noise class. The consistently high recall, particularly for the CE-based configurations, shows that the EW-ANN models successfully identify the vast majority of injected signals, even at early in-band stages. The precision values, which measure the fraction of predicted foreground cases that truly contain an injection, are also high for all detectors considered. This indicates that, when the corresponding ANN classifier issues an early-warning trigger, it is very likely to correspond to a genuine signal of astrophysical relevance rather than to a spurious fluctuation. Furthermore, the F1-score, which balances precision and recall, remains close to the overall accuracy for all configurations, demonstrating that the models do not achieve high performance merely by favoring one class but instead maintain a well-balanced trade-off between correctly detecting signals and avoiding false positives.

Complementarily, the specificity values measure the true-negative rate for the noise-only class. The high specificities obtained for all detectors indicate that noise-only segments are very rarely misclassified as containing a signal, which is reflected in the correspondingly low FAP and FPR values.  For the employed ET-based configurations, the reported FAP and FPR remain at most at the level of a few percent, while for the CE1 and CE2 configurations, they drop to well below the percent level. In practical terms, this means that spurious early-warning triggers caused by noise fluctuations are relatively infrequent for ET and even rarer for CE, where the lower noise floor makes it more difficult for noise to mimic the learned signal patterns. This behavior confirms that the EW-ANN models can operate with stringent false-alarm requirements without severely compromising their detection efficiency.

In all cases, the employed training and test evaluation measures are very close, with differences at the level of a few tenths of a percent for most quantities. This small train–test gap suggests that the EW-ANN models generalize well to unseen data and do not exhibit significant overfitting. Moreover, the reported FAP and FPR values remain at the sub-percent level for CE1 and CE2, and at only a few percent for ET1 and ET2, confirming that the proposed classifiers achieve low false-alarm probabilities and false-positive rates while maintaining high detection efficiency across the range of datasets and detectors considered.

\subsubsection{Confusion matrices and EW-ANN detections}

\begin{figure*}[!thb]
    \centering
    \subfloat[ET1 Detector: Confusion matrix for the EW-ANN model 1 predictions on the associated $\mathcal{D}_{1,\mathrm{ET1}}$ test subset.]{%    
    \includegraphics[width=0.51\textwidth]{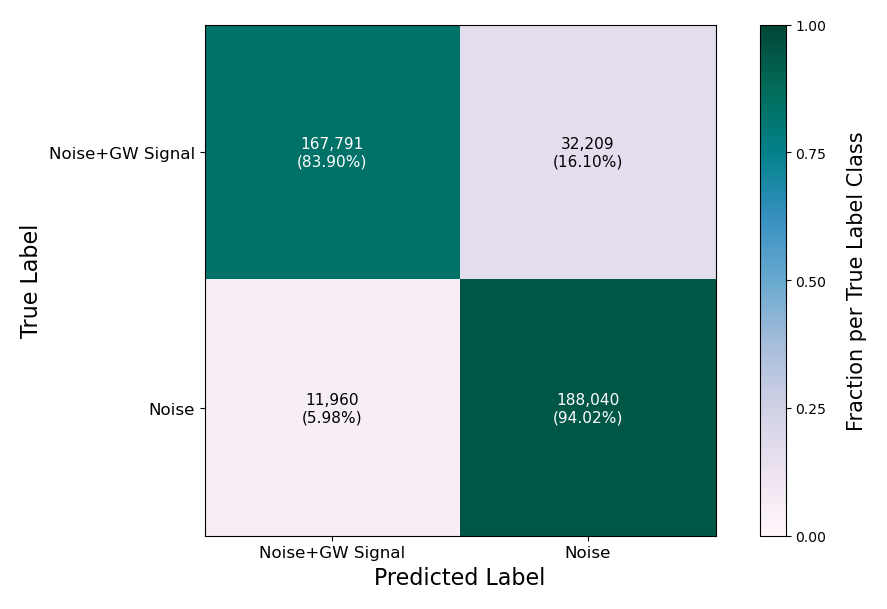}
    \label{fig:ETV_confmat_fmax17}}
    \subfloat[ET2 Detector: Confusion matrix for the EW-ANN model 1 predictions on the associated $\mathcal{D}_{1,\mathrm{ET2}}$ test subset.]{%
    \includegraphics[width=0.51\textwidth]{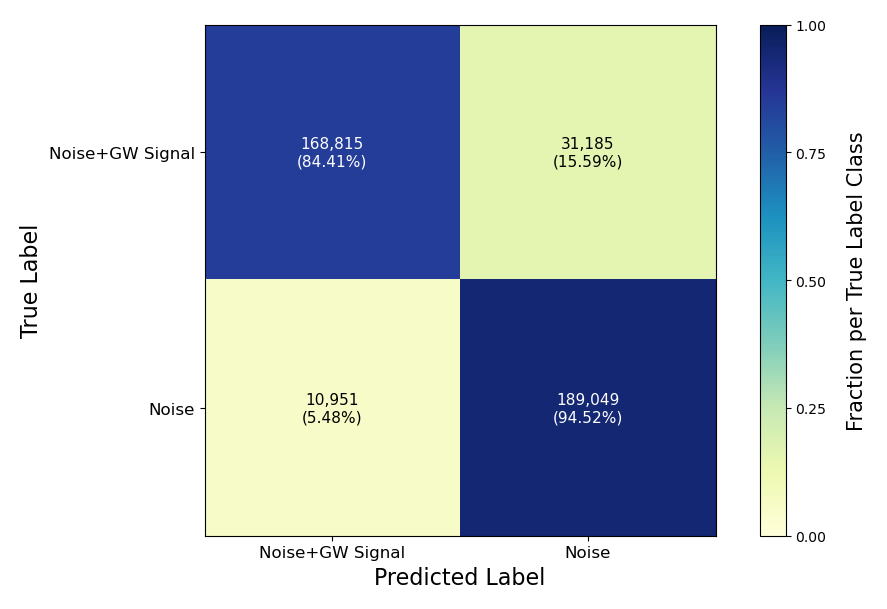}
    \label{fig:ETVG_confmat_fmax17}}\\
    \subfloat[CE1 Detector: Confusion matrix for the EW-ANN model 1 predictions on the associated $\mathcal{D}_{1,\mathrm{CE1}}$ test subset.]{%    
    \includegraphics[width=0.51\textwidth]{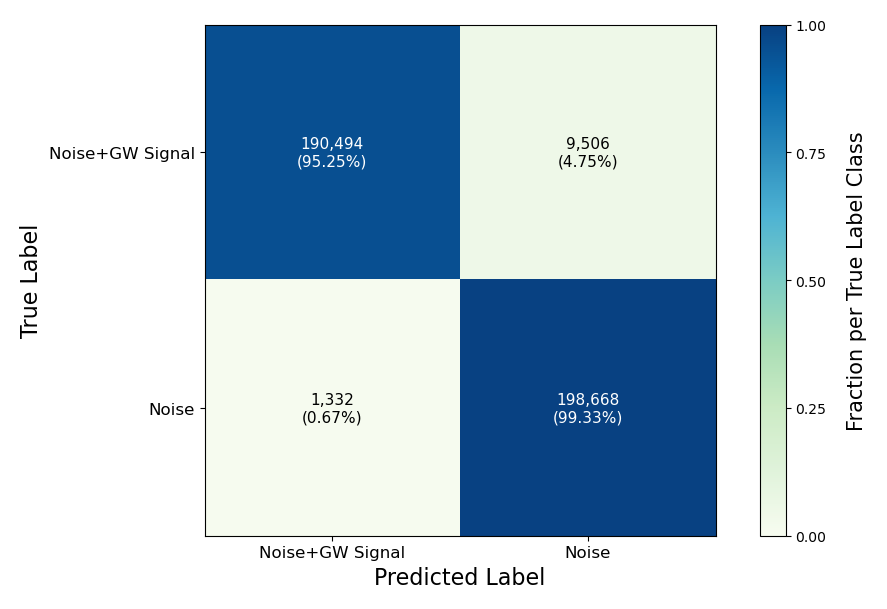}
    \label{fig:CEH_confmat_fmax17}}
    \subfloat[CE2 Detector: Confusion matrix for the EW-ANN model 1 predictions on the associated $\mathcal{D}_{1,\mathrm{CE2}}$ test subset.]{%
    \includegraphics[width=0.51\textwidth]{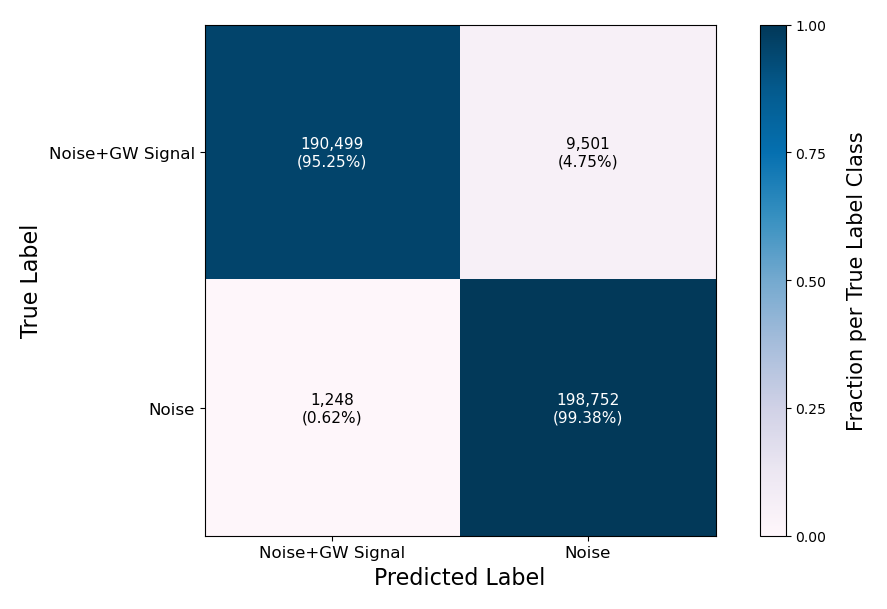}
    \label{fig:CEL_confmat_fmax17}}\\
    \caption{\label{fig:confusion_matrices_fmax17}
    Confusion matrices for the $\mathcal{D}_1$ test datasets, showing the predictions of the associated EW-ANN model 1 for each detector configuration. In each panel, the diagonal entries (true positives and true negatives) clearly dominate over the off-diagonal ones (false positives and false negatives), illustrating the strong separation achieved between foreground (GW signal+noise) and background (noise-only) examples. In each panel, the associated color bar shows the fraction of samples belonging to each true-label class. For the CE1 and CE2 configurations, the trained ANN models exhibit better confusion-matrix patterns than for ET1 and ET2, consistent with the higher-quality feature representations provided by the greater sensitivity and lower noise levels of the CE detectors.
    }
\end{figure*}

To complement these summary statistics and provide a more intuitive view of the classifiers' behavior, we also present their confusion matrices. As an illustrative example, Fig. \ref{fig:confusion_matrices_fmax17} shows the confusion matrices on the associated test datasets for all four detector configurations, for the EW-ANN models trained on the $\mathcal{D}_1$ datasets with maximum frequency $f_{\max} = 17\,\mathrm{Hz}$. 
In each case, the number of true positives and true negatives clearly dominates over the misclassified examples, while the ET-based configurations exhibit slightly higher counts of false positives and false negatives than the CE-based ones. This visual representation is consistent with the trends reported in Table \ref{tab:detector_dataset_metrics} for the reported models and illustrates how the higher sensitivity of the CE detectors translates into cleaner separation between foreground and background examples at the level of individual predictions. Analogous confusion matrices can be constructed for all remaining detector–dataset pairs, and they display the same qualitative behavior, with a strong dominance of the diagonal entries and systematically fewer misclassifications for the CE-based configurations than for the ET-based ones.

In addition, based on the confusion-matrix outcomes, Fig. \ref{fig:BNS_conf_matrix_elements_hists} in Appendix \ref{app:classification_eval_measures} shows the distributions of the EW-ANN output probability for samples in each category (TP, FP, TN, FN). More specifically, we report results for the corresponding test subsamples drawn from $\mathcal{D}_1$ and $\mathcal{D}_8$ datasets for the ET1, ET2, CE1, and CE2 detector configurations. The chosen ANN classifiers illustrate earlier and later early-warning alert regimes, respectively. The first and last models in the sequence correspond to progressively higher PI SNR in the foreground samples, leading to improved performance (a higher true-positive rate and fewer false alarms). The models also maintain good separation for background (noise-only) samples, which are predominantly assigned low probability values, while the number of false negatives remains small.

\begin{figure}[!thb]
    \includegraphics[width=0.47\textwidth]{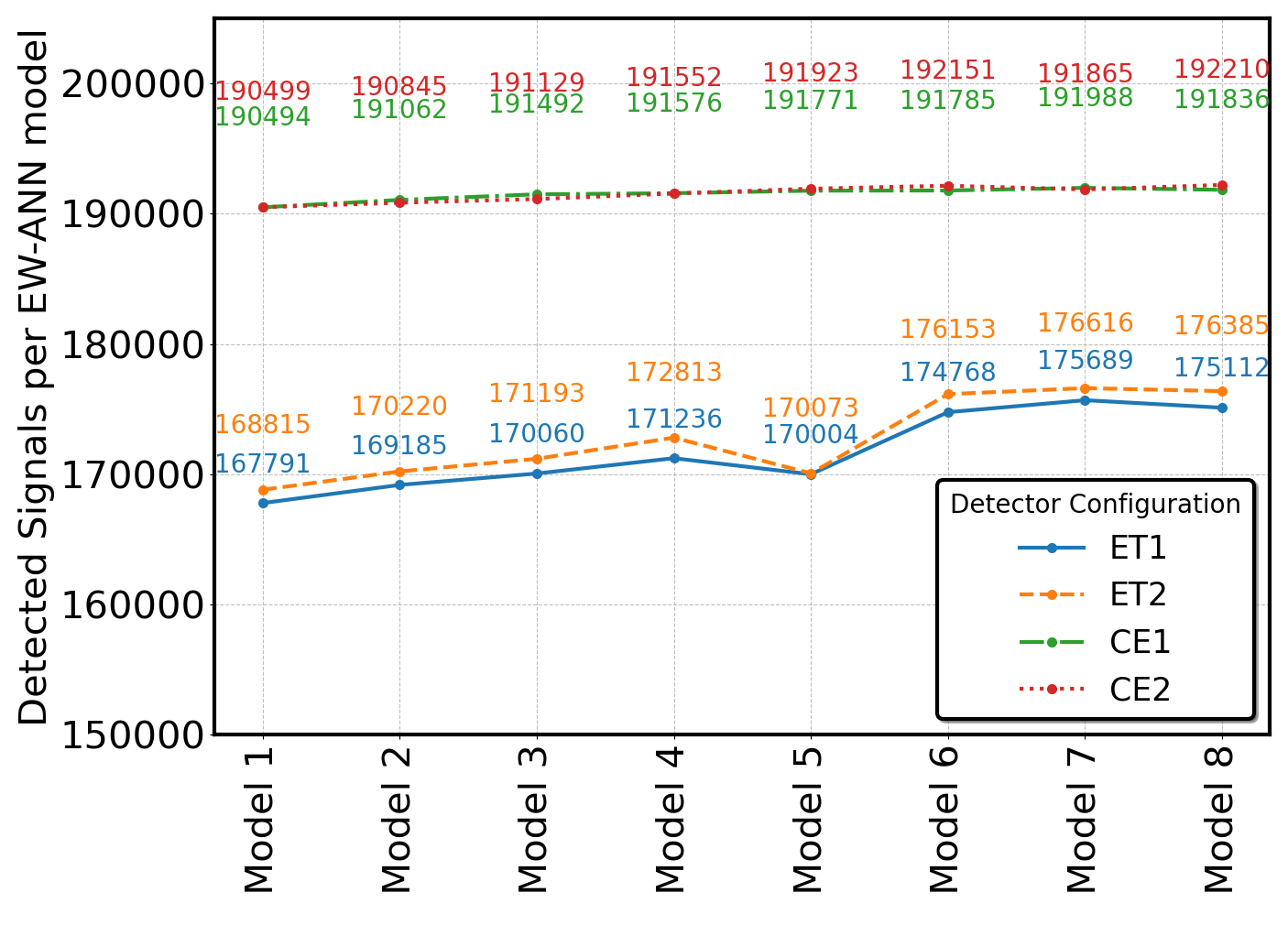}\hfill
    \caption{\label{fig:BNS_detections_per_model}
    Number of detected BNS injections in the test datasets for each model and detector configuration. The horizontal axis labels the EW-ANN models ($1–8$), while the vertical axis shows the number of signals correctly classified as detections (true positives) by each individual model. Colored curves correspond to different detectors, and the annotated values indicate the exact number of detected injections for each model-detector pair. For the ET1 and ET2 configurations, the detected-signal counts increase with model index, with a noticeable improvement around models 4-6 and only modest changes for models 7 and 8, indicating the onset of performance saturation. For the CE1 and CE2 configurations, the detected-signal counts are systematically higher and vary only weakly with model index, consistent with the fact that each CE-based model already recovers nearly all detectable injections in the test datasets.
    }
\end{figure}

To summarize the impact of these detection trends across the whole ensemble of models, Fig. \ref{fig:BNS_detections_per_model} shows, for each detector configuration, the number of BNS injections in the test datasets that each EW-ANN correctly identifies. For the ET1 and ET2 detectors, the detected-signal counts generally increase as the model index grows: they rise from models 1 to 4, show a slight dip at model 5, and then jump again for models 6 and 7, with only marginal variation for model 8. This behavior indicates that the ET-based classifiers continue to benefit from extending the OTWs to higher instantaneous maximum frequency, but approach a saturation regime at the highest model indices. More specifically, for models 7 and 8, the additional signal content gained by extending the OTWs to higher $f_{\max}$ lies in a frequency band with slightly increased ET noise levels. As a result, the features extracted from these windows are quite less informative for the ANNs, and the overall change in performance remains modest (see, e.g., Fig. \ref{fig:design_psds} for a review). In contrast, the CE1 and CE2 configurations yield substantially higher detection counts for all models and exhibit only a very mild dependence on model index: the curves are nearly flat, with small incremental gains from one model to another. This is consistent with the picture that each CE-based EW-ANN already recovers almost all detectable BNS signals in the test datasets.

\subsubsection{Detection efficiency as a function of BNS PI SNR}

\begin{figure*}[!thb]
    \centering
    \subfloat[ET1 detector]{%    
    \includegraphics[width=0.50\textwidth]{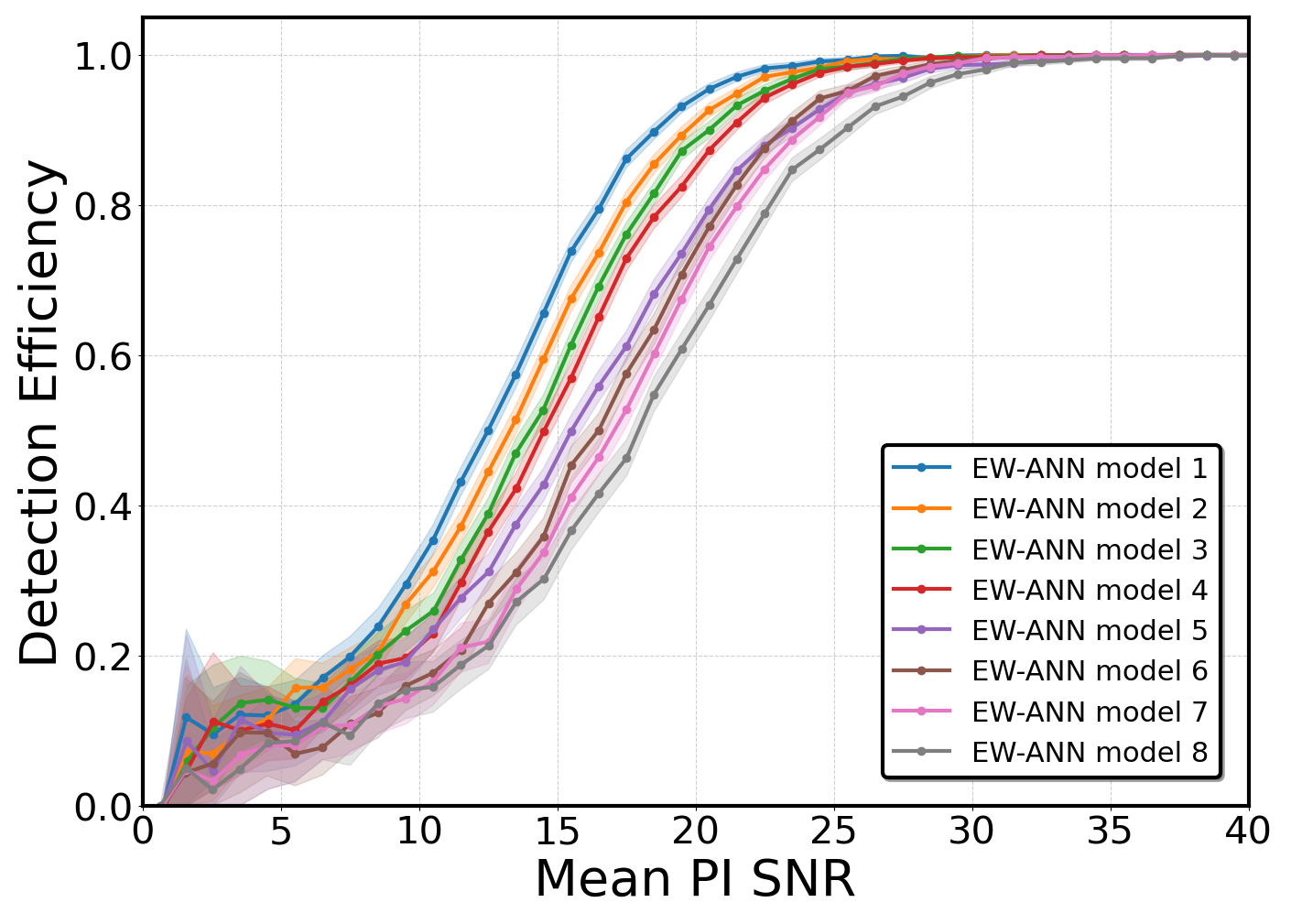}
    \label{fig:ETV_BNS_det_efficiency}}
    \subfloat[ET2 detector]{%
    \includegraphics[width=0.50\textwidth]{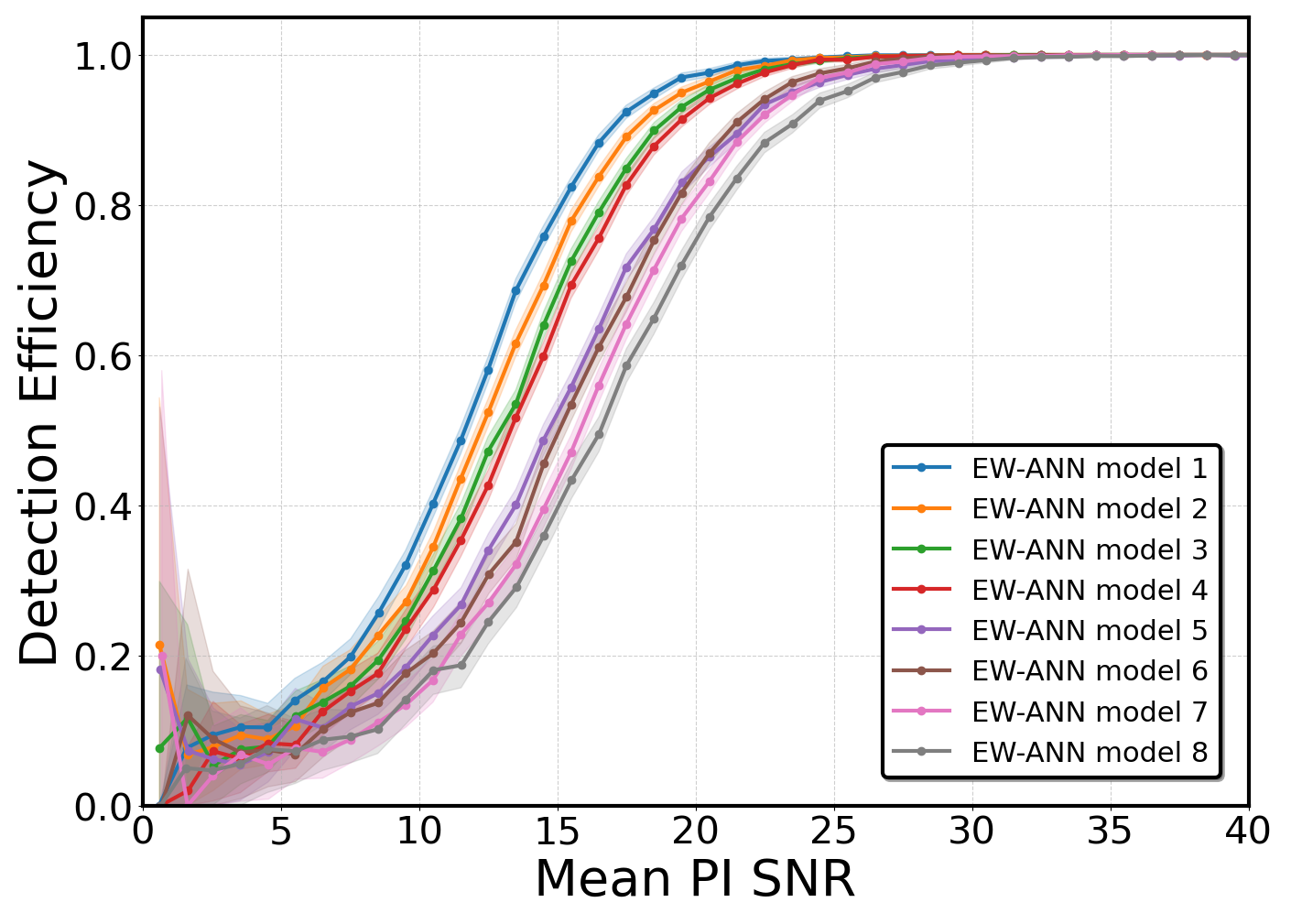}
    \label{fig:ETVG_BNS_det_efficiency}}\\
    \subfloat[CE1 detector]{%    
    \includegraphics[width=0.50\textwidth]{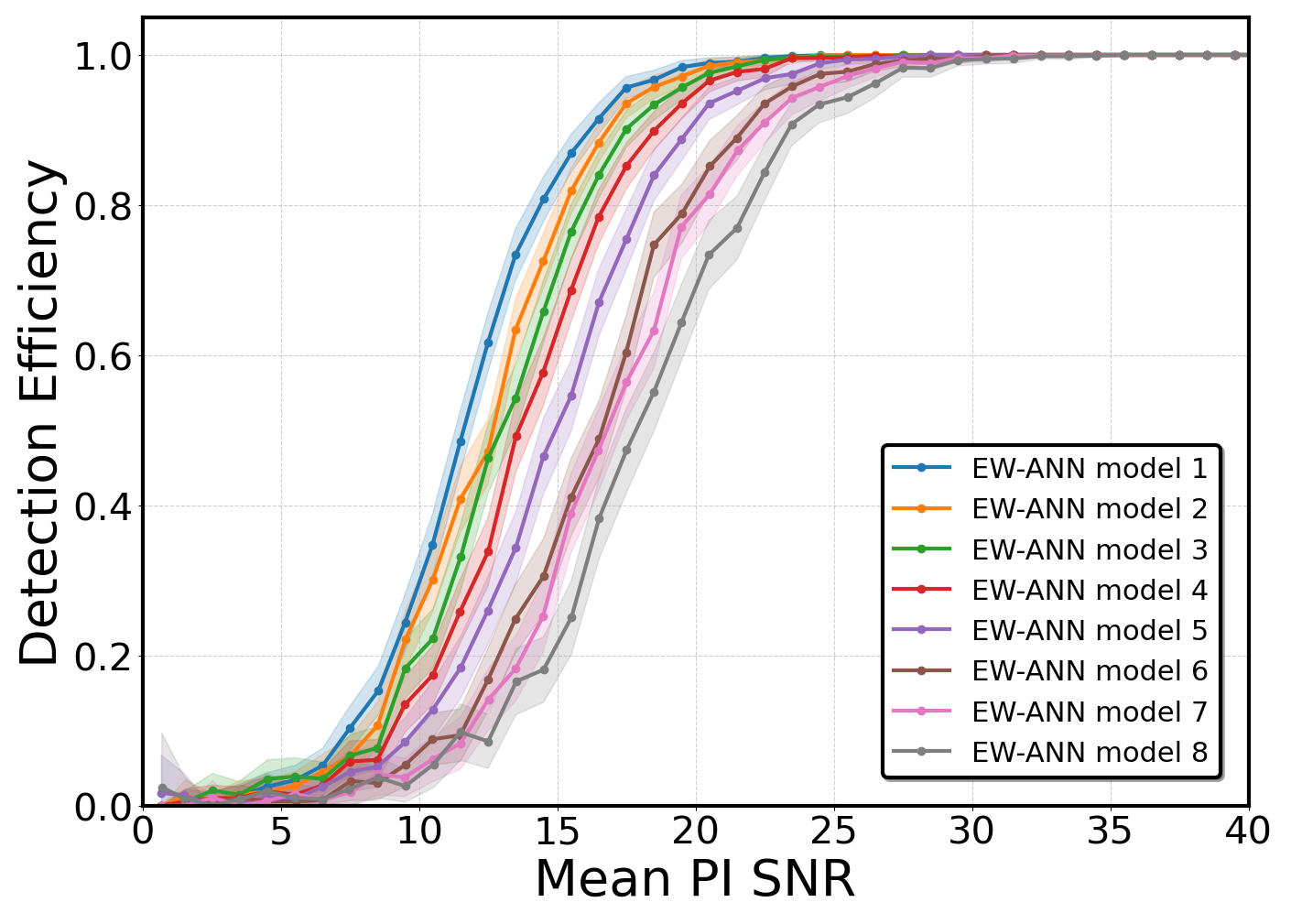}
    \label{fig:CEH_BNS_det_efficiency}}
    \subfloat[CE2 detector]{%
    \includegraphics[width=0.50\textwidth]{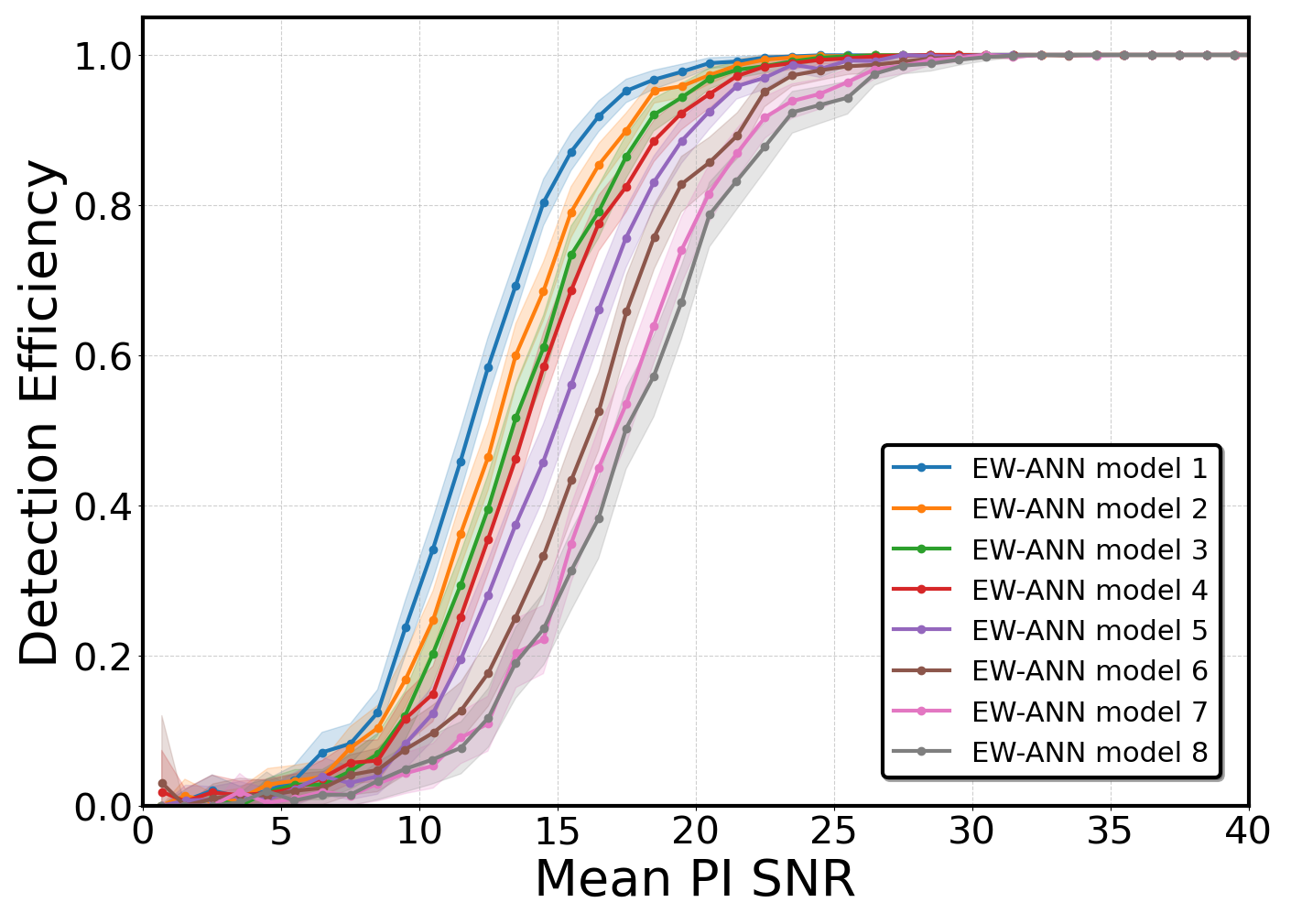}
    \label{fig:CEL_BNS_det_efficiency}}\\
    \caption{\label{fig:BNS_det_efficiency_vs_mean_PI_snr}
    Detection efficiency as a function of mean PI SNR for the BNS EW-ANN models. Each panel corresponds to one detector configuration. The colored curves show the detection efficiency on the test datasets for the different EW-ANN models (models 1–8), while the shaded bands denote the associated $3\sigma$ uncertainty regions. The curves illustrate the gradual transition from low efficiency at small mean PI SNR to near-unity efficiency at high mean PI SNR, with the CE configurations reaching high efficiencies at lower PI SNR values than the ET configurations.
    }
\end{figure*}

To further assess the behavior of the EW-ANNs, we also examine how the models’ detection efficiency varies with the loudness of the incoming BNS signals. For each detector configuration and test dataset considered, we estimate the detection efficiency (recall) as a function of the mean PI SNR. In practice, injection waveforms in the test set are grouped into narrow PI SNR bins (with a bin width of 1 in PI SNR); for each bin, we compute the mean PI SNR of all injections (true positives and false negatives) and the associated detection efficiency, defined as the fraction of those injections that are correctly identified as belonging to the signal+noise class. The resulting detection-efficiency curves are shown in Fig. \ref{fig:BNS_det_efficiency_vs_mean_PI_snr}. Each panel corresponds to one detector configuration and displays, for the eight EW-ANN models, the recall as a function of the mean PI SNR on the associated test dataset. The solid lines trace the mean efficiency in each PI SNR bin, while the shaded bands indicate the corresponding $\pm 3\sigma$ uncertainty. In each case, we should note that the horizontal axis is restricted to mean PI SNR values up to 40; for larger mean PI SNR, all ANN models for all detectors already operate at essentially perfect detection efficiency, so the corresponding curves coincide.

For all detectors, the curves exhibit the expected sigmoidal behavior: the efficiency is close to zero at very low mean PI SNR, rises steeply through an intermediate regime as the signal becomes more prominent in the data, and then saturates near unity once the mean PI SNR is sufficiently large. Within a given detector, networks trained on features from OTWs that extend closer to merger (higher $f_{\max}$) tend to ``turn on’’ at slightly higher PI SNR but still converge to high efficiencies at large PI SNR, reflecting the trade-off involved in using later-time windows that accumulate more signal power. Comparing detectors, the CE1 and CE2 panels show a noticeably earlier and steeper rise in efficiency than the ET1 and ET2 panels, together with narrower $\pm 3\sigma$ uncertainty bands at intermediate PI SNR. In particular, the CE configurations achieve essentially perfect detection already for mean PI SNRs of about $22$–$27$, whereas the ET configurations require somewhat larger values, typically mean PI SNRs of about $30$–$35$, for the curves to plateau at unity. This indicates that, for a fixed PI SNR threshold, the CE configurations provide more reliable and stable early-warning triggers for BNS signals, consistent with their improved sensitivity for the employed BNS waveforms and the more informative feature representations entering the EW-ANNs.

Above all, the behavior of these efficiency curves is consistent with the PI SNR distributions of the signal portions of injections used for each dataset. As illustrated for the CE1 case in the left panel of Fig.~\ref{fig:CEH_PI_SNR_distributions} (with analogous distributions for the other detectors shown in Appendix~\ref{app:PI_SNR_distributions}), most BNS signal portions have PI SNR values well above $20$, while only a small fraction occupy the low-PI SNR tail below $10$. Consequently, at a mean PI SNR within the range $5-10$, the ANNs operate in a regime where signal power is comparable to the noise level and test examples are relatively scarce, so the detection efficiency remains low and the uncertainty bands are wide. As the mean PI SNR increases into the $10-20$ range and beyond, an increasingly large fraction of signals contributes, the extracted features become more distinctive, and the classifiers transition to the high-efficiency plateau seen in Fig. \ref{fig:BNS_det_efficiency_vs_mean_PI_snr}. Because the CE1 and CE2 detectors yield PI SNR distributions that are systematically shifted to higher values than those of ET1 and ET2, the corresponding ANNs are trained on a larger proportion of loud, well-resolved signals; this naturally explains the earlier and steeper rise of the CE efficiency test datasets' curves and their approach to unity at lower mean PI SNR.

Finally, within each detector configuration, the relative performance of the individual EW-ANN models can also be interpreted in terms of the PI SNR content of their datasets. As $f_{\max}$ increases from dataset~1 to dataset~8, the OTWs from which features are extracted include a progressively larger fraction of the inspiral, and the associated PI SNR distributions shift toward higher values, so that a larger proportion of BNS injections lies in the intermediate-and high-PI SNR regimes where the networks operate with high efficiency. At the same time, the substantial overlap between the PI SNR distributions of neighboring datasets implies that adjacent models are trained on many signals of comparable loudness, which naturally leads to similar detection efficiencies and smooth transitions between their efficiency curves. In the ET-based configurations, this pattern is modulated by the slight rise of the ET noise at the chosen higher frequencies, which can partially offset the benefits of extending the OTWs in the investigated frequency band. By contrast, for the CE detectors, the lower noise floor allows the increased signal content at higher $f_{\max}$ to translate more directly into improved or saturated detection performance. Furthermore, in terms of absolute advance-warning alerts, the partial-inspiral strain windows considered here correspond to signals being observed well before coalescence. For the most favorable BNS configurations and for the chosen OTWs, this implies that the \texttt{GW-FALCON} pipeline can deliver robust early-warning triggers from several tens up to a few hundred seconds before merger, depending on the detector's noise level for the respective frequency band, the GW signal's PI SNR, the corresponding source parameters of the binary, and the associated sky localization (for the relative times distributions see, e.g., the left panel of Fig. \ref{fig:delta_t_distributions} for an indicative review).

\subsection{Performance of EW-ANNs for NSBH mergers}

\subsubsection{Overall classification performance}

Having assessed the performance of the EW-ANNs for BNS early-warning alerts, we now investigate the corresponding models trained for NSBH systems. The classification results for this case are summarized in Table \ref{tab:detector_dataset_metrics_II}, which presents, for each detector–dataset pair, the same set of performance measures as in the BNS analysis.

\begin{table*}[!t]
  \caption{\label{tab:detector_dataset_metrics_II} Same for Table \ref{tab:detector_dataset_metrics} for the NSBH early alerts investigation.}
  \centering
  \resizebox{\textwidth}{!}{%
  %\begin{tabular}{lll*{16}{c}}
  \begin{tabular}{cc|c|cc|cc|cc|cc|cc|cc|cc|cc}
    \toprule
    \multirow{2}{*}{Detector}& \multirow{2}{*}{Dataset Subsets}& \multirow{2}{*}{EW-ANN}&
    \multicolumn{2}{c}{Accuracy [\%]} &
    \multicolumn{2}{c}{Precision [\%]} &
    \multicolumn{2}{c}{Recall [\%]} &
    \multicolumn{2}{c}{F1-score [\%]} &
    \multicolumn{2}{c}{Specificity [\%]} &
    \multicolumn{2}{c}{MCC [\%]} &
    \multicolumn{2}{c}{FAP [\%]} &
    \multicolumn{2}{c}{FPR [\%]} \\
    \cmidrule(lr){4-5}
    \cmidrule(lr){6-7}
    \cmidrule(lr){8-9}
    \cmidrule(lr){10-11}
    \cmidrule(lr){12-13}
    \cmidrule(lr){14-15}
    \cmidrule(lr){16-17}
    \cmidrule(lr){18-19}
    & &  
    & Train & Test    % Accuracy
    & Train & Test    % Precision
    & Train & Test    % Recall
    & Train & Test    % F1
    & Train & Test    % Specificity
    & Train & Test    % MCC
    & Train & Test    % FAP
    & Train & Test    % FAR
    \\
    \midrule
    \multirow{6}{*}{ET1}
      & $\mathcal{D}_{1,\mathrm{ET1}}$ & model 1 & 89.06 & 89.00 & 95.34 & 95.28 & 82.14 & 82.06 & 88.25 & 88.18 & 95.98 & 95.93 & 78.89 & 78.75 & 4.02  & 4.07 & 4.66 & 4.72 \\
      & $\mathcal{D}_{2,\mathrm{ET1}}$ & model 2 & 89.66 & 89.53 & 95.49 & 95.39 & 83.26 & 83.08 & 88.96 & 88.81 & 96.07 & 95.98 & 79.99 & 79.73 & 3.93 & 4.02 & 4.51 & 4.61 \\
      & $\mathcal{D}_{3,\mathrm{ET1}}$ & model 3 & 90.04 & 89.90 & 95.70 & 95.54 & 83.84 & 83.71 & 89.38 & 89.23 & 96.23 & 96.09 & 80.69 & 80.41 & 3.77 &  3.91 & 4.30 & 4.46 \\
      & $\mathcal{D}_{4,\mathrm{ET1}}$ & model 4 & 90.41 & 90.25 & 95.84 & 95.68 & 84.48 & 84.31 & 89.80 & 89.63 & 96.33 & 96.19 & 81.39 & 81.07 & 3.67 & 3.81 & 4.16 & 4.32 \\
      & $\mathcal{D}_{5,\mathrm{ET1}}$ & model 5 & 90.97 & 90.80 &  96.11 & 95.86 & 85.40 & 85.27 & 90.44 & 90.26 & 96.54 & 96.32 & 82.45 & 82.10 & 3.46 & 3.68 & 3.89 & 4.14 \\
      & $\mathcal{D}_{6,\mathrm{ET1}}$ & model 6 & 91.48 & 91.33 & 96.11 & 95.97 & 86.47 & 86.28 & 91.03 & 90.87 & 96.50 & 96.38 & 83.39 & 83.09 & 3.50 & 3.62 & 3.89 & 4.03 \\
    \midrule
    \multirow{6}{*}{ET2}
      & $\mathcal{D}_{1,\mathrm{ET2}}$ & model 1 & 90.13 & 90.08 & 96.14 & 96.12 & 83.63 & 83.53 & 89.45 & 89.39 & 96.64 & 96.63 & 80.95 & 80.86 & 3.36 & 3.37 & 3.86 & 3.88 \\
      & $\mathcal{D}_{2,\mathrm{ET2}}$ & model 2 & 90.65 & 90.54 & 96.15 & 95.97 & 84.69 & 84.62 & 90.06 & 89.94 & 96.61 & 96.45 & 81.88 & 81.64 & 3.39 & 3.55 & 3.85 & 4.03 \\
      & $\mathcal{D}_{3,\mathrm{ET2}}$ & model 3 & 91.02 & 90.99 & 96.27 & 96.18 & 85.34 & 85.37 & 90.48 & 90.45 & 96.70 & 96.61 & 82.57 & 82.51 &  3.30 & 3.39 & 3.73 & 3.82 \\
      & $\mathcal{D}_{4,\mathrm{ET2}}$ & model 4 & 91.35 & 91.26 & 96.34 & 96.26 & 85.96 & 85.85 & 90.86 & 90.76 & 96.74 & 96.66 & 83.19 & 83.00 & 3.26 & 3.34 & 3.66 & 3.74 \\
      & $\mathcal{D}_{5,\mathrm{ET2}}$ & model 5 & 91.86 & 91.81 & 96.59 & 96.54 & 86.79 & 86.72 & 91.42 & 91.37 & 96.93 & 96.89 & 84.15 & 84.05 & 3.07 & 3.10 & 3.41 & 3.46 \\
      & $\mathcal{D}_{6,\mathrm{ET2}}$ & model 6 & 92.29 & 92.24 & 96.63 & 96.57 & 87.63 & 87.59 & 91.91 & 91.86 & 96.94 & 96.89 & 84.94 & 84.85 & 3.06 & 3.11 & 3.37 & 3.43 \\
    \midrule
    \multirow{6}{*}{CE1}
      & $\mathcal{D}_{1,\mathrm{CE1}}$ & model 1 & 97.11 & 97.11 & 99.29 & 99.28 & 94.91 & 94.90 & 97.05 & 97.04 & 99.32 & 99.31 & 94.32 & 94.31 & 0.68 & 0.69 & 0.71 & 0.72 \\
      & $\mathcal{D}_{2,\mathrm{CE1}}$ & model 2 & 97.35 & 97.34 & 99.49 & 99.46 & 95.20 & 95.19 & 97.30 & 97.28 & 99.51 & 99.48 & 94.80 & 94.76 & 0.49 & 0.52 & 0.51 & 0.54 \\
      & $\mathcal{D}_{3,\mathrm{CE1}}$ & model 3 & 97.52 & 97.52 & 99.46 & 99.46 & 95.56 & 95.56 & 97.47 & 97.47 & 99.49 & 99.48 & 95.12 & 95.12 & 0.51 & 0.52 & 0.54 & 0.54 \\
      & $\mathcal{D}_{4,\mathrm{CE1}}$ & model 4 & 97.64 & 97.63 & 99.44 & 99.41 & 95.83 & 95.82 & 97.60 & 97.59 & 99.46 & 99.44 & 95.35 & 95.33 & 0.54 & 0.56 & 0.56 & 0.59 \\
      & $\mathcal{D}_{5,\mathrm{CE1}}$ & model 5 & 97.79 & 97.77 &  99.43 &  99.42 & 96.13 & 96.09 & 97.75 & 97.73 & 99.45 & 99.44 & 95.64 & 95.59 & 0.55 & 0.56 & 0.57 & 0.58 \\
      & $\mathcal{D}_{6,\mathrm{CE1}}$ & model 6 & 97.91 & 97.90 & 99.54 & 99.53 & 96.27 & 96.25 & 97.88 & 97.86 & 99.55 & 99.54 & 95.88 & 95.85 & 0.45 & 0.46 & 0.46 & 0.47 \\
    \midrule
    \multirow{6}{*}{CE2}
      & $\mathcal{D}_{1,\mathrm{CE2}}$ & model 1 & 97.10 & 97.03 & 99.38 & 99.34 & 94.79 & 94.69 & 97.03 & 96.65 & 99.41 & 99.37 & 94.30 & 94.16 & 0.59 & 0.63 & 0.62 & 0.66 \\
      & $\mathcal{D}_{2,\mathrm{CE2}}$ & model 2 & 97.33 & 97.29 & 99.40 & 99.37 & 95.23 & 95.19 & 97.27 & 97.24 & 99.43 & 99.40 & 94.74 & 94.67 & 0.57 & 0.60 & 0.60 & 0.63 \\
      & $\mathcal{D}_{3,\mathrm{CE2}}$ & model 3 & 97.51 & 97.43 & 99.48 & 99.47 & 95.51 & 95.36 & 97.45 & 97.37 & 99.50 & 99.49 & 95.09 & 94.94 & 0.50 & 0.51 & 0.52 & 0.53 \\
      & $\mathcal{D}_{4,\mathrm{CE2}}$ & model 4 & 97.63 & 97.59 & 99.54 & 99.53 & 95.70 & 95.63 & 97.58 & 97.54 & 99.56 & 99.56 & 95.32 & 95.25 & 0.44 & 0.46 & 0.46 & 0.47 \\
      & $\mathcal{D}_{5,\mathrm{CE2}}$ & model 5 & 97.77 & 97.71 & 99.55 & 99.54 & 95.98 & 95.87 &  97.73 &  97.67 & 99.57 & 99.56 & 95.61 & 95.49 & 0.43 & 0.44 & 0.45 & 0.46 \\
      & $\mathcal{D}_{6,\mathrm{CE2}}$ & model 6 & 97.90 & 97.87 & 99.66 & 99.66 & 96.12 & 96.08 & 97.86 & 97.84 & 99.67 & 99.67 & 95.85 & 95.81 & 0.33 & 0.33 & 0.34 & 0.34 \\
    \bottomrule
  \end{tabular}
  }% end resizebox
\end{table*}

Overall, the NSBH EW-ANNs display consistently strong classification performance across all ET- and CE-based configurations, with trends broadly similar to those obtained in the BNS early-warning investigation. For the ET1 and ET2 detectors, the train and test sets exhibit accuracies in the range $89\%$-$92\%$, with precision values around $95\%$-$96.6\%$, recall in the $82\%$-$87.6\%$ range, and specificities close to $96\%$-$97\%$. The corresponding MCC values lie between about $79\%$ and $85\%$, and the reported FAP and FPR have values below $5\%$, indicating robust signal-detection efficiency and a satisfactory separation between foreground and background samples. For the CE1 and CE2 detectors, the performance is even more uniform and closer to ideal: the models' train and test set accuracies exceed $97\%$, precision and specificity are above $99\%$, recall values cluster around $95$-$96\%$, and MCC values are near $95\%$, while FAP and FPR are well below the percent level. These results reflect the cleaner data provided by the CE noise curves, which yield more informative feature representations for the employed datasets and enable the networks to distinguish GW signal+noise from noise-only class examples with greater confidence.

Within each detector configuration, there is again a clear and monotonic trend across the datasets indexed by $i=1,\dots,6$. As one moves from $\mathcal{D}_1$ to $\mathcal{D}_6$, the reported evaluation performance indicators for the test sets improve systematically. This behavior is consistent with the underlying NSBH injection properties: increasing $f_{\max}$ implies that a larger and more rapidly evolving portion of the inspiral is captured within the observational time window, leading to higher PI SNRs and more distinctive feature patterns for the EW-ANNs (see, e.g., the PI SNR distributions shown in Figs. \ref{fig:CEH_PI_SNR_distributions} and \ref{fig:other_PI_SNR_distributions}). As in the BNS case, louder NSBH signals translate directly into easier discrimination between the two categories.

\begin{figure}[!thb]
    \includegraphics[width=0.47\textwidth]{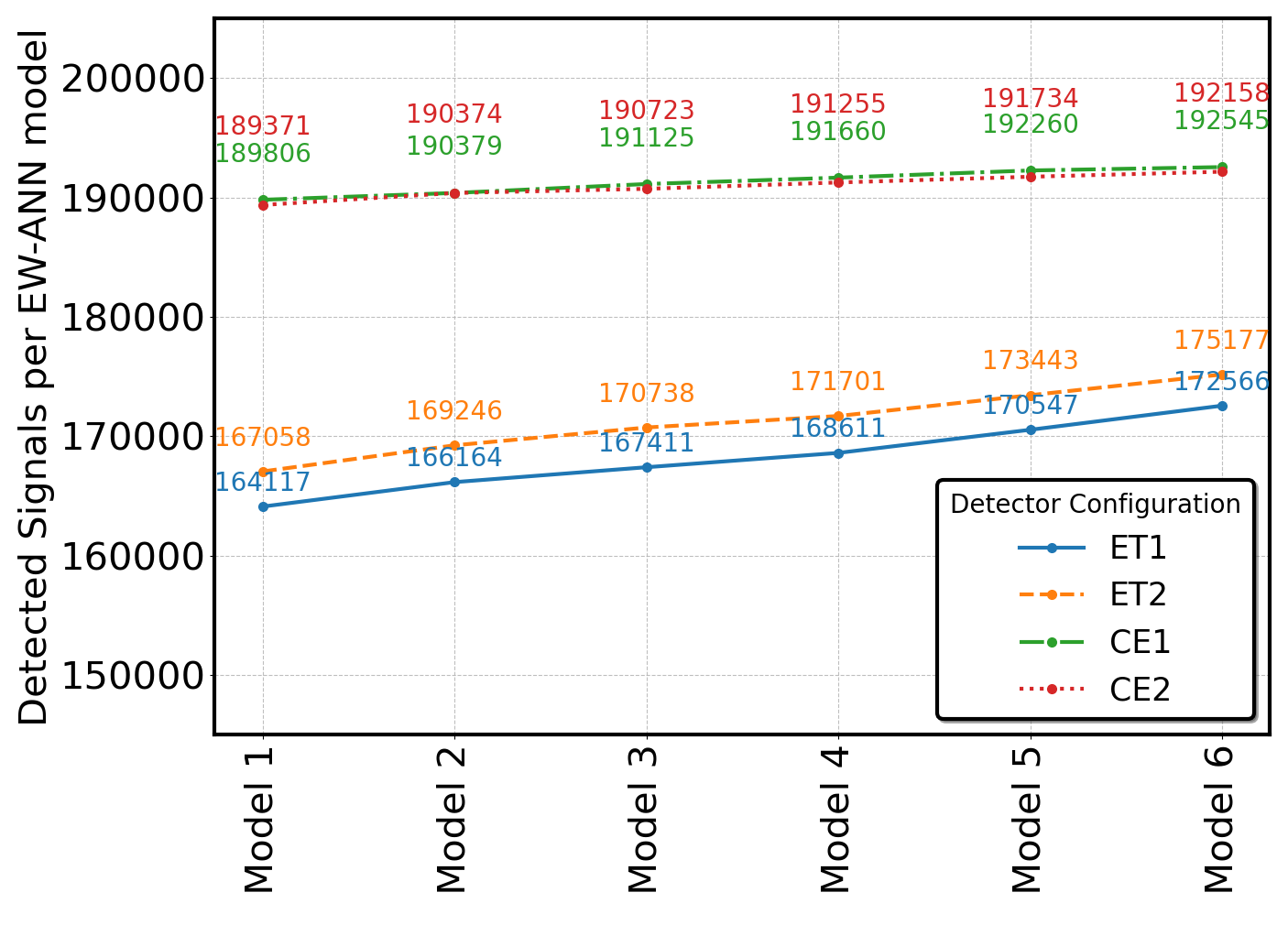}\hfill
    \caption{\label{fig:NSBH_detections_per_model}
    Number of detected NSBH injections in the test datasets for each model and detector configuration. The horizontal axis labels the EW-ANN models ($1–6$), while the vertical axis shows the number of signals correctly classified as detections (true positives) by each individual model. Colored curves correspond to different detectors, and the annotated values indicate the exact number of detected injections for each model–detector pair. For the ET1 and ET2 configurations, the detected-signal counts increase steadily with model index, reflecting the improved performance obtained when the OTWs extend to higher $f_{\max}$. For the CE1 and CE2 configurations, the detected-signal counts are systematically higher and vary only weakly with model index, consistent with the fact that the CE-based ANN models already recover nearly all detectable NSBH injections across the test datasets considered.
    }
\end{figure}

To complement these summary statistics, Fig.~\ref{fig:NSBH_detections_per_model} shows, for each detector configuration, the number of NSBH injections in the test datasets that are correctly identified by each individual EW-ANN. For the ET1 and ET2 detectors, the detected-signal counts increase steadily from model~1 to model~6, reflecting the improved performance obtained as the OTWs extend to higher $f_{\max}$ and capture a larger, more rapidly evolving portion of the inspiral. The CE1 and CE2 configurations yield systematically higher detection counts for all network indices and exhibit only a weak dependence on the specific model, with the corresponding curves being nearly flat. This behavior is consistent with the high PI SNRs achieved in the CE-based datasets: the associated feature representations are already very informative, so that each CE EW-ANN recovers nearly all detectable NSBH injections in the employed test datasets.

In addition, in contrast to the BNS early-warning classifiers, the ET-based NSBH EW-ANNs do not show a noticeable degradation in the evaluation measures for model 5. In this case, the OTWs associated with the dataset $\mathcal{D}_5$ (with $f_{\max} = 22 \ \mathrm{Hz}$) still contain comparatively strong NSBH signals: NSBH systems exhibit more pronounced amplitude and frequency evolution within the analysed windows than BNS systems. As a result, the features extracted from these OTWs remain sufficiently informative, allowing the networks for the associated ET1 and ET2 detectors to discern characteristic patterns in the data and to maintain stable performance despite the increased ET noise level in this frequency band (see, e.g., Fig \ref{fig:design_psds} for a review).

The values listed in Table \ref{tab:detector_dataset_metrics_II} also show that, for all detectors, the signal-detection efficiency remains high while the rate of false positive triggers is kept under control. For the ET-based configurations, the false-alarm probability and false-positive rate remain at the level of only a few percent, whereas for CE1 and CE2, they are reduced to well below the $1\%$ level. This confirms that the NSBH EW-ANNs can operate under stringent false-alarm requirements, particularly in the CE case, without sacrificing their ability to recover the majority of injected signals at early inspiral stages. To further illustrate these trends at the level of individual predictions, Fig. \ref{fig:NSBH_conf_matrix_elements_hists} in Appendix \ref{app:classification_eval_measures} shows the distribution of the EW-ANN output probability for events in each confusion-matrix class (TP, FP, TN, FN), using the corresponding test subsamples from $\mathcal{D}_1$ and $\mathcal{D}_6$ across the ET1, ET2, CE1, and CE2 detector configurations. The selected ANN classifiers are representative of earlier and later early-warning alert regimes, respectively. Consistent with the respective evaluation measures, background (noise-only) samples are predominantly assigned low network probability values, while the false-negative population remains small; moreover, moving from the first to the last model in the sequence (higher PI SNR in the foreground samples population) yields a clearer separation and fewer false-positive triggers.

Finally, the proximity between the training and test-set values across all entries in Table \ref{tab:detector_dataset_metrics_II} indicates a small train-test gap, typically at the level of a few tenths of a percent. This behavior mirrors what was found for BNS and suggests that the NSBH EW-ANN models generalize well to unseen data, with no evidence of significant overfitting. Taken together with the BNS results, this provides strong support for the robustness of the proposed \texttt{GW-FALCON}-based early warning classification framework across different compact binary source populations.

%EM followup

\subsubsection{Detection efficiency as a function of NSBH PI SNR}

\begin{figure*}[!thb]
    \centering
    \subfloat[ET1 detector]{%    
    \includegraphics[width=0.50\textwidth]{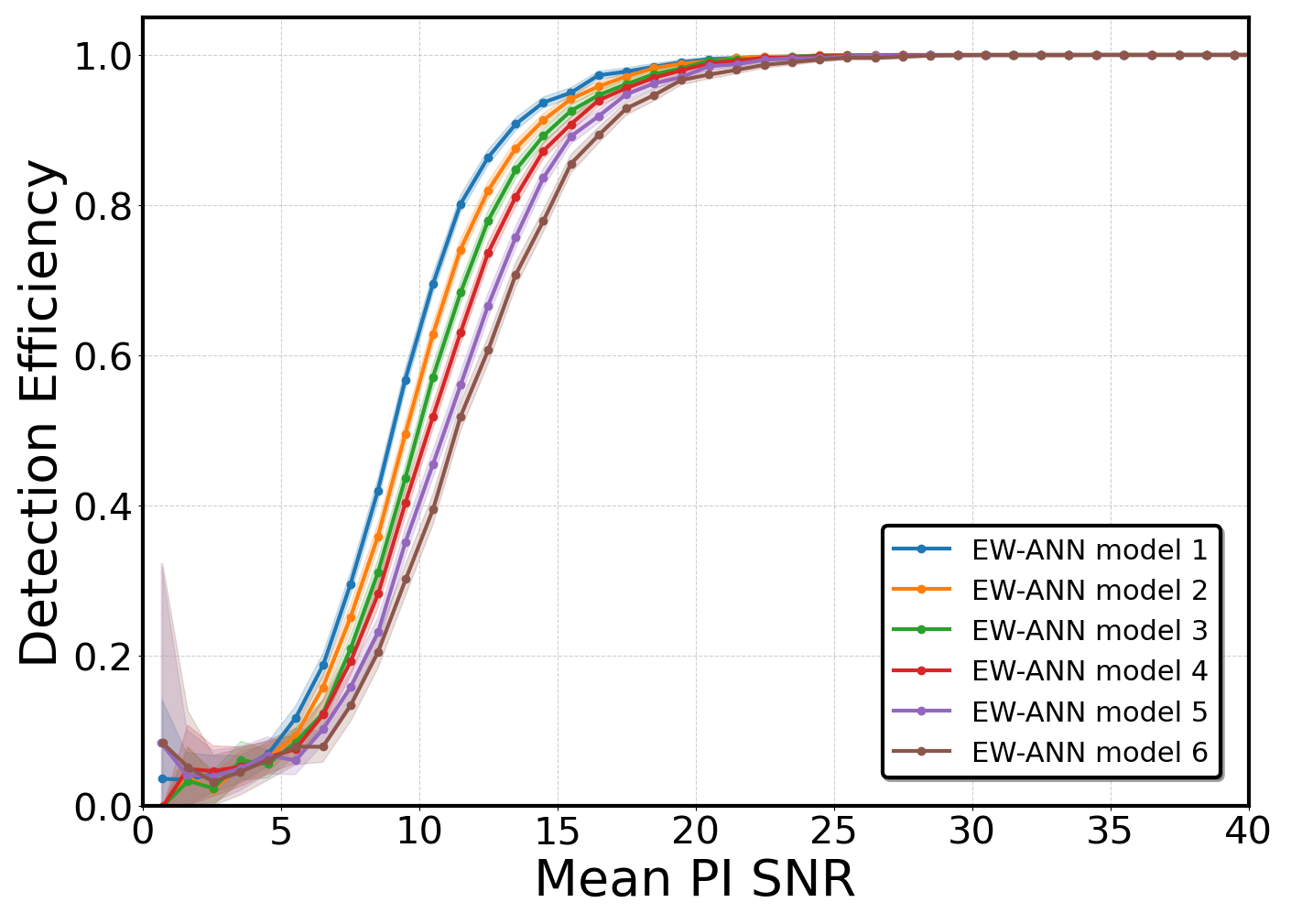}
    \label{fig:ETV_NSBH_det_efficiency}}
    \subfloat[ET2 detector]{%
    \includegraphics[width=0.50\textwidth]{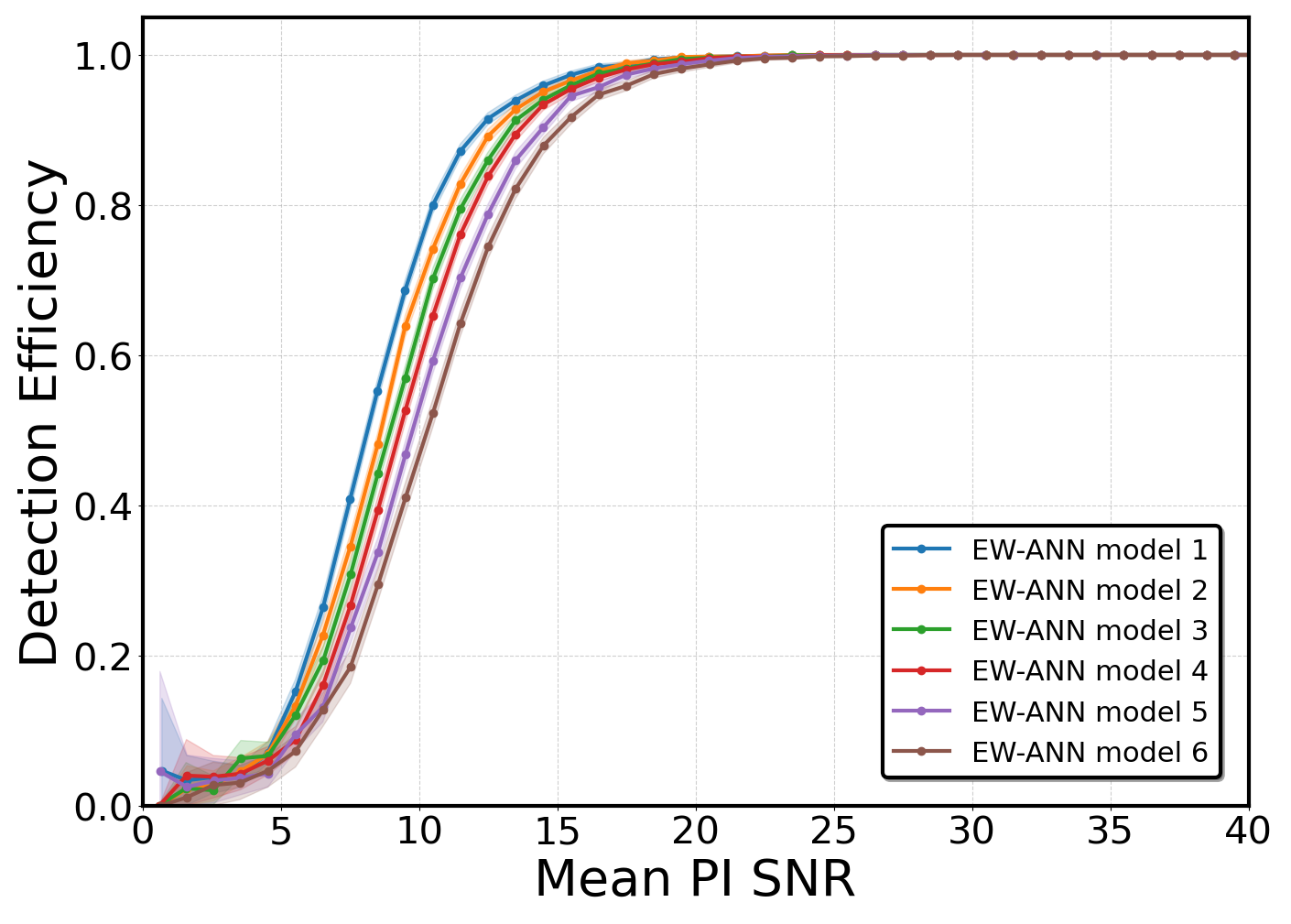}
    \label{fig:ETVG_NSBH_det_efficiency}}\\
    \subfloat[CE1 detector]{%    
    \includegraphics[width=0.50\textwidth]{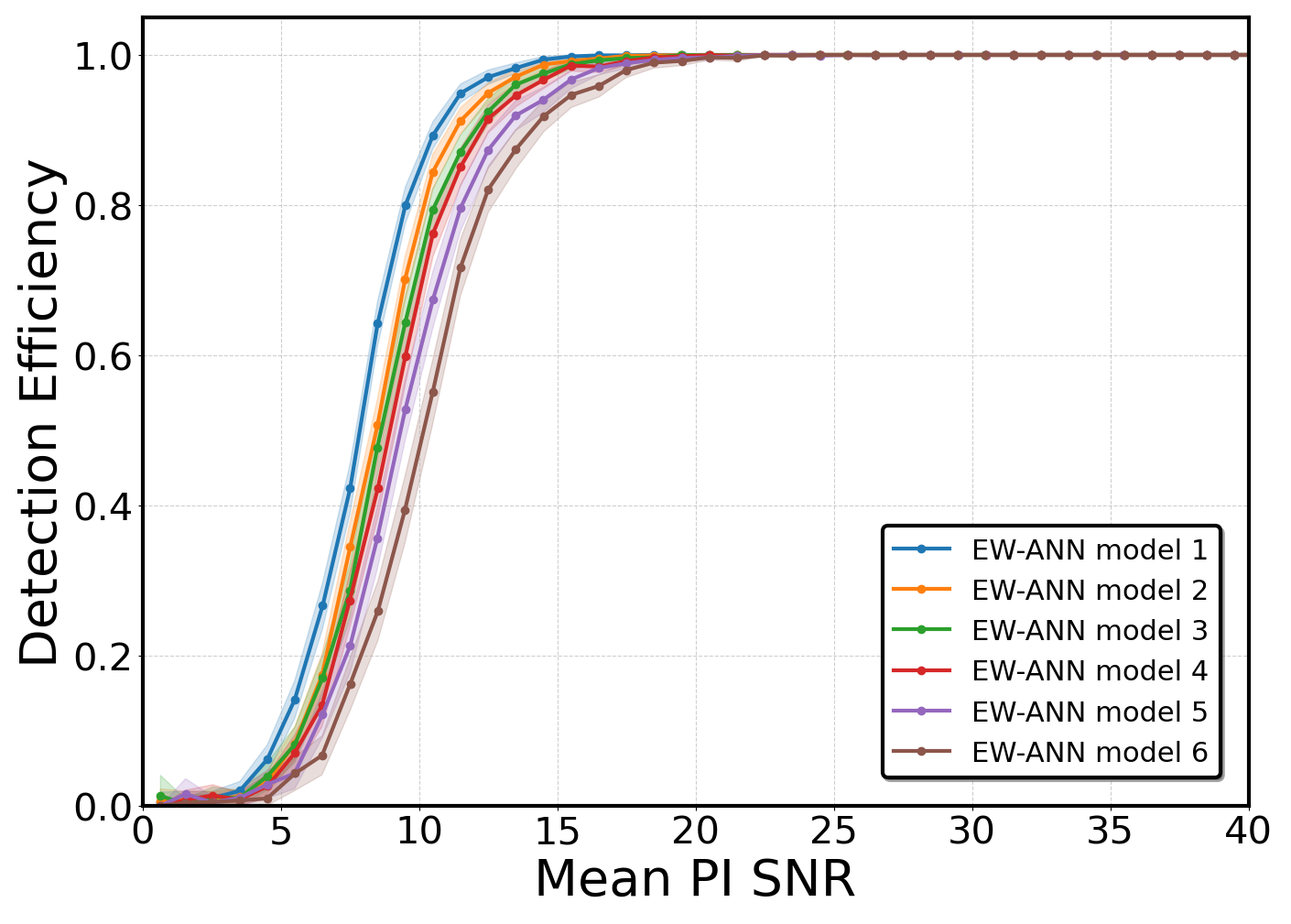}
    \label{fig:CEH_NSBH_det_efficiency}}
    \subfloat[CE2 detector]{%
    \includegraphics[width=0.50\textwidth]{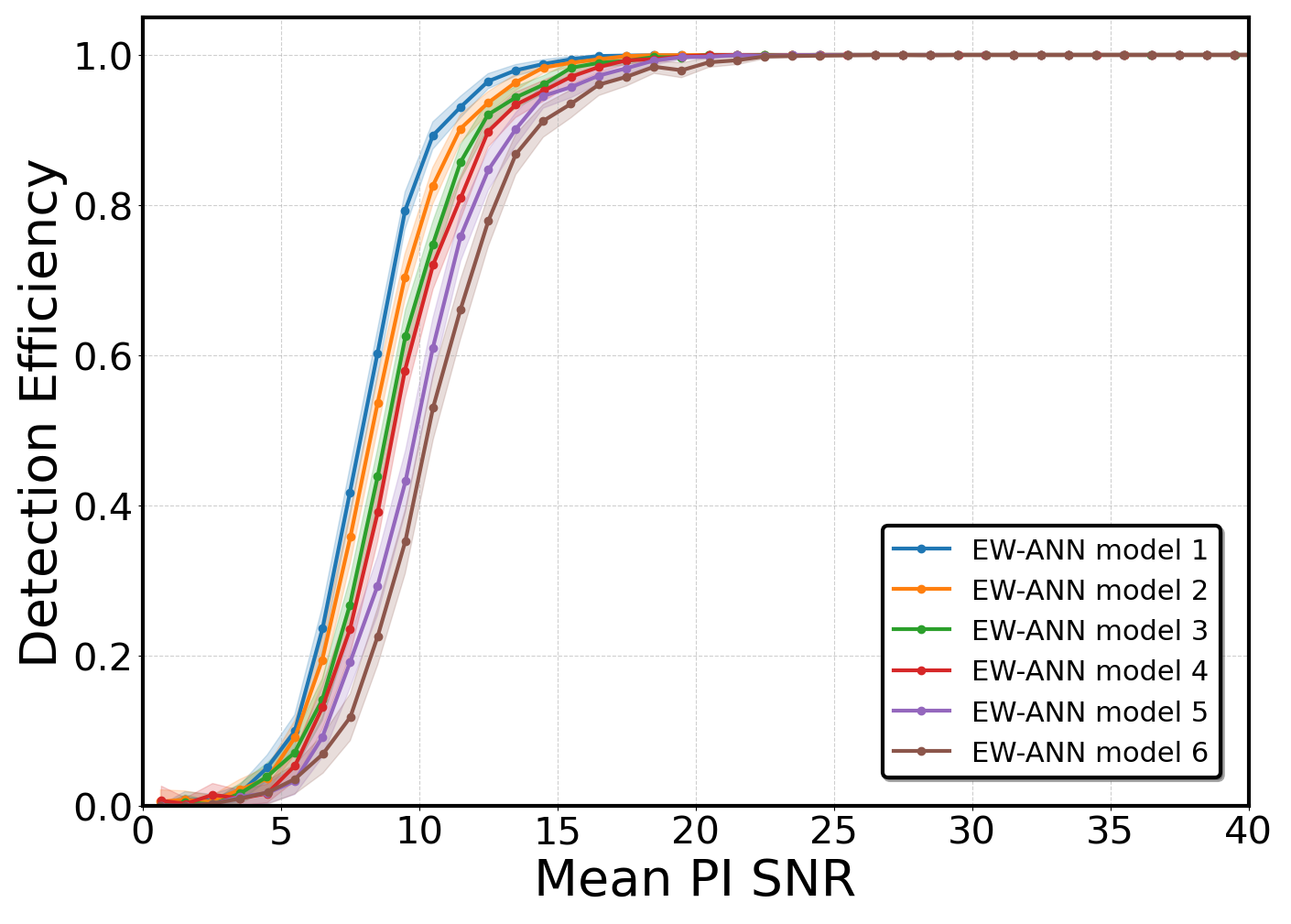}
    \label{fig:CEL_NSBH_det_efficiency}}\\
    \caption{\label{fig:NSBH_det_efficiency_vs_mean_PI_snr}
    Detection efficiency as a function of mean PI SNR for the NSBH EW-ANN models. Each panel corresponds to one detector configuration. The colored curves show the detection efficiency on the test datasets for the different EW-ANN models (models 1-6), while the shaded bands denote the associated $3\sigma$ uncertainty regions. The curves illustrate the gradual transition from low efficiency at small mean PI SNR to near-unity efficiency at high mean PI SNR, with the CE configurations reaching high efficiencies at lower PI SNR values than the ET configurations.
    }
\end{figure*}

Analogously to the BNS case, we also investigate how the detection efficiency of the NSBH EW-ANNs depends on the loudness of the incoming signals. For each detector configuration and NSBH test dataset, we compute the recall as a function of the mean PI SNR using the same PI SNR binning and uncertainty estimation procedure described in the previous section. The resulting efficiency curves are shown in Fig. \ref{fig:NSBH_det_efficiency_vs_mean_PI_snr}. Each panel corresponds to one detector configuration and displays, for the six NSBH EW-ANN models, the detection efficiency on the test set as a function of the mean PI SNR, with solid lines and shaded bands indicating the mean values and associated $\pm 3\sigma$ uncertainties, respectively. Across all detectors employed, the curves exhibit the expected transition from low efficiency at very small mean PI SNR to a high-efficiency plateau once the signals become sufficiently loud. For the ET1 and ET2 configurations, the EW-ANN models show a gradual increase in detection efficiency as the mean PI SNR grows. In contrast, the CE1 and CE2 configurations reach high efficiencies at comparatively lower PI SNR values and exhibit narrower uncertainty bands, reflecting the cleaner data and more informative feature representations afforded by the CE noise curves. As in the BNS analysis, the horizontal axis refers to mean PI SNR values up to 40; for louder signals, all NSBH EW-ANN models for all detectors already operate at essentially perfect detection efficiency, and the corresponding curves are indistinguishable.

For all detectors, the NSBH efficiency curves in Fig. \ref{fig:NSBH_det_efficiency_vs_mean_PI_snr} again display a sigmoidal behavior: the efficiency is very low at mean PI SNRs below $4-5$, rises rapidly through the intermediate range $5-12$ as the associated GW signal portion becomes more prominent in the OTWs, and saturates close to unity once the mean PI SNR exceeds $\simeq 15-18$. In the ET1 and ET2 configurations, the transition from low to high efficiency is somewhat more gradual and extends to slightly higher PI SNR, whereas in the CE1 and CE2 configurations, the rise is sharper and shifted toward lower PI SNR, with visibly tighter uncertainty bands, reflecting the cleaner CE noise curves and the more informative feature representations available to the corresponding models. For all four detectors, the efficiencies are essentially perfect for mean PI SNRs above $20$, indicating that NSBH injections of this loudness are almost always recovered by the suggested EW-ANNs.

In addition, the interpretation of NSBH detection efficiency curves is closely linked to the PI SNR distributions of the corresponding signal portions. As shown for the CE1 detector in the right panel of Fig. \ref{fig:CEH_PI_SNR_distributions} (with the corresponding PI SNR distributions for NSBH injections in the other detectors given in Appendix \ref{app:PI_SNR_distributions}), the employed NSBH waveform portions still carry substantial signal power: most events have PI SNR values above $15$, while only a small fraction populate the low-PI SNR tail below $8-10$. At the mean PI SNR $\lesssim 5-8$, the ANNs therefore operate in a regime where the signal is only marginally above the noise, and relatively few training and test examples are available, so the detection efficiency remains modest and the uncertainty bands are comparatively wide. As the mean PI SNR increases into the $10-20$ range and beyond, a growing fraction of NSBH signals contributes, the extracted features become more distinctive, and the classifiers rapidly transition to the high-efficiency plateau seen in Fig. \ref{fig:NSBH_det_efficiency_vs_mean_PI_snr}. Because the CE1 and CE2 detectors yield NSBH PI SNR distributions that are systematically shifted to higher values than those of ET1 and ET2, the corresponding ANNs are trained on a larger proportion of loud, well-resolved NSBH signals; this naturally explains the earlier and steeper rise of the CE efficiency curves and their approach to unity at lower mean PI SNR.

Within each detector configuration, the differences in detection performance between the NSBH EW-ANN models can be traced back to the PI SNR content of the underlying datasets on which they are trained. As $f_{\max}$ increases from dataset 1 to dataset 6, the OTWs used for feature extraction contain a progressively larger portion of the late inspiral. The corresponding PI SNR distributions shift toward higher values, so that a growing fraction of NSBH injections falls in the intermediate- and high-PI SNR regimes where the classifiers already operate with high efficiency. At the same time, the substantial overlap between the PI SNR distributions of neighboring datasets means that adjacent models are trained on many signals of comparable loudness, which naturally yields similar detection efficiencies and smooth changes in their efficiency curves. For the ET-based configurations, the modest rise of the ET noise at the higher frequencies considered slightly tempers the gains from extending the OTWs, but the comparatively loud NSBH signals employed ensure that performance remains uniformly high across models. For the CE detectors, the qualitative features extracted from a lower noise floor lead the EW-ANN models to near-saturated efficiencies over much of the PI SNR range; therefore, extending the OTWs primarily consolidates, rather than dramatically improves, the already excellent detection performance. In addition, because NSBH systems spend less time in the detector's band than BNSs, the corresponding advance-warning times are shorter. Within the present setup, however, the trained EW-ANNs can still provide reliable pre-merger alerts typically tens of seconds before coalescence in the most favorable cases (for the relative time distributions see, e.g., the right panel of Fig. \ref{fig:delta_t_distributions} for an indicative review).

\subsection{\label{sec:triggers_from_EW_ANNa}Assessing EW Triggers on Embedded GW Signals}

\subsubsection{Sliding-window analysis for detector's strain data}

We next move beyond the supervised datasets employed in the previous sections and assess, as an indicative example, how the EW-ANN models perform when applied to continuous, detector-like data streams containing simulated BNS and NSBH mergers. This step serves as a proof-of-concept validation of the networks’ triggering behavior and further illustrates how the models respond to signals embedded in long stretches of noise. In particular, it demonstrates the performance of the trained ANNs under more realistic, low-latency conditions that mimic an operational early-warning pipeline.

For this purpose, we process independent detector frames that each contain a single injected BNS or NSBH signal, using essentially the same preprocessing and feature-extraction steps as in the supervised study. We analyze the data in $44~\mathrm{s}$ analysis windows. Starting from the beginning of the frame, we select a $44~\mathrm{s}$ segment of strain, whiten it using a power spectral density (PSD) estimated from a $400~\mathrm{s}$ stretch of independent background data, and then apply the same high- and low-pass filters used in the previous analysis. To mitigate edge effects introduced by filtering, we discard the first and last $2~\mathrm{s}$ of the filtered segment, retaining a central $40~\mathrm{s}$ interval of processed time-series data. In this way, we obtain eight OTWs for the BNS case and six OTWs for the NSBH case, which subsequently serve as input to the TSFEL feature-extraction pipeline \cite{barandas2020tsfel}. For the standardization step, we rescale each feature using the mean and standard deviation estimated from the corresponding test dataset feature distributions, ensuring that the resulting feature vectors are suitable inputs for the EW-ANNs.

The initial $44~\mathrm{s}$ analysis window is then shifted forward by $4~\mathrm{s}$, and the procedure is repeated so that successive OTWs overlap by $40~\mathrm{s}$ until the entire data stream is covered. For whitening, we reuse the same background PSD across all analysis windows, which provides a representative noise estimate while remaining computationally efficient. This choice avoids repeatedly loading independent $400~\mathrm{s}$ stretches of background data and recomputing the PSD for each $44~\mathrm{s}$ segment during the whitening phase, a procedure that would be significantly more time-consuming. Note that the $4~\mathrm{s}$ step is arbitrary and could be reduced; in a realistic early-warning pipeline, however, the minimum step should not be shorter than the time required to load $44~\mathrm{s}$ of data, apply the preprocessing, and obtain predictions from the models' evaluation.

In its current implementation, the \texttt{GW-FALCON} framework is computationally inexpensive. For each $44~\mathrm{s}$ analysis window, TSFEL-based feature construction on the CPU requires an average characteristic time of $t_f \sim 0.1~\mathrm{s}$ per $40~\mathrm{s}$ OTW, while inference with the trained EW-ANNs on an NVIDIA RTX~A6000 GPU is essentially negligible in comparison. Using a precomputed PSD obtained with PyCBC, whitening and band-pass filtering of each $44~\mathrm{s}$ strain segment add only a modest overhead relative to $t_f$, so that the dominant contribution to the pipeline latency is the time required to load $44~\mathrm{s}$ of detector data. In practice, whitening, filtering, feature extraction, standardization, and ANN evaluation can be organized so that several stages run in parallel, and the total processing time per analysis window remains comfortably below the $4~\mathrm{s}$ step adopted for the sliding analysis, making the method compatible with low-latency early-warning operation.

%These timing considerations also motivate the chosen stride: since the construction of one feature vector requires, on average, a time $t_f$, the interval between successive evaluations should be at least $t_f + \epsilon$ (for a small safety margin $\epsilon$) to preserve the online character of the pipeline. At the same time, the employed OTWs span $40~\mathrm{s}$ of filtered data within each $44~\mathrm{s}$ analysis window, so using a much smaller step would produce highly overlapping windows with largely redundant information while unnecessarily increasing the computational cost. A $4~\mathrm{s}$ step, therefore, provides a reasonable compromise, respecting the latency implied by $t_f$ while avoiding an oversampling of nearly identical OTWs.

%\gv{It should be noted that the generation of the feature vector for each OTW requires, on average, (to be defined) $t_f$ seconds, whereas the prediction time of the classifier is negligible. Consequently, to preserve the online nature of the methodology, the sliding window must be advanced by at least $t_f + \epsilon$ seconds between successive evaluations. Moreover, for such large observation time windows, using a sliding interval that is small relative to the OTW length does not introduce additional informative content. Therefore, advancing the window with a smaller interval not only compromises the online operation but also fails to provide meaningful gains in information quality.}

\subsubsection{Illustrative BNS and NSBH triggering examples}

\begin{figure*}[!thb]
    \includegraphics[width=1.0\textwidth]{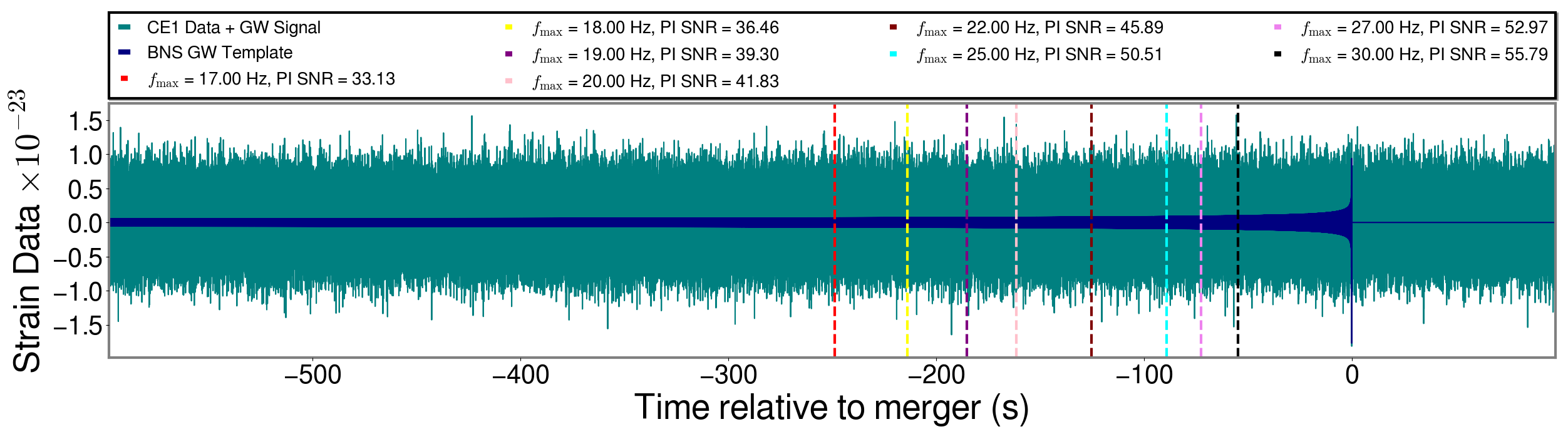}\hfill
    \includegraphics[width=1.0\textwidth]{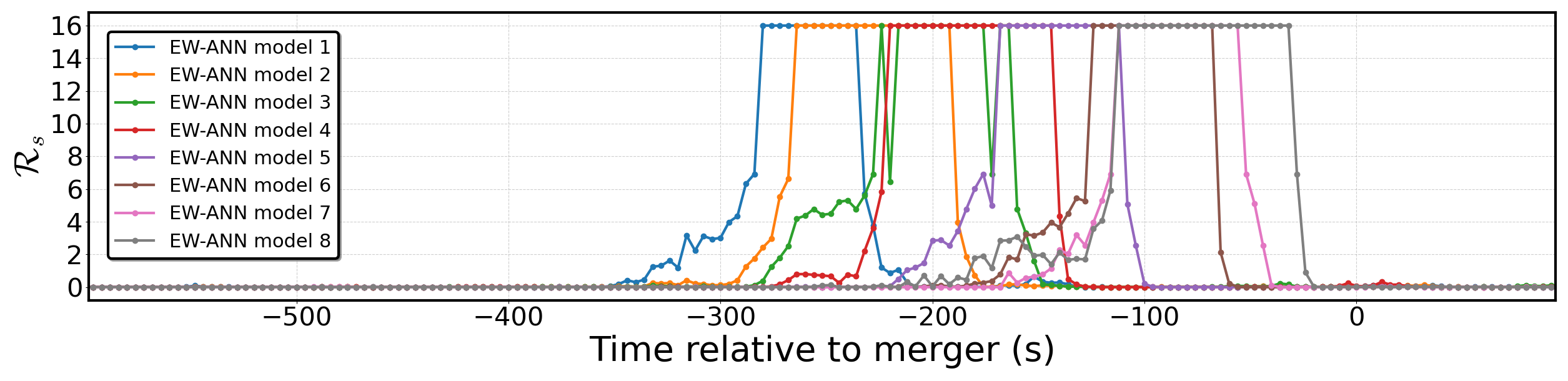}\hfill
    \caption{\label{fig:simulated_BNS_example_and_Ranking_statistic} Top panel: CE1 strain data containing an injected BNS signal with detector-frame component masses $m_{1,\mathrm{det}} = 1.5 \ M_{\odot}$ and $m_{2,\mathrm{det}} = 1.3 \ M_{\odot}$, located at a luminosity distance $d_L = 400\,\mathrm{Mpc}$. Colored vertical lines mark the times (relative to merger) at which the waveform’s instantaneous frequency reaches the corresponding $f_{\max}$ values used in the analysis; for each $f_{\max}$, the associated PI SNR is also presented, illustrating how the accumulated SNR in band increases as the system approaches merger. The coalescence time corresponds to $t_c = 0\,\mathrm{s}$. Bottom panel: evolution of the ranking statistic $\mathcal{R}_s$ across the sliding $4\,\mathrm{s}$ analysis windows, showing how the trained eight EW-ANNs, evaluated simultaneously on the corresponding feature vectors, respond to the continuous CE1 detector data stream and highlight the presence (or absence) of the injected signal. As expected, in several intervals, their predictions overlap and produce multiple high-probability trigger points in a row, illustrating sequences of consistent early-warning responses.}
\end{figure*}

\begin{figure*}[!thb]
    %\resizebox{0.5\textwidth}{!}{
    \includegraphics[width=1.0\textwidth]{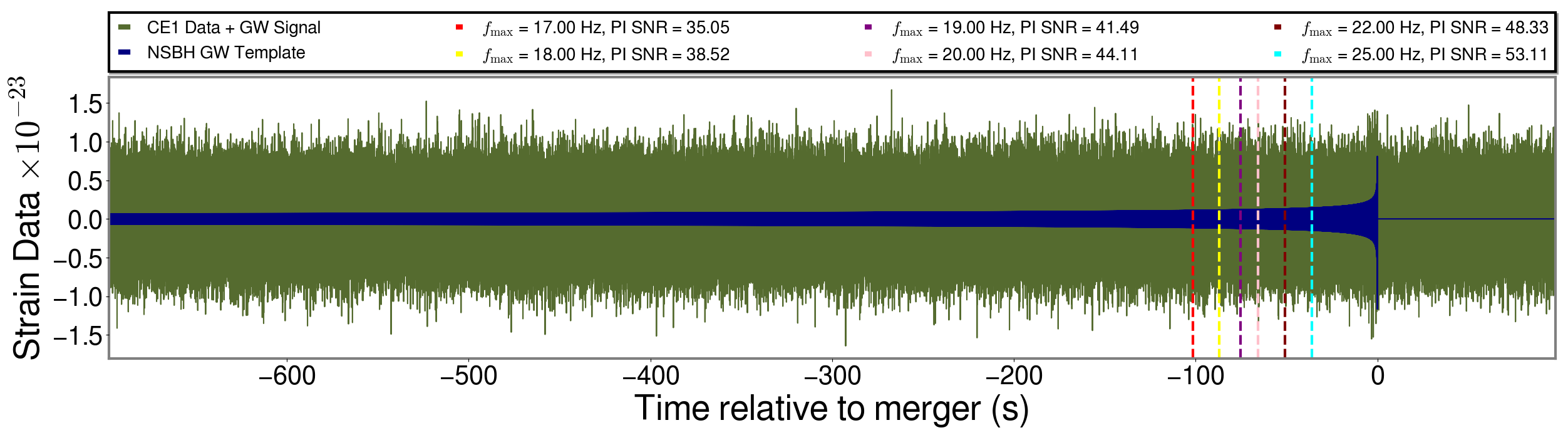}\hfill
    \includegraphics[width=1.0\textwidth]{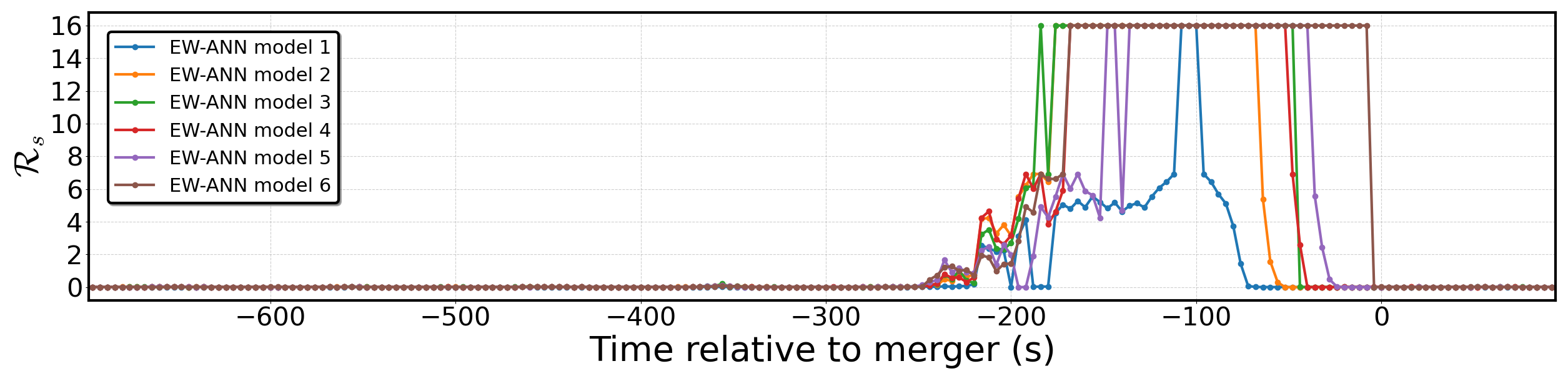}\hfill
    \caption{\label{fig:simulated_NSBH_example_and_Ranking_statistic} Top panel: CE1 strain data containing an injected NSBH signal with detector-frame component masses $m_{1,\mathrm{det}} = 4 \ M_{\odot}$ and $m_{2,\mathrm{det}} = 1.5 \ M_{\odot}$, located at a luminosity distance $d_L = 600\,\mathrm{Mpc}$. Colored vertical lines mark the times (relative to merger) at which the waveform’s instantaneous frequency reaches the corresponding $f_{\max}$ values used in the analysis; for each $f_{\max}$, the associated PI SNR is also presented, illustrating how the accumulated SNR in band increases as the system approaches merger. The coalescence time corresponds to $t_c = 0\,\mathrm{s}$. Bottom panel: evolution of the ranking statistic $\mathcal{R}_s$ across the sliding $4\,\mathrm{s}$ analysis windows, showing how the trained six EW-ANNs, evaluated simultaneously on the corresponding feature vectors, respond to the continuous CE1 detector data stream and highlight the presence (or absence) of the injected signal. As expected, in several intervals, their predictions overlap and produce multiple high-probability trigger points in a row, illustrating sequences of consistent early-warning responses.}
\end{figure*}

Following the procedure outlined above for sliding-window analysis and low-latency inference, we now illustrate the operation of the suggested framework on two representative CE1 data streams. In the first case, the stream contains a single BNS injection with detector-frame component masses $m_{1,\mathrm{det}} = 1.5 \ M_\odot$ and $m_{2,\mathrm{det}} = 1.3 \ M_\odot$, while in the second it contains a single NSBH injection with detector-frame masses $m_{1,\mathrm{det}} = 4.0 \ M_\odot$ (black hole, within the lower mass-gap region \cite{ligo2024observation}) and a NS of $m_{2,\mathrm{det}} = 1.5 \ M_\odot$. For each stream, we process the data using the $44~\mathrm{s}$ overlapping analysis windows described above, whiten and band-pass filter the strain, construct the corresponding sequence of $40~\mathrm{s}$ OTWs, and extract TSFEL features that are standardized and passed to the associated CE1 EW-ANNs. To monitor the models’ output in a way that emphasizes confident triggers, we follow the definition presented in Ref. \cite{koloniari2025new} and transform the ANN-predicted probability $p_{\theta^\star}$ for the signal+noise class into the logarithmic ranking statistic,
\begin{align}
    \mathcal{R}_s = -\log_{10}(1-p_{\theta^\star} +10^{-16}),
\end{align}
which spreads out the values of highly confident predictions (with $p_{\theta^\star}$ close to 1), making them easier to distinguish, while keeping the statistic finite as $p_{\theta^\star} \to 1$. In Figs. \ref{fig:simulated_BNS_example_and_Ranking_statistic} and \ref{fig:simulated_NSBH_example_and_Ranking_statistic}, we show, for each of these CE1 examples, the input detector-frame strain time series containing the injected CBC waveform, together with the chosen maximum instantaneous frequencies and the corresponding PI SNR values (top panels). The bottom panels display the evolution of the ranking statistic $\mathcal{R}_s$ across the sliding $4\ \mathrm{s}$ analysis windows, illustrating how the corresponding EW-ANNs respond to an isolated BNS or NSBH signal embedded in a continuous, detector-like data stream and how potential low-latency early-warning triggers appear in this setup.

The behavior of the ranking statistic in Figs. \ref{fig:simulated_BNS_example_and_Ranking_statistic} and \ref{fig:simulated_NSBH_example_and_Ranking_statistic} (bottom panels) show that the ANN classifiers begin to respond significantly well before the merger. In these examples, the earliest model (EW-ANN model~1, trained on features extracted from OTWs with the lowest $f_{\max}$) starts to produce elevated values of $\mathcal{R}_s$ already a few hundred seconds before coalescence, with these values increasing towards their maxima as the merger approaches. Subsequent models “light up’’ progressively closer to merger, as the features extracted from the portions of the signal contained in their corresponding OTWs enter the sensitive band and accumulate SNR. Over a substantial time interval, as predictions are obtained from the trained models, several EW-ANNs simultaneously yield elevated values of $\mathcal{R}_s$, producing overlapping sequences of triggers that reflect the coherent growth of the signal across the different partial-inspiral windows.

In a triggering context, it is natural to require not just a single high value of $\mathcal{R}_s$, but several consecutive “triggers in a row’’ from at least one EW-ANN—ideally from the earliest model—to mitigate the influence of transient noise excursions.
%As an indicative choice, one could, for example, demand activations of the same for example EW-ANN model with $\mathcal{R}_s > 1$ (network probability $p_{\theta^{\star}}>0.9$) or multiple models if possible in at least five consecutive analysis windows before issuing an alert; 
As an indicative choice, one could, for example, require that the same EW-ANN model (or several of the models evaluated simultaneously on the data stream) yield positive activations with $\mathcal{R}_s > 1$ (network probability $p_{\theta^\star} > 0.9$) in at least $k=5$ consecutive analysis triggers before issuing an alert; with the $4~\mathrm{s}$ sliding window step adopted here, this corresponds to a span of about $16~\mathrm{s}$ between the first and last trigger in the sequence. The use of multiple consecutive triggers inevitably increases the waiting time before an alert is generated and therefore reduces the remaining time to merger, so in practice, one must balance the expectation for very early warnings against the need to control the false-alarm rate. The NSBH case exhibits analogous behavior, although the shorter in-band duration of NSBH inspirals leads to a correspondingly shorter interval between the first sustained sequences of triggers and the merger time.

In addition, the common choice of a $4~\mathrm{s}$ sliding step for both the BNS and NSBH early-warning trigger examples was adopted for simplicity and to facilitate a direct comparison between the two source classes. For BNS systems, which spend a comparatively long time in band and exhibit a more gradual build-up of SNR, such a stride appears adequate: successive analysis windows still probe overlapping portions of the inspiral, and the evolution of the ranking statistic can be tracked with sufficient temporal resolution. For NSBH systems, however, the inspiral progresses more rapidly through the sensitive band, so a shorter sliding step may be preferable to resolve more finely the rise of the network response and to maximize the available warning time. In those cases, the family of EW-ANNs, each trained on a different partial-inspiral window, is particularly valuable: as the signal sweeps upward in frequency, different classifiers become active in turn, providing multiple, temporally ordered opportunities to issue a trigger. Reducing the step size would increase the number of analysis windows and, consequently, the computational cost of whitening, feature extraction, and ANN evaluation, making the real-time requirements more demanding and motivating the development of even faster implementations or hardware acceleration. A detailed optimization of the sliding-window step size, balancing temporal resolution, computational resources, and early-warning performance, is therefore a significant direction for future work.

\subsubsection{False-alarm rate of the suggested EW-ANN models}

Within such a sliding-window analysis and models' inference, it is also significant to quantify not only how early and how often the EW-ANNs trigger on true signals, but also how frequently they produce spurious triggers on segments of pure noise. In the context of low-latency searches, it is therefore natural to complement the usual classification evaluation measures with a time-based measure of false positives. In analogy with matched-filter searches, the false-alarm rate (FAR) is defined as \cite{messick2017analysis,nitz2018rapid,cannon2021gstlal,usman2016pycbc,davies2020extending, nitz2022dfinstad, kumar2024optimized},
\begin{align}
   \mathrm{FAR}(>\mathcal{R}_s) =  \frac{N_{\mathrm{FP}}}{T_{\mathrm{obs}}},
\end{align}
where at this point $N_{\mathrm{FP}}$ is the number of spurious triggers produced by the ANN classifier and $T_{\mathrm{obs}}$ is the total observation time over which the analysis is performed. Throughout this work, we use only colored Gaussian noise generated from the ET and CE design power spectral densities, rather than realistic non-Gaussian, non-stationary detector noise.

\begin{figure*}[!thb]
    \centering
    \subfloat[\label{fig:CEH_FAR_BNS} $\mathrm{FAR}$ vs $\mathrm{R}_s$ for BNS EW-ANN models]{%
        \includegraphics[width=0.50\textwidth]{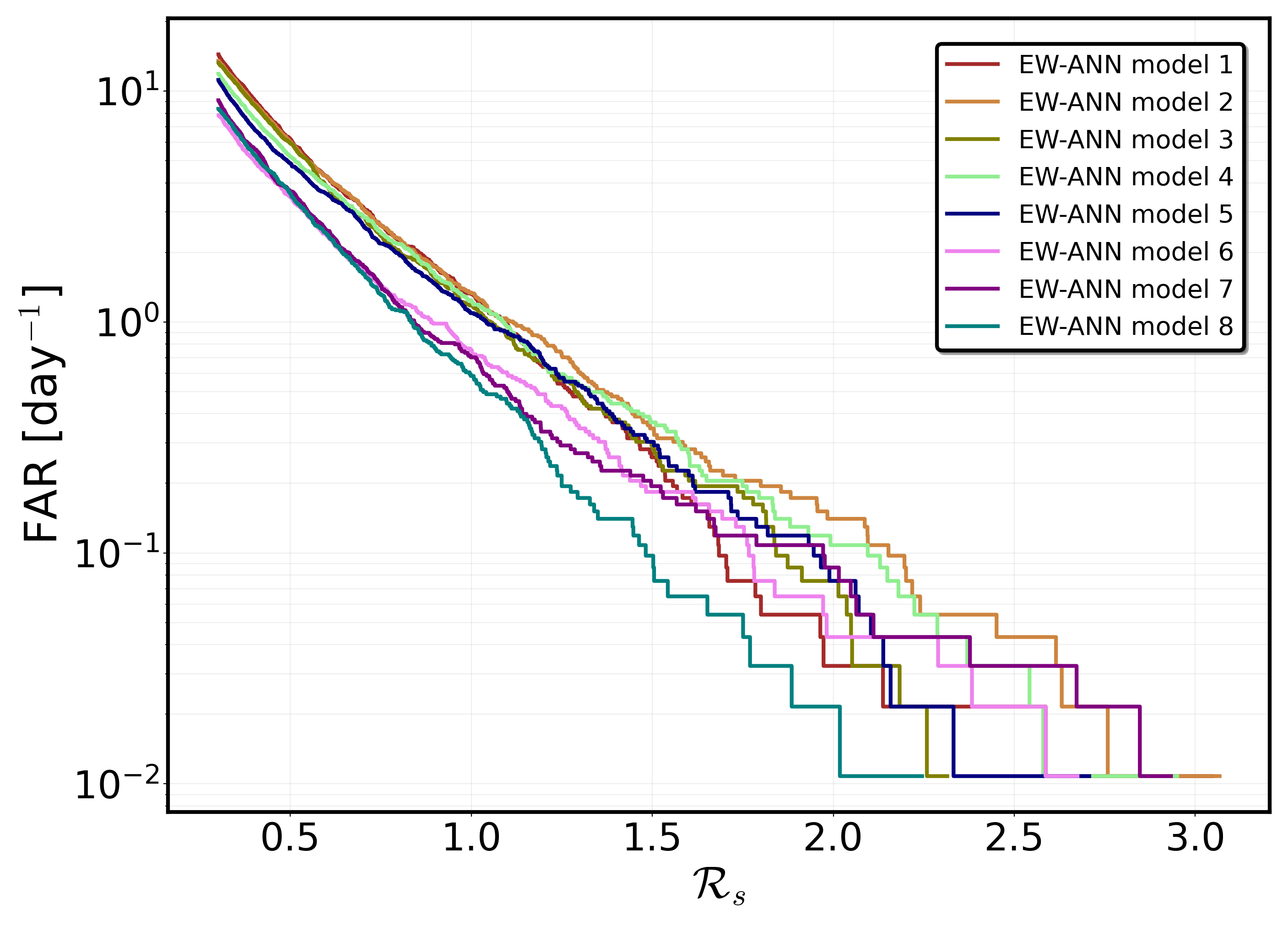}}
    \subfloat[$\mathrm{FAR}$ vs $\mathrm{R}_s$ for NSBH EW-ANN models]{%
        \includegraphics[width=0.50\textwidth]{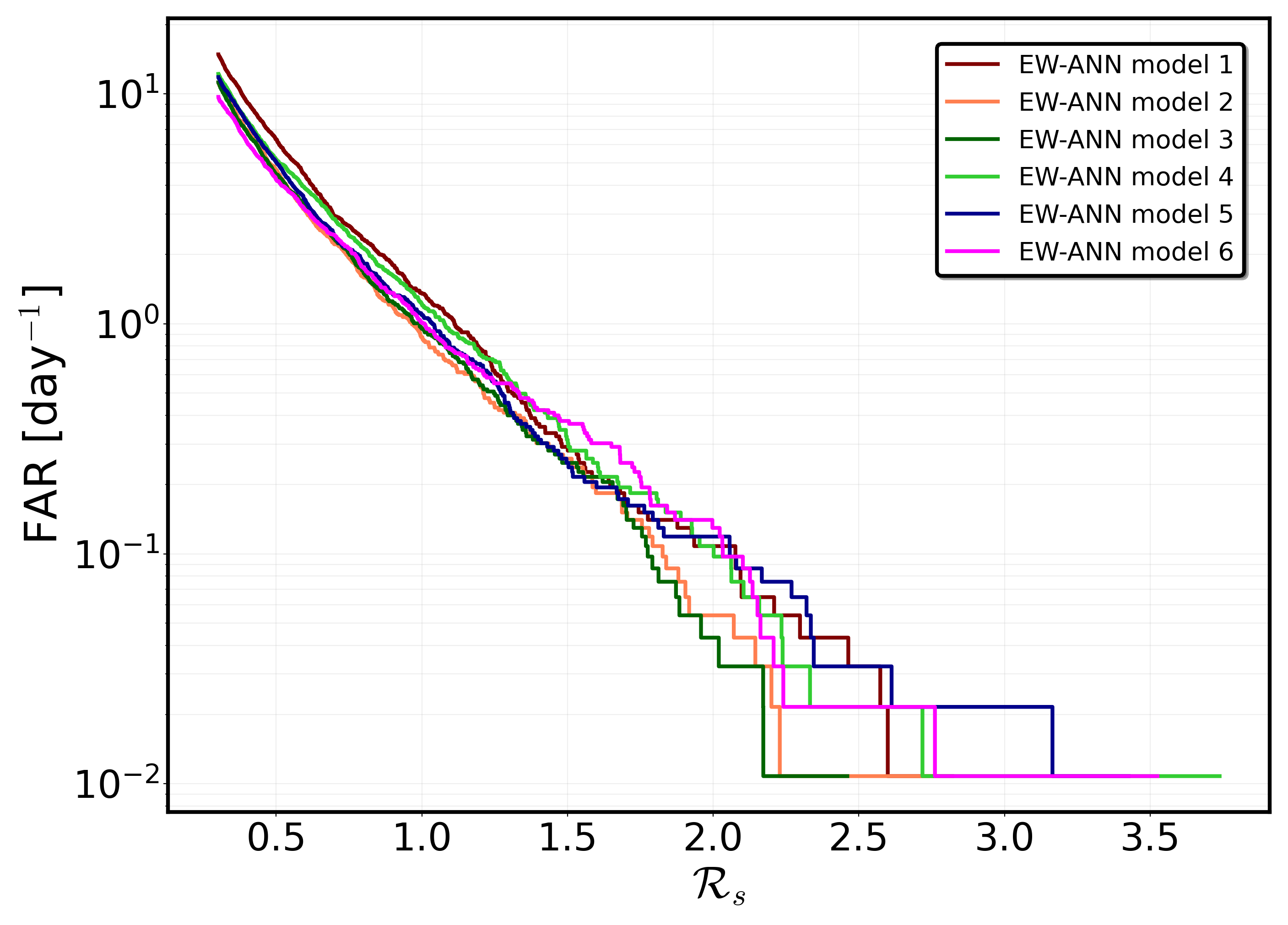}
        \label{fig:CEH_FAR_NSBH}}
    \caption{\label{fig:CEH_FAR_distributions} 
    CE1 detector: False-alarm rate as a function of the ranking statistic $\mathcal{R}_s$. Each curve corresponds to a different EW-ANN model evaluated on background-only test data. The left panel corresponds to the BNS ANN models, whereas the right panel shows the NSBH models. At high $\mathcal{R}_s$ thresholds, the curves become step-like and approach a finite-duration floor set by the available background livetime.}
    %CE1 detector: False-alarm probability $\mathrm{FAP}_t$ as a function of the ranking-statistic $\mathcal{R}_s$. Each curve corresponds to a different EW-ANN model evaluated on background-only test data. Left panel: BNS classifiers. Right: NSBH classifiers. At high thresholds, the curves exhibit step-like behavior and approach a finite-duration floor set by the available background livetime ($\sim 1/T_{\mathrm{obs}}$).}
   %The optimal SNR distribution related to the time of merger $t_c$ is also shown in black. 
\end{figure*}

%%%%%%%%%%%%%%%%%%%%%%%%%%%%%%%%%%%%%%%%%%
In our investigation, $\mathrm{FAR}$ for each EW-ANN is estimated directly from the corresponding background-only test data. Specifically, we use $2\times10^{5}$ non-overlapping OTWs of duration $40\,\mathrm{s}$, corresponding to a total background livetime of $T_{\mathrm{obs}} \simeq 8\times10^{6}\,\mathrm{s} \approx 92.6\,\mathrm{days}$ for each test dataset. For a given threshold on the ranking statistic, we compute the cumulative $\mathrm{FAR}(>\mathcal{R}_s)$ by counting the number of background windows with values $\ge \mathcal{R}_s$ and dividing by the total observational duration $T_{\mathrm{obs}}$. Figure \ref{fig:CEH_FAR_distributions} shows the resulting $\mathrm{FAR}$ curves as a function of $\mathcal{R}_s$ for the CE1 EW-ANNs, for BNS (left panel) and NSBH (right panel) early-warning classification models; corresponding results for the ET1, ET2, and CE2 configurations are provided in Appendix \ref{app: FAP_t_for_other_det_cases}. Because third-generation observatories are expected to yield very high detection rates per day \cite{iacovelli2022forecasting, kalogera2021next, gupta2023characterizing}, we report the false-alarm rate in units of $\mathrm{day}^{-1}$ (rather than, e.g., per month) to provide an operationally relevant scale for low-latency early-warning alerts. In the regime relevant for our analysis, the inferred false-alarm rates for representative high-confidence triggers remain low, typically $\lesssim \mathcal{O}(10^{-2}-10^{-1})\,\mathrm{day}^{-1}$, indicating that spurious early-warning notifications are expected to be rare even in the presence of frequent true alerts.

For the CE1 BNS case (left panel of Fig. \ref{fig:CEH_FAR_distributions}), $\mathrm{FAR}$ decreases steeply and monotonically with increasing $\mathcal{R}_s$, spanning more than three orders of magnitude across the plotted range. In particular, requiring $\mathcal{R}_s \gtrsim 2$ yields $\mathrm{FAR} \lesssim \mathcal{O}(10^{-2}\text{--}10^{-1})\,\mathrm{day}^{-1}$ for all models. At intermediate-to-high thresholds ($\mathcal{R}_s \gtrsim 1.3$), the spread among the different EW-ANN models becomes more apparent: while the curves are largely consistent at low $\mathcal{R}_s$, $\mathrm{FAR}$ can differ by up to about an order of magnitude between models at fixed $\mathcal{R}_s$. In particular, the later models in the sequence (especially model~8, and to a lesser extent models~6-7) tend to yield lower $\mathrm{FAR}$ in the high-$\mathcal{R}_s$ tail, indicating stronger background rejection. In contrast, some earlier models retain comparatively higher $\mathrm{FAR}$ at the same threshold. Consequently, for a fixed false-alarm requirement (e.g., $\mathrm{FAR} \lesssim 10^{-1}\,\mathrm{day}^{-1}$), the corresponding $\mathcal{R}_s$ threshold varies modestly across models. At the highest thresholds, the curves exhibit the expected step-like behavior and approach a floor near $ 10^{-2}\,\mathrm{day}^{-1}$, set by the employed background duration.

Furthermore, the NSBH CE1 classifiers (right panel of Fig. \ref{fig:CEH_FAR_distributions}) show the same overall trend: $\mathrm{FAR}$ decreases with increasing $\mathcal{R}_s$, with only modest variation among the different EW-ANN models. Already at $\mathcal{R}_s \simeq 1$, $\mathrm{FAR}$ is reduced to the $\mathcal{O}(1)\,\mathrm{day}^{-1}$ level, while thresholds $\mathcal{R}_s \gtrsim 2$ suppress spurious triggers to $\lesssim \mathcal{O}(10^{-1})\,\mathrm{day}^{-1}$ across all models. The inter-model spread is most evident at intermediate thresholds ($\mathcal{R}_s \simeq 1.5-2.2$), whereas it becomes negligible once the curves approach the background-limited floor. As for BNS, the curves become step-like at high thresholds and approach the livetime-limited floor near $\sim 10^{-2}\,\mathrm{day}^{-1}$; small differences in the $\mathcal{R}_s$ value at which this floor is reached indicate mildly improved background rejection for some EW-ANN models.

To approximate more realistic low-latency operating conditions, we impose a persistence requirement: the early alert is issued only if at least one individual EW-ANN model (i.e., the same classifier) produces $k=5$ consecutive above-threshold triggers with $\mathcal{R}_s>1$. In the sliding-window analysis with the $4\,\mathrm{s}$ stride that was discussed previously, each ANN evaluation uses as input standardized features extracted from a $40\,\mathrm{s}$ OTW. Successive OTWs are therefore strongly overlapping, and the resulting $\mathcal{R}_s$ values are temporally correlated; in particular, consecutive triggers should not be interpreted as statistically independent events, but rather as clustered triggers associated with a single underlying GW signal. Temporal correlations introduced by overlapping windows primarily affect trigger clustering, which is mitigated by the adopted persistence requirement. In this framework, the background-derived cumulative $\mathrm{FAR}(>\mathcal{R}_s)$ curves remain directly applicable as an estimate of the expected rate of above-threshold background exceedances. Moreover, since $\mathrm{FAR}$ is defined per unit time (rather than per window), it provides a portable calibration that can be mapped to a sliding-window setting to quantify the expected incidence of spurious triggers for a given $\mathcal{R}_s$ threshold.

For the BNS injection shown in Fig \ref{fig:simulated_BNS_example_and_Ranking_statistic}, EW-ANN model 1 is the first classifier to produce $k=5$ consecutive triggers with $\mathcal{R}_s>1$. We therefore evaluate $\mathrm{FAR}$ for the specific model that issues the alert, while subsequent high-confidence triggers from the remaining models provide a consistency check. Using the corresponding background-derived $\mathrm{FAR}$ curve, we estimate the false-alarm rate associated with this persistence rule by mapping the average ranking statistic $\langle \mathcal{R}_s\rangle$ of the five-trigger sequence to $\mathrm{FAR}$. For this event, the alert condition is satisfied with $\langle \mathcal{R}_s\rangle = 1.7$ over the five consecutive triggers, which corresponds to $\mathrm{FAR} = 7.56\times10^{-2}\,\mathrm{day}^{-1}$. In this simulated BNS case, the resulting early-warning notification should be issued $316\,\mathrm{s}$ before the merger. In addition, for the NSBH example shown in Fig. \ref{fig:simulated_NSBH_example_and_Ranking_statistic}, four EW-ANN models (models 2-5) satisfy the persistence criterion, each producing $k=5$ consecutive triggers within a time span of $16\,\mathrm{s}$. In this case, each of the models 2-5 would independently issue an early-warning alert, and we interpret this simultaneous set of alerts as a single joint early-warning notification. For each reported triggering model, we compute the average ranking statistic $\langle \mathcal{R}_s\rangle$ over the five-trigger sequence, and map it to the corresponding background-derived $\mathrm{FAR}$ curve. We then summarize the joint notification by taking the mean $\mathrm{FAR}$ across the four triggering ANN models, obtaining $\langle \mathrm{FAR}\rangle = 2.43\times10^{-2}\,\mathrm{day}^{-1}$ for this event. In this case, the early-alert notification should be issued $200\,\mathrm{s}$ before the merger.

%%%%%%%%%%%%%%%%%%%%%%%%%%%%%%%%%%%%%%%%%%%
Finally, we have to note that the $\mathrm{FAR}$ values inferred here primarily reflect idealized stationary Gaussian-noise conditions and should therefore be interpreted as baseline estimates. While this setting enables controlled comparisons between detector configurations and EW-ANN models, realistic third-generation data will also contain non-Gaussian transients and non-stationary noise artifacts that can increase the effective false-alarm probability trigger rate. Accordingly, we place primary emphasis on standard classification measures, which provide a robust assessment of foreground-background separability within our simulated datasets. Extending the present $\mathrm{FAR}$ analysis to more realistic noise realizations (or to ET/CE-like mock data when available) will be an important step toward translating these baseline curves into operational false-alarm rate predictions.

%\change{Finally, we have to note that any $\mathrm{FAP}_t$ value inferred from such simulations would mainly reflect the idealized Gaussian noise conditions and would therefore underestimate the false-alarm trigger rate expected in real third-generation data.} 
%For this reason, we restrict our quantitative assessment to standard classification evaluation measures, and a physically meaningful estimation of $\mathrm{FAP}_t$ is left for future work based on more realistic mock data or actual ET/CE-like detector noise.

\section{\label{sec:summary_conclusions}Summary and conclusions}

In this work, we have presented \texttt{GW-FALCON} (\texttt{G}ravitational-\texttt{W}ave Feature-based deep-learning \texttt{A}pproach for \texttt{L}ow-latency \texttt{C}lassificati\texttt{ON}), a novel feature-driven deep-learning framework for early-time discrimination between GW signal+noise and noise-only data in third-generation ground-based detectors, focusing on the Einstein Telescope (ET) and Cosmic Explorer (CE) configurations \cite{hild2011sensitivity,punturo2010einstein,abbott2017exploring,maggiore2020science,abac2025science,reitze2019cosmic,branchesi2023science,evans2021horizon}. Motivated by the prospects for systematic early-warning alerts of BNS and NSBH mergers that are expected in these observatories \cite{li2022exploring,hu2023rapid,abbott2020guide,sachdev2020early,tsutsui2022early,chan2018binary}, and by the demonstrated importance of such alerts for multimessenger astronomy following GW170817 \cite{cannon2012toward,abbott2017gw170817,coulter2017swope,Abbot2017multi,abbott2017gravitational,cowperthwaite2017electromagnetic,soares2017electromagnetic,nicholl2017electromagnetic,margutti2017electromagnetic,chornock2017electromagnetic,abbott2019properties}, we have explored an alternative to end-to-end time series deep-learning approaches \cite{cuoco2020enhancing,benedetto2023ai,zhao2023dawning,cuoco2025applications}, based on supervised classification in a suitably engineered feature space. To the best of our knowledge, this is the first work to develop a feature-based deep-learning framework for GW signal–noise discrimination in third-generation detectors, explicitly linking the detector's strain data feature extraction to robust early-warning classification.

In this proof-of-concept study, we constructed foreground and background datasets by injecting simulated BNS and NSBH inspirals into colored Gaussian noise generated from the ET-D and wideband CE design power spectral densities \cite{hild2011sensitivity,maggiore2020science,abac2025science,abbott2017exploring,reitze2019cosmic,evans2021horizon}. For each detector configuration considered, we defined a sequence of partial-inspiral observational time windows (OTWs) characterized by different maximum instantaneous frequencies $f_{\max}$, which serve as the basis for constructing the supervised-learning datasets. In total, we generated large ensembles of simulated signals for each source class and detector configuration, producing $10^6$ injections per detector for both BNS and NSBH configurations. For each detector–dataset pair, we then constructed balanced supervised-learning sets by combining $10^6$ foreground examples (OTWs containing an injected signal plus noise) with $10^6$ background examples (noise-only OTWs), as detailed in Sec. \ref{sec:datasets_feature_extraction}.

Rather than feeding long stretches of raw strain data directly to convolutional or other complex neural-network architectures, in the \texttt{GW-FALCON} framework, we map each observational time window (OTW) of detector strain into a high-dimensional feature vector using the Time Series Feature Extraction Library (TSFEL) \cite{barandas2020tsfel}. The resulting statistical, temporal, and spectral quantities are collected into fixed-length feature vectors, which define the supervised-learning datasets for each detector configuration and for each choice of maximum instantaneous frequency $f_{\max}$. These feature-based datasets are then standardized and split into training and test subsets using the 80:20 partition described in Sec. \ref{sec:GWFDA_framework}. In this form, the feature vectors serve as inputs to compact feed-forward artificial neural networks (ANNs) trained to discriminate between early-inspiral GW signal+noise and pure noise examples, with a dedicated EW-ANN trained on the corresponding feature set for each detector configuration and dataset. 
%In online operation, the feasibility of early-warning triggering is additionally conditioned by the per-window processing and feature-extraction latency, which sets the minimum update step of a sliding-window implementation (see the discussion in Sec.~\ref{sec:results_discussion}).

Across all detector and dataset combinations constructed, the resulting ANN classifiers exhibit high classification performance as quantified by standard machine-learning evaluation measures, computed from the associated confusion matrices \cite{bishop2006pattern,goodfellow2016deep,lecun2015deep,prince2023understanding} and summarized in Tables \ref{tab:detector_dataset_metrics} and \ref{tab:detector_dataset_metrics_II}. For the ET-based configurations, test-set accuracies for BNS and NSBH typically lie around $90\%$, with detection efficiency values in the mid-$80\%$ range and false-alarm probabilities of only a few percent. For the CE-based configurations considered, the performance is even closer to ideal: test-set accuracies exceed $97\%$, recall values cluster around $95$-$96\%$, and the reported false-alarm probabilities (FAP) and rates (FPR) are well below the percent level for both source classes. These trends are reflected directly in the confusion matrices, where the diagonal entries (true positives and true negatives) strongly dominate over the off-diagonal ones, particularly for CE1 and CE2, demonstrating robust separation between signal+noise and noise-only examples at the level of individual predictions.

%In a realistic low-latency search, the EW-ANN output would naturally be treated as a ranking statistic whose operating point is set by background behavior, analogous to how matched-filter pipelines report candidate significance through a false-alarm rate (FAR) per unit time and, increasingly, an astrophysical probability $p_{\mathrm{astro}}$ \cite{messick2017analysis,usman2016pycbc,aubin2021mbta,farr2015counting}. In this setting, the EW-ANN score threshold could be calibrated on extended noise-only streams (and, in network operation, with coincidence requirements) to achieve a target FAR while maximizing detection efficiency, enabling the \texttt{GW-FALCON} framework to serve as an online potential trigger process that is interoperable with the established low-latency pipelines \cite{sachdev2020early, sachdev2019gstlal,biwer2019pycbc,hanna2020fast,adams2016low,aubin2021mbta,chu2022spiir, hanna2020fast, cannon2021gstlal, nitz2018rapid,  usman2016pycbc, nitz2017detecting,davies2020extending, nitz2022dfinstad, kumar2024optimized,klimenko2016method, drago2021coherent, klimenko2021cwb, klimenko2011localization}.

By examining the detection efficiency as a function of the partial-inspiral signal-to-noise ratio (PI SNR), we have further characterized how the classifiers respond as the GW signal accumulates in band. For both BNS and NSBH populations employed, the detection efficiency curves display the expected sigmoidal behavior: they are close to zero at very low mean PI SNR, rise steeply through an intermediate regime as the signal becomes more prominent in the OTWs, and saturate near perfect detection once the mean PI SNR is sufficiently large. ANN models for CE-based detectors consistently reach high efficiencies at lower PI SNR than ET-like ones, with narrower uncertainty bands, in line with their superior strain data noise sensitivity and the more informative feature representations available to the corresponding EW-ANNs \cite{hild2011sensitivity,maggiore2020science,abac2025science,abbott2017exploring,reitze2019cosmic,evans2021horizon}. We have shown that these efficiency curves are quantitatively consistent with the PI SNR distributions of the injected waveform portions, and that the relative behavior of the different EW-ANN models within each detector configuration can be interpreted in terms of how the PI SNR content of the datasets shifts as we move from earlier to later OTWs that capture a progressively larger fraction of the inspiral.

Within this setup, the partial-inspiral windows are chosen such that, for the most favorable BNS systems and the highest-frequency OTWs, the corresponding signal portions can lie from several tens up to a few hundred seconds before merger, depending on the detector and source parameters \cite{li2022exploring,hu2023rapid,abbott2020guide}. Our results, therefore, indicate that, in such cases, the suggested ANNs are capable of delivering robust pre-merger triggers in time to support rapid electromagnetic follow-up of BNS events \cite{hild2011sensitivity,hu2023rapid,maggiore2020science,abbott2020guide,abac2025science, sachdev2020early,tsutsui2022early,nitz2020gravitational,nitz2018rapid,chan2018binary,li2022exploring,kovalam2022early,magee2022observing,chaudhary2024low,kang2022electromagnetic,banerjee2023pre}. Because NSBH systems typically spend less time in band than BNSs, the associated advance-warning times are shorter. Nevertheless, for the configurations investigated here, the trained NSBH EW-ANNs can still provide reliable early-warning alerts tens of seconds before coalescence in the most favorable scenarios, particularly for CE-like sensitivities.

In addition, in a realistic low-latency search, the EW-ANN output would naturally be treated as a ranking statistic whose operating point is set by background behavior, analogous to how matched-filter pipelines report candidate significance through a false-alarm rate (FAR) per unit time and, increasingly, an astrophysical probability $p_{\mathrm{astro}}$ \cite{messick2017analysis,usman2016pycbc,aubin2021mbta,allene2025mbta,farr2015counting}. In this setting, the EW-ANN score threshold could be calibrated on extended noise-only streams (and, in network operation, with coincidence requirements) to achieve a target FAR while maximizing detection efficiency, enabling the \texttt{GW-FALCON} framework to serve as an online potential trigger process that is interoperable with the established low-latency pipelines \cite{sachdev2020early, sachdev2019gstlal,biwer2019pycbc,hanna2020fast,adams2016low,aubin2021mbta,chu2022spiir, hanna2020fast, cannon2021gstlal, nitz2018rapid,  usman2016pycbc, nitz2017detecting,davies2020extending, nitz2022dfinstad, kumar2024optimized,klimenko2016method, drago2021coherent, klimenko2021cwb, klimenko2011localization}.

Taken together, these findings demonstrate that feature-based deep-learning methods can provide a viable and competitive complement to more complex end-to-end approaches for GW detection \cite{cuoco2020enhancing,benedetto2023ai,zhao2023dawning,cuoco2025applications}. While the present analysis is dedicated to simulated inspirals with simplified source modeling choices (e.g., the restricted spin/tidal assumptions described in Sec. \ref{sec:methodology}) in Gaussian noise and to relatively simple feed-forward architectures, the structure of the \texttt{GW-FALCON} pipeline---from feature extraction and standardization to supervised classification---is modular and readily extensible to more realistic data, multi-detector networks, and richer neural-network architectures. Future work will include the incorporation of non-Gaussian and non-stationary noise, the exploration of multi-class setups that distinguish between different source types, and the integration of local, feature-level interpretability tools to quantify how individual features contribute to the ANN decisions, building on recent interpretable deep-learning studies of rapidly rotating NSs \cite{papigkiotis2025assessing}. Ultimately, such extensions will be crucial to deploy robust, low-latency early-warning pipelines within the data-analysis frameworks of ET and CE \cite{messick2017analysis,sachdev2019gstlal,biwer2019pycbc}.

%Beyond single-detector demonstrations, the strongest early-warning returns will come from network deployment, where feature vectors from multiple instruments can be combined to exploit independent antenna responses and noise realizations, reducing susceptibility to single-detector artifacts while increasing effective network sensitivity. In third-generation networks, this is also directly tied to multimessenger utility: improved geographical coverage and detector multiplicity can significantly sharpen sky localization, so that early-warning triggers become more actionable for electromagnetic follow-up campaigns \cite{branchesi2023science,abac2025science,evans2021horizon,cannon2012toward,sachdev2020early}.

Furthermore, while our results focus on single-detector classifiers, the suggested methodology naturally extends to network operation in ET-, CE-, or joint ET+CE configurations, where detector-specific feature vectors can be concatenated into a unified classifier input.  In that setting, early-warning classification can benefit from multiple, independent views of the same signal, improving robustness near threshold and reducing vulnerability to single-detector noise artifacts, while remaining computationally lightweight and therefore compatible with hierarchical low-latency workflows \cite{cannon2012toward,sachdev2020early,messick2017analysis,sachdev2019gstlal,biwer2019pycbc}. Crucially, the scientific value of such networked early-warning triggers extends beyond detection: the sky-localization capability of third-generation networks improves markedly with detector number and geographic distribution \cite{branchesi2023science,abac2025science,evans2021horizon}, so combining higher-confidence premerger triggers with network information is expected to yield smaller early-warning localization regions and more actionable targets for electromagnetic follow-up—thereby strengthening the prospects for prompt counterpart identification and multimessenger astronomy.

Finally, in parallel with the present investigation focused on third-generation observatories, we are developing extensions of the proposed framework for the upgraded ground-based interferometers of the LVK network, including Advanced LIGO A+ \cite{LIGO_A_plus} and the planned Virgo NExt configuration \cite{garufi2024advanced,abac2025gwtc}. These studies aim to quantify the performance of the feature-based DL approach for signal detection under the anticipated noise characteristics of these facilities and to investigate whether a unified analysis pipeline can support both low-latency early-warning alerts and offline searches, including searches for BBH mergers. In this context, we are benchmarking the methodology on more realistic, non-Gaussian noise datasets and plan to apply the \texttt{GW-FALCON} framework to archival LVK observations from previous observing runs. A full exploration of these extensions, including detailed multi-detector training strategies and the impact of realistic detector glitches and data-quality issues, will be presented in future work.

%To the best of our knowledge, we introduce the first comprehensive feature-based gravitational-wave discrimination framework, providing an end-to-end pipeline from feature extraction to robust signal–noise classification.

%Taken together, these elements define a comprehensive feature-based gravitational-wave discrimination pipeline, from feature extraction to robust signal–noise classification, which is modular and readily extensible to real data and more advanced classifier architectures.

%However, detections of the early inspiral phase by the gravitational wave detectors would allow the observation of the earlier stages of the merger in the electromagnetic band, improving multimessenger astronomy and giving access to new information.

%Within this setup it is possible to produce an early alert from tens to hundreds seconds before the merger for the best-case scenario.

\section*{\label{sec:acknowledgements} Acknowledgements}
We are grateful to Melissa Lopez and Reem ALfaidi for a careful reading of the manuscript and for comments that improved it. G. P.
wants to thank Christos Paschalidis for useful discussions. G.~P. acknowledges financial support from the Hellenic Foundation for Research and Innovation (H.F.R.I.) under the 5th Call for H.F.R.I. PhD Fellowships (Fellowship No.~20450), as well as partial funding from the project “3rd Call for H.F.R.I. Research Projects to support Faculty Members and Researchers” (Project No.~26254). G.~V. and N.~S. also acknowledge funding from the same H.F.R.I. project (Project No.~26254). In addition, this publication is part of a project that has received funding from the European Union’s Horizon Europe Research and Innovation Programme under Grant Agreement No 101131928. The data collection and analysis were carried out at the Department of Physics, Aristotle University of Thessaloniki (AUTh), Greece. Training of the proposed deep-learning models was performed on the Department’s computational facilities. We gratefully acknowledge the AUTh IT Center for providing access to the university’s High-Performance Computing infrastructure, on which the results of this work were obtained. Virgo is funded, through the European Gravitational Observatory (EGO), by the French Centre National de Recherche Scientifique (CNRS), the Italian Istituto Nazionale di Fisica Nucleare (INFN) and the Dutch Nikhef, with contributions by institutions from Belgium, Germany, Greece, Hungary, Ireland, Japan, Monaco, Poland, Portugal, Spain. KAGRA is supported by Ministry of Education, Culture, Sports, Science and Technology (MEXT), Japan Society for the Promotion of Science (JSPS) in Japan; National Research Foundation (NRF) and Ministry of Science and ICT (MSIT) in Korea; Academia Sinica (AS) and National Science and Technology Council (NSTC) in Taiwan.

%Finally, we are grateful to the anonymous referee for their insightful comments and suggestions that improved the final version of this manuscript.

\section*{Data Availability}
The BNS and NSBH datasets, the derived foreground/background feature representations, and the trained ANN weights that support the findings of this study are not publicly available. The trained model parameters $\theta^\star$ can be obtained from the authors upon reasonable request.

\appendix

\section{\label{app:PI_SNR_distributions}PI SNR Data distributions}
\begin{figure*}[!thb]
    \centering
    \subfloat[\label{fig:ET_triangle_PI_snr_BNS} ET1 Detector: PI SNR histograms for BNS signals. ]{%    
    \includegraphics[width=0.50\textwidth]{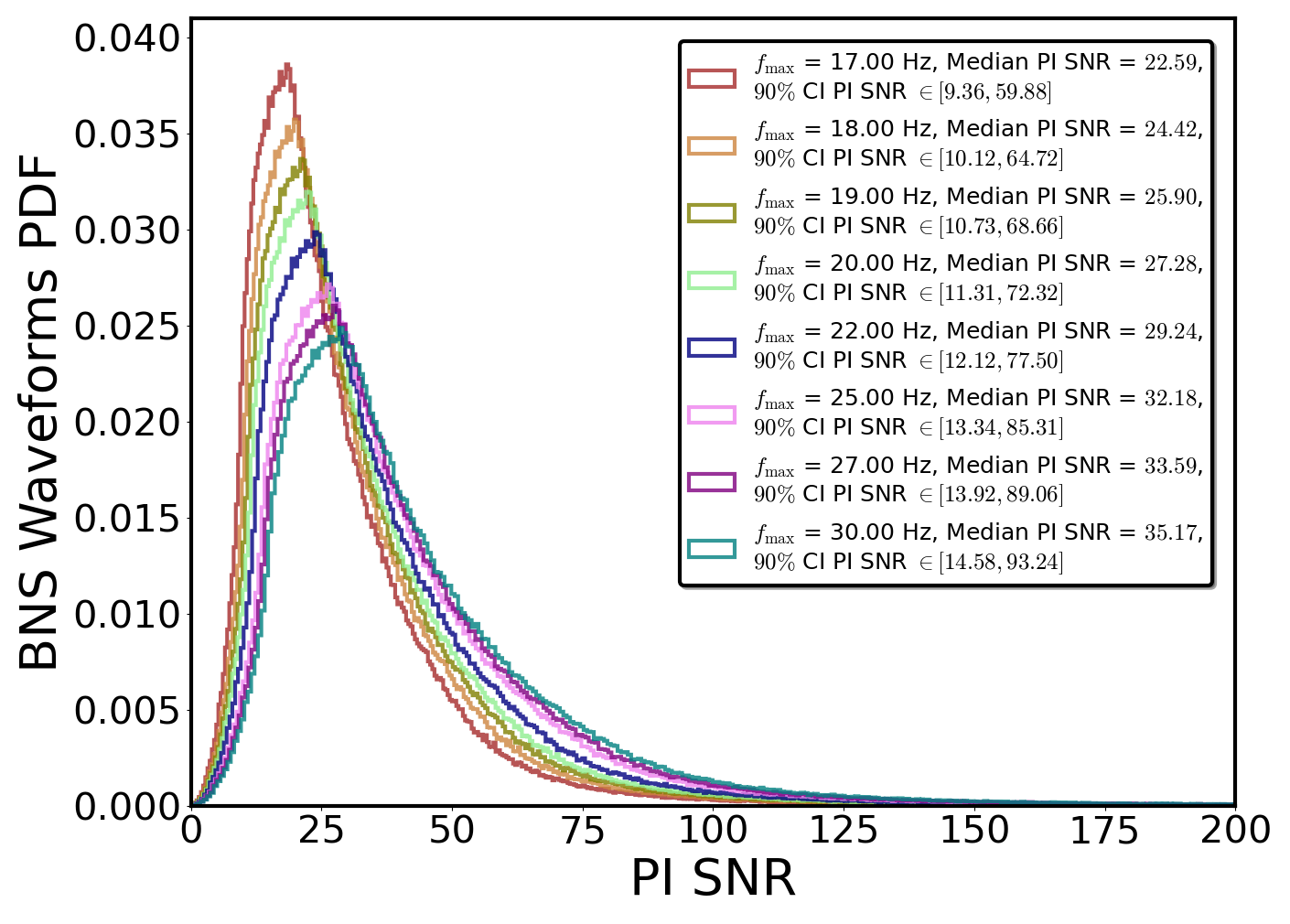}}
    \subfloat[ET1 Detector: PI SNR histograms for NSBH signals.]{%
    \includegraphics[width=0.50\textwidth]{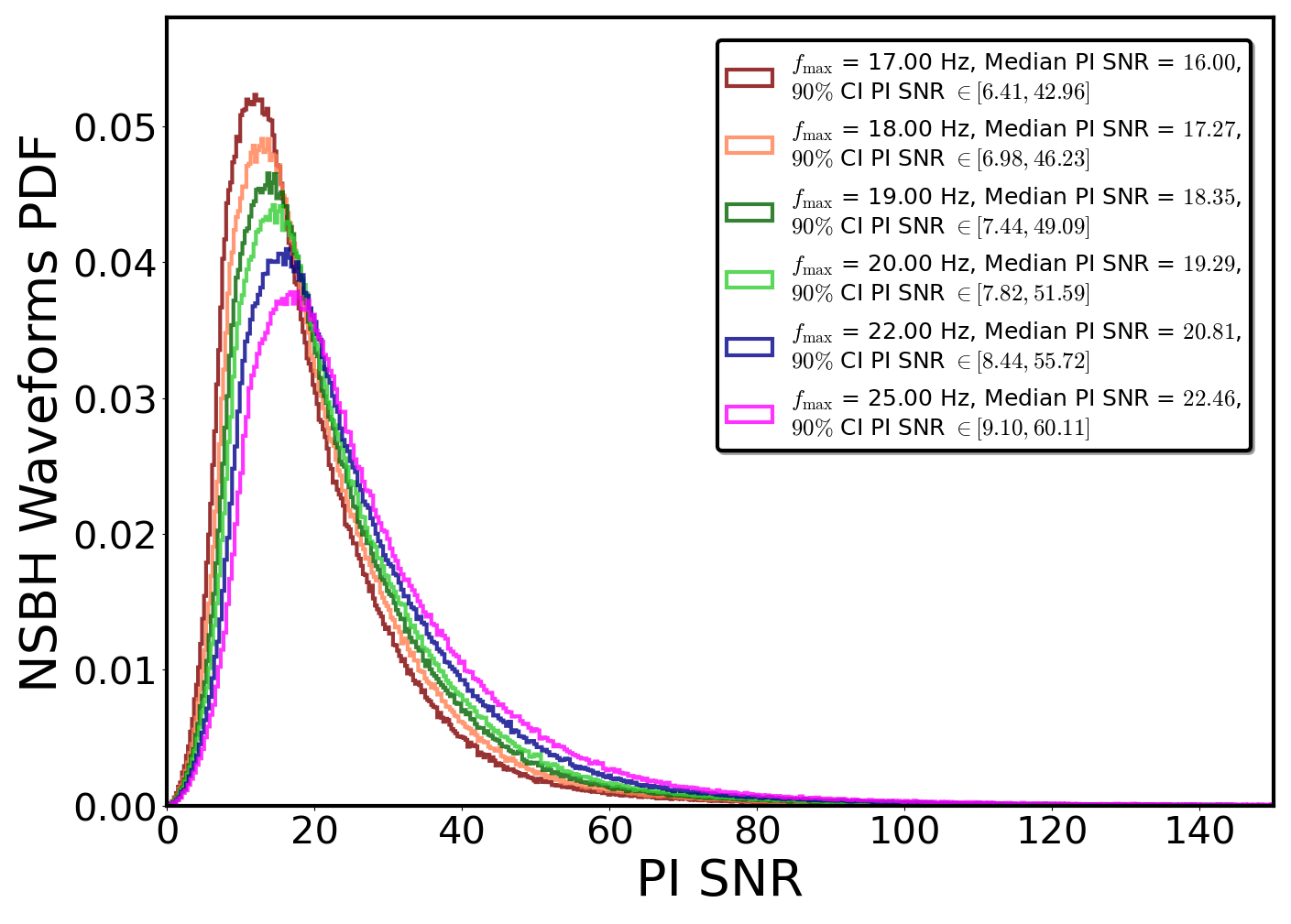}
    \label{fig:ET_triangle_PI_snr_NSBH}}\\
    \subfloat[\label{fig:ET_2L_45_PI_snr_BNS} ET2 Detector: PI SNR histograms for BNS signals. ]{%    
    \includegraphics[width=0.50\textwidth]{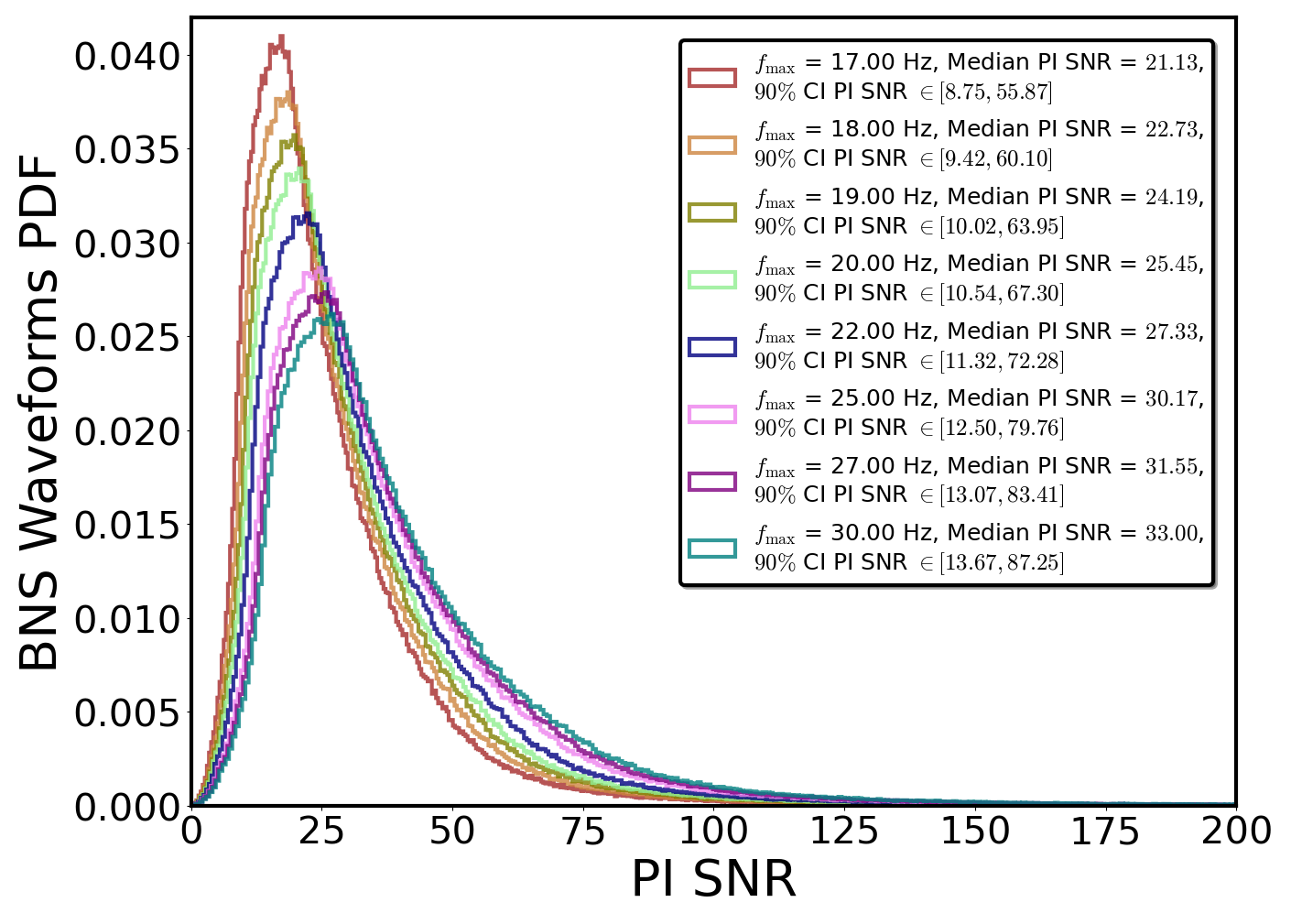}}
    \subfloat[ET1 Detector: PI SNR histograms for NSBH signals.]{%
    \includegraphics[width=0.50\textwidth]{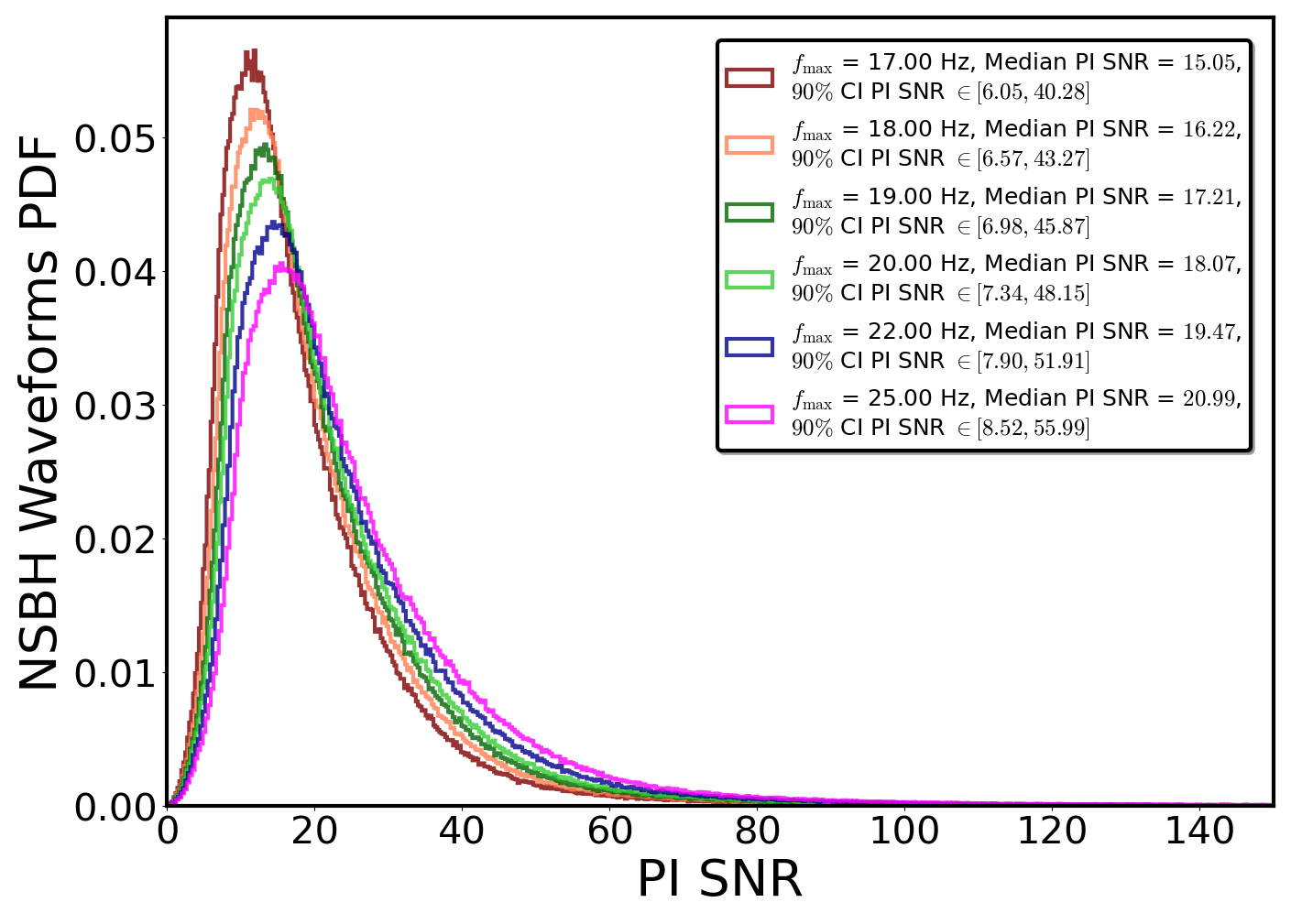}
    \label{fig:ET_2L_45_PI_snr_NSBH}}\\
    \subfloat[\label{fig:CEL_PI_snr_BNS} CE2 Detector: PI SNR histograms for BNS signals. ]{%    
    \includegraphics[width=0.50\textwidth]{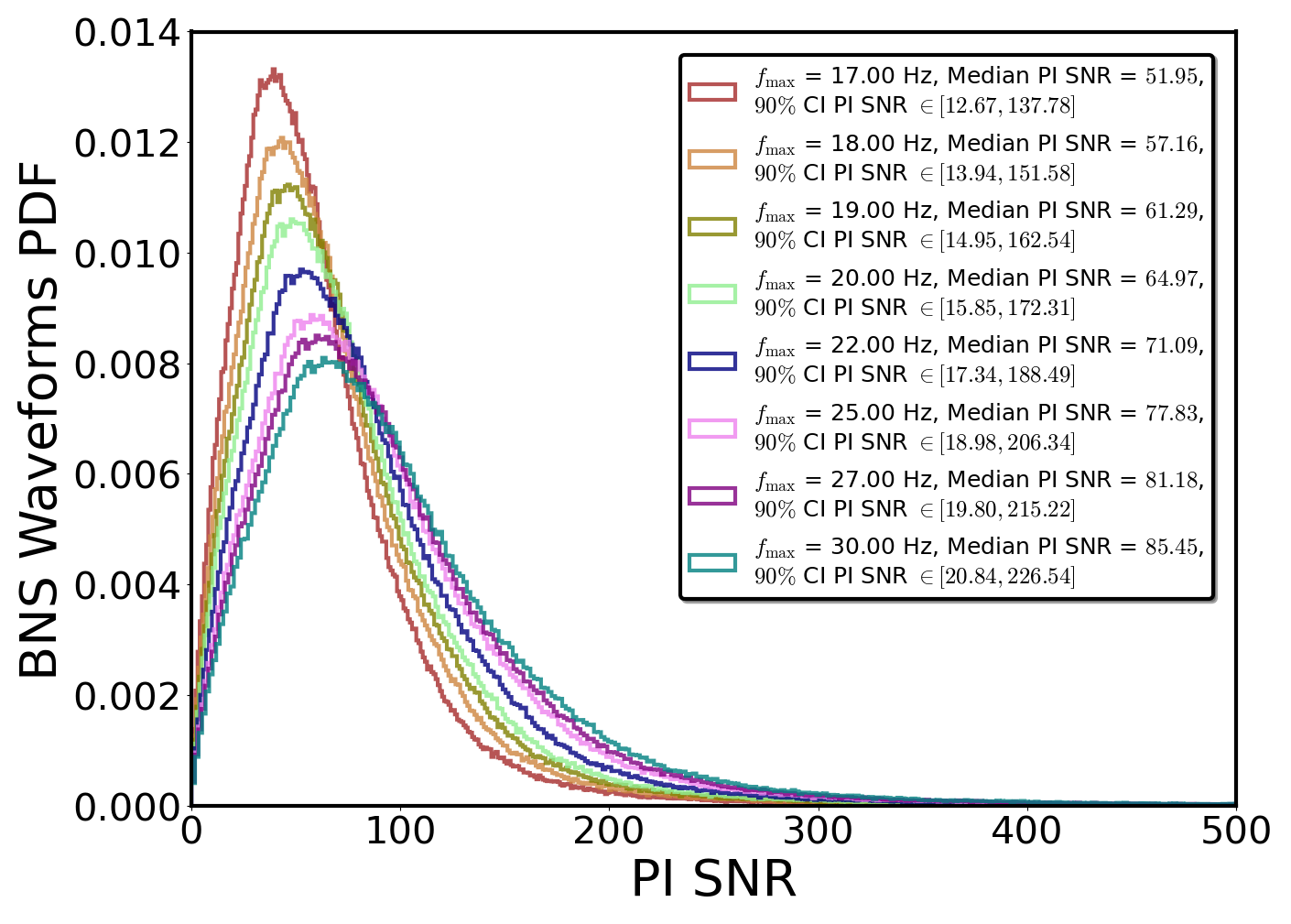}}
    \subfloat[CE2 Detector: PI SNR histograms for NSBH signals.]{%
    \includegraphics[width=0.50\textwidth]{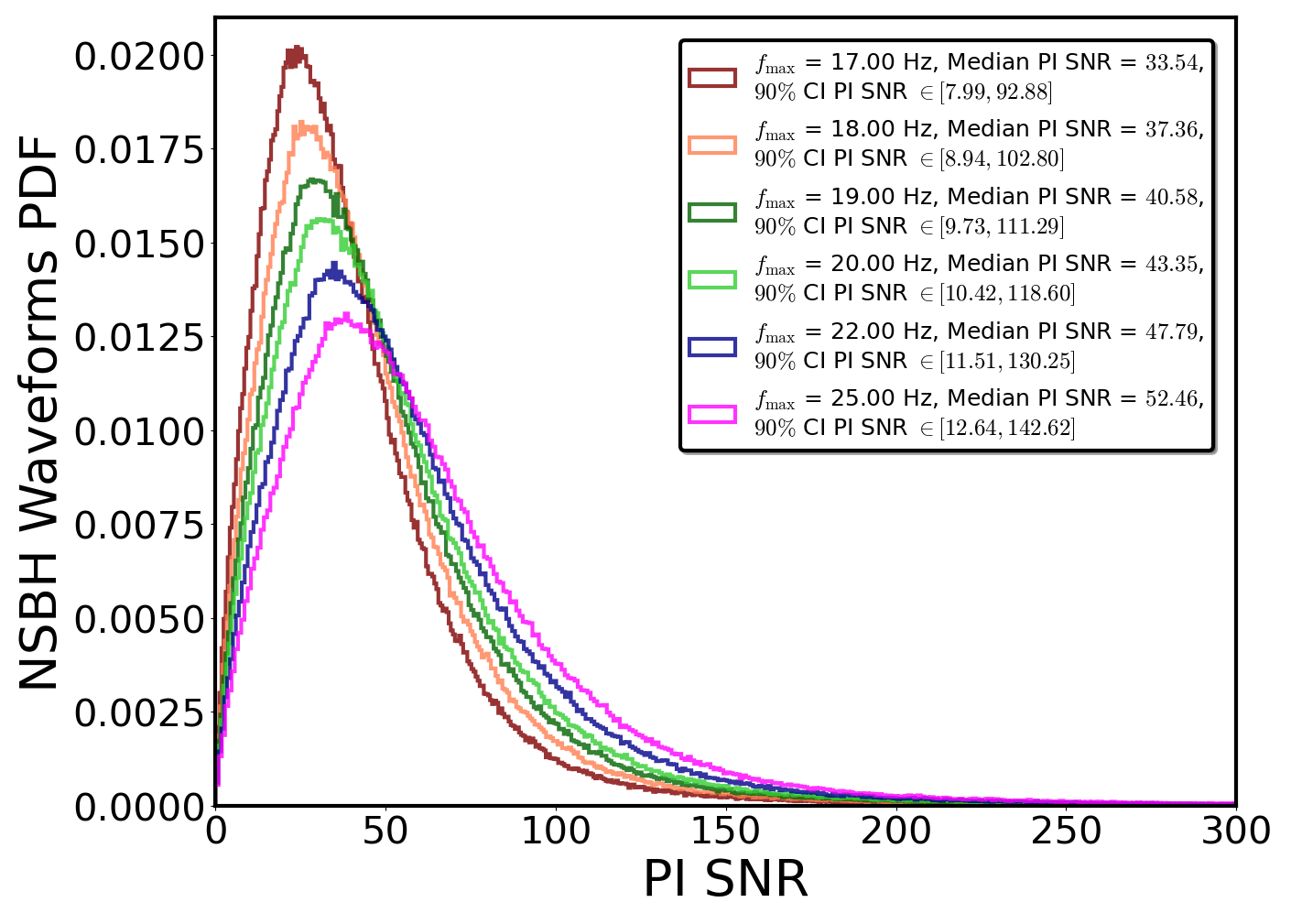}
    \label{fig:CEL_PI_snr_NSBH}}
    \caption{\label{fig:other_PI_SNR_distributions} PI SNR distributions for BNS (left column) and NSBH (right column) injections embedded in the employed detectors' strain data. Each histogram is color-coded by the annotated maximum instantaneous frequency $f_{\mathrm{max}}$. For each histogram, we report the median PI SNR along with the corresponding $90\%$ credible interval, which together characterize the central value and spread of the distribution, thus enabling a direct comparison of the typical signal strength and its variability across the different cases.
   %The optimal SNR distribution related to the time of merger $t_c$ is also shown in black. 
   }
\end{figure*}

In this section, Fig.~\ref{fig:other_PI_SNR_distributions} provides the PI SNR distributions for the BNS and NSBH injections associated with the ET1, ET2, and CE2 detectors discussed in the main text, as supplementary material.

%%%%%%%%%%%%%%%%%%%%%%%%%%%%%%%%%%%%%%%%%%%%%%%%
%%%%%%%%%%%%%%%%%%%%%%%%%%%%%%%%%%%%%%%%%%%%%%%%%%%

\section{\label{app:Ann_training_prop} Training properties of EW-ANN models}

In this subsection, we describe the training procedure for each ANN classifier that yields the final parameter sets $\theta^\star$ discussed in the main text. In total, we trained 56 models, covering all cases considered in this work. Table \ref{tab:model_train_specifics} summarizes the main training settings and the learning strategy adopted in each case. Specifically, it lists the batch size, the total number of training epochs, and the learning rate used.

\begin{table*}[!t]
\caption{\label{tab:model_train_specifics}
Training characteristics per EW-ANN model index. Each table entry is illustrated as $(B,\eta,E)$, where
$B$ is the batch size, $\eta$ the learning rate, and $E$ the number of epochs.}
\centering
\resizebox{\textwidth}{!}{%
\begin{tabular}{c| c|c c c c}
\toprule
\multirow{2}{*}{\textbf{Case}} &
\multirow{2}{*}{\textbf{EW-ANN Model}} &
\multicolumn{4}{c}{\textbf{Model Training Hyperparameters: $(B,\eta,E)$}} \\
\cmidrule(lr){3-6}
& & \textbf{ET1} & \textbf{ET2} & \textbf{CE1} & \textbf{CE2} \\
\midrule

\multirow{8}{*}{BNS}
& 1 & $(512,2\times10^{-4},80)$ & $(512,3\times10^{-4},100)$ & $(1024,2.5\times10^{-4},300)$ & $(1024,2.5\times10^{-4},300)$ \\
& 2 & $(512,2\times10^{-4},80)$ & $(512,3\times10^{-4},100)$ & $(1024,2.5\times10^{-4},300)$ & $(1024,2.5\times10^{-4},300)$ \\
& 3 & $(512,2\times10^{-4},80)$ & $(512,3\times10^{-4},100)$ & $(1024,2.5\times10^{-4},300)$ & $(1024,2.5\times10^{-4},300)$ \\
& 4 & $(512,2\times10^{-4},80)$ & $(512,3\times10^{-4},100)$ & $(1024,2.5\times10^{-4},300)$ & $(1024,2.5\times10^{-4},300)$ \\
& 5 & $(512,2\times10^{-4},200)$ & $(512,2\times10^{-4},300)$ & $(1024,2.5\times10^{-4},300)$ & $(1024,2.5\times10^{-4},300)$ \\
& 6 & $(512,2\times10^{-4},80)$ & $(512,2\times10^{-4},100)$ & $(1024,2.5\times10^{-4},300)$ & $(1024,2.5\times10^{-4},300)$ \\
& 7 & $(512,2\times10^{-4},200)$ & $(512,2\times10^{-4},100)$ & $(1024,2.5\times10^{-4},300)$ & $(1024,2.5\times10^{-4},300)$ \\
& 8 & $(512,2\times10^{-4},200)$ & $(512,3\times10^{-4},200)$ & $(1024,2.5\times10^{-4},300)$ & $(1024,2.5\times10^{-4},300)$ \\
\midrule

\multirow{6}{*}{NSBH}
& 1 & $(512,3\times10^{-4},80)$ & $(512,3\times10^{-4},100)$ & $(1024,2.5\times10^{-4},300)$ & $(1024,2\times10^{-4},300)$ \\
& 2 & $(512,3\times10^{-4},80)$ & $(512,3\times10^{-4},100)$ & $(1024,2.5\times10^{-4},300)$ & $(1024,2\times10^{-4},300)$ \\
& 3 & $(512,3\times10^{-4},80)$ & $(512,3\times10^{-4},100)$ & $(1024,2.5\times10^{-4},300)$ & $(1024,2\times10^{-4},300)$ \\
& 4 & $(512,3\times10^{-4},80)$ & $(512,3\times10^{-4},100)$ & $(1024,2.5\times10^{-4},300)$ & $(1024,2\times10^{-4},300)$ \\
& 5 & $(512,3\times10^{-4},80)$ & $(512,3\times10^{-4},100)$ & $(1024,2.5\times10^{-4},300)$ & $(1024,2\times10^{-4},300)$ \\
& 6 & $(512,3\times10^{-4},80)$ & $(512,3\times10^{-4},100)$ & $(1024,2.5\times10^{-4},300)$ & $(1024,2\times10^{-4},300)$ \\
\bottomrule
\end{tabular}%
}
\end{table*}

\section{\label{app:classification_eval_measures} Classification Evaluation Measures}

The standard classification measures \cite{goodfellow2016deep,lecun2015deep, bishop2006pattern, prince2023understanding} used in machine learning and adopted in this work to evaluate the performance and robustness of the ANN models on the training and test sets are presented in Table \ref{tab:classification_measures}.

\begin{widetext}

\begin{table}[!t]
  \caption{\label{tab:classification_measures}
  Machine learning classification measures that are used to evaluate the performance and robustness of the deployed ANN classifiers.}
  \begin{ruledtabular}
    \begin{tabular}{lll}
      \textbf{Evaluation Measure} & \textbf{Notation} & \textbf{Mathematical Definition} \\
      \hline
      &  &  \\[2pt]

      Accuracy & $\mathrm{ACC}$ &
      $\displaystyle \frac{TP + TN}{TP + TN + FP + FN}$ \\[6pt]

      Precision (positive predictive value) & $\mathrm{Prec}$ &
      $\displaystyle \frac{TP}{TP + FP}$ \\[6pt]

      Recall (true positive rate) & $\mathrm{Rec}$ &
      $\displaystyle \frac{TP}{TP + FN}$ \\[6pt]

      Specificity (true negative rate) & $\mathrm{Spec}$ &
      $\displaystyle \frac{TN}{TN + FP}$ \\[6pt]
      
      F1-score & $\mathrm{F1}$ &
      $\displaystyle 2\,\frac{\mathrm{Prec} \times \mathrm{Rec}}{\mathrm{Prec} + \mathrm{Rec}}$ \\[6pt]

      Matthews correlation coefficient & $\mathrm{MCC}$ &
      $\displaystyle \frac{TP \times TN - FP \times FN}{
      \sqrt{(TP + FP)(TP + FN)(TN + FP)(TN + FN)}}$ \\[6pt]

      False alarm probability & $\mathrm{FAP}$ &
      $\displaystyle \frac{FP}{FP + TN}$ \\[6pt]

      False positive rate & $\mathrm{FPR}$ &
      $\displaystyle \frac{FP}{FP + TP}$ \\
    \end{tabular}
  \end{ruledtabular}
\end{table}
In the above expressions, \(TP\), \(FP\), \(FN\), and \(TN\) denote the entries of the confusion matrix presented in Fig \ref{fig:confusion_matrix}: 
\(TP\) (true positives) is the number of foreground samples correctly classified as foreground, 
\(FP\) (false positives) is the number of background samples incorrectly classified as foreground, 
\(FN\) (false negatives) is the number of foreground samples incorrectly classified as background, 
and \(TN\) (true negatives) is the number of background samples correctly classified as background.

\begin{figure}[!t]
\centering
\begin{tikzpicture}[
    font=\small,
    cell/.style={minimum width=3.0cm, minimum height=1.2cm, draw, align=center},
    tp/.style={cell, fill=green!15},
    tn/.style={cell, fill=green!15},
    fp/.style={cell, fill=red!10},
    fn/.style={cell, fill=red!10},
    header/.style={minimum width=3.0cm, minimum height=1.2cm, draw, align=center, fill=gray!15},
    axislabel/.style={font=\small}
]

% === Confusion matrix cells (2x2), tightly packed ===
% Centers chosen so widths/heights match and cells touch.
\node[tp] (tp) at (0,0)          {True Positive\\(TP)};
\node[fn] (fn) at (3.0,0)        {False Negative\\(FN)};
\node[fp] (fp) at (0,-1.2)       {False Positive\\(FP)};
\node[tn] (tn) at (3.0,-1.2)     {True Negative\\(TN)};

% Outer frame around the 2x2 block
\draw[thick] ($(tp.north west)$) rectangle ($(tn.south east)$);

% === Header row for predicted classes ===
\node[header] (pred_pos) at (0,1.2) {GW signal + Noise};
\node[header] (pred_neg) at (3.0,1.2) {Noise};

% Box around header row
\draw[thick] ($(pred_pos.north west)$) rectangle ($(pred_neg.south east)$);

% === Header column for true classes ===
\node[header] (true_pos) at (-3,0) {GW signal + Noise};
\node[header] (true_neg) at (-3,-1.2) {Noise};

% Box around header column
\draw[thick] ($(true_pos.north west)$) rectangle ($(true_neg.south east)$);

% === Axis titles ===
\node[axislabel] at (1.5, 2.4) {Predicted class};
\node[axislabel, rotate=90] at (-5.0, -0.6) {True class};
\end{tikzpicture}
\caption{Confusion matrix for binary classification, with true positives (TP), false negatives (FN), false positives (FP), and true negatives (TN) shown in the central $2\times 2$ block, and separate headers indicating true and predicted classes. Positives correspond to samples that correspond to features that contain a GW signal embedded in noise, whereas negatives denote the associated ones consisting of pure background noise only.}
\label{fig:confusion_matrix}
\end{figure}

Accuracy quantifies the overall fraction of correctly classified samples and provides a first indication of model performance. Precision (positive predictive value) measures the fraction of samples predicted as positive that are truly positive, thus characterizing the reliability of detections and penalizing false alarms. Recall (true positive rate or sensitivity) or detection efficiency measures the fraction of truly positive samples that are correctly identified, reflecting the ability of the classifier to avoid missed detections. The F1-score, defined as the harmonic mean of precision and recall, offers a single summary statistic that balances these two aspects. Specificity (true negative rate) measures the fraction of truly negative samples that are correctly classified as negative and therefore quantifies the ability of the classifier to reject pure background noise. The Matthews correlation coefficient (MCC) provides a global measure of binary classification quality by jointly summarizing all four entries of the confusion matrix; it takes values in the interval [-1,1], with $+1$ indicating perfect prediction, $0$ performance no better than random guessing, and $-1$ complete disagreement between predictions and true labels. Finally, the false alarm probability (FAP) denotes the fraction of negative samples misclassified as positive, while the false positive rate (FPR) denotes the fraction of predicted positive samples that are in fact negative, both providing complementary characterizations of spurious ``signal'' declarations. All metrics except MCC take values in the range $[0,1]$, with values closer to $1$ indicating the model's better performance.

Finally, based on the confusion-matrix outcomes, Figs. \ref{fig:BNS_conf_matrix_elements_hists} and \ref{fig:NSBH_conf_matrix_elements_hists} show the distributions of the EW-ANN output probability $p$ for samples classified as TP, FP, TN, and FN. For the BNS case, we report results for the associated test subsamples drawn from $\mathcal{D}_1$ and $\mathcal{D}_8$, while for the NSBH case, we use the corresponding test subsamples from $\mathcal{D}_1$ and $\mathcal{D}_6$ as detailed in the main text. In both figures, panels display results for the ET1, ET2, CE1, and CE2 detector configurations. The selected ANN classifiers illustrate earlier and later early-warning alert regimes, respectively; increasing model index corresponds to progressively higher PI SNR for the foreground samples, which improves performance by increasing the true-positive rate while reducing false alarms. In addition, the models maintain good separation for background (noise-only) samples, yielding predominantly low probability values.

\begin{figure*}[!thb]
    \centering
    \subfloat[ET1 Detector: EW-ANN model 1]{%
    \includegraphics[width=0.50\textwidth]{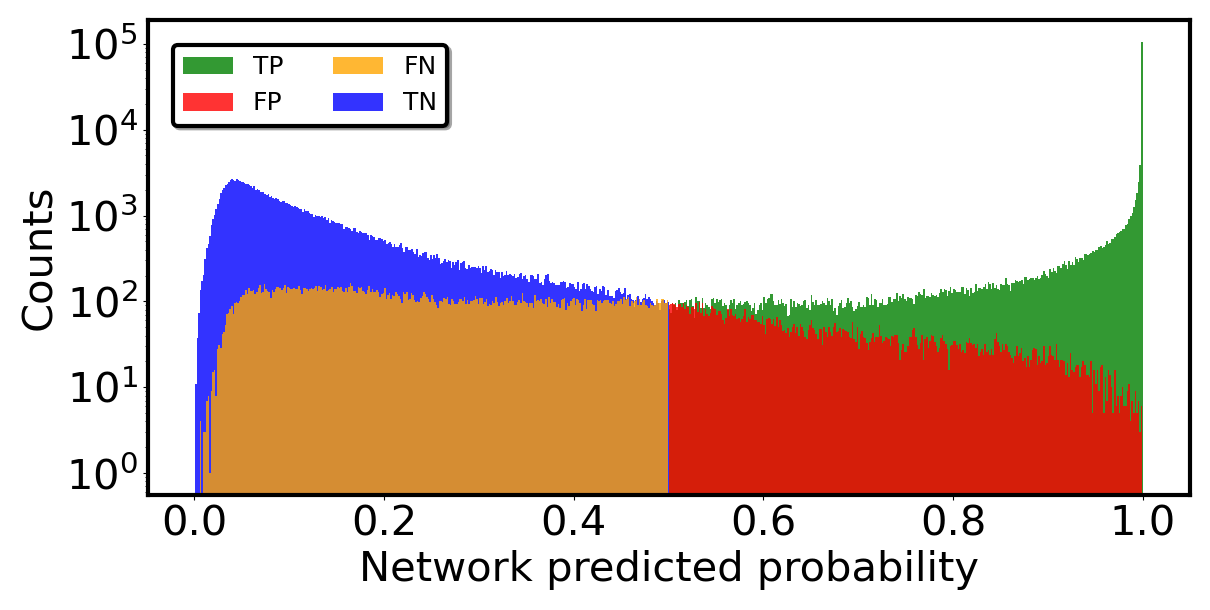}
    \label{fig:BNS_ETV_conf_matrix_elements_model_1}}
    \subfloat[ET1 Detector: EW-ANN model 8]{%
    \includegraphics[width=0.50\textwidth]{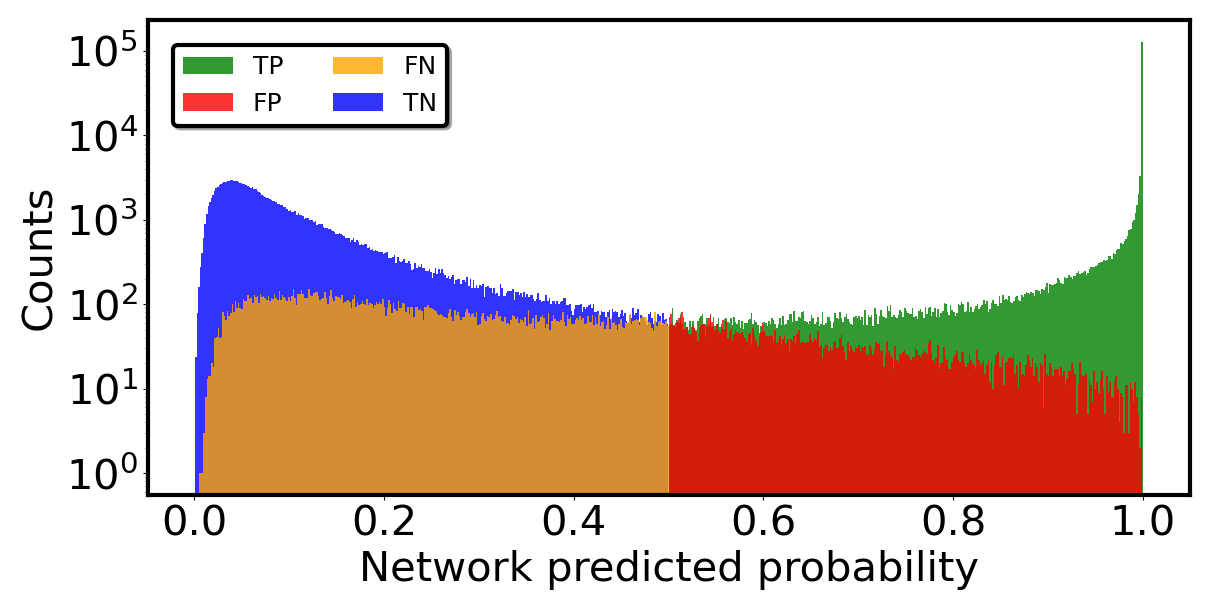}
    \label{fig:BNS_ETV_conf_matrix_elements_model_8}}\\
    
    \subfloat[ET2 Detector: EW-ANN model 1]{%
    \includegraphics[width=0.50\textwidth]{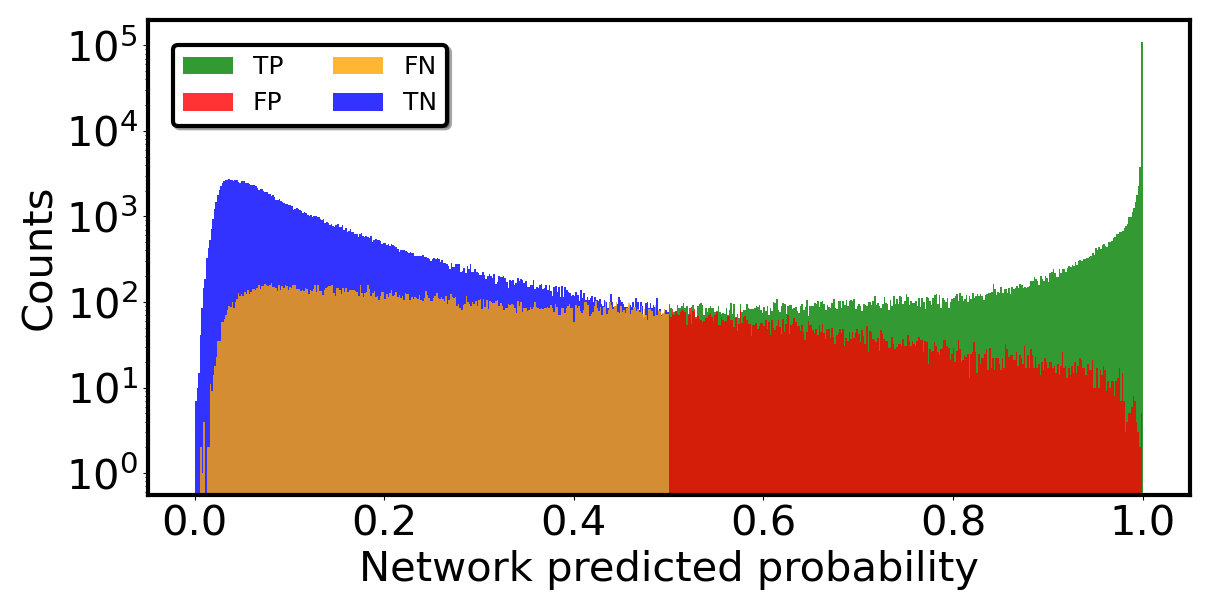}
    \label{fig:BNS_ETVG_conf_matrix_elements_model_1}}
    \subfloat[ET2 Detector: EW-ANN model 8]{%
    \includegraphics[width=0.50\textwidth]{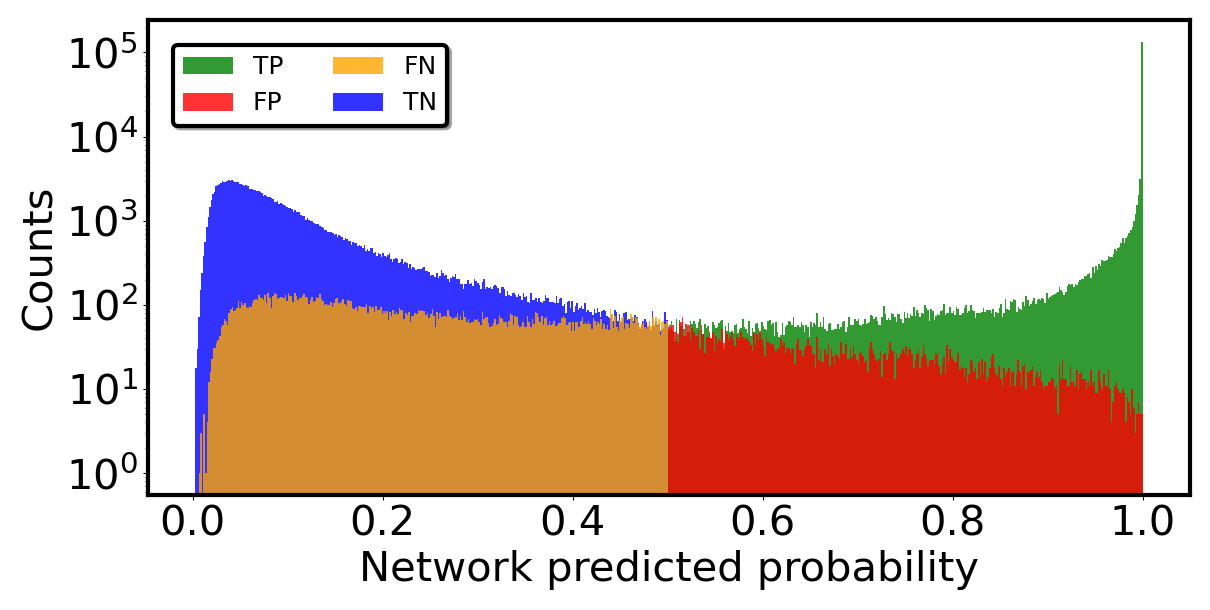}
    \label{fig:BNS_ETVG_conf_matrix_elements_model_8}}\\

    \subfloat[CE1 Detector: EW-ANN model 1]{%
    \includegraphics[width=0.50\textwidth]{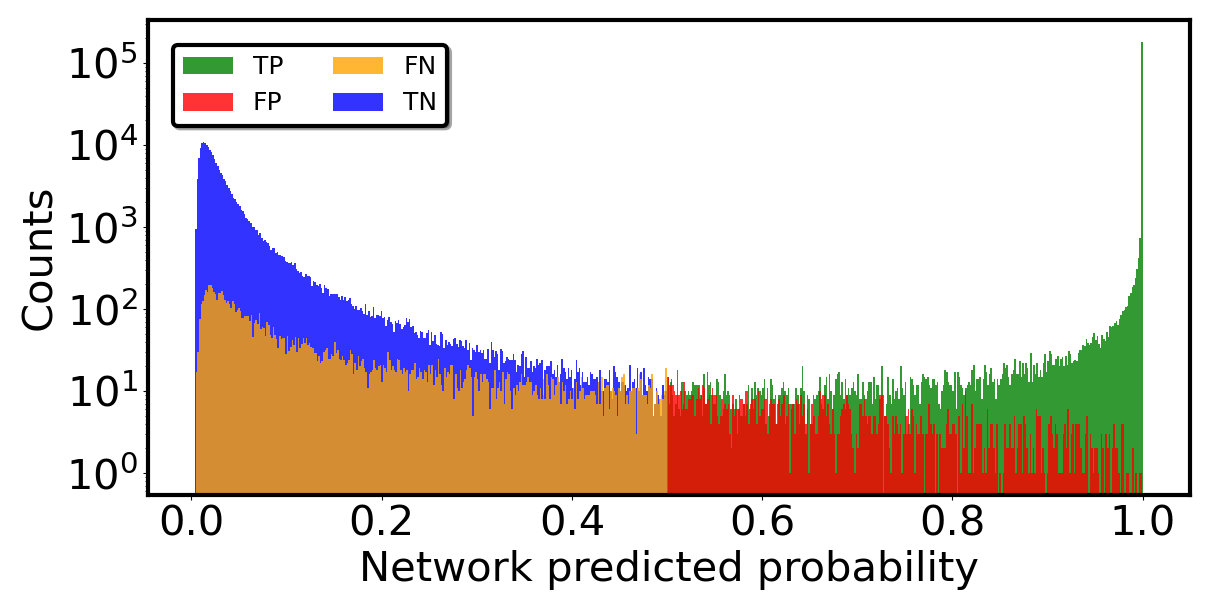}
    \label{fig:BNS_CE1_conf_matrix_elements_model_1}}
    \subfloat[CE1 Detector: EW-ANN model 8]{%
    \includegraphics[width=0.50\textwidth]{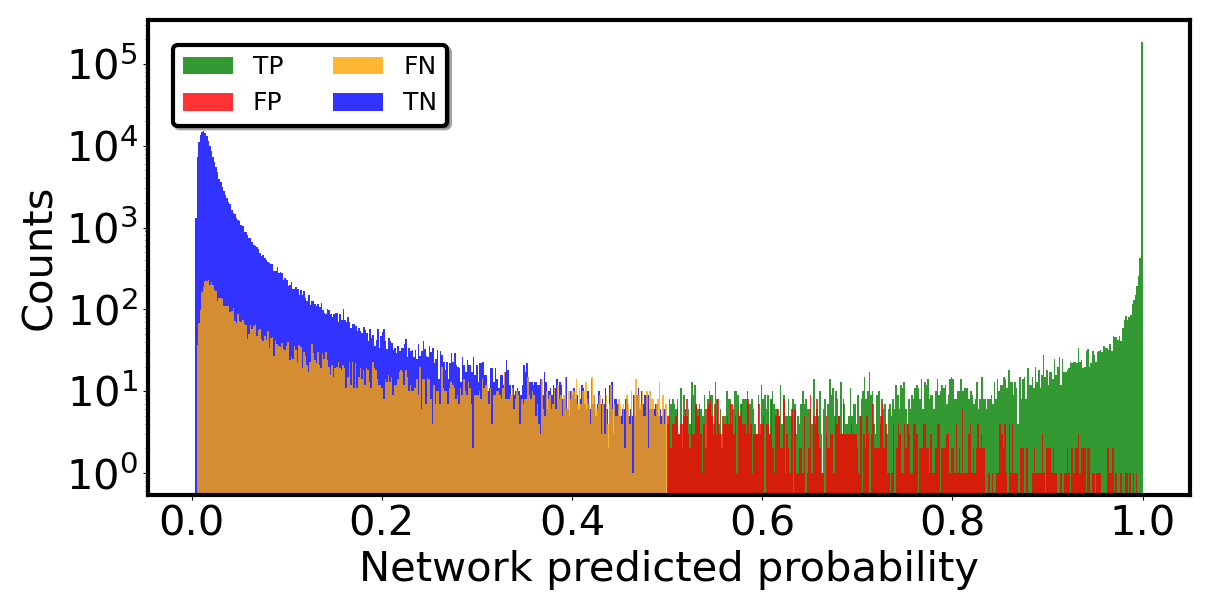}
    \label{fig:BNS_CE1_conf_matrix_elements_model_8}}\\

    \subfloat[CE2 Detector: EW-ANN model 1]{%
    \includegraphics[width=0.50\textwidth]{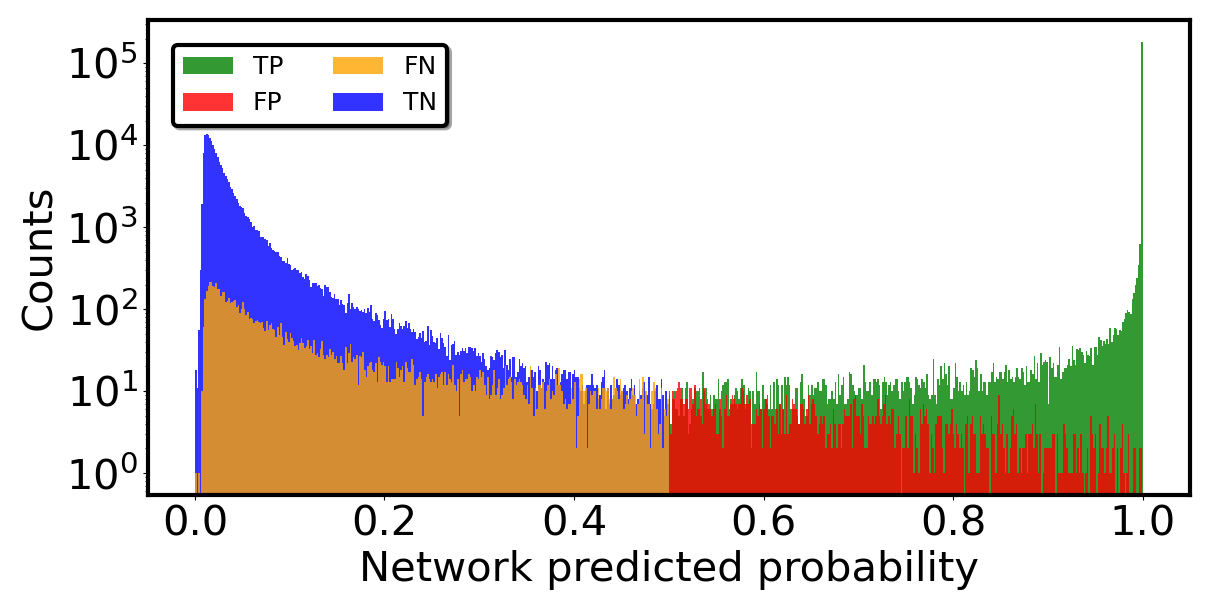}
    \label{fig:BNS_CE2_conf_matrix_elements_model_1}}
    \subfloat[CE2 Detector: EW-ANN model 8]{%
    \includegraphics[width=0.50\textwidth]{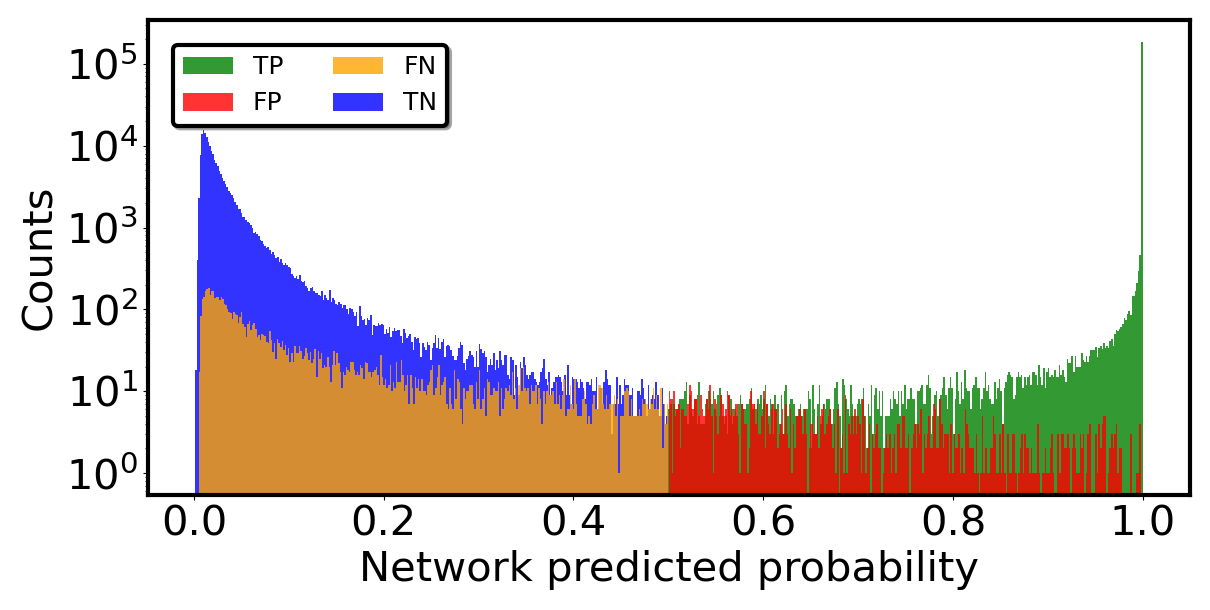}
    \label{fig:BNS_CE2_conf_matrix_elements_model_8}}
    \caption{\label{fig:BNS_conf_matrix_elements_hists}  Distributions of the EW-ANN output probability $p$ on the associated test subsamples for BNS mergers in the early-warning classification task. Histograms show the counts (on a logarithmic scale) of samples in each confusion-matrix category (TP, FP, TN, FN) as a function of $p$, using a decision threshold of $p=0.5$. Panels correspond to the ET1, ET2, CE1, and CE2 detector configurations and to a representative subset of EW-ANN models (ANNs 1 and 8). Here, classifier models 1 and 8 illustrate earlier and later early-warning alert regimes, respectively, with increasing model index corresponding to progressively higher PI SNR values for the foreground samples.}
\end{figure*}

\begin{figure*}[!thb]
    \centering
    \subfloat[ET1 Detector: EW-ANN model 1]{%
    \includegraphics[width=0.50\textwidth]{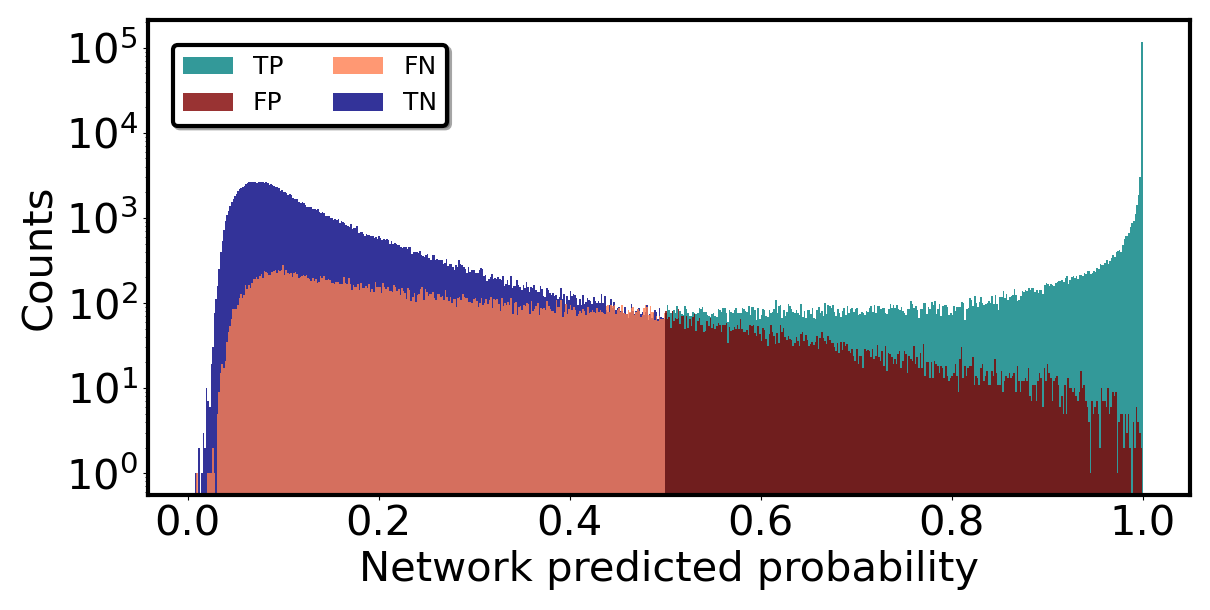}
    \label{fig:NSBH_ETV_conf_matrix_elements_model_1}}
    \subfloat[ET1 Detector: EW-ANN model 8]{%
    \includegraphics[width=0.50\textwidth]{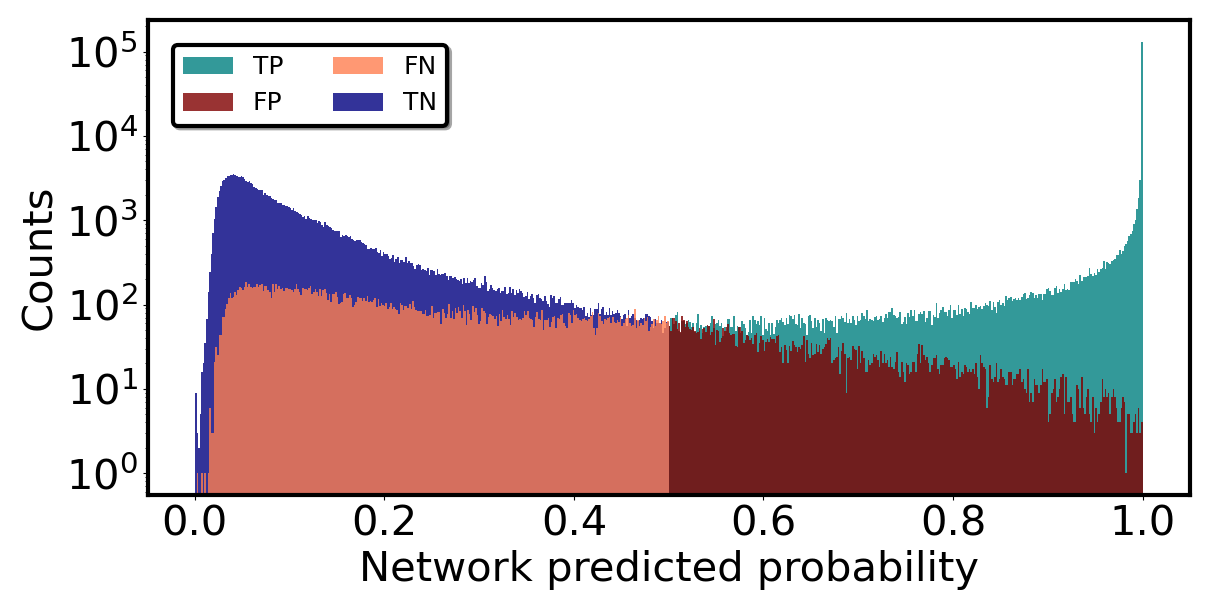}
    \label{fig:NSBH_ETV_conf_matrix_elements_model_8}}\\ 
    \subfloat[ET2 Detector: EW-ANN model 1]{%
    \includegraphics[width=0.50\textwidth]{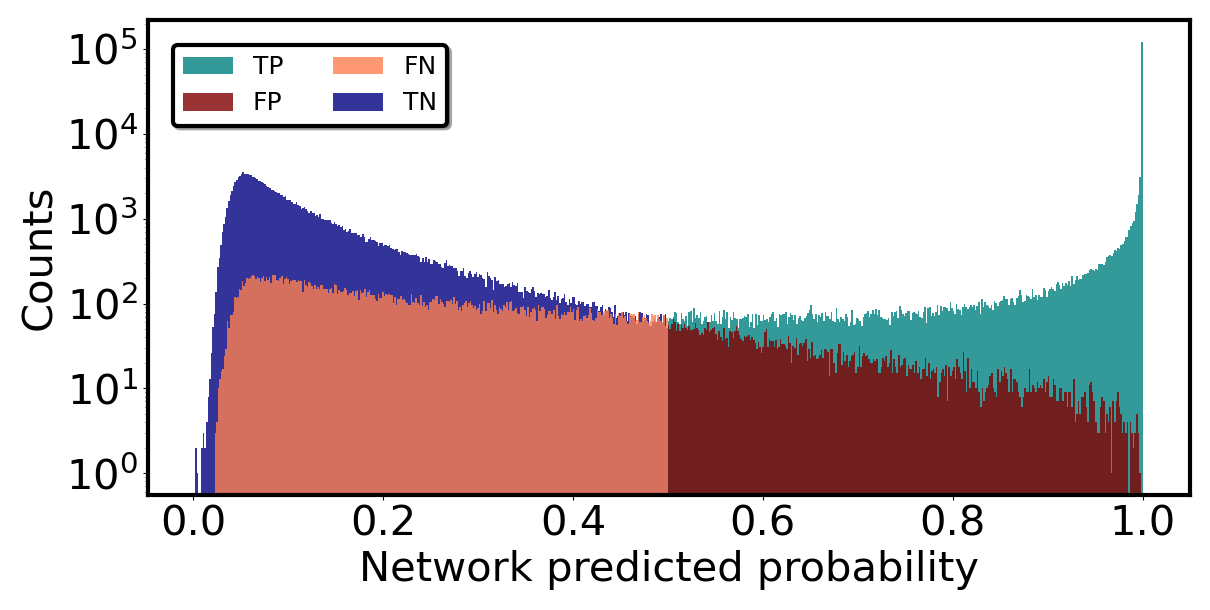}
    \label{fig:NSBH_ETVG_conf_matrix_elements_model_1}}
    \subfloat[ET2 Detector: EW-ANN model 8]{%
    \includegraphics[width=0.50\textwidth]{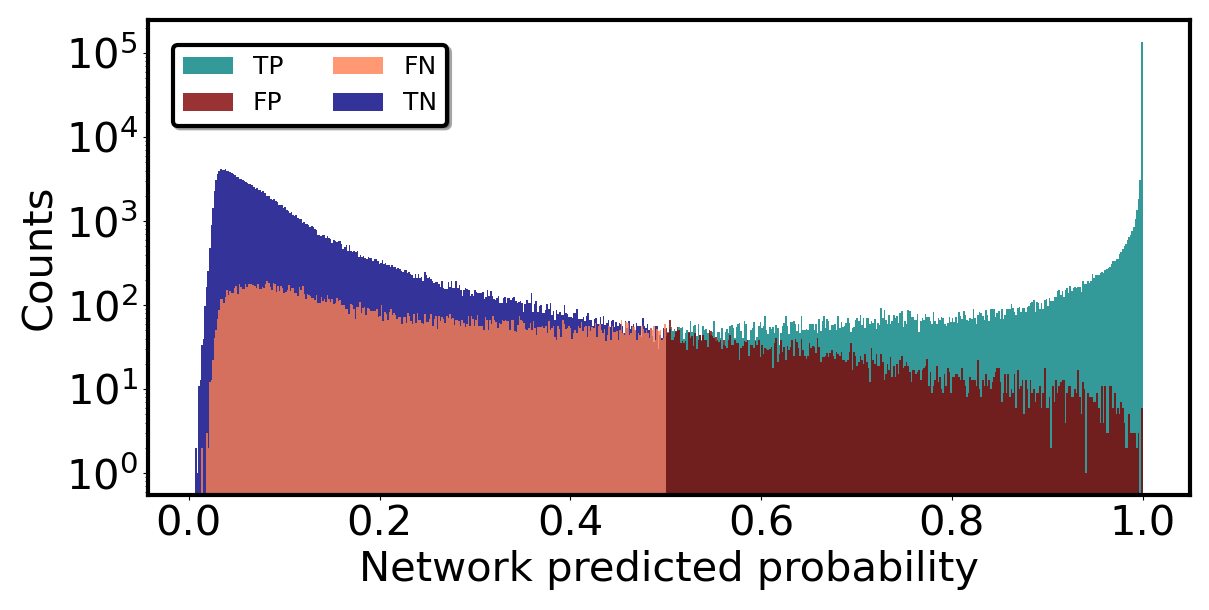}
    \label{fig:NSBH_ETVG_conf_matrix_elements_model_8}}\\
    \subfloat[CE1 Detector: EW-ANN model 1]{%
    \includegraphics[width=0.50\textwidth]{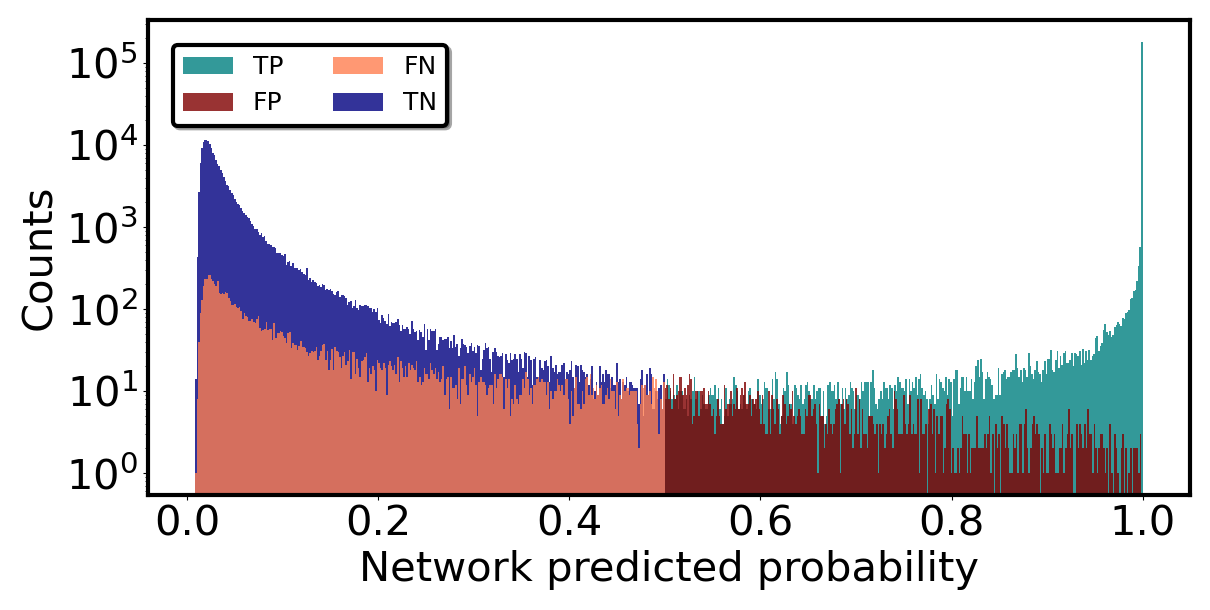}
    \label{fig:NSBH_CE1_conf_matrix_elements_model_1}}
    \subfloat[CE1 Detector: EW-ANN model 8]{%
    \includegraphics[width=0.50\textwidth]{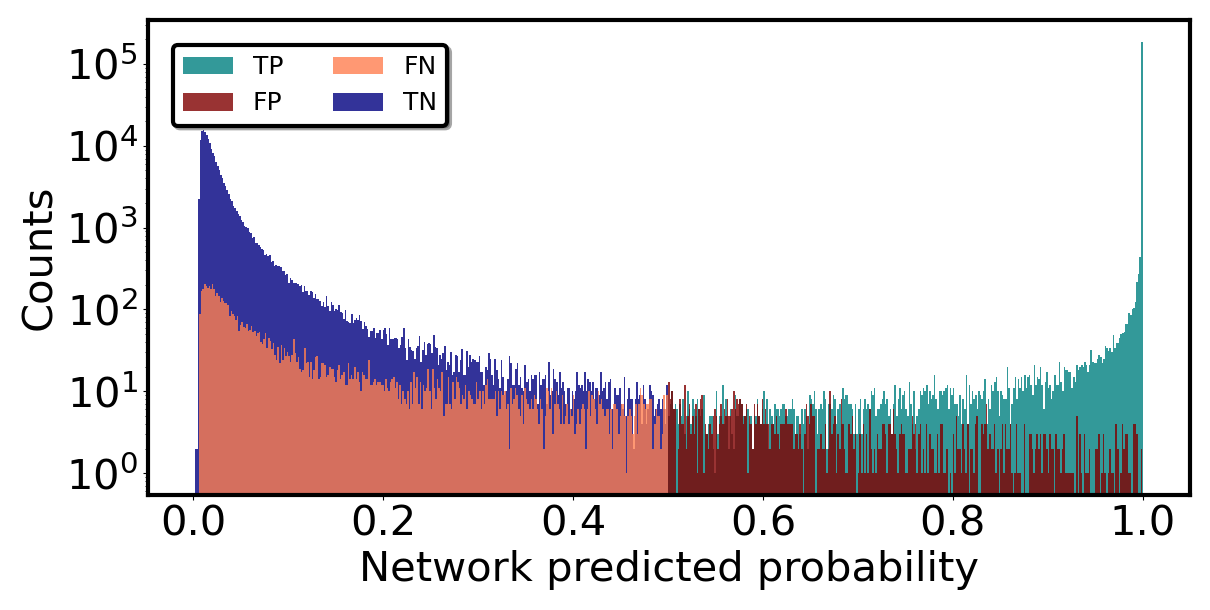}
    \label{fig:NSBH_CE1_conf_matrix_elements_model_8}}\\
    \subfloat[CE2 Detector: EW-ANN model 1]{%
    \includegraphics[width=0.50\textwidth]{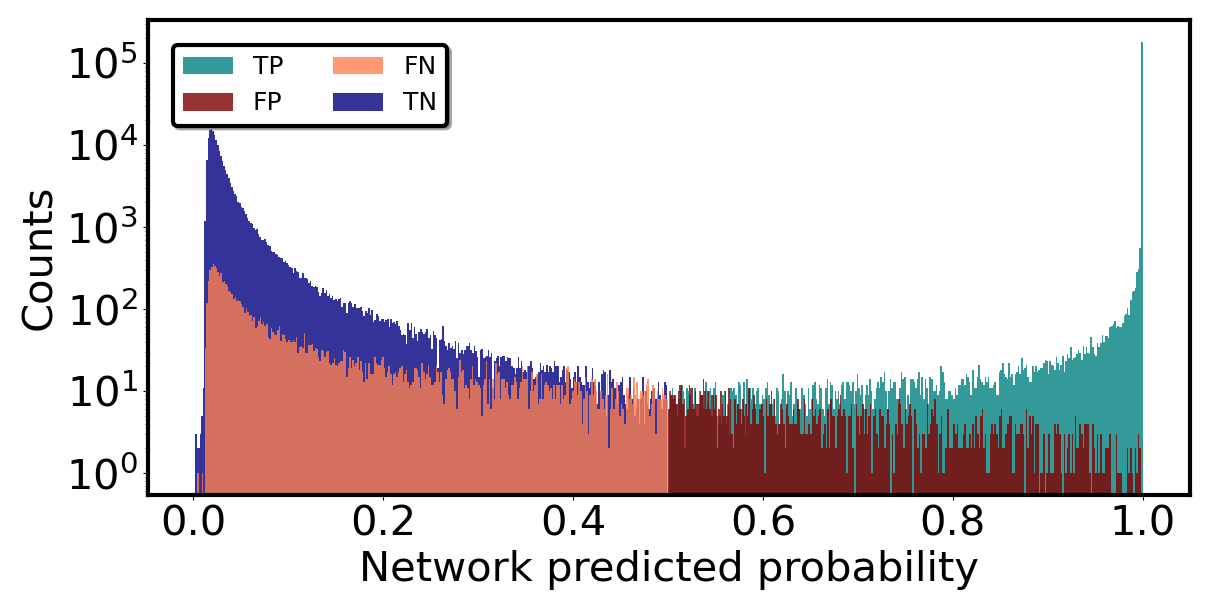}
    \label{fig:NSBH_CE2_conf_matrix_elements_model_1}}
    \subfloat[CE2 Detector: EW-ANN model 8]{%
    \includegraphics[width=0.50\textwidth]{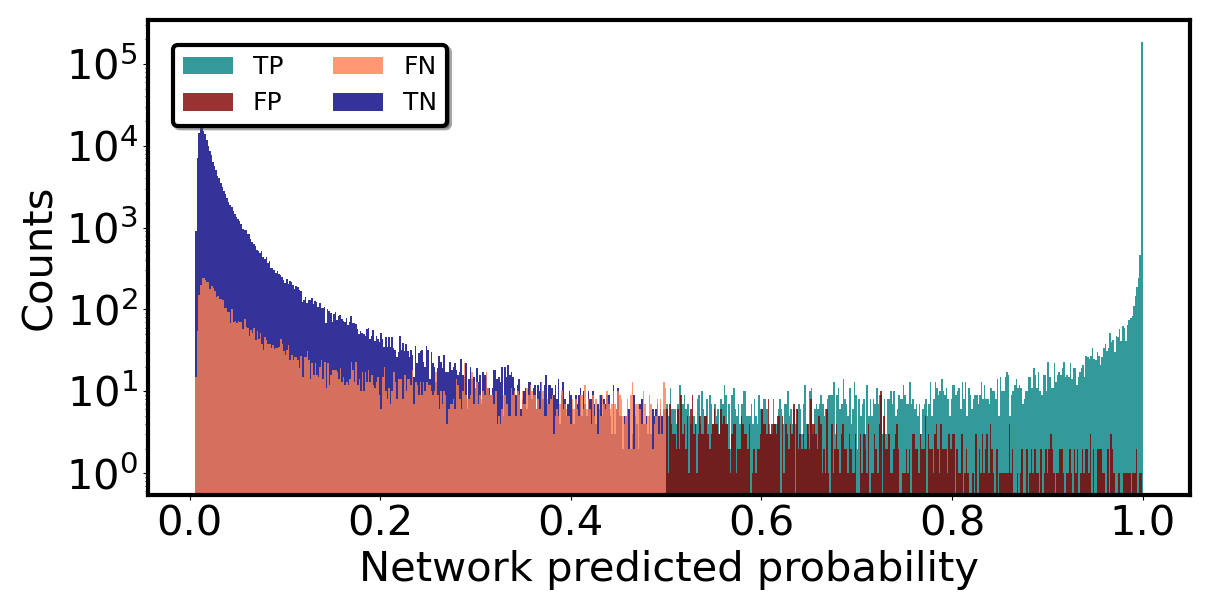}
    \label{fig:NSBH_CE2_conf_matrix_elements_model_8}}
    \caption{\label{fig:NSBH_conf_matrix_elements_hists} Distributions of the EW-ANN output probability $p$ on the associated test subsamples for NSBH mergers in the early-warning classification task. Histograms show the counts (on a logarithmic scale) of samples in each confusion-matrix category (TP, FP, TN, FN) as a function of $p$, using a decision threshold of $p=0.5$. Panels correspond to the ET1, ET2, CE1, and CE2 detector configurations and to a representative subset of EW-ANN models (ANNs 1 and 6). Here, classifier models 1 and 6 illustrate earlier and later early-warning alert regimes, respectively, with increasing model index corresponding to progressively higher PI SNR values for the foreground samples.}
\end{figure*}

\end{widetext}

\section{\label{app: FAP_t_for_other_det_cases} EW-ANN $\mathrm{FAR}$ representations}

\begin{figure*}[!thb]
    \centering
    
    \subfloat[ET1: $\mathrm{FAR}$ vs $\mathcal{R}_s$ for BNS EW-ANN models]{%
    \includegraphics[width=0.50\textwidth]{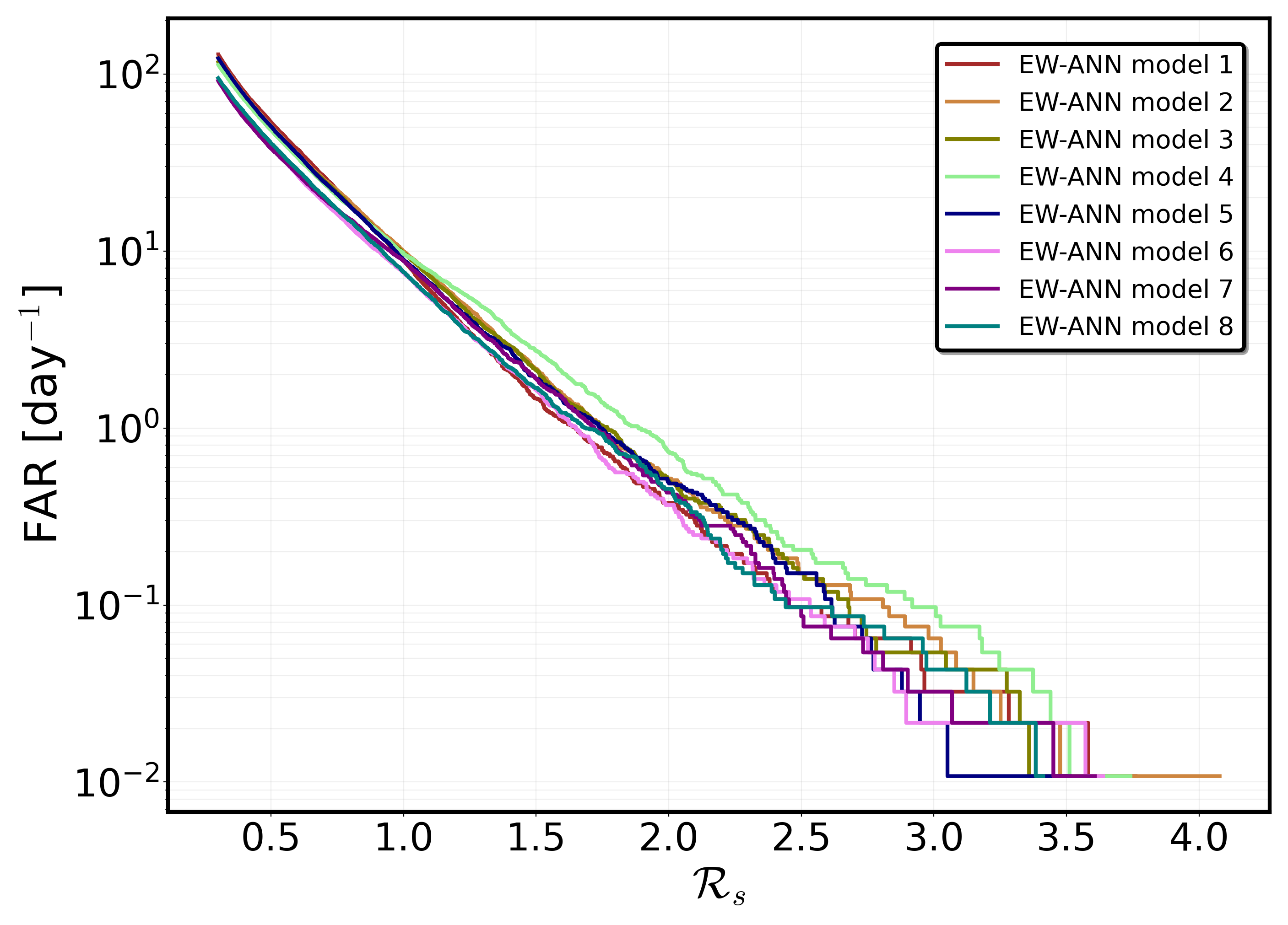}
    \label{fig:ETV_FAR_BNS}}
    \subfloat[ET1: $\mathrm{FAR}$ vs $\mathcal{R}_s$ for NSBH EW-ANN models]{%
        \includegraphics[width=0.50\textwidth]{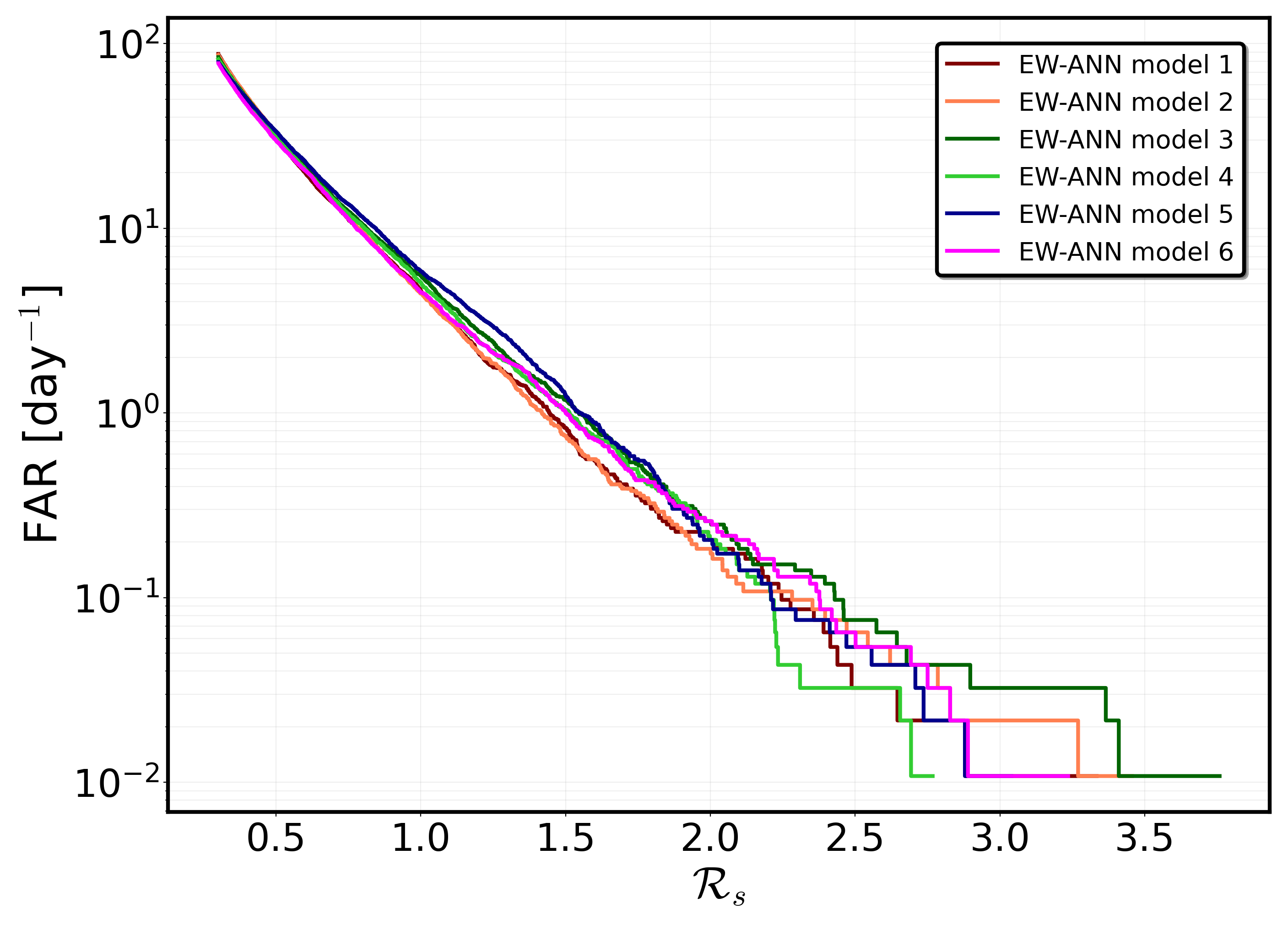}
        \label{fig:ETV_FAR_NSBH}}\\

    \subfloat[ET2: $\mathrm{FAR}$ vs $\mathcal{R}_s$ for BNS EW-ANN models]{%
    \includegraphics[width=0.50\textwidth]{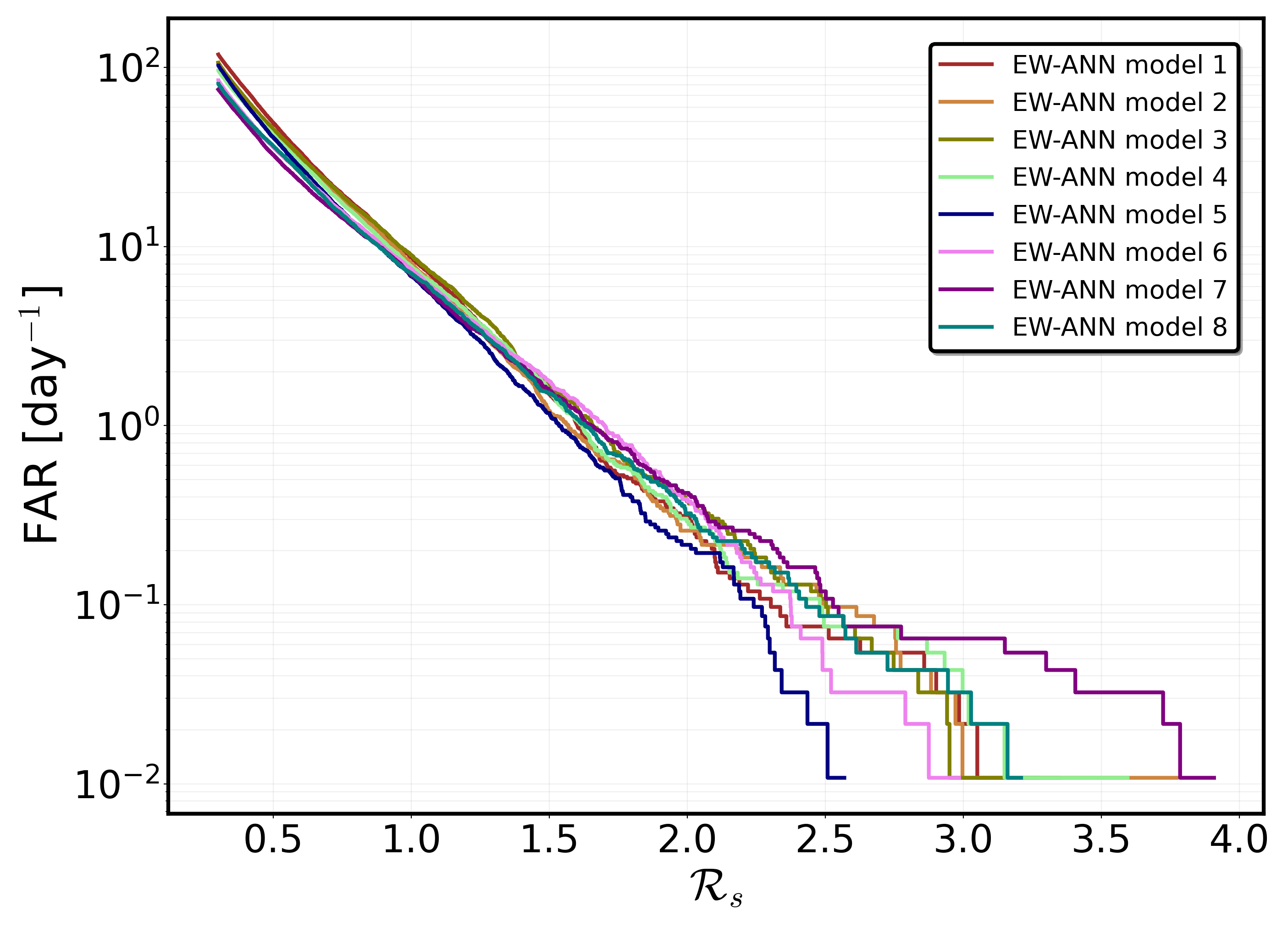}
    \label{fig:ETVG_FAR_BNS}}
    \subfloat[ET2: $\mathrm{FAR}$ vs $\mathcal{R}_s$ for NSBH EW-ANN models]{%
        \includegraphics[width=0.50\textwidth]{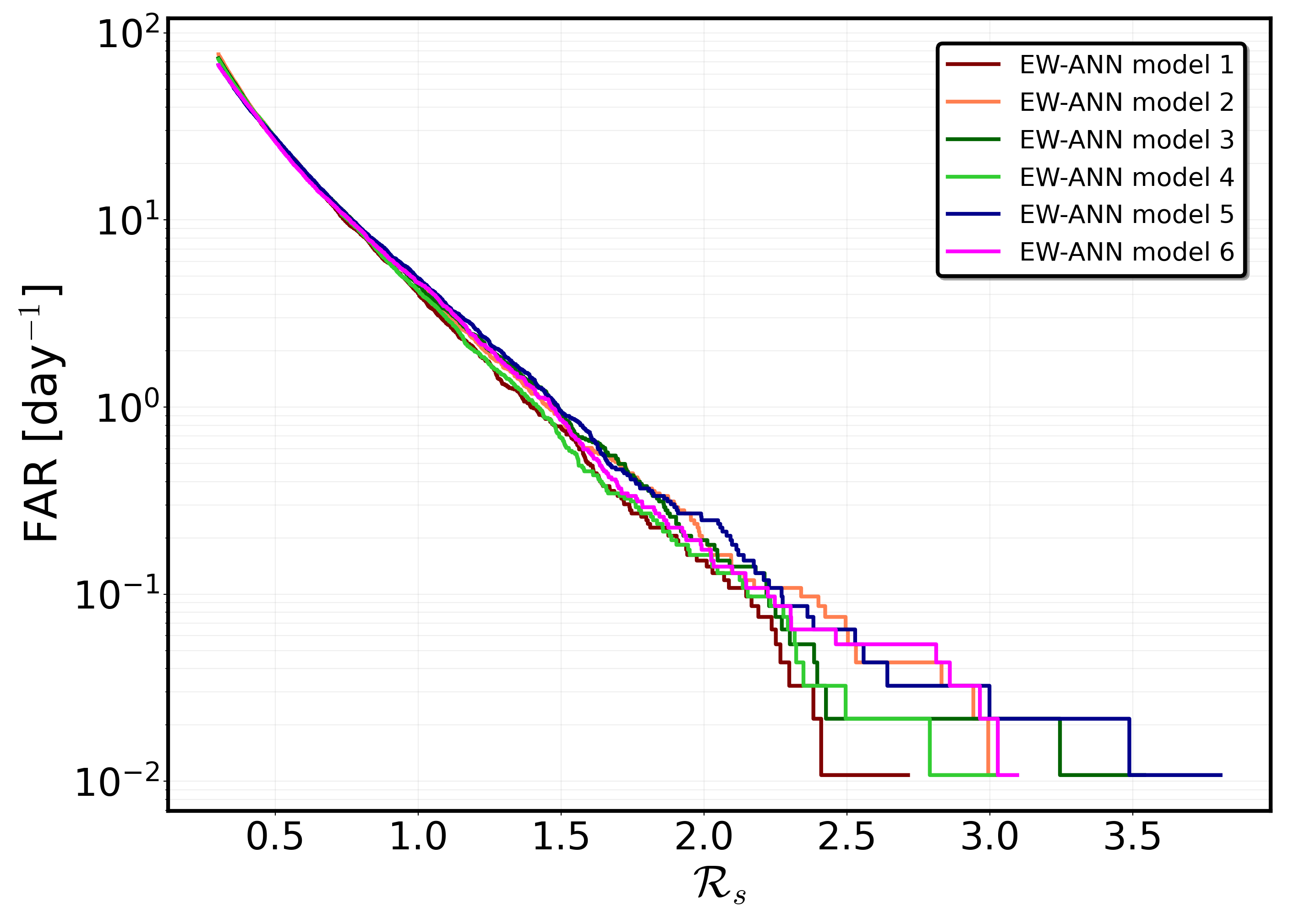}
        \label{fig:ETVG_FAR_NSBH}}\\

    \subfloat[CE2: $\mathrm{FAR}$ vs $\mathcal{R}_s$ for BNS EW-ANN models]{%
        \includegraphics[width=0.50\textwidth]{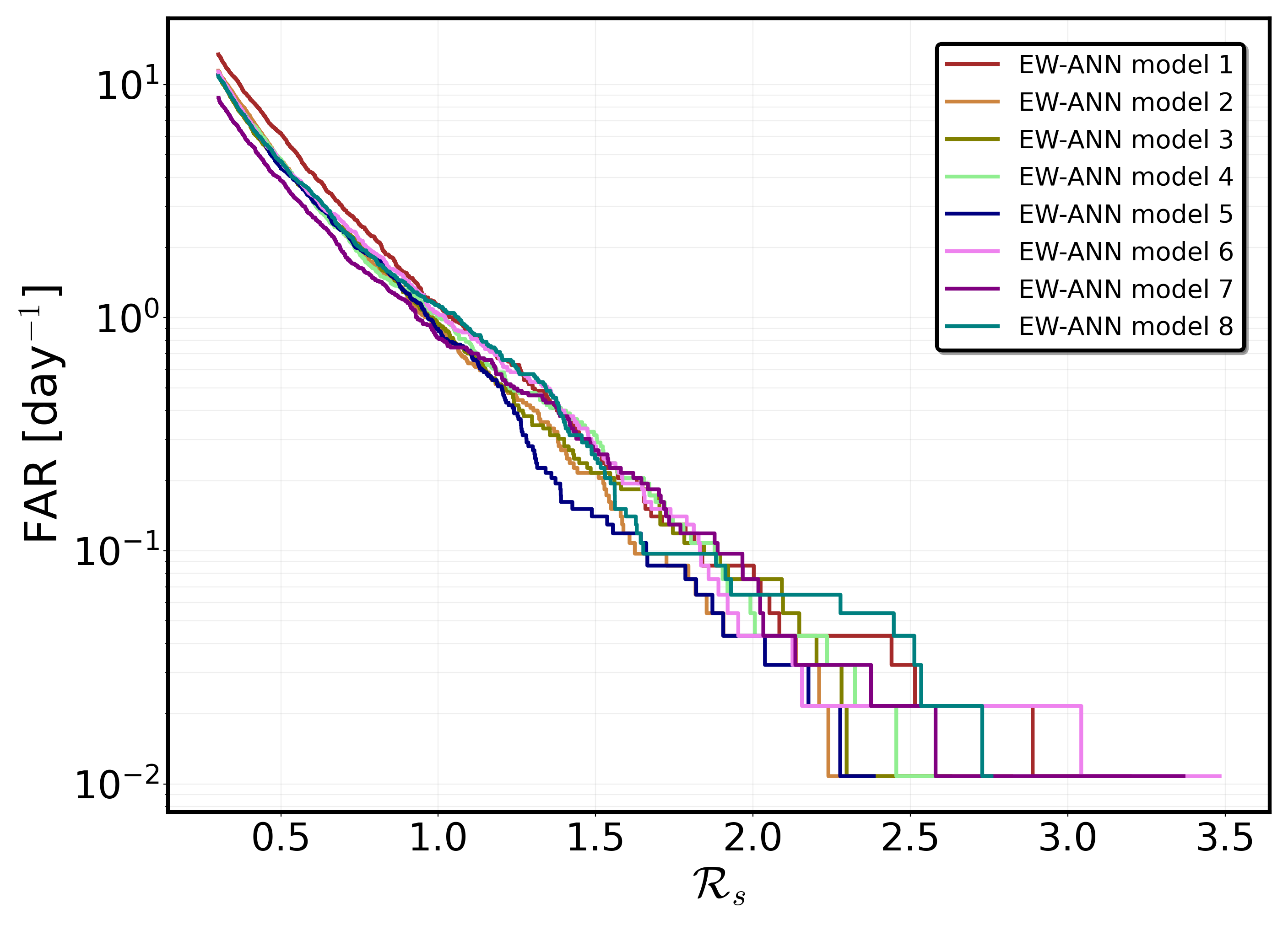}
        \label{fig:CEL_FAR_BNS}}
        \subfloat[CE2: $\mathrm{FAR}$ vs $\mathcal{R}_s$ for NSBH EW-ANN models]{%
            \includegraphics[width=0.50\textwidth]{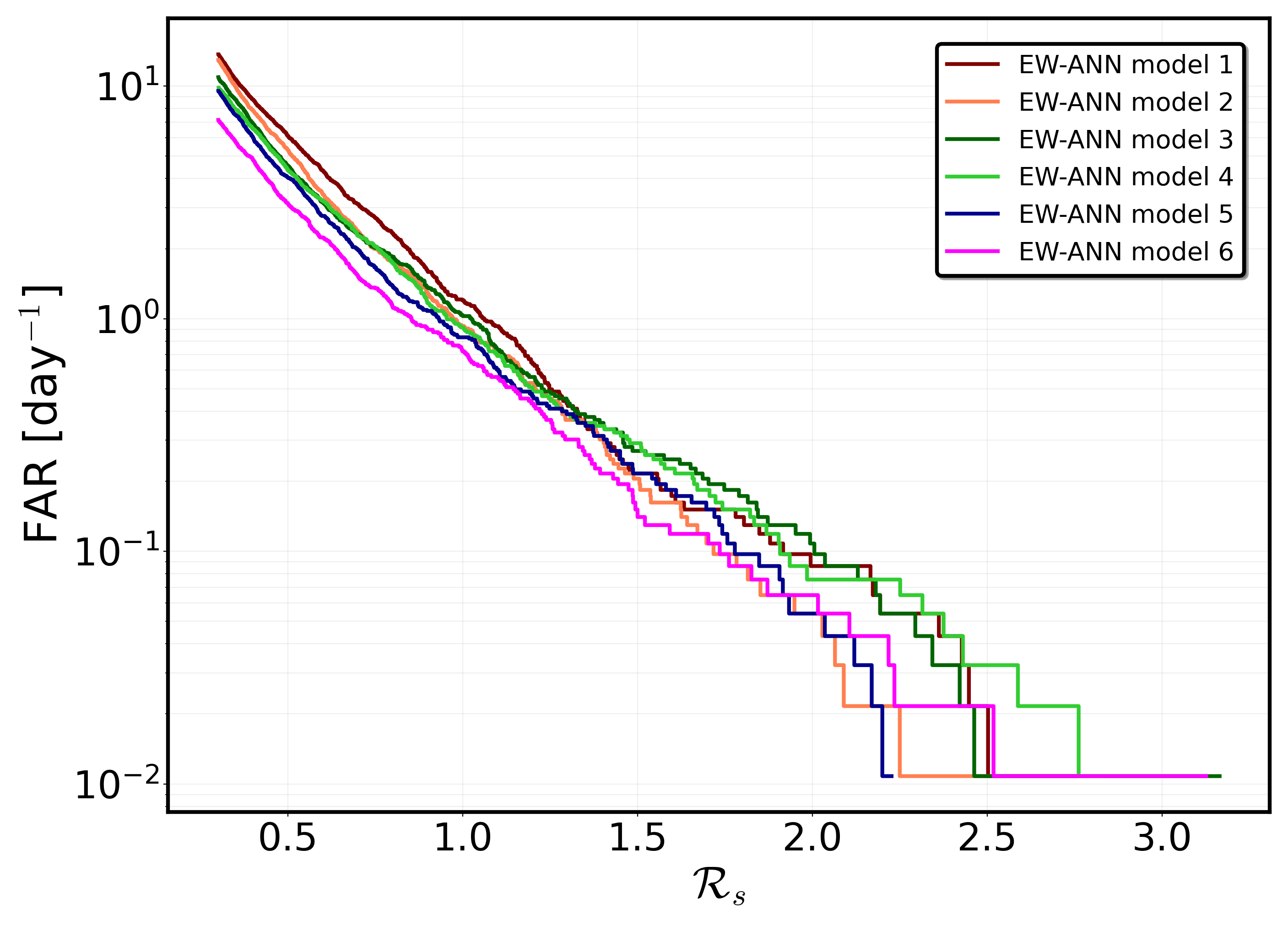}
            \label{fig:CEL_FAR_NSBH}}
    \caption{\label{fig:FAP_t_distributions} False-alarm rate as a function of the ranking statistic $\mathcal{R}_s$ for the ET1, ET2, and CE2 detector configurations, evaluated on background-only test data. Within each panel, each curve corresponds to a different EW-ANN model. Left panels show the false-alarm behavior of the BNS ANN models, while right ones show that of the NSBH models. In each case, at high $\mathcal{R}_s$ thresholds, the curves become step-like and approach a finite-duration floor set by the available background duration.}
\end{figure*}

In this section, Fig. \ref{fig:FAP_t_distributions} presents, as supplementary material, the false-alarm rate as a function of ranking statistic $\mathcal{R}_s$ for the BNS and NSBH EW-ANN models associated with the ET1, ET2, and CE2 detector configurations discussed in the main text. In all panels, for sufficiently high $\mathcal{R}_s$ thresholds (for each EW-ANN model), the inferred false-alarm rate is very small, corresponding to a negligible expected number of spurious triggers per day.

%\begin{comment}
    
%\end{comment}

%%%%%%%%%%%%%%%%%%%%%%%%%%%%%%%%%%%%%%%%
%%%%%%%%%%%%%%%%%%%%%%%%%%%%%%%%%%%%%%%%%
%%%%%%%%%%%%%%%%%%%%%%%%%%%%%%%%%%%%%%%%%

\bibliography{apssamp}% Produces the bibliography via BibTeX.

\end{document}